# Influence of Culture on e-Government Acceptance in Saudi Arabia

Ibrahim Khalil Abu Nadi

*Bachelor of Science in Computer Science, Master of Information Systems*

School of Information and Communication Technology

Science, Environment, Engineering and Technology Group

Griffith University

Submitted in fulfilment of the requirements of the degree of

Doctor of Philosophy

June 2012



# ABSTRACT


The main purpose of this thesis is to determine the influence of culture on e-government (electronic government) acceptance in a developing nation namely the Kingdom of Saudi Arabia (KSA). Citizens' transactional interactions (electronic transactions or e-transactions) with the government via the Internet are examined. To this end, the following research question is addressed: 'How does culture influence the acceptance of e-transactions?' Defining and understanding cultural factors influencing e-transactions provides an insight into the actual requirements of citizens. The findings of this study include design and implementation strategies that can serve as guidance for the Saudi government, as well as for the developers and implementers of e-transactions in the KSA.

Numerous models and theories were referred to in identifying the research context requirements that enabled the analysis of e-transaction acceptance. A research model that fits the research context was developed to predict and elucidate acceptance. A sample of 671 Saudi citizens was recruited using an online survey. Structural equation modelling was used to assess the relationship between intention to use e-transactions and perceptions of e-transactions, trust, preferences for using e-transactions as a communication method with the government, social influence and cultural values. Preference for using e-transactions as a communication method, perceptions of the compatibility of e-transactions with values and citizens' needs, communicability of the results of using e-transactions, trust in the Internet as a medium of communication with the government, and conservation values are positive significant determinants of e-transaction acceptance. Conversely, trust in government agencies, as well as motivation towards gaining prestige and possessing dominance over people and resources (i.e. power value) exhibit a negative significant effect on acceptance.




**STATEMENT OF ORIGINALITY**

"This work has not previously been submitted for a degree or diploma in any university. To the best of my knowledge and belief, the thesis contains no material previously published or written by another person except where due reference is made in the thesis itself."

_______________________________

Ibrahim Khalil Abu Nadi



# TABLE OF CONTENTS





















## LIST OF FIGURES









# LIST OF TABLES







## LIST OF ABBREVIATIONS

| | |
|---|---|
| A | Achievement |
| AGFI | Adjusted Goodness of Fit Index |
| B | Benevolence |
| BHV | Basic Human Values |
| BPV | Basic Personal Values |
| C | Conformity |
| CFA | Confirmatory Factor Analysis |
| CFI | Comparative Fit Index |
| CMIN | Minimum discrepancy |
| CMX | Complexity |
| CON | Conservation values |
| CR | Critical Ratio |
| CT | Compatibility |
| CVR | Content Validity Ratio |
| df | Degree of Freedom |
| EFA | Exploratory Factor Analysis |
| GFI | Goodness of Fit Index |
| H | Hedonism |
| ICT | Information Communication Technology |
| IFI | Incremental Fit Index |
| IS | Information Systems |
| IT | Information Technology |
| KMO | Kaiser-Meyer-Olkin |
| KSA | Kingdom of Saudi Arabia |
| MI | Modification Index |
| ML | Maximum Likelihood |
| P | Power |
| PC | Personal Computer |
| PCET | Perceived Characteristics of e-Transactions |
| PCI | Perceived Characteristics of Innovation |
| POC | Perspective on communication |
| PVQ | Portrait Values Questionnaire |
| RED | Result demonstrability |
| RMSEA | Root Mean Square Error Approximation |
| SD | Self-direction |



| SE | Security |
|----|----------|
| SEM | Structural Equation Modelling |
| SGA | Saudi Government Agencies |
| SI | Social influence |
| SRW | Standardised regression weight |
| SRMR | Standardised Root Mean Square Residual |
| ST | Stimulation |
| T | Tradition |
| TAM | Technology Acceptance Model |
| TG | Trust in government agencies |
| TI | Trust in the Internet |
| TPB | Theory of Planned Behaviour |
| TRA | Theory of Reasoned Action |
| U | Universalism |
| UK | United Kingdom |
| UNDESA | United Nation Department of Economic and Social Affairs |
| USA | United States of America |
| UTAUT | Unified Theory of Acceptance and Use of Technology |
| WTO | World Trade Organisations |



**ACKNOWLEDGEMENTS**

My abilities, inspiration and hope during this PhD journey come from Allah who has showered me with many blessings in this life.

Sincere thanks go to my wife Alya and my daughter Ayat for their motivation, patience, and for sharing this journey with me. Also, I am very thankful to my beloved mother, father (may Allah bless his soul), brothers, father-in-law, and mother-in-law for their continuous prayers and encouragement.

I am very grateful to my principle supervisor Dr. Steve Drew for his sincere advice and constant guidance. Also, special thanks go to Dr. Kuldeep Sandhu for his constructive opinions, wise words, and support. I would like to extend my thanks to Dr. Steve and Dr. Kuldeep for reviewing the publications which were submitted from this thesis. There are many in the Griffith School of Information and Communication Technology and the Griffith Business School who assisted me in this journey including academics, staff members, and PhD colleagues. I am indebted to Dr. Louis Sanzogni and Dr. Geoffrey Carter, my supervisors at an initial stage of this study, for what I learned from them. I am also very grateful to Professor Peter Green from the University of Queensland and Professor Viswanath Venkatesh from the University of Arkansas for their directions in this journey. I would like to acknowledge the statistical advice of Dr. Peter Grimbeek. Also I would like to thank to Alexa Grunner for providing editorial services which included proofreading and advice on presentation and use of clear language.

Prince Sultan University, the Ministry of Higher education, and Saudi Cultural Mission have provided the opportunity and support that enabled me to continue this degree. Last but not least, I would like to thank all of those friends who accompanied me on the PhD journey and who were there when I needed them.



## DEDICATION

I dedicate this work to my late father may Allah bless his soul and to my mother for she is the source of light in my life.



# 1    INTRODUCTION

Continued globalisation has prompted many countries to move towards increased use of new technologies. The drive to shifting to digital technologies is so pronounced that lack of acceptance would almost certainly guarantee the loss of competitive advantage. The increasing demand for acceptance has also been observed at all levels of government given that numerous nations provide services to citizens via electronic means (including computers, digital communication channels, and the Internet). This platform of service provision is known as electronic government, or e-government. e-Government can be defined as a means for providing government services to citizens through online communication channels (Sharifi & Zarei, 2004).

The Kingdom of Saudi Arabia (KSA) continues to experience rapid growth in terms of economy, education, population, and technology. Such progress stems from the constantly increasing oil revenues earned by the country. Nevertheless, completion in the context of globalisation and its membership in the World Trade Organization (WTO) have prompted the KSA government to pursue global-scale developments in the quest to elevate the country to the status of developed nations (WTO, 2008). As a concrete step towards this objective, the government has created comprehensive overall development plans that feature a national information technology programme, which includes the implementation of e-government.

In 2005, the KSA initiated its e-government project, focusing on implementing e-government platforms in most of its government agencies by the end of 2010 (Ministry of Communications and Information Technology, 2006). The launch of an ambitious e-government programme indicates that the country is keeping pace with developed nations. The Saudi government hopes that effective implementation and acceptance of e-government services will extensively improve the internal effectiveness and efficiency of its agencies (Ministry of Communications and Information Technology, 2007b). A recent report by the United Nations



Department of Economic and Social Affairs (UNDESA) indicates that Saudi Arabia is considered an emerging leader in e-government development; as a developing nation, it has achieved a status on par with that of developed countries. The report nonetheless also discusses the lack of acceptance of e-government services at the global scale which indicates that acceptance of e-government is problematic and thus requires further research (UNDESA, 2012).

This chapter first provides the background of the thesis and then discusses culture as a determinant of technology acceptance in order to demonstrate its significance as a research topic. The research questions are also presented, along with the research methodology, and overview of the thesis structure.

## 1.1    Research Background

Saudi culture is a combination of Islamic and Arabic beliefs and traditions. Seeking prestige, adhering to tribalism, acknowledging hierarchy, and maintaining conservatism characterise Saudi societal–cultural values (Bhuian, 1998; Gallagher & Searle, 1985; Hofstede, Hofstede, & Minkov, 2010). Because Saudi society is conservative, its traditions and culture affect every aspect of life. Saudis have a strong affinity for their heritage, which exerts a considerable influence on the manner in which they live and work (AL-Shehry, Rogerson, Fairweather, & Prior, 2006). Although these cultural attributes affect many Saudis positively, they also impose negative effects (Al-Yahya, 2009). The Saudi government recognises these drawbacks and has consequentially introduced e-government to KSA citizens in the hopes of maximising its advantages. e-Transaction acceptance is therefore crucial because it would induce meaningful change in Saudi society. For example, prestige in the country is gained by using personal connections; a common practice is to circumvent rules so that certain individuals are able to complete transactions faster than others. These unfair methods of acquiring government services through tribal connections, family kinships, or personal relationships are considered a form of corruption (Abu Nadi, 2010; Smith, Huang, Harb, & Torres, 2011). The acceptance of e-transactions is expected to reduce such practices within government agencies. Another important factor in e-transaction acceptance is facilitating the diversification of the



KSA economy, which is, freeing it from its dependence on the production and export of oil to make way for establishing a knowledge-based economy. The widespread acceptance of e-transactions would facilitate Saudi Arabia's into transition to an information society, which would in turn advance its transition to a knowledge-based economy (Ramady, 2010). Information societies and knowledge-based economies rely on the use of knowledge to create a competitive advantage and to earn economic benefits for a country (Webster, 2002). Nevertheless, the changes that arise from the introduction of e-transactions will be rejected if e-transaction acceptance is not studied from a cultural perspective and carefully implemented in accordance with cultural nuances (AL-Shehry, et al., 2006). Research has shown that socio-cultural factors dictate the acceptance of computer usage (M. Ali & Alshawi, 2004) and the Internet (Loch, Straub, & Kamel, 2003), but studies that focus only on the effect of culture on technology acceptance, or more specifically, e-transaction acceptance, are lacking.

## 1.2    Rationale for the Cultural Approach

       Culture is one of the most important and abstract factor that determines human behaviour (Gong, Li, & Stump, 2007). Despite its importance, culture has been accorded limited attention in research on the relationship between culture and information technology acceptance (Thatcher & Foster, 2003). A technological innovation reflects the culture of the developer and the socio-cultural needs of the country where it was developed (Straub, Loch, & Hill, 2003). Many researchers (Gefen, Karahanna, & Straub, 2003; Karahanna, Straub, & Chervany, 1999; Straub, et al., 2003) agreed that the use of information technology varies across different cultures. Technology can be rejected on the grounds of its incompatibility with cultural practices, values, and traditions (C. Hill, Straub, Loch, Cotterman, & El-Sheshai, 1994). An example of technology incompatibility is the time required for acceptance and the difference in usage patterns for electronic meeting systems across different cultures (Raman & Wei, 1992). Hill, Loch, Straub, and El-Sheshai (1998) found that Arab cultures prefer face-to-face communication over communication using technological means. This example shows one kind of cultural difference that has implications for technology use. Zakaria, Stanton, and Sarker-



Barney (2003) indicated that the Internet and related applications have been spreading throughout different cultures. Many of these developed applications facilitate personal communications, an activity closely related to cultural values and beliefs (Zakaria, et al., 2003). Given that the use of Internet communication applications can enhance information sharing, a particular culture can encourage or inhibit this behaviour over the Internet (Thatcher & Foster, 2003).

Al-Gahtani, Hubona and Wang (2007) asserted that, given the high levels of uncertainty avoidance in cultures such as the KSA, a technology is scrutinised before accepted. Because collectivism dominates KSA society, its mainstream citizens reflect on the perspectives of first adopters of a technology. Mainstream citizens tend to avoid the uncertainties associated with new technologies until others (first adopters) have accepted and recommended such innovations or technologies. Given this situation, some researchers have raised concerns that collectivism will affect e-transaction acceptance in the country (Abu Nadi, 2010; Al-Gahtani, et al., 2007).

No study found to use Schwartz's theory of Basic Personal Values (BPV) to explain the cultural values of the KSA. Hofstede's cultural dimensions, on the other hand, have been explored by a number of researchers for the purpose of understanding the mechanisms that underlie culture and communication in KSA (Al-Gahtani, et al., 2007; Alshaya, 2002; Bjerke & Al-Meer, 1993; Hall & Hall, 1990; Würtz, 2005). These studies provide insight into Saudi society, and serve as basis for constructing models that predict acceptance and for formulating hypotheses designed to validate assumptions about relationships between culture and technology acceptance.

## 1.3    Research Questions

The primary objective of this study is to determine the cultural factors that may affect the acceptance of e-government. The main research question is 'How does culture influence the acceptance of e-transactions?' Cultural factors include perceptions, trust, social influences, and cultural values. Perceptions of a technology are shaped by culture, in which people evaluate the



use of technology against their cultural perspectives (Leidner & Kayworth, 2006; Moore & Benbasat, 1991). Trust in the medium (the Internet) through which e-transactions are provided as well as in the provider of such services (government agencies) are also expected to play significant roles in the acceptance of e-transactions (Carter & Bélanger, 2005). This influence is especially relevant given that Saudi society is characterised by lack of trust and avoidance of uncertainties (Hofstede, et al., 2010). The use of e-transactions as a medium of communication between citizens and the government is a main element of this research inquiry, and the influence of the opinions of others (social influence) was also considered (Al-Gahtani, et al., 2007; Aoun, Vatanasakdakul, & Li, 2010; Venkatesh, Morris, Davis, & Davis, 2003). Cultural values are the most influential component of behaviour and it was expected play a role in the behavioural intention towards e-transaction usage (Schwartz, 2003).

The secondary research questions attend to issues that are related to the main question. These questions are as follows:

- **Research Question 1:** How do perceived characteristics of e-transactions affect e-transaction acceptance?

- **Research Question 2:** How does trust in the Internet and government agencies influence acceptance?

- **Research Question 3:** How does the social influence of existing e-transaction users affect the acceptance of e-transactions?

- **Research Question 4:** How does using e-transactions as a communication method affect acceptance of e-transactions?

- **Research Question 5:** How do cultural values influence the acceptance of e-transactions?

## 1.4    Research Methodology

In carrying out this research, the perspective of soft positivism's ontology was adopted; that is, the research procedure was designed to capture pre-existing phenomena and study the



relationships among them, with careful consideration of context (Kirsch, 2004). This ontological position facilitates the study's examination of culture and e-government acceptance through the comprehensive analysis of the cultural nuances of Saudi society.

Quantitative survey questionnaires are commonly used instruments for determining the perceptions, cultural values, and beliefs that affect the acceptance and adoption of technology. Because this study concentrates on citizens' behavioural intention toward e-transactions, online surveys were used (Leidner & Kayworth, 2006). Using an online survey restricts sample selection to individuals with Internet access because they are able to respond to the survey in a timely manner. Additionally, Internet users are more experienced with and aware of online transactions, making them more capable of completing and accurately responding to the survey questions (Alomari, Woods, & Sandhu, 2009). An online survey also addresses the issue of geographical access given that citizens are scattered across cities and villages in this large country. Compared with paper-and-pencil surveys, an online survey is more economical and a more efficient method of data collection, especially because travelling (or sending mail surveys) throughout the KSA and other countries where citizens might be located to distribute questionnaires would be expensive and time consuming (Van Selm & Jankowski, 2006). This study focuses on Saudi citizens to facilitate the examination of the national culture of Saudis and its influence; the inclusion of other nationalities may have distorted the findings.

The study was carried out in three stages: The first stage involved an extensive literature review, which served as the foundation for developing the research model. The second stage involved questionnaire development and the third was devoted to data analysis. As shown in Figure 1.1 below, the second stage was further divided into six phases: after the literature review and development of the research model, associated constructs (concepts) and related items (questions) were determined for inclusion in the instrument. The questionnaire was then translated into Arabic and compared with the English version multiple times until the translation was deemed accurate. The developed questionnaire was reviewed and pre-tested by nine Saudi participants to evaluate the clarity and accuracy of the items' intended meaning. At this stage,



the participants tested the usability of Qualtrics.com as online survey software. The fourth phase was intended to ascertain the content validity of the contextualised items and constructs. Lewis et al.'s (1995) questionnaire development and content validity procedures (described later in section 6.2.3) was applied. Fifth, the resultant instrument was pilot-tested with 113 participants; feedback was collected and feasible recommendations were adopted. Finally, a full-scale survey was sent to the sampling frame (100,000 online users).



Figure 1.1
*Overview of the research methodology.*

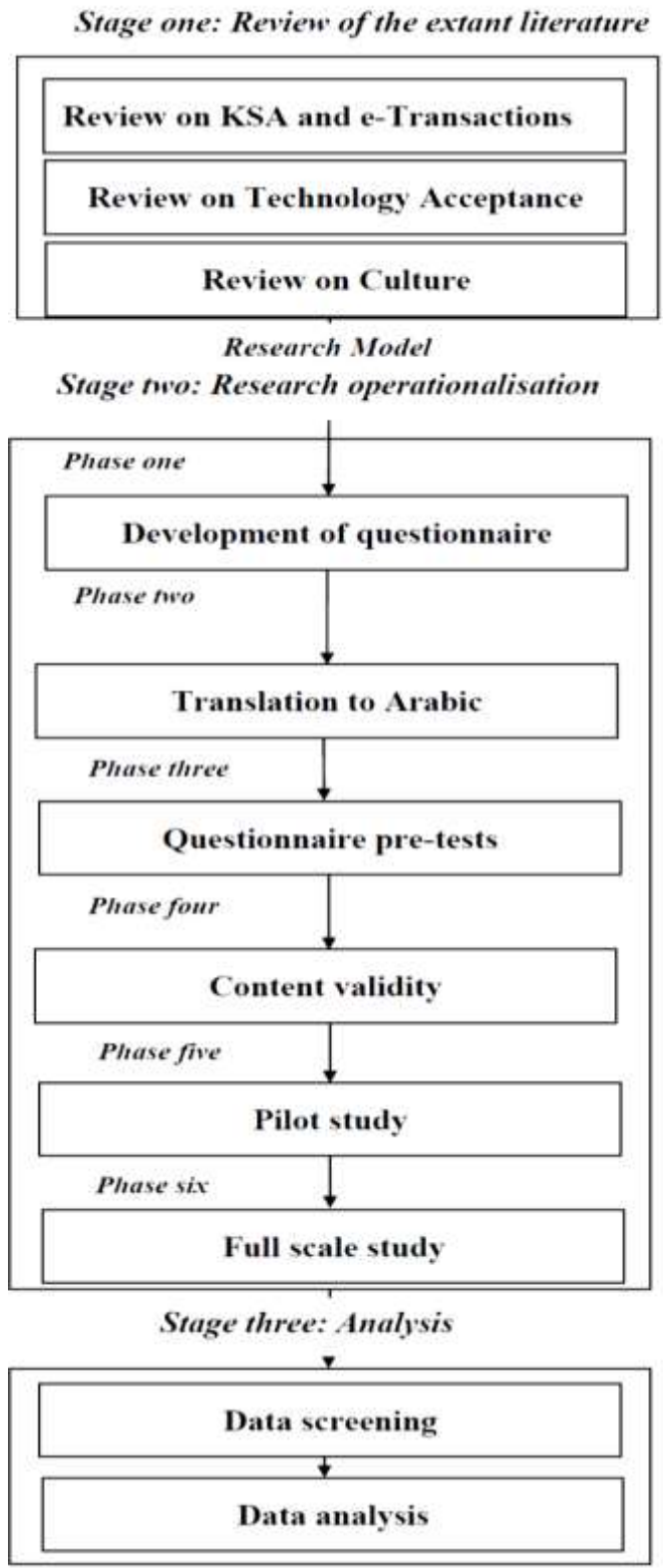

In the third stage, the data was first screened for possible elimination of outliers, and then the demographics of 671 Saudi citizens were described. After screening the collected data,



it was assessed for suitability of use by Structural Equation Modelling (SEM). SEM was employed to assess the hypothesised relationships.

## 1.5    Thesis Structure

Excluding this chapter, the thesis comprises nine chapters. Chapter 2 provides an overview of the KSA and its e-government programme. Necessary information on the KSA as a country, economy, and culture, as well as a discussion of e-transaction development and implementation in the country, is presented. Additionally, a review of the literature on the acceptance of technology in the KSA is discussed.

Chapter 3 reviews the relevant theories on the acceptance of technology and discusses the theories most suitable for this research. Other research models on e-government adoption and acceptance are also presented. The tasks described in this chapter served as the stepping stone to the development of the research model and hypotheses.

Chapter 4 discusses and reviews the theories and conceptual models for explaining culture. Culture is defined and its importance is re-emphasised. Cultural values are explained and different models of culture are elucidated. A more detailed explanation is devoted to the cultural model chosen for this study: the theory of Basic Human Values (BHV) and the instrument used to capture the essence of Saudi culture.

Chapter 5 presents the conceptual development of the research model. The model of the perceived characteristics of e-transactions is first discussed, in which the constructs used in previous studies are redefined and contextualised. The chapter presents the overall research model, which includes the following factors: perceived characteristics of e-transactions, trust in the government and the Internet, social influence, perspective on communication, cultural values, and intention to use e-transactions. The research hypotheses, which were developed on the basis of the research questions, are presented.



Chapter 6 explains the research methodology. The philosophy and paradigm that underpin this study are discussed. The research design used for data collection is also presented, followed by a detailed explanation of the questionnaire development process and the sampling techniques used. Here, the initial test and pilot test on the questionnaire are discussed, as well as the design and approach used to collect data for analysis (full-scale study). The study's ethical considerations are discussed, and finally, the manner by which the research model was quantitatively designed using the developed questionnaire is explained.

Chapter 7 presents the data preparation and examination. The demographic characteristics of the Saudi population with Internet access are compared with those of the sample recruited for this study to determine the similarity between the two groups. The sample is screened for any extreme or invalid responses from the participants. The sampled data was then assessed for its suitability for SEM. This chapter advances the understanding of the features of the sampled data and enabled the determination of a suitable analysis method.

Chapter 8 reports the results of the data analysis. The reliability of the constructs is evaluated to determine the level to which the items are internally consistent with their underlying measured concept. Furthermore, the validity of the measured constructs is assessed to determine the level to which the items measure the intended concept. The methods used to ascertain construct validity are exploratory factor analysis and confirmatory factor analysis. All of the previously mentioned steps enabled the assessment of the hypothesised relationships between constructs, thereby facilitating the identification of the direction and significance of the relationships. Ascertaining such relationships provided answers to the research questions.

Chapter 9 comprehensively discusses the results. Research questions are revisited to establish the link with the findings. Based on the research questions and hypotheses, both the significant and non-significant relationships are explained.

Chapter 10 concludes the thesis. A brief summary of the thesis is provided. The theoretical and practical contributions of this study are also presented. Using the findings as



bases, design strategies for the enhancement of the e-government programme in the KSA are proposed. The chapter ends with a discussion of limitations and future research directions.

## 1.6    Summary

This chapter introduced the thesis, presenting the research background and overview of the culture and e-transactions of the KSA. The emphasis on cultural inquiry was presented to provide a justification for the chosen approach. The research questions and their interrelationships were discussed. The chapter also briefly described the research procedure and methodology, after which the structure of the thesis was presented.



## 2    SAUDI ARABIA AND E-GOVERNMENT

The economy of the KSA depends heavily on production and exports of oil (Niblock, 2006), which are unsustainable sources of revenue because oil and fossil fuels are non-renewable energy resources. Some experts predict that oil and fossil fuel reserves will be depleted by 2050, especially given increasing global demand (Bentley, 2002). Therefore, the Saudi Arabian government plans to shift the economy from an oil production-based structure to a knowledge-based one. A pillar of this plan is the introduction of e-government transactions (e-transactions) which simplify the delivery of services between the government and different stakeholders, including citizens using the Internet. The acceptance of e-transactions is a step towards transforming the country into an electronic society, which will enhance the chances of a successful shift towards a knowledge-based economy. However, Saudi Arabia is characterised by a conservative society, where cultural values and perceptions of technology underlie acceptance (AL-Shehry, et al., 2006; Mackey, 2002). Chapter 2 discusses the focus of the thesis: the KSA as a country, economy, society, and culture. Studies and reports related to the KSA and e-governments are reviewed and e-transactions are defined. Finally, this chapter discusses how Saudis' acceptance of technology is similar to that of e-transactions.

### 2.1    KSA Profile

The KSA is the largest country in the Arabian Peninsula, occupying about 2,240,000 square kilometres (AL-Shehry, et al., 2006; Bowen, 2008). It is bordered by Kuwait, Iraq, and Jordan to the north; Yemen and Oman to the south; the Arabian Gulf, the United Arab Emirates, and Qatar to the east; and the Red Sea to the west. According to the latest census, which was carried out in 2010, the population (including all residents citizens and non-citizens) of the KSA is 27,136,977 while the number of Saudi citizens is 18,707,576 (Central Department of Statistics & Information, 2010). Riyadh is the capital and largest city of the KSA; Jeddah and Dammam are considered major ports, located on the Red Sea and Arabian Gulf, respectively.



The major spoken and written language in the KSA is Arabic, and most of the residents are Muslims (Bowen, 2008; Central Department of Statistics & Information, 2010).

### 2.1.1    Brief History of KSA Cultural Heritage

The KSA inherited a long history of civilisation that comes from to the culture and society of the Arabian Peninsula (Al-Rāshid, 1986). Nomadic tribes inhabited the Peninsula and developed a high level of independence and deep connection to the land over the years, especially to Mecca. Mecca was considered a trade and pilgrimage destination. It is the place where the Prophet of Islam Mohammad began preaching about Islam, making this land the birthplace of this religion. The nomadic tribes were unified in the seventh century under one religion, worshiping one god as Muslims.

Mecca and Medina are the holiest Muslim cities and are located in the western province of the KSA. The modern KSA, as it is now known, was established and unified as one kingdom by King Abdul-Aziz Al-Saud in 1932. The Saudi constitution is based on the Muslims' holy book of Quran and Sunnah (speech and teachings of the Prophet of Islam, Mohammad). Up to the present, a strong connection has existed between Muslims and Islam, affecting how people live, especially in the KSA (Bowen, 2008; Niblock, 2006).

### 2.1.2    Economic and Technological Developments

The Saudis discovered oil reserves in 1938, boosting the Saudi economy and enabling the modernisation of the KSA. The KSA government and society moved towards having high standards of living and modern lifestyles (Bowen, 2008; Niblock, 2006; Ochsenwald & Fisher, 2010; Simmons, 2005). Although uprisings in the Middle East (the Arab Spring) have recently changed the political status of many neighbouring countries, such as Tunisia, Egypt, and Libya, the economy and monarchy of the KSA remain stable (Gause, 2011). This secure environment has improved KSA governance, infrastructure, technology, and society over the years and these improvements represent the potential milestones of a transition to a knowledge-based economy.



A step towards the improvement of the Information and Communications Technology (ICT) infrastructure in the KSA is the liberalisation of the Ministry of Telecommunications, which transitioned from a public entity to a large private company called Saudi Telecom Company in 2002 (Shoult & Consulting ASA, 2006). This step was followed in 2004 by the opening of the telecom market to competition from other mobile phone services providers. These initiatives have increased the adoption of mobile technologies within the population by reducing the price and improving the quality of telecom services. According to a government report intended to measure the indicators of improvements contributing in information society vision, the liberalisation of the telecom sector increased mobile phone dissemination for the Saudi population from about 100,000 in 2000 to about 9 million in 2004 (Ministry of Communications and Information Technology, 2007b). The latest census from the Ministry of Communication and Information Technology indicated that mobile subscriptions in the KSA reached 53.7 million by the end of 2011 (Ministry of Communications and Information Technology, 2011).

Other developments in the Saudi economy include the regulation and monitoring of the stock exchange market by the Capital Market Authority (CMA) in July 2003. The enhanced operations of the stock exchange market facilitated local and foreign exchange, as well as the trading of Saudi stocks. The initialisation of *Tadawul* (the Saudi stock exchange) and the CMA legalised online stock market transactions, but the introduction of the electronic stock (e-stock) exchange by banks and other financial institutions met a lack of acceptance by Saudi society. A study in 2008 showed that only 17% of financial transactions were conducted online via the e-stock exchange (Abu Nadi, 2010; Khan, 2008).

The above-mentioned developments facilitated KSA membership in the World Trade Organisation (WTO). The country's membership, approved in November 2005 by the WTO General Council, followed the KSA government's adoption of the Council's terms of accession, which include creating a suitable environment for international investment and world trade by liberalising trade and accelerating the country's integration with the world economy (Niblock &



Malik, 2007). In 2007, the Saudi Ministry of Communication and Information Technology (IT) concluded planning for the national Plan of Communication and IT, whose main goal is to implement measures that enable the transformation of the KSA into an information society and knowledge-based economy (Ministry of Communications and Information Technology, 2007b). Implementing e-government is also one of the primary targets that the national plan aimed to comprehensively achieve by 2010, with Saudi government agencies as frontliners of the initiative (Gartner Group, 2007). As a supplementary effort, the KSA government created a strategic plan called 'The Long-Term Strategy 2025' to enable diversification of the economy, that is, a transition from an oil production-based to a knowledge-based economy. One of the main goals of this strategy is e-government implementation (Alothman & Busch, 2009; Ramady, 2010), which is discussed in the following section.

## 2.2    Saudi e-Government Initiative

The Saudi e-government initiative is a part of the national IT Plan, which focuses on the use of ICT in reforming Saudi public agencies (AL-Shehry, et al., 2006). The objectives of the IT plan are to build a strong and reliable ICT infrastructure, transform Saudi society into an electronic or information society, and satisfy the requirements for a knowledge-based economy. Information societies rely on information acquired through technology to guarantee critical operations. A knowledge economy is formed through such a society, in which knowledge production and transfer are adopted for economic advantage (Ministry of Communications and Information Technology, 2007b; Webster, 2002).

Within the same setting of the national IT plan, the *Yesser* (Arabic for 'facilitating' and within this context 'facilitating government transactions') e-government initiative was officially founded in 2003 by a supreme royal decree. However, this decree was implemented only in 2005 through the establishment of the online *Yesser* programme, www.yesser.gov.sa (Sahraoui, Gharaibeh, & Al-Jboori, 2006). The objectives of the programme (regarded as the foundation of the Saudi e-government initiative; Ministry of Communications and Information Technology, 2007a) are (1) to increase the productivity and efficiency of the public sector; (2) to provide



better and simplified services for individuals and the business sector; (3) to increase returns on

investment; and (4) to provide required information for stakeholders, including citizens,

businesses, and other government agencies in a timely and accurate manner (Abu Nadi, 2010).

## 2.3     e-Government Status in KSA

Some studies and research organisations have evaluated the Saudi e-government

programme and provided good insights into the programme's form and status. Abanumy and

Mayhew (2005a) indicated that the KSA e-government initiative began in 2001, and

summarised its objectives as follows:

> The main objectives of this program were to enhance productivity of
>
> public organizations; to provide government services to citizens and business
>
> in a simple and convenient way; and to provide the required information in a
>
> timely and highly accurate style. (p. 4)

Bawazir (2006) declared that 2001 was not the first time the KSA initiated e-

government implementation, and that e-government applications were available in the country

as early as 1995, as represented by a project called the Saudi Electronic Data Interchange

(SaudiEDI). SaudiEDI served as a link between businesses and government agencies (Bawazir,

2006). Al-Elaiwi (2006), whose research focused on the Saudi Ministry of Labour, discussed

another early e-government application (automation of labour information and employment

processing systems), indicating that the ministry has initially failed many times to provide

online services to the public. He stated that implementing e-government in the KSA is

confronted with many challenges (including management of processes, technologies and

people) that government officials need to be aware of and actively address. Kostopoulos (2006)

mentioned that the KSA government developed a website for providing information to pilgrims;

this website later became a major portal that offers many other e-government services. In a

thesis presented by Alharbi (2006), the author agreed with Kostopoulos assertion that the Saudi

e-government has visibly improved in a very short period. Alharbi, however, did not detail the



magnitude of improvement, and criticised the weakly articulated content of government websites, especially the services provided by educational institutions. Alharbi (2006) stated that each society has its own needs and requirements for readiness for e-government acceptance after such an initiative is implemented.

Sahraoui, Gharaibeh, and Al-Jboori (2006) critically analysed the status of Saudi e-government, praising the level of the country's progress towards the information society aim. The authors pointed out that starting in 1998, the Saudi government has thoroughly improved the ICT infrastructure and telecom sector (Sahraoui, et al., 2006). These improvements include the 'Home Personal Computers Initiative', which is primarily a public-to-private partnership intended to provide a million home personal computers (Sahraoui, et al., 2006) to Saudi citizens at a very low price (Abanumy & Mayhew, 2005b; Alsabti, 2005). The 'EasyNet' initiative simplifies Internet access and reduce barriers to usage such as reducing access cost for users (Sahraoui, et al., 2006). The third improvement, 'e-awards', aims to confer recognition and advance the promotion of local initiatives towards the advancement of e-services (Sahraoui, et al., 2006). These improvements and initiatives represented only the initial phases of the e-government programme (Sahraoui, et al., 2006). The United Nations Department of Economic and Social Affairs (UNDESA) e-government survey in 2012 ranks Saudi Arabia in the e-government development index as 41st out of 190 countries. This progress positions the KSA as a leader in the delivery of e-government services within Asia and, according to the report, is also an indicator of the productivity and efficiency of the public sector in the KSA. However, the report states that the acceptance and usage of e-government in the KSA and around the world is generally low. For example, only two out of five Australians use the Internet to contact the government, and the average e-government usage rate in European countries is only 32%. No specific percentage or details on the number of e-government users in the KSA are provided. According to the UNDESA report, 60% of all Saudi government services and transactions can be completed online via e-government transactions (UNDESA, 2012).



**2.4    e-Government Transactions**

Heeks (2006) defined e-government as the automation of internal government processes with information systems (IS) and Internet technologies (e.g., websites) for the purpose of providing services to stakeholders, including citizens, businesses, and other government agencies. Lenk and Traunmüller (2002) suggested that: "e-government focuses upon relatively simple transactions between identifiable customers (citizens, enterprises), on one side, and a multitude of government organisations in charge of registering objects, issuing passports, collecting taxes or paying benefits, on the other" (p. 147). In the current study, e-government transactions are regarded as operations that take place between citizens and governments through government Internet services. Online e-government transaction, as a simple, direct, and important contact point between the government and citizens, is the basis of this research. e-Transactions can be conducted between citizens and the government (G2C), between businesses and the government (B2G), or internally within the government (G2G). This thesis focuses on G2C transactions. The acceptance of technology by Saudi society is discussed under the premise that e-transaction acceptance is similar to technology acceptance (Abu Nadi, 2010; Al-Gahtani, 2011).

**2.5    Saudi Society and Technology Acceptance**

The KSA is characterised by a considerable transformation over the past 50 years that changed it from an isolated desert land into the modern KSA, which represents a rare paradox of technological proficiency and traditional social conservatism. Such a combination cannot be achieved without a conflict of interest. A clear example of this conflict is the adoption of the Internet (Gallagher & Searle, 1985; Sait, Al-Tawil, & Hussain, 2004). Whereas a study conducted by Internet World Stats (2008) calculated an Internet usage growth rate of 75.8% between 2007 and 2008, a recent scan in 2011 determined that only 44% of the population uses the Internet (Communications and Information Technology Commission, 2011). One of the reasons for this relatively low usage is that the ICT infrastructure is still under continual improvement (Abanumy, Al-Badi, & Mayhew, 2005). This issue is not the focus of the present



study, however, because other socio-cultural issues that affect the acceptance of the Internet may also be related to the acceptance and usage of e-government (Weerakkody, 2008). Loch, Straub, and Kamel (2003) declared that the essence of this issue is that technological innovation carries the culture-specific beliefs and values of its innovator, which if adopted as it is, will conflict with the culture of the hosting environment. Some scholars have explained that the culture of the hosting environment is often disregarded by the innovators (Khalil & Elkordy, 2001; Straub, et al., 2003). Loch et al. (2003) argued that the low rate of Internet acceptance in the Arab world is due to cultural inconsistencies with this invention. Straub et al. (2003) revealed that the reluctance to accept the technologies introduced in the Arab world stems from the strong affinity of Arabs for their cultural beliefs and values. Rejection is directed not towards the technology itself, but towards the culture that it carries within it—a culture that clashes with Arab cultural beliefs and values. Indeed, the reluctance to accept technology is a difficulty faced by governments and organisations in the Arab world; as an Arab country, the KSA shares many characteristics, such as language, culture, and religion, with other Arab states (Al-Yahya, 2009). Hill, Loch, Straub, and El-Sheshai (1998) stated that Arab countries and organisations devote millions of dollars to technology transfer. Weerakkody (2008) pointed out that Internet acceptance in the KSA is affected by the socio-cultural attributes of adopters. These same factors may also influence the acceptance and usage of e-services, such as e-commerce and e-government. Adopting e-commerce and e-banking initiatives in the KSA has been characterised by numerous difficulties (Al-Somali, Gholami, & Clegg, 2009; Sait, et al., 2004). Sait et al. (2004), who examined the factors that influence the acceptance of e-commerce in the KSA, discussed the lack of trust in Internet security and privacy, lack of computer and Internet education, and lack of exposure to and awareness of Internet services. These negatively affect not only the acceptance of e-commerce but also that of e-banking. Despite the aforementioned studies, research that explains the acceptance of the Internet and Internet services in the KSA is rare.



e-Government acceptance is a cultural issue because the implementation of this innovation will bring about cultural changes within Saudi society; its acceptance therefore depends on cultural norms. For instance, tribal systems affect government workplaces because some government officials exhibit a preference while providing services for the members of certain tribes. Such a practice reflects negatively on the efficiency with which they accomplish their duties and the soundness of their work ethic because of the unfair emphasis on specific tribes rather than others. Having an e-government would reduce corruption because it enables the direct access of citizens to government services. However, the disintermediation caused by e-government may be unaccepted, causing rejection trends within these particular tribes (AL-Shehry, et al., 2006; Mackey, 2002).

## 2.6 Summary

This chapter provided an overview of the KSA and e-government. The literature review included background on KSA culture, economy, and technology. This chapter also presented the sub-focus of this study (e-government), discussing the Saudi e-government initiative, e-transactions, and the current e-government situation. Finally, studies most relevant to culture and technology acceptance in the KSA were discussed. In the next chapter, innovation and technology acceptance and adoption-related theories are reviewed in order to determine the design of the research model.



## 3    E-GOVERNMENT ACCEPTANCE

Cultural differences can affect how people interact with technology (Straub, et al., 2003). Any new technology carries within it the culture of the inventor, which is not necessarily compatible with receiver's culture (Loch, et al., 2003). This study has utilised different theories to explain e-government acceptance, namely the Diffusion of Innovation (DOI) theory, the Unified Theory of Acceptance and Use of Technology (UTAUT), and Perceived Characteristics of Innovation (PCI) (Moore & Benbasat, 1991; Rogers, 2003; Venkatesh, et al., 2003). Additionally, this study has adopted elements (perceptions of trustworthiness and perspective on communication) from two acceptance models (Aoun, et al., 2010; Carter & Bélanger, 2005) which are discussed in this chapter.

To enable development of the research model and hypothesis, the relevant literature is reviewed. Frequently cited theories of acceptance of technology are discussed, including the Technology Acceptance Model (TAM), DOI, UTAUT, perceptions of trustworthiness and PCI. Furthermore, the related e-government acceptance models are examined. Finally, relevant constructs are identified, with justification of their selection for the research model.

### 3.1    Theory of Reasoned Action (TRA)

TRA, which is originally drawn from social psychology, is one of the most influential theories in the behavioural and social sciences and information systems (Sheppard, Hartwick, & Warshaw, 1988; Venkatesh, et al., 2003). As shown in Figure 3.1 below, TRA is concerned with predicting behaviour on the basis of the posited associations between behaviour, behavioural intentions, and attitudes.



Figure 3.1
*Theory of reasoned action (adapted from Fishbein & Ajzen, 1975).*

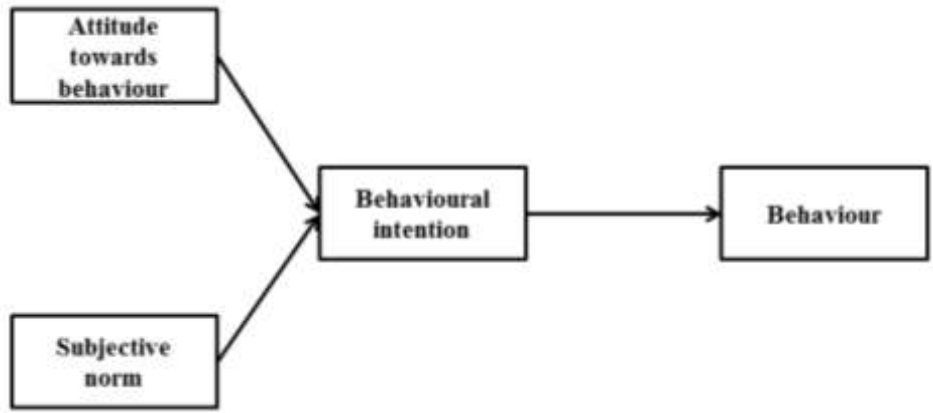

One of the most significant tenets of TRA is the proposed relationship between behavioural intention and behaviour. Behavioural intention is defined as a "person's subjective probability that he or she will perform some behaviour" (Fishbein & Ajzen, 1975, p. 288). Behavioural intention is determined by the attitude towards behaviour and subjective norm. The attitude towards a behaviour is a bipolar (positive or negative) feeling about performing a behaviour (Fishbein & Ajzen, 1975). A subjective norm is "a person's perception that most people who are important to [her or] him think [she or] he should or should not perform the behaviour in question" (Fishbein & Ajzen, 1975, p. 302). TRA suggests that attitudes, whether positive or negative, arise as a result of beliefs about the perceived consequences of a given action or behaviour. A subjective norm is more related to a person's motivation or normative beliefs about complying with the perceived normative standards (Ajzen, 1991; Fishbein & Ajzen, 1975). In technology acceptance research, the use of TRA has been prevalent, whether it is used directly, as a basis for explaining acceptance, or used as a springboard to advance new models or theories (Venkatesh, et al., 2003). The latter use of TRA is discussed in the following sections.



**3.2     Theory of Planned Behaviour (TPB)**

The importance of TRA and TPB resides in their applicability to a variety of settings and their successful projections of behavioural intention and behaviour (Ajzen, 1991; Taylor & Todd, 1995). TPB is a descendent of TRA where there is always a need to provide a more detailed explanation for the complex human behaviour (Ajzen, 1991). Figure 3.2 illustrates the addition of the construct of perceived behaviour control, which influences both behaviour, and behavioural intention and the addition of the correlations between the antecedents of behavioural intention.

Figure 3.2
*Theory of planned behaviour (adapted from Ajzen, 1991, p. 182).*

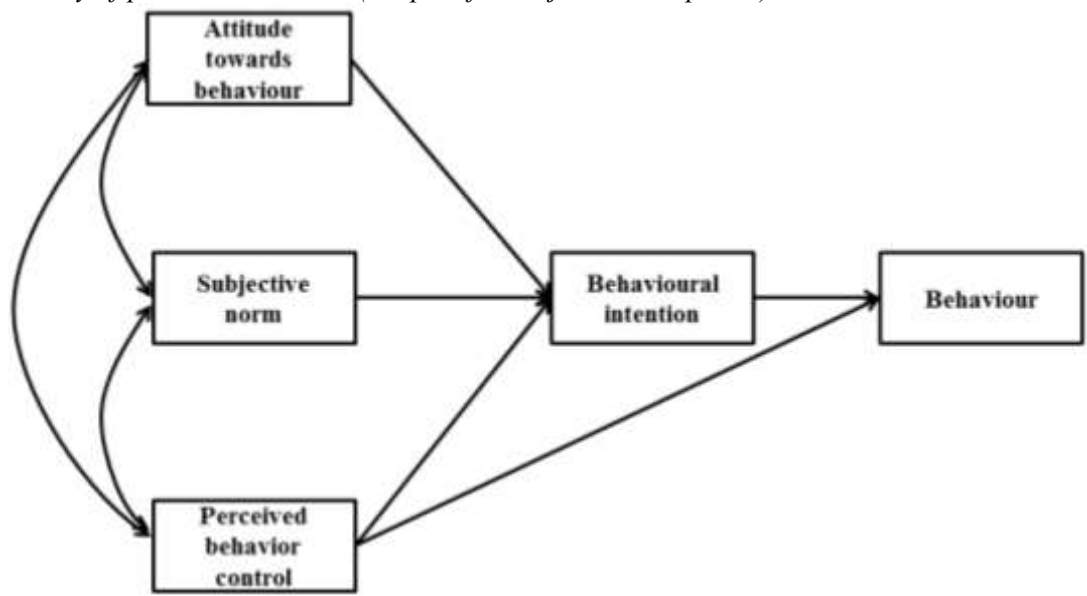

Perceived behavioural control represents the extent to which "the resources and opportunities available to a person … dictate the likelihood of behavioral achievement" (Ajzen, 1991, p. 183). The Decomposed Theory of Planned Behaviour (DTPB) is an adoption of TPB to the field of information systems (Taylor & Todd, 1995). The main components of TPB (attitudes, subjective norms, and perceived behavioural control) are decomposed into other related factors. Attitude was influenced by relative advantage, complexity and compatibility, subjective norms was influenced by normative influences and perceived behaviour control was influenced by efficacy and facilitating conditions (Taylor & Todd, 1995). DTPB was also



influential in the emergence of widely cited theories such as UTAUT (Al-Gahtani, et al., 2007; Venkatesh, et al., 2003). DTPB was not only an extension of TPB, but also of TAM, which is discussed in the following section (Taylor & Todd, 1995).

### 3.3    Technology Acceptance Model (TAM)

Of the different approaches concerning the adoption of new technology, a major approach in the field of information systems is the Technology Acceptance Model (TAM) (Abu Nadi, 2010; D'Angelo & Little, 1998; Davis, 1989). TAM is an adaptation and technology-oriented contextualisation of the social psychological TRA (Davis, 1986; Fishbein & Ajzen, 1975).

The majority of the research on TAM (Gefen, et al., 2003; Moon & Kim, 2001) has been conducted on organisation employees' perceptions of technology, not on the perceptions of consumers or citizens (who are the focus of this study). Nonetheless, Burton-Jones and Hubona (2005) concluded that the TAM constructs are insignificant in determining system usability. Lu, Yu, Liu, and Yao (2003) argued that the TAM, as a result of its generality, is unable to provide detailed information on users' opinions of a system. Another major criticism mentioned by Legris et al. (2003) is that TAM should have included social and organisational factors which are considered the most important factors for determining technology acceptance. Figure 3.3 below shows the TAM model as identified by Davis (1986, 1989).

Figure 3.3
*TAM model (adapted from Davis, 1986, 1989).*

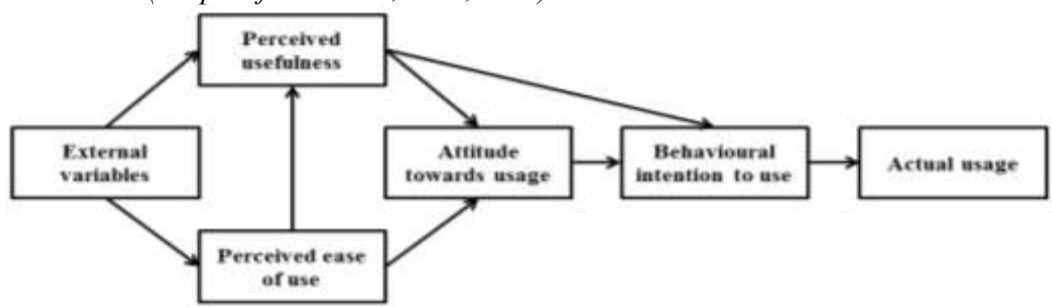

The TAM was theoretically extended by Venkatesh and Davis (2000) to contain social and organisational factors, as shown in Figure 3.4. The TAM constructs of perceived ease of use



and perceived usefulness were the basis of the model. The model also included social influence (subjective norm, voluntariness and image) and cognitive instrumental processes (job relevance, output quality and result demonstrability). Image, job relevance, output quality and result demonstrability were considered determinants of the dependent variables of perceived usefulness. Perceived usefulness and usage intention were postulated to influence actual usage.

Figure 3.4
*TAM2 model (adapted from Venkatesh & Davis, 2000).*

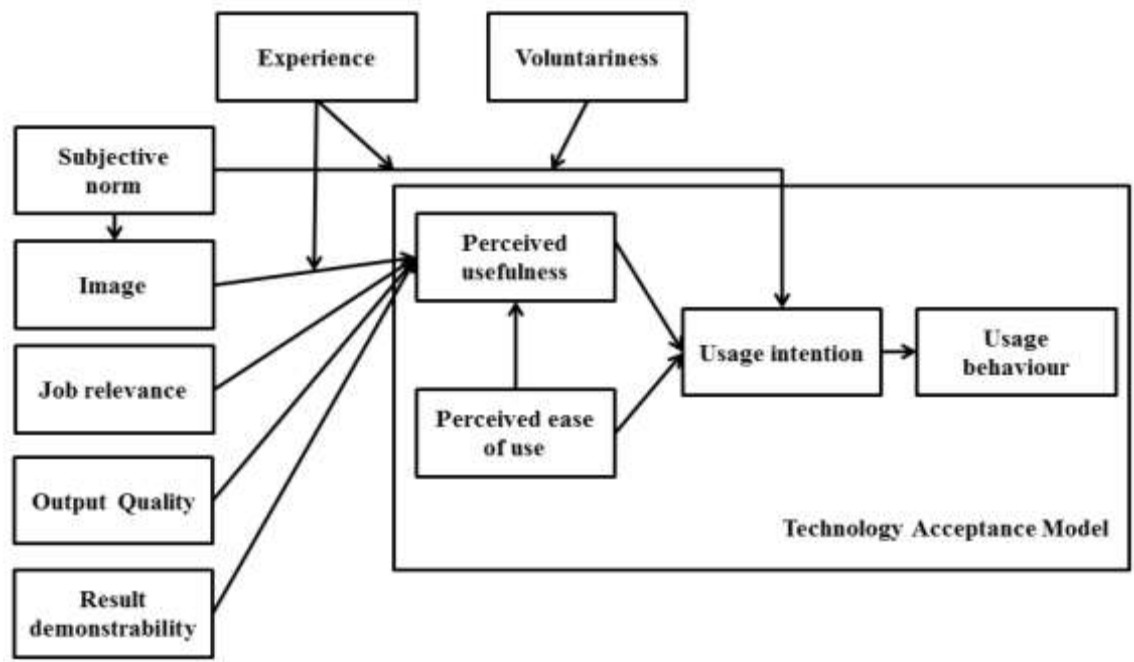

TAM continued to be extended and developed (Venkatesh & Brown, 2001), and eventually included the influence of utilitarian and hedonic outcomes perceptions, until it reached the current UTAUT form, which is considered in the next section. Nevertheless, it should be noted that McCoy, Galletta and King (2007) indicated that TAM does not fully apply for individuals who scored highly in Hofstede's cultural dimensions of uncertainty avoidance, power distance, masculinity and collectivism (discussed in section 4.3.1.1). These cultural attributes match those in Saudi society. This raises many questions regarding the influence of culture on technology acceptance, at least in the context of the Saudi society (Al-Gahtani, et al., 2007; Alshaya, 2002; Bjerke & Al-Meer, 1993).



**3.4      Unified Theory of Acceptance and Use of Technology (UTAUT)**

The UTAUT model was created by Venkatesh, Morris, Davis, and Davis (2003) and aimed to explain the behavioural intention and consequent usage of technology. The authors formulated a unified model which integrates elements from eight models (social cognitive theory, TRA, TPB, model of PC utilisation, motivational model, DOI, TAM, combined TAM and TPB) covered in the acceptance literature. The UTAUT was validated on four business organisations in different industries and then cross-validated using two additional organisations. The model was able to explain 70% of the variance in intention to use technology, which is significantly higher than previous acceptance models (ranging between 17 and 41 percent).

Venkatesh et al. (2003) developed their research model by (1) reviewing and discussing eight models of acceptance, (2) empirically comparing these models and their extensions with each other, (3) formulating a unified model that integrated elements from the eight models, and (4) empirically validating the final unified model. Basically, the method used in this thesis to develop a research model is similar to the method used by Venkatesh et al. (2003).



Figure 3.5
*UTAUT (adapted from Venkatesh, et al., 2003).*

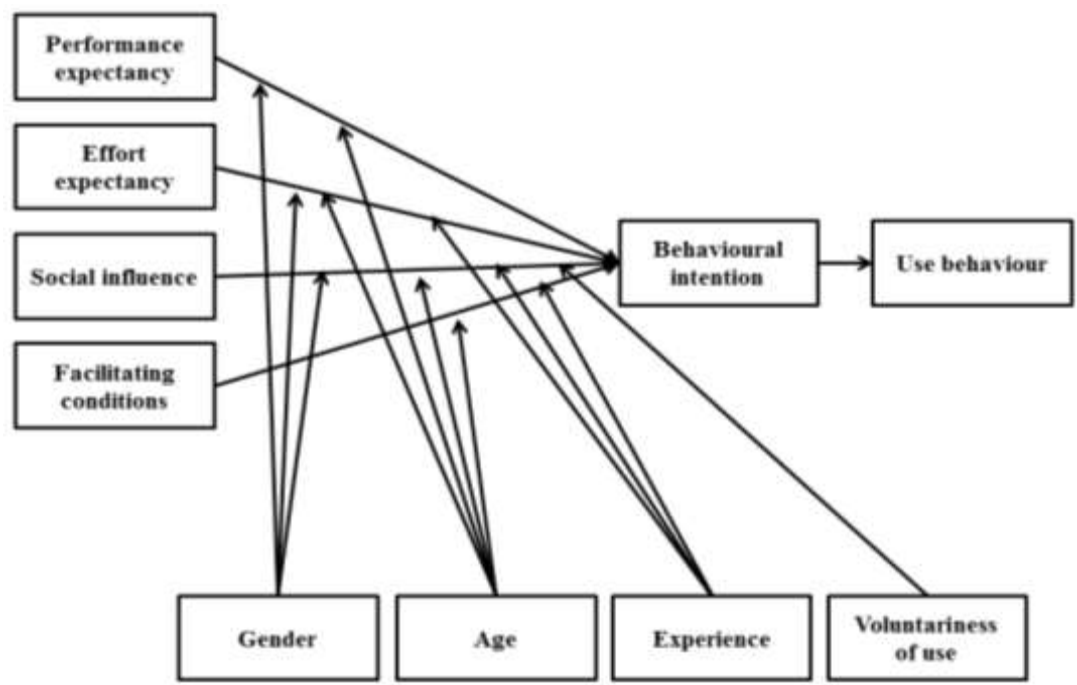

As Figure 3.5 shows, intentions for accepting technology are determined with four main constructs (performance expectance, effort expectance, social influence and facilitating conditions) and four moderators (gender, age, experience and voluntariness of use). Performance expectancy was defined as the "degree to which an individual believes that using the system will help him or her to attain gains in job performance" (Venkatesh, et al., 2003, p. 447). Venkatesh et al. (2003, p. 450) defined effort expectance as "the degree of ease associated with the use of the system." Social influence was identified as: "the degree to which an individual perceives that important others believe he or she should use the new system" (Venkatesh, et al., 2003, p. 451). Finally, facilitating conditions was defined as "the degree to which an individual believes that an organisational and technical infrastructure exists to support the use of the system" (Venkatesh, et al., 2003, p. 453). The first three constructs (performance expectancy, effort expectancy and facilitating conditions) had an influence on the intention of acceptance behaviour, which was defined by Fishbein and Ajzen (1975) as "the person's subjective probability that he will perform the behaviour in question" while the facilitating



conditions construct had an influence only on the behavioural intention construct. Behavioural intention construct influence the actual usage of technology construct (Venkatesh, et al., 2003).

The UTAUT study showed that the model is able to explain the acceptance of technology in a more realistic and complete way than earlier models. However, UTAUT has been criticised for its inability to measure acceptance of technology outside the boundaries of organisations and working environments (S. Hill & Troshani, 2010). Indeed, e-government acceptance is not limited to these boundaries. Users of these electronic services are not necessarily affected by the organisational mindset captured by UTAUT and TAM. Furthermore, since this study focuses on the demand side and not the provider and organisational context, some elements of UTAUT is selected according to their relevance to this study.

## 3.5    Diffusion of Innovation (DOI)

The DOI model provided a conceptual background that was used by many disciplines and researchers to explain the acceptance of innovations or new technologies (Rogers, 2003). Rogers (2003) conceptualised the process of innovation acceptance and distribution by creating a framework that includes definitions and attributes of DOI. Rogers (2003, p. 5) defined diffusion as "the process in which an innovation is communicated though certain channels over time among the members of social system," adding that "[a]n innovation is an idea, practice, or object that is perceived as new by an individual or other unit of acceptance" (p. 12). DOI has four elements derived from the definition of the diffusion process, namely "(1) an innovation (2) is communicated though certain channels (3) over time (4) among the members of a social system" (Rogers, 2003, p. 11).

According to the DOI model, the evaluation of an innovation is based on the five attributes of relative advantage, compatibility, complexity, trialability, and observability. Rogers (2003, p. 229) defined relative advantage as "the degree to which an innovation is seen as being better than the idea is supersedes"; compatibility as the consistency of an innovation to a potential adopter's needs, past experiences, and values; complexity as the level to which an



innovation is perceived as acceptable and effortless in terms of usage; trialability as "the degree to which an innovation may be experimented with on a limited basis" (p. 258); and observability as "the degree to which the results of an innovation are visible to others" (p. 258).

The strong advantage of DOI is that it is able to explain the acceptance of innovations at the levels of societies, organisations, and individuals (Rogers, 2003). Van Dijk et al. (2007) noted that although DOI provides a theoretically rich explanation of acceptance, the generality of the theory makes specification difficult. By contrast, the theory of Perceived Characteristics of Innovation (PCI) is a more technology-specific model. It was originally developed by Moore and Benbasat (1991) from Rogers' (1983) DOI model.

**3.6      Perceived Characteristics of Innovation (PCI)**

Moore and Benbasat (1991) extended DOI, focusing only on perceived characteristics of IT innovation. They included eight perceived innovation characteristics, namely relative advantage, compatibility, complexity, ease of use, result demonstrability, trialability, visibility, and voluntariness. Venkatesh et al. (2003) mention that there is a strong conceptual resemblance between PCI model and Davis's (1986) technology acceptance model in particularly relative advantage and perceived usefulness.  Nevertheless, PCI is based on assumptions about technology acceptance and how this process occurs, which is considered more relevant to technology than generic DOI assumptions about innovations (Lyytinen & Damsgaard, 2001). Thus, PCI would provide a more focused explanation of the acceptance of e-government. The following section is concerned with perceptions of trustworthiness, which are relevant to the usage of e-transactions but not theoretically covered by PCI.

**3.7      Perceptions of Trustworthiness**

In many cultures, trust is a concern for people when it comes to intentions to engage with technology (Jarvenpaa & Tractinsky, 1999). Carter & Bélanger's (2005) e-government acceptance model (discussed in section 3.9) included trustworthiness perceptions of the Internet and government. Rotter (1971, p. 444) defined trust as "[e]xpectancy held by an individual or a



group that the word, promise, verbal, or written statement of another individual or group can be relied on." The definition of trustworthiness adopted by Carter and Bélanger (2005) was originally found in Bélanger, Hiller, & Smith (2002) e-commerce acceptance study as follows: "the perception of confidence in the electronic marketer's reliability and integrity" (p. 252). In e-government transactions, citizens could be concerned with the government's reliability and integrity to conduct their errands and provide services using the Internet. McKnight, Choudhury, and Kacmar (2002) contended that individuals formulate trust perceptions by utilising any available information on the service provider. An individual's final decision on the usage of e-government transactions depends on whether the service-providing entity is trusted or not (Carter & Bélanger, 2005). Carter and Bélanger (2005) reported that acceptance decisions would not be made if the enabling technology was not trustworthy. Taking into consideration that the Saudi culture has high uncertainty avoidance (80) in comparison to the world average (65) according to Hofstede's index (Al-Gahtani, et al., 2007; Alshaya, 2002; Bjerke & Al-Meer, 1993; Hofstede, 2001a), it is important to consider trustworthiness. Thus, both constructs (government and Internet trustworthiness) are included in this study to examine the influence of trustworthiness perceptions. The additionally significant element of communication for acceptance by the Saudi culture, based on Aoun, et al. (2010), is discussed in the following section.

## 3.8    Perspective on Communication

Based on Edward Hall's (1973) intercultural communication theory, Aoun et al. (2010) extended the UTAUT by adapting a construct based originally on Aoun, Vatanasakdakul, and Yu (2009). Before elaborating on the construct of perspective on communication, it is worthwhile to briefly discuss Hall's theory. Hall's cultural theory proposed a classification of cultures based on their communication styles. Hall distinguished between high-context cultures (such as Japan, Arab countries, and China) and low-context cultures (such as German-speaking countries, Scandinavian countries, the USA, the UK, and Australia) (Hall, 1990, 2000). According to Hall (1976, p. 91):



A high-context communication or message is one in which most of the
information is either in the physical context or internalized in the person,
while very little is in the coded, explicit, transmitted part of the message. A
low-context communication is just the opposite; i.e., the mass of the
information is vested in the explicit code.

The meanings of messages in high-context cultures are implied rather than articulated
by words. Many meanings are derived from the status quo and the environment when and where
the message is being delivered. Furthermore, the interlocutors' behaviour, paraverbal cues, and
closeness of relationships tend to affect how the message is understood. For example, a close
relationship between interlocutors enables a commonly shared history of information which
might help to establish the context in which messages are understood. On the other hand, low-
context cultures focus on the explicit content of messages (Hall, 1976, 1990, 2000; Würtz,
2005).

The perspective on communication construct employs Hall's theory to measure the
ability of technology to enable adequate communication (Aoun, et al., 2010; Aoun, et al., 2009).
Aoun et al. (2010) proposed that there is a direct link between usage of technology and
communication orientation in two studies. The first study (Aoun, et al., 2009) tested enterprise
resource planning in post-implementation performance in Chinese manufacturing companies. It
concluded that many factors affect performance and that the high-context culture of Chinese
employees was a negative factor. Although, performance is different from intention of usage,
and enterprise resource planning systems are different from e-transactions, the Chinese culture
of communication is similar to that of the Saudis as both are considered highly contextual (Hall,
1976, 1990, 2000; Würtz, 2005). Therefore, high-context communication would be relevant.
The second study (Aoun, et al., 2010) concluded that in Australia (considered a low-context
culture), accounting practitioners' acceptance intentions for accounting information systems are
positively affected by it as a communication medium. Aoun et al. (2010) asserted that low-
context cultures (such as Western cultures, including Australia) welcome the usage of



technology as a communication medium. This is attributed to the preference for the explicit communication of information. On the other hand, Eastern cultures (high-context cultures) utilise information during communication differently by acquiring indirect details. Thus, it might be concluded at first glance that usage of technologies which only provide hard facts is not welcomed by high-context cultures, but preferred by low-context cultures.

However, these results cannot be generalised to the Saudi case. The context of this study is different as the Saudi culture is not similar to the Australian or Chinese culture in terms of national culture and communication (Hall, 2000; Hofstede & Hofstede, 2005). Furthermore, e-transactions are different from accounting information systems and enterprise resource planning. The perspective on communication construct has the ability to capture perceptions of the adequacy of e-transactions as a communication medium between Saudi government agencies and citizens; thus, it was included in the research model. The researcher contacted the main author of the study by Aoun et al. (2009) and acquired the items of this reflective construct, which are as follows:

- My ability to communicate is enhanced when using accounting information systems.
- Communications through the systems enhance my ability to interpret business issues.
- Textual, verbal, and visual information is important for business communication.

The 'reflective' nature of this construct and these items means that an increase in the perspective on communication construct would entail an increase in all other items (Aoun, et al., 2009; Petter, Straub, & Rai, 2007). The following section provides a summary of the e-government adoption models available in the literature.

### 3.9     e-Government Adoption Models

Many researchers (Carter & Bélanger, 2005; Gefen & Straub, 2003; Warkentin, Gefen, Pavlou, & Rose, 2002) have commented that the usage of e-government actually depends on the willingness of citizens to adopt these online services. Therefore, the focus of this study is to



understand the factors that would affect the willingness of Saudi citizens to use e-government services. To enable an understanding of the factors that are related to the acceptance of e-government, related models outlined in the literature are discussed, taking into consideration cultural factors. Most models were used for investigation in developed countries such as the UK and the USA; few were implemented for emerging economies such as Thailand, and even fewer for Arab countries such as Oman, Jordan and Saudi Arabia.

An early model developed by Carter and Bélanger (2005) explained the adoption of e-government in the USA. Their model originated from Moore and Benbasat's (1991) PCI model and was used to determine relevant factors in e-government adoption. However, this model was later improved to include the following factors: compatibility, relative advantage, and complexity (from DOI), image (from PCI), and perceived ease of use, perceived usefulness (from the TAM), trust in the Internet and trust in government agencies. These were hypothesised to influence intention to use e-government in the context of an online voting system, as shown in Figure 3.6.



Figure 3.6
*Adoption of e-government initiative (adapted from Carter & Bélanger, 2005).*

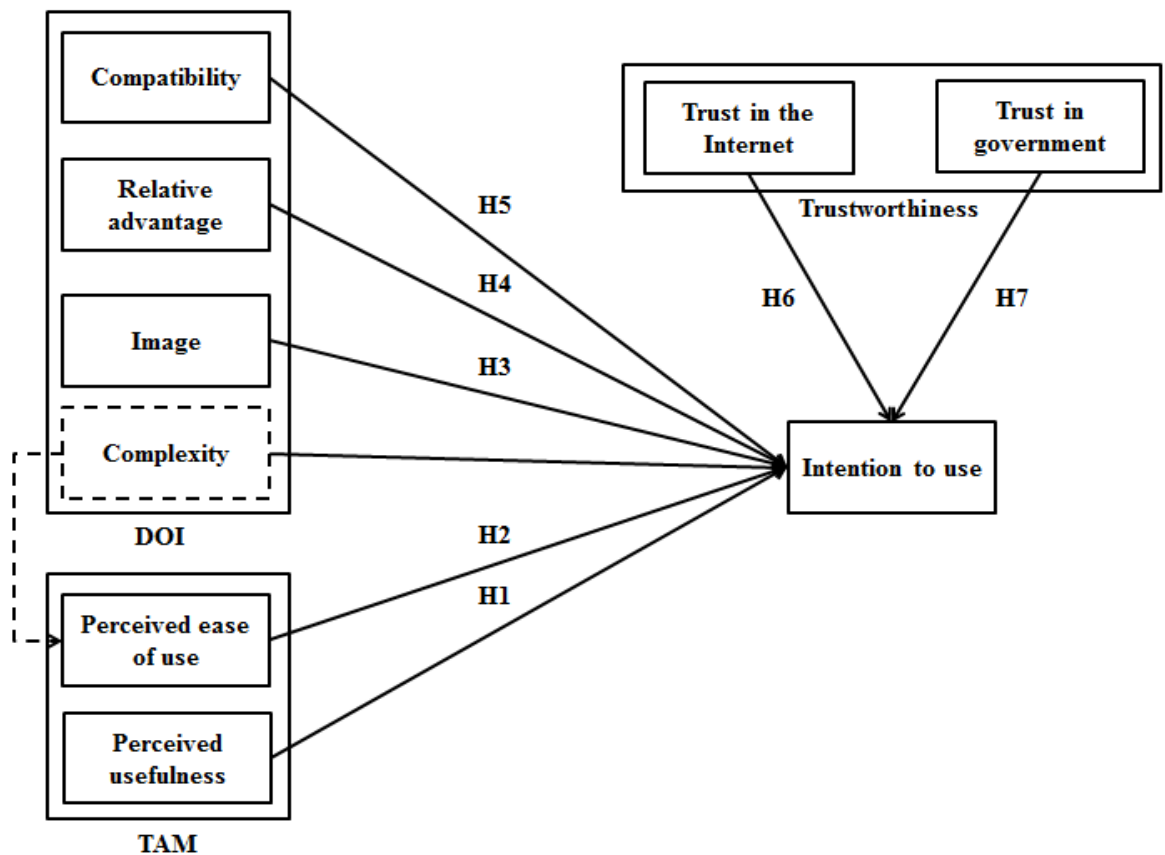

The authors chosen the ease of use construct over complexity construct (the reverse of ease of use) considering that the two concepts overlap. They argued that ease of use is well-tested and would better represent the concept of difficulty or ease of understanding and using e-government services. The study results showed that all of the abovementioned factors were significant, except for image and relative advantage for the intention to use e-voting services (Carter & Bélanger, 2005).

In a later study, Bélanger and Carter (2008) analysed the role of risk perceptions and trust in citizens' intention to use e-government services in the USA. Their results indicated that trust in the Internet, trust in the government and perceived risk positively affected the Intention to Use (IU). The model, as shown in Figure 3.7, also included disposition to trust which is indirectly related to IU.



Figure 3.7
*Trust and risk in e-government adoption (adapted from Bélanger & Carter, 2008).*

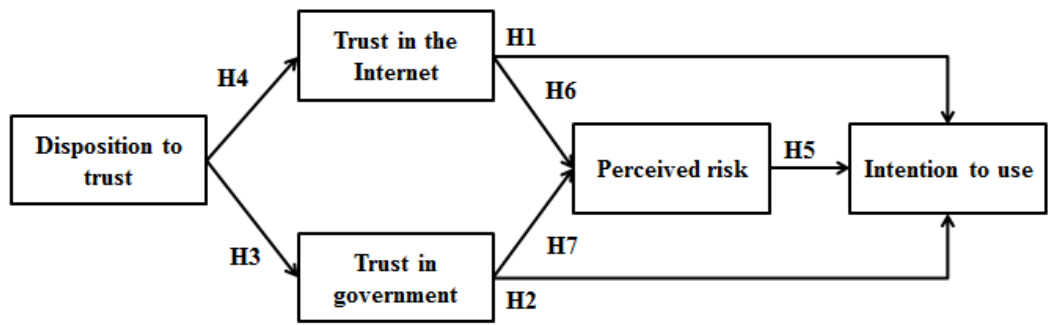

Another similar model was developed by Alsaghier et al. (2009, 2010), with the focus on antecedent factors of trust and trust beliefs regarding e-government in Saudi Arabia, as shown in Figure 3.8 below. The authors included trust factors such as perceived risk, trust in e-government and perceived website attributes including perceived ease of use, perceived usefulness, perceived website quality and website familiarity. These factors were suggested to have an influence on trust in e-government, but not on intention to use. About 400 surveys were collected in five major Saudi cities, and focus group studies were conducted to verify the model. The authors found that e-government trust was affected by the following: institution based trust, familiarity, perceived website quality and perceived ease of use. Additionally and more importantly, intention to engage with e-government was influenced by trust in e-government, perceived usefulness, and perceived risk. Although Alsaghier et al. (2009, 2010) discussed the importance of cultural factors for e-government acceptance in the KSA, they did not study its effect empirically.



Figure 3.8

*e-Government adoption with focus on trust in Saudi Arabia (adapted from Alsaghier, et al., 2010).*

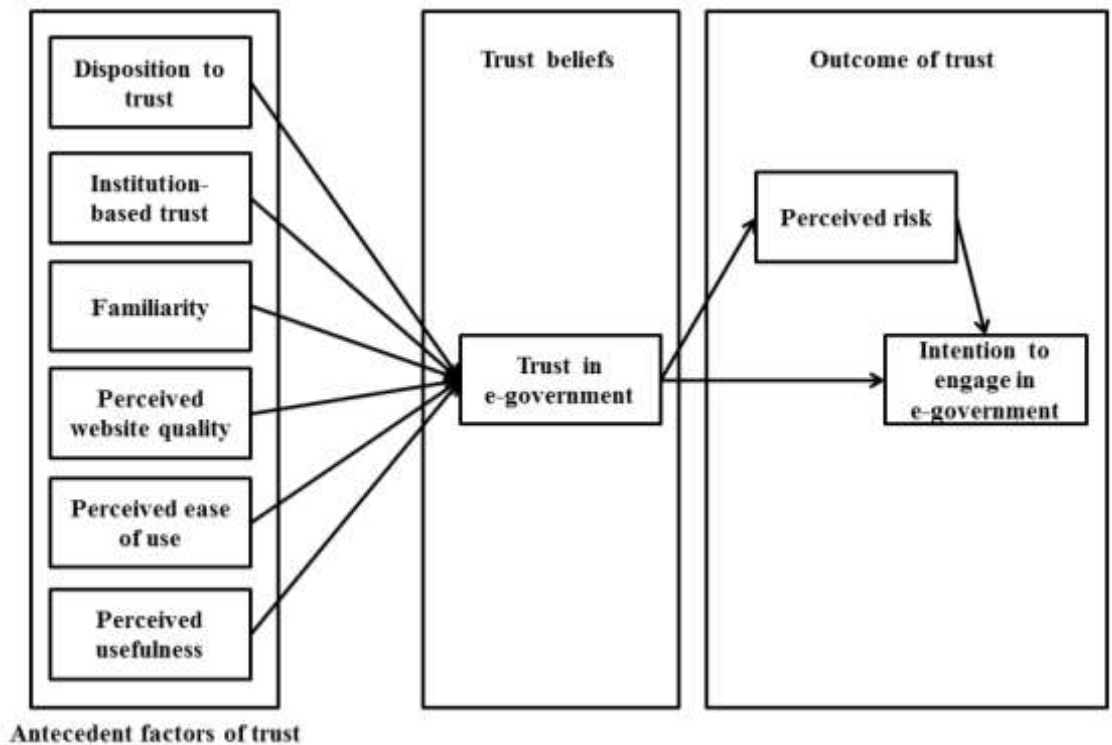

 Warkentin et al. (2002) suggested a conceptual model, also based on trust, for the online taxing system in the USA; but, this model was not empirically tested. Nevertheless, compared to previous models, this one had ties with Hofstede's cultural dimensions, as shown in Figure 3.9 below. The authors proposed that there is a direct link between power distance and intention to receive and request e-government services. Moreover, they used uncertainty avoidance as a moderator between the relationship of perceived ease of use and intention to receive and request e-government services.



Figure 3.9
*Encouraging citizen's adoption proposed model (adapted from Warkentin, et al., 2002).*

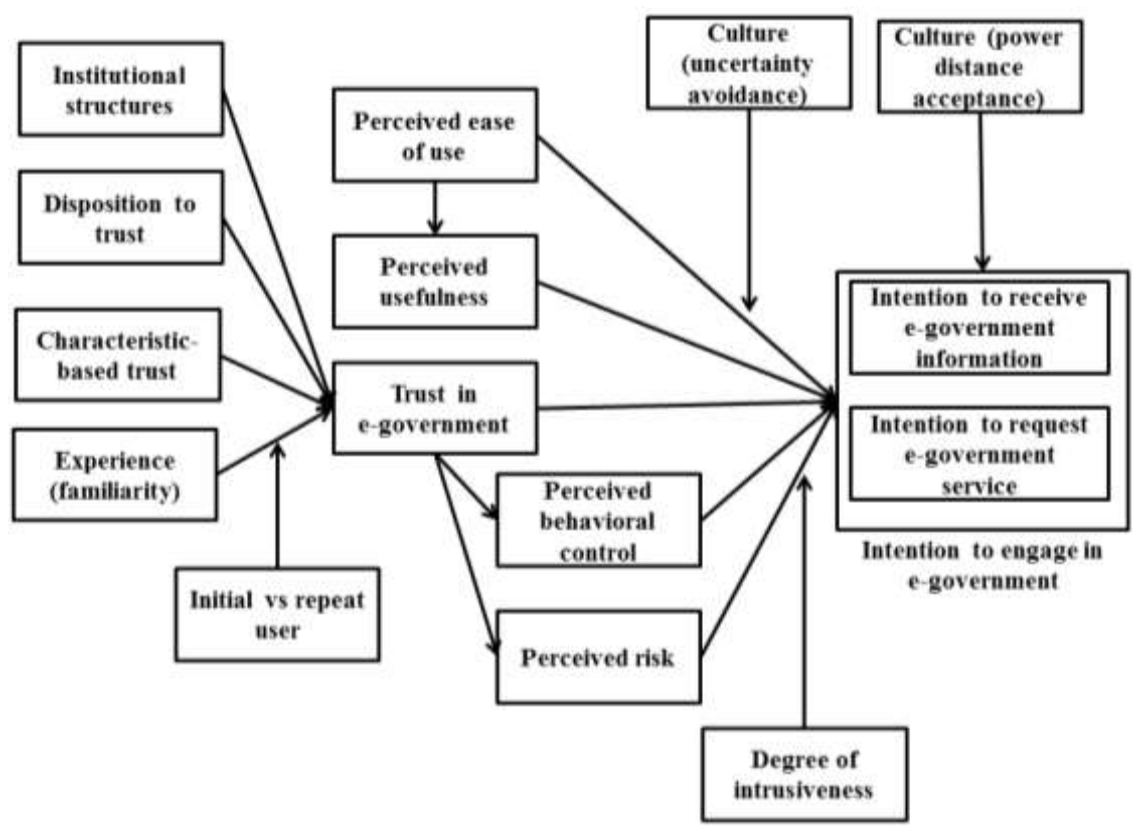

Warkentin et al. (2002) indicated that culture is a probable contributor to or inhibitor of e-government adoption, since culture is a determinant of how people act. Using Hofstede's (Hofstede, 2001a) dimensions of national culture values, they claimed that the power distance value would have an effect on e-government adoption and usage. They explained that societies with a lower power distance (which view the government as a serving entity), such as the USA, would not accept e-government as easily as a societies with higher power distance (which obey government instructions). Therefore, they proposed that high power distance would positively influence intentions to engage in e-government (Warkentin, et al., 2002). The second factor related to culture was uncertainty avoidance, about which they suggested that the higher the likelihood to avoid uncertainty, the greater the influence of trust on e-government adoption. On the other hand, the lower the uncertainty avoidance within a culture, the less hesitant the citizens would be to trust e-government. Thus, they suggested that "higher uncertainty avoidance will reinforce the positive effect of citizen trust on intentions to engage in e-government"



(Warkentin, et al., 2002, p. 161). As a methodology to measure these factors, they suggested distributing a pilot-tested survey to citizens within different countries and cultures. However, they did not include the actual results of this survey, if it was conducted in the first place (Warkentin, et al., 2002).

Another non-empirically tested conceptual model was developed by Kumar et al. (2007). They divided perception of e-government which has an influence on e-government adoption into three categories: user characteristics (perceived risk and perceived control), website design (perceived usefulness and perceived ease of use) and user satisfaction, as shown in Figure 3.10 below.

Figure 3.10
*Factors for successful e-government adoption proposed model (adapted from Kumar, et al., 2007).*

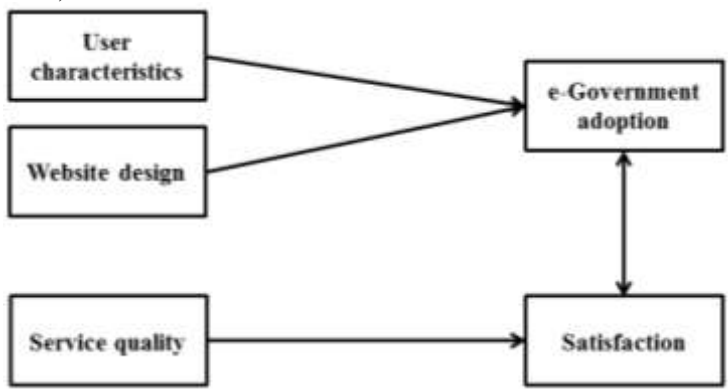

The authors discussed whether service quality, website experience and users' perceived content, such as ease of navigation, accessibility and personalisation, have an effect on their satisfaction and adoption. They also assumed that the higher levels of citizens' satisfaction with e-government quality of service, the higher the probability that they will adopt e-government.

Studying Taiwanese citizens, Hung et al. (2006) developed an e-government acceptance model using TPB and by studying the Online Tax Filing and Payment System (OTFPS). The following factors were important determinants for the Taiwan case: perceived usefulness, perceived ease of use, perceived risk, trust, compatibility, external influences, interpersonal



influence and facilitating conditions, as shown in Figure 3.11.The only factor that was not found significant was personal innovativeness.

Figure 3.11
*Results of the e-government acceptance of OTFPS in Taiwan (adapted from Hung, et al., 2006).*

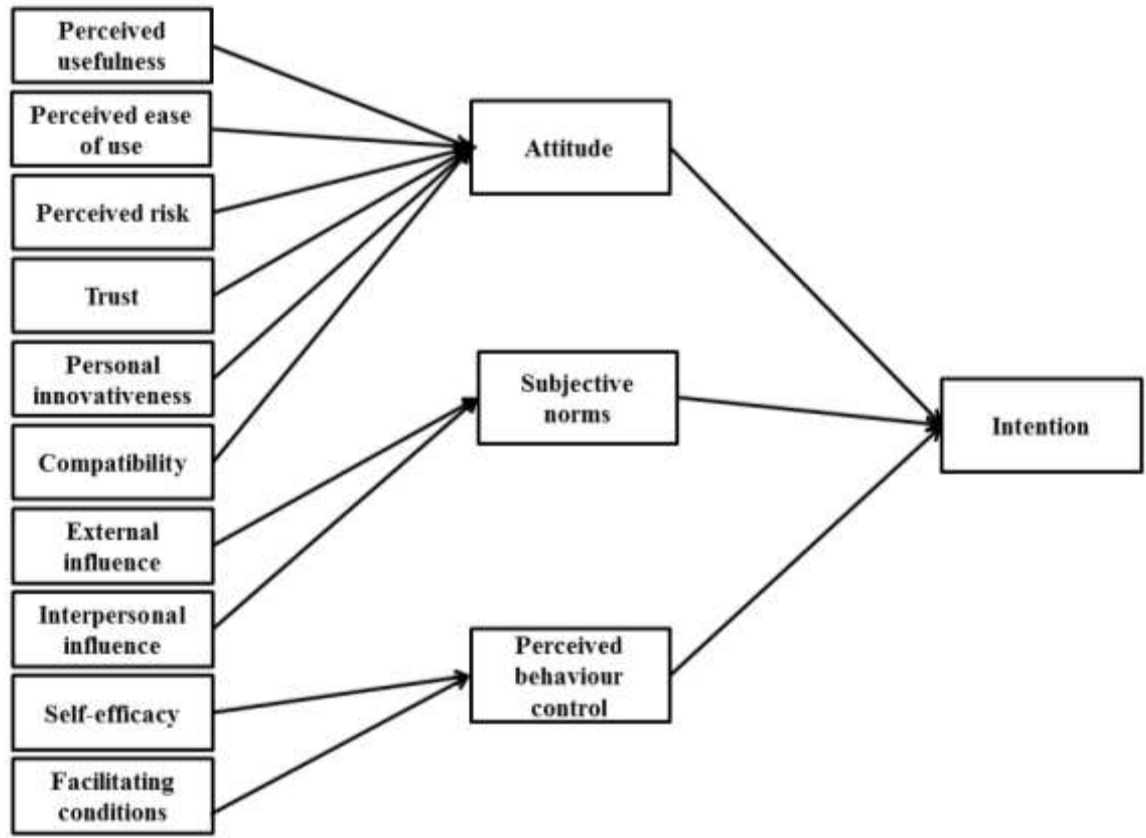

Although Hung et al.'s (2006) study was focused on the Taiwanese culture, no consideration was given for cultural factors. It was noticed, however, that Taiwan has a very low individuality score (17) according to Hofstede's Individualism Index (Hofstede & Hofstede, 2005). This might explain the strong influence exerted by the subjective norm on the adoption of e-government in Taiwan. As Hung et al. (2006, p. 113) mentioned: "Subjective norm significantly affects non-adopters' intention to not use." Their final contribution was a set of recommendations to Taiwan Government Agencies based on the relative importance of the previously mentioned factors (Hung, et al., 2006).

Using a culturally specific approach, Carter and Weerakkody (2008) studied adoption of e-government by comparing the two different but very similar cultures of the UK and USA. The



model (shown in Figure 3.12) was adopted from Carter and Bélanger (2005), who had

previously conducted their survey in the USA. Specifically, Carter and Weerakkody (2008)

compared the results of their study which was conducted in the UK with Carter and Bélanger's

(2005) results. Carter and Weerakkody (2008) found that relative advantage and trust (in the

Internet and government) were for both cultures the only significant influences on intention to

use e-government.

Figure 3.12
*e-Government adoption in the UK (adapted from Carter & Weerakkody, 2008).*

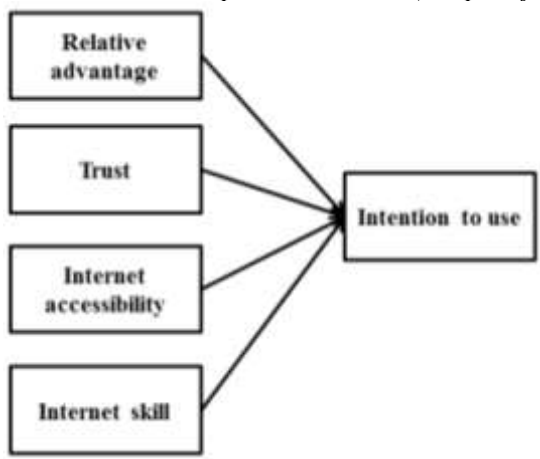

Although the study's title described it as a cultural comparison and its main focus was

the UK, the constructs and main elements of the model and research were not synthesised or

chosen based on British culture. Furthermore, Internet accessibility and Internet skills are

factors related to the ICT digital divide and might be more relevant to infrastructure and

education than to culture.

Another model was tested in Jordan by Alomari, Sandhu, and Woods (2009) who

studied the social factors affecting e-government adoption. Alomari, Sandhu, and Woods's

(2009) model was adapted from that of Carter and Weerakkody (2008) and Carter and Bélanger

(2005), but it added "attitudes and beliefs," as shown in Figure 3.13 below. This construct was

adapted from West (2004), who conducted surveys on the state e-government services in the

USA. Although West's (2004) construct was beliefs in effectiveness of state e-government



services, Alomari et al.'s (2009) construct focused on the preference to use traditional methods over e-government.

Figure 3.13
*Deployment of e-government in the Jordan model (adapted from Alomari, Woods, et al., 2009).*

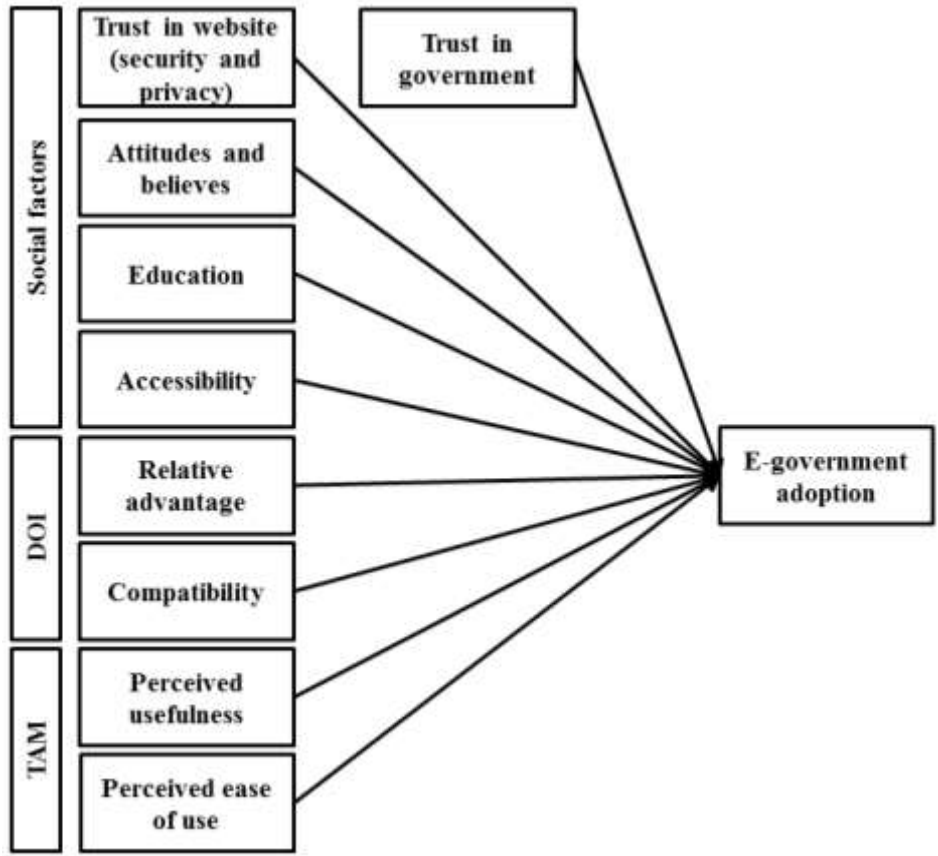

Alomari et al. (2009) proposed that these constructs would be relevant for the case of Jordan, yet gave no detailed explanation of why these any of factors, with the exception of attitudes and beliefs, would be related to this specific culture. The authors mentioned that religious beliefs are an important determinant of e-government adoption in Jordan since the country was a former socialist nation. Additionally, they explained that the lack of e-government centricity in Jordan was one of the main reasons for trust issues to arise for citizens, including trust in the website and trust in the provider of the service (the government). Internet usage in Alomari et al.'s (2009) model has taken factors from the digital divide in Jordan (education and accessibility to the Internet) and perceived website design aspects (privacy, security, usefulness and ease of use, complexity, and relative advantage). The following



constructs (Education, relative advantage, compatibility and perceived ease of use) were dropped from the study due to lack of discriminant validity owning to the conceptual similarity of these constructs with the retained constructs (beliefs, accessibility, complexity, and perceived usefulness). Results indicated that beliefs, accessibility, complexity, and perceived usefulness were significant. Although this model yielded noteworthy results, Alomari, Sandhu, and Woods (2009) and Alomari, Woods, and Sandhu (2009) focussed on the digital divide and perceived website design, with no detailed elaboration of cultural factors except for the attitudes and beliefs construct which was discussed as a social factor. Furthermore, neither of these studies focused on specific elements of e-government such as transactions, services or simply the acquisition or submission of information.

A very important model in terms of its relevance to this thesis it that of Alhujran (2009). Alhujran (2009) studied the influence of Arab national culture on TAM, but extended TAM by considering the influence of national culture, trustworthiness, and perceived public value on e-government acceptance, as shown in Figure 3.14 below.

Figure 3.14
*Determinants of the e-government adoption model in Jordan (adapted from Alhujran, 2009).*

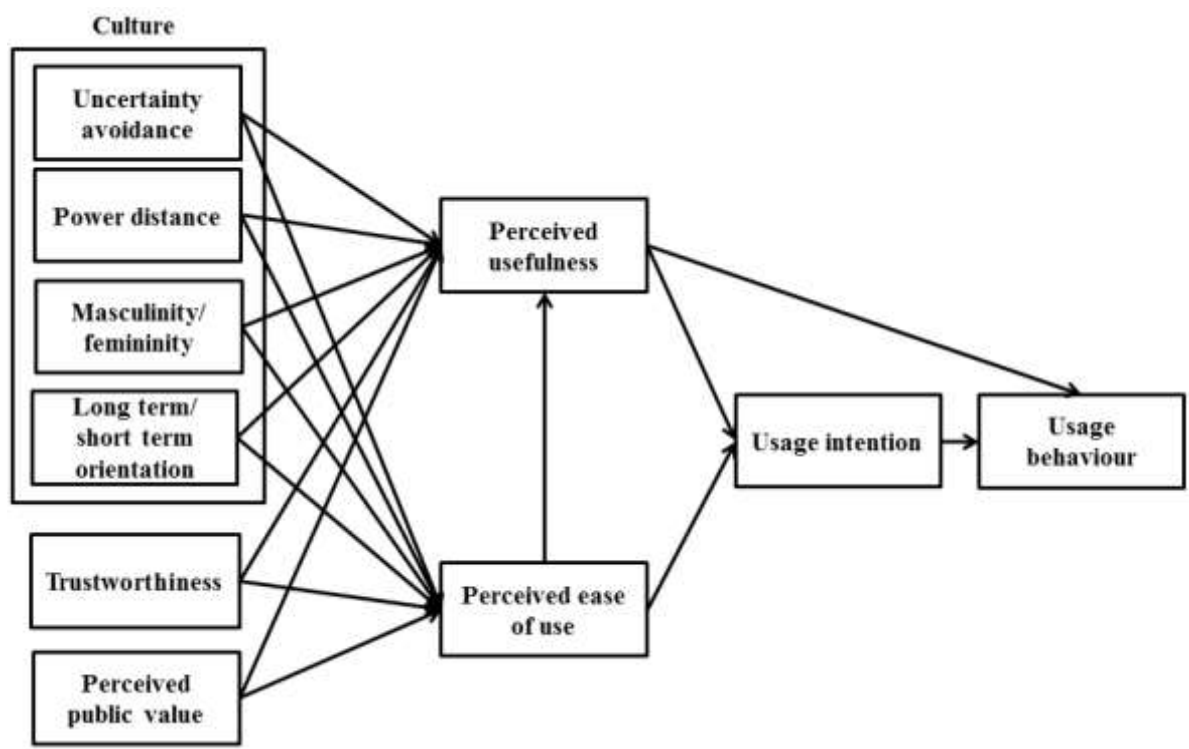



Alhujran (2009) considered the influence of five of Hofstede's National Culture Dimensions (that is, uncertainty avoidance, power distance, masculinity/femininity, individualism vs. collectivist and long term orientation/short term orientation) as well as of trustworthiness and perceived public value on perceived usefulness and perceived ease of use. The author defined trustworthiness as what the government does to make itself trustworthy in terms of e-government services delivery. Therefore, trustworthiness of the government is an antecedent of citizens' trust in e-government as postulated by Alhujran (2009). The trustworthiness construct was actually adopted from a previously described constructs in Carter and Bélanger (2005): trust in government agencies and the Internet. Alhujran (2009) hypothesised a relationship between national culture and how e-government websites are perceived in terms of usefulness and ease of use. Perceived public value was defined by Alhujran (2009) as how citizens perceive e-government services. Although this construct might not be directly related to the Arab culture, Alhujran (2009) hypothesised that perceived public value would have an influence on e-government perceived ease of use and perceived usefulness.

Specifically, Alhujran (2009) conducted a self-administered paper survey to which 335 Jordanian citizens responded. In terms of national culture and TAM, there was mainly no significant relationship between perceived usefulness and the five values of national culture, except for uncertainty avoidance. Furthermore, perceived ease of use was significantly influenced only by the uncertainty avoidance and power distance values. All other hypotheses were supported by the empirical study, except for the hypothesised relation between perceived usefulness and behavioural intention.

Other studies on e-government acceptance are summarised in Table 3.1 below. Although this table is inclusive, it is not exhaustive. The table highlights the factors used for each model, the methods used to empirically test and validate the model, and the findings (significant and non-significant factors for acceptance).



Table 3.1
*e-Government Acceptance Models*

| Reference | Examined factors | Method | Influence on acceptance |
|---|---|---|---|
| (Kanat & Ozkan, 2009) | **TAM** (*perceived ease of use, perceived usefulness*)<br><br>**TPB** (*perceived behaviour control, subjective norm, intentions of usage*)<br><br>**Trust factors** (*trust in the Internet, trust in government*) | Online pilot study; Quantitative surveys of Turkish citizens<br><br>Quantitative student surveys | Not available |
| (Carter, 2008) | **TAM** (*perceived ease of use, perceived usefulness)*<br><br>**Computer usage** (*computer self-efficacy)*<br><br>**Experience with e-government** (*previous e-government transaction*)<br><br>**Trust factors** (*trust in the Internet, trust in government*) | Quantitative surveys of USA citizens | Significant factors:<br><br>Perceived ease of use<br><br>Perceived usefulness<br><br>Perceptions of previous e-government transaction<br><br>Trust in the Internet<br><br>Insignificant factors:<br><br>Trust in the government<br><br>Computer self-efficacy |
| (Al-adawi, Yousafzai, & Pallister, 2005) | Acquiring Information:<br><br>**TAM** (*perceived ease of use, perceived usefulness*)<br><br>**Trustworthiness** (*perceived risk, Trust*)<br><br>Transaction:<br><br>**TAM** (*perceived ease of use, perceived usefulness*)<br><br>**Trustworthiness** (*perceived risk, trust*)<br><br>**TPB** (*behavioural intention, adoption* | Quantitative surveys of citizens who have nationalities from the middle eastern countries | Not available |



| | *behaviour*) | | |
|---|---|---|---|
| (Carter & Bélanger, 2004b) | **TAM** (*perceived ease of use*)<br><br>**DOI** (*compatibility, relative advantage and image*) | Quantitative pilot study of citizens from the USA<br><br>Quantitative surveys of citizens from the USA | <u>Significant factors:</u><br><br>Compatibility<br><br>Relative advantage<br><br>Image<br><br><u>Perceived ease of use not of significant influence</u> |
| (Carter & Weerakkody, 2008) | **DOI** (*relative advantage*)<br><br>**Trust** (*Trustworthiness of e-government*) | Expert opinion on surveys<br><br>Quantitative surveys of UK citizens from different cultural backgrounds | <u>Significant factors:</u><br><br>Relative advantage<br><br>Trustworthiness of e-government |
| (Gilbert, Balestrini, & Littleboy, 2004) | <u>Perceived benefits:</u><br><br>Avoid personal communications with government officials<br><br>Control of service<br><br>Convenience<br><br>Cost of service<br><br>Personalisation<br><br>Time needed to complete transaction<br><br><u>Perceived barriers:</u><br><br>Confidentiality<br><br>Difficulty of use<br><br>Lack of enjoyably<br><br>Expected reliability<br><br>Expected safeness<br><br>Visual appeal of website | Quantitative online surveys of UK citizens | <u>Significant factors:</u><br><br>Trust<br><br>Financial security (Confidentiality)<br><br>Information quality (visual appeal and reliability)<br><br>Time<br><br>Money<br><br><u>All other factors were insignificant</u> |



| (Ozkan & Kanat, 2011) | **TAM** *(perceived ease of use, perceived usefulness)*<br><br>User skills<br><br>Access to services<br><br>**TBP** *(attitude towards e-government, subjective norm, perceived behavioural control)*<br><br>**Trust factors** *(Trust in government and Internet)* | Quantitative surveys for Turkish students in Turkey | Significant factors:<br><br>Trust<br><br>Perceived behavioural control<br><br>Attitudes towards e-government<br><br>Other factor were mediated by the above constructs and not directly related to intention of e-government usage |

## 3.10  Summary

This chapter provided the necessary background on the acceptance of e-government and used theories to identify the gap in the literature. Furthermore, the discussion included the definitions and criticisms of innovation and technology acceptance theories in the literature. Finally, e-government models were discussed to enable identification of the gap in the e-government acceptance literature. This gap is the lack of consideration of cultural values when investigating e-government acceptance. The next chapter focuses on culture values, leading to the synthesis of a new research model for technology acceptance in Chapter Five.



# 4  CULTURAL VALUES

It is important to provide a theoretical background and literature review for the cultural values that are relevant to this study. Cultural values are a major component of this study, and explaining them would enable an understanding of culture and its effects on the acceptance of technology and e-transactions. The cultural unit of analysis for a study can be individual, organisational, national, or ethnic (Hofstede, 2001a; House, Hanges, Javidan, Dorfman, & Gupta, 2004; Srite, Straub, Loch, Evaristo, & Karahanna, 2003). Relevant cultural models is reviewed, including Trompenaars' Cultural Dimensions (1998), Triandis' Cultural Syndromes (1994), Hofstede's National Culture Dimensions (2001a), and Schwartz's theory of Basic Human Values (1992, 1994b). The instruments used to capture data relating to Schwartz's theory are also reviewed. There are many other cultural models, such as Hall's (1976) theory of intercultural communication, which focus on the culture of communications rather than on the conceptualisation of a holistic view of culture. The previously mentioned cultural models are comprehensive and provide concepts relevant to this study, and they are discussed in this chapter, with a particular focus on this study's adopted attitude on cultural values: Schwartz's Basic Human Values theory. Research on culture's importance for and effect on technology acceptance and e-transaction acceptance in particular, is discussed in the following section.

## 4.1  Impact of Culture

The study of culture originated from anthropology and sociology and has been used by many disciplines as an explanation of why people behave in particular ways (Davison & Martinsons, 2003). The literature has indicated the significance of studying the relationship between culture and technology, particularly e-government and culture (AL-Shehry, et al., 2006; Warkentin, et al., 2002). Literature on technology acceptance and adoption has also shown that culture is a key determinant in the acceptance of technology (Leidner & Kayworth, 2006).



Successful e-government utilisation cannot occur without a close modelling of cultural influences on the acceptance of e-government (Sharifi & Zarei, 2004). The influence of culture has become evident for many applied disciplines, including information systems and technology (M. Ali, Weerakkody, & El-Haddadeh, 2009; Davison & Martinsons, 2003). Additionally, scholars have found a significant correlation between cultural factors and the adoption of ICT (Erumban & de Jong, 2006; Zhang & Maruping, 2008), IS (Min, Li, & Ji, 2009; Twati, 2008) and IT (Srite & Karahanna, 2006).

Many researchers have argued that the diffusion of technology across cultures occurs in a highly culture-specific manner (Al-Gahtani, et al., 2007; Erumban & de Jong, 2006; C. Hill, et al., 1994; Karahanna, et al., 1999; Straub, et al., 2003). Straub et al. (2003) posited that these differences in technology diffusion are due to the strong relation between culture and technology acceptance. In fact, Straub et al. (2003) explained that the success of technologies developed in one culture and then transferred to another requires more than just technical instructions. Given that culture is a collection of values and beliefs which differentiates one culture from another, culture affects how technological systems are designed and received. Therefore, lack of acceptance occurs because individuals carry cultural biases, beliefs, and values which affect their perceptions of the technology. Thus, understanding and communicating with the receiving cultures would enable a better and more successful transfer of technology systems (Straub, et al., 2003). Espoused cultural values are considered a powerful explanation of the socio- psychological phenomenon of technology acceptance (Al-Gahtani, et al., 2007; Carter & Weerakkody, 2008; Ford, Connelly, & Meister, 2003; Srite & Karahanna, 2006; Straub, et al., 2003; Zakaria, et al., 2003). Hence, before studying this relationship, a better understanding of culture is required and is explored in the following section.

## 4.2     Definition of Culture

Defining culture is a challenge, considering the numerous definitions, dimensions, and theorisations used to describe this concept (Leidner & Kayworth, 2006; Straub, Loch, Evaristo, Karahanna, & Srite, 2002). Definitions of culture are to be found in many disciplines, including



anthropology, physiology, sociology, history, economics, management, business, technology, and information systems (Leidner & Kayworth, 2006; Srite, et al., 2003; Twati, 2006). A review of the literature on culture identified a wide range of contradictory concepts and opinions, of which beliefs, values, and norms should be considered the differentiating attributes of culture (Srite, et al., 2003).

Hofstede (1980) defined culture as "the collective programming of the mind which distinguishes the members of one human group from another" (p. 9). Previously, Kroeber (1952) defined culture as "the historically differentiated and variable mass of customary ways of functioning of human societies" (p. 157). According to Samovar, Porter, and McDaniel (1998), culture is a collection of beliefs, values, attitudes, religion, philosophy of time, roles, spatial relations, understanding of the universe and material objects, knowledge, experience, and belongings gained over generations by the group and the individuals within it. One anthropologist, Mead (1953), defined culture as simply "shared patterns of behaviour" (Davison & Martinsons, 2003, p. 3).

Coming from the social psychology discipline, Schwartz (2006) identified culture as "the rich complex of meanings, beliefs, practices, symbols, norms, and values prevalent among people in a society" (p. 138). Schein (2010) defined culture from a sociological point of view as a set of basic common assumptions that defines an interpretation of the world; what is an acceptable emotional reaction to what is going on; and what actions are required in response to an event. From the discipline of business, Trompenaars and Hampden-Turner (1998) defined culture as the way a group of individuals solve problems. A cross-disciplinary (sociology and anthropology) definition of culture by Kroeber and Parsons (1958) is as follows: "transmitted and created content and patterns of values, ideas, and other symbolic-meaningful systems as factors in the shaping of human behaviour and the artefacts produced through behaviour" (p. 583).



Even though any comprehensive and exact definition of culture is considered debatable and difficult to achieve (Davison & Martinsons, 2003; Triandis, Bontempo, Villareal, Asai, & Lucca, 1988), the literature described above has some common characteristics. Most scholars have agreed that culture is a pattern of values, attitudes, and behaviours shared by two or more individuals (Davison & Martinsons, 2003; Leidner & Kayworth, 2006; Srite, et al., 2003; Twati, 2008). Moreover, they have commonly made a distinction between the objective and subjective elements of culture and described them as being causal of each other (Leidner & Kayworth, 2006; Triandis, 1994).

Indeed, culture has been framed in the literature as a tacit set of beliefs and basic assumptions and the collective software of the mind (Hofstede, 1980) and as shared values or agreed-upon ideologies (Sackmann, 1992). Others view culture in terms of languages, symbols, ceremonies, myths, rituals, norms, and common practices (Hofstede, 1998a; Leidner & Kayworth, 2006) or include inventions, tools, and technologies in the definition of culture (Triandis, 1994). Kroeber and Kluckhohn (1952), who are widely cited in the anthropology literature, critically analysed the concepts of culture established in the early literature and concluded with the following definition:

> Culture consists of patterns, explicit and implicit, of and for behaviour acquired and transmitted by symbols, constituting the distinctive achievements of human groups, including their embodiments in artefacts; the essential core of culture consists of traditional (i.e., historically derived and selected) ideas and especially their attached values; culture systems may on the one hand, be considered as products of action, and on the other as conditioning elements of further action. (p. 181)

Focusing on the adaptability of an individual's belief system, Herskovits (1955) argued that "there is a general agreement that culture is learned; that it allows man to adapt himself to



his natural and social setting; that it is greatly variable; that it is manifested in institutions, thought patterns, and material objects" (p. 305).

Slocum, Fry, and Gaines (1991) distinguished between implicit (ideational) and explicit (material) components of culture. On the other hand, Parsons and Shils (1951) discussed the connection between these components. They claimed that both implicit and explicit components of culture which includes cultural beliefs, values, and norms have a direct effect on an individual's behaviour. Kluckhohn (1951) also emphasised this association:

> [C]ulture consists in patterned ways of thinking, feeling and reacting, acquired and transmitted mainly by symbols, constituting the distinctive achievements of human groups, including their embodiments in artefacts the essential core of culture consists of traditional (i.e., historically derived and selected) ideas and especially their attached values. (p. 86)

An earlier definition of culture that fits within this focus is provided by Redfield (1948): "shared understandings made manifest in act and artifact" (p. vii). Kluckhohn and Strodtbeck (1961) and Hall (1976) have concurred that commonly held beliefs and values within a group define how this group acts and what is and is not considered acceptable. They mentioned that humans manage and adapt to their environments by creating a system of values and beliefs that dictates how people behave, think, solve problems, make decisions, and organise their economic, political, and transportation systems (Davison & Martinsons, 2003; Hall, 1976; F. Kluckhohn & Strodtbeck, 1961; Markus & Kitayama, 1991).

According to Altman and Chemers (1984), culture is not isolated, but affects and is affected by the natural environment. They posit that the environment shapes, changes, or reinforces cultural traits within a cultural group or nation. Originating and living in different environments shapes how people perceive the world and act accordingly (Altman & Chemers, 1984; Davison & Martinsons, 2003; Herskovits, 1955). For example, people living in environments with extreme weather tend to avoid risk and uncertainty more than people living



in environments with mild weather (Boholm, 2003). Nevertheless, culture can change when environments and circumstances change (Hendry, 1999). Giddens (1979) claimed that immigrants or travellers are influenced by the cultures they interact with, causing some changes to their national native culture when they return to their original country. Although the cultural value system of individuals is relatively stable, it can change in response to adopted national and organisational cultural values (M Ali & Brooks, 2008). Even though culture contains many elements (e.g., assumptions, beliefs, norms, artefacts), most definitions and discussions of culture emphasise values due to their significance in determining behaviour (Berry, 2007).

**4.3     Cultural Values**

Cultural values have been the main element and distinctive feature of culture for many scholars (Hofstede, 2001a; Hunt & At-Twaijri, 1996; Srite & Karahanna, 2006; Zakaria, et al., 2003; Zhang & Maruping, 2008). Schwartz and Bardi (2001) argued that many factors determine the values of individuals. Values are prioritised based on enculturation, social locations, personal experiences, and genetic heritage (Schwartz & Bardi, 2001). Straub et al. (2002) stressed the role of values in shaping culture, claiming that culture is a representation of core values. They also stressed on the influence of core values on technology adoption.

Rokeach (1973) defined cultural values as "an enduring belief that a specific mode of conduct or end-state of existence is personally or socially preferable to an opposite or converse mode of conduct or end-state of existence. A value system is an enduring organization of beliefs concerning preferable modes of conduct or end-states of existence along a continuum of relative importance" (p. 5). According to Kluckhohn (1951), a value refers to "a conception, explicit or implicit, distinctive of an individual or characteristic of a group, of the desirable which influences the selection from available modes, means and ends of actions" (p. 395). Schwartz et al. (1997), in citing and synthesising Kluckhohn's (1951) and Rokeach's (1973) definitions of values, stated that values are "desirable, trans-situational goals, varying in importance, that serve as guiding principles in people's lives" (p. 6). Hofstede (2001a, p. 6) described values as



feelings that have positive or negative indications and argued that values address a number of psychological concerns:

- Evil versus good
- Dirty versus clean
- Dangerous versus safe
- Decent versus indecent
- Ugly versus beautiful
- Abnormal versus normal
- Paradoxical versus logical
- Irrational versus rational
- Moral versus immoral

Values are gained (Karahanna, Evaristo, & Srite, 2005) or mentally programmed (Hofstede, 2001a) mainly through family, education, and social environment. Moreover, values are mostly, but not completely, acquired at an early stage of life and provide basic assumptions about life and how the world is perceived. An organised system of values is unconsciously created by integrating and prioritising learned values and adapting to current environmental and social standards. A system of values, although fairly stable, can be changed, which in turn leads to changes in culture due to various personal experiences and causing distinctions in individual personalities (M Ali & Brooks, 2008; Bagchi, Hart, & Peterson, 2004; Karahanna, et al., 2005; Srite, et al., 2003; Straub, et al., 2002). Distinctions in individual's values can be noticed in the actual behaviour of individuals, where the relationship between values (the implicit component of culture) and behaviour (the explicit component of culture) may be apparent (Hofstede, 2001a).

Hofstede (2001a) created a cultural manifestation model which shows the importance of values. The model illustrates various levels of depth and values that are at the core with direct influence on practices, while symbols, which occupy the outer level, are preceded by heroes and rituals, as shown in Figure 4.1.



Figure 4.1

*Hofstede's cultural manifestation model at different levels of depth (adapted from Hofstede, 2001a, p. 11).*

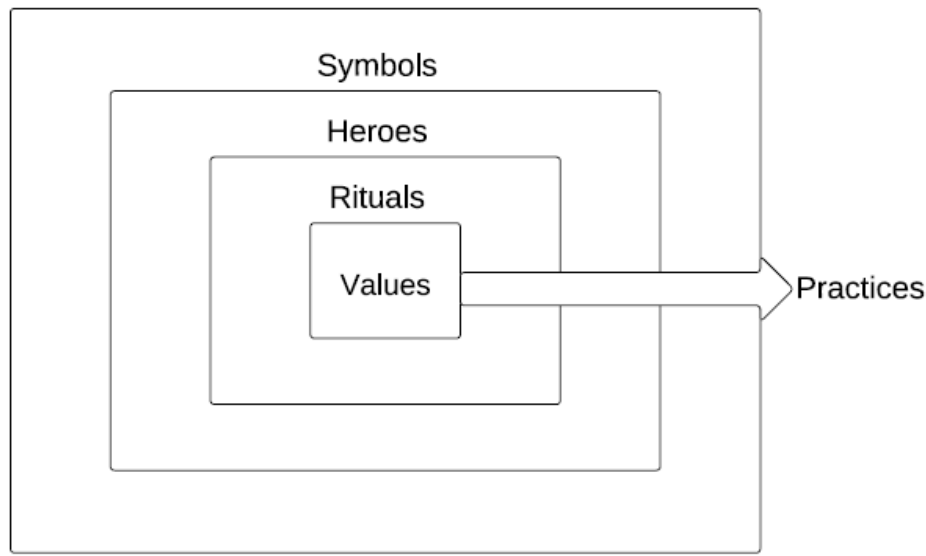

Although values are hidden, they represent the deepest manifestation of culture and are considered a determinant of preferences. Symbols, heroes, and rituals are the visible part of culture and are included under practices in Hofstede's model. Symbols are words, gestures, pictures, and objects that hold meaning and can be understood only by a specific culture. Heroes are individuals who are alive or dead, real or imaginary, but are highly esteemed in a culture and are a model for behaviour. Rituals are a group of actions that are not necessarily required to achieve an end, such as forms of greetings. In this model, values have a direct influence on practices, which in turn have an effect on the social environment

### 4.3.1.1   Cultural models

There are many models of culture which focus on different values and present them as the most relevant and important (Leidner & Kayworth, 2006). Examples are Trompenaars' Cultural Dimensions (1998), Triandis' Cultural Syndromes (1994), Hofstede's National Cultural Dimensions, and Schwartz's Basic Human Value theory (1992, 1994b). From these models, this study adopted Schwartz's BHV theory as a representation of culture and uses its instrument, the Portrait Values Questionnaire, to capture the cultural values of Saudi citizens who have Internet



access. These models are briefly discussed and the justification for adopting Schwartz's theory is presented.

Trompenaars and Hampden-Turner (1998) created their cultural model based on cultural issues of business and management executives. Categorising culture as a combination of behavioural and values patterns, Trompenaars and Hampden-Turner (1998) offered a different description of cultural values, which focuses only on dialectic opposites of cultural values, namely universalism versus particularism, affective versus neutral relationships, specificity versus diffuseness, achievement versus ascription, and internal versus external control. These five cultural dimensions are very similar to those of Parson's (1951) theory of social systems, which included affectivity versus affective neutrality, self-orientation versus collective-orientation, universalism versus particularism, ascription versus achievement, and specificity versus diffuseness. These cultural values are used to solve problems that are related to the environment, relationships with others, and time (Trompenaars, 2006). Trompenaars and Hampden-Turner's (1998) cultural dimensions are extensive in describing culture, but they are not exclusive (for instance, attitude to the environment and individuality are very similar to each other). However, in terms of applying these concepts and testing them in reality, they significantly lacked the simplicity and precision in describing behaviour which is important for this thesis (Chanchani & Theivananthampillai, 2002). Another cultural model is Triandis' (1994) Cultural Syndromes, where he defines culture as:

> a set of human-made objective and subjective elements that in the past have increased the probability of survival and resulted in satisfaction of the participants in an ecological niche, and thus became shared among those who could communicate with each other because they had a common language and lived in the same time and place (p. 22).

Furthermore, Triandis (1994) distinguishes between objective elements (such as tools, roads, and radio) and subjective elements (that is, categorisations, associations, norms, roles,



and values). The original model included only the four cultural syndromes of complexity, tightness, individualism, and collectivism (Triandis, 1994), to which Triandis added the following syndromes in his later work: vertical versus horizontal, active versus passive, universal versus particular, ascription versus achievement of status, diffuse versus specific, instrumental versus specific, and emotionally expressive versus suppressive (Triandis, 2001). Although these syndromes may seem to be comprehensive, there is a lack of provision for a specified methodology and guidelines to measurement. Additionally, because the current form of the syndromes does not have structured metrics for measurement, applying a quantitative methodology to the syndromes would not be feasible. Like many other cultural models, even though Triandis' Cultural Syndromes are conceptually unique, interesting, and rich in content, they are not completely developed in terms of instrumentalisation and operationalisation (Chanchani & Theivananthampillai, 2002).

Most of the previously discussed cultural models lacked a complete and simple foundation for testing the concepts of culture they represent. Nonetheless, Hofstede's research (1980, 1997, 1998b, 1999, 2001a, 2001b) has been widely cited in technology acceptance research (Vatanasakdakul, Tibben, & Cooper, 2004). Hofstede's work is based on an extensive series of surveys that took place from 1968 to 1972 with 116,000 IBM employees in 72 international divisions. To explain the differences between the respondents using factor analysis, Hofstede established the four overall dimensions of uncertainty avoidance, power distance, masculinity vs. femininity, and individualism vs. collectivism, which explain the variations between half of the participants. Hofstede claimed that these results are universal and stable across time since people are highly resistant to changing values learned at an early age. Nevertheless, as adults, people might slightly adapt and change their core values when exposed to opposing ones (Gould, 2005). A fifth dimension called "time orientation" (also called Confucian dynamism) was added with the help of Michael Bond (Hofstede & Bond, 1988). Bond suggested adding this value using the Chinese Value Survey, which was developed by scholars from Taiwan and Hong Kong in the early 1980s. This fifth dimension was added after



analysis in order to address Hofstede's lack of explanation for Asian values. Hofstede's dimensions were used as a basis for many studies because of their simplicity, precision, and strong explanation of cultural differences and because of the size and scope of the distributed surveys; yet there were many critics on the theory (Gould, 2005). Gould (2005) summarised the critiques of Hofstede's dimensions, which covered the following three issues:

1.  IBM company employees do not represent a national culture. Moreover, IBM had a very conservative organisational culture when the survey was conducted, affecting the culture of its employees. Hofstede argued that although the IBM sample may not represent a national culture, it could be used to identify national cultural values. He explained that focussing on one company would enable a functionality equivalence sample. Hofstede added that concentrating on the distinct organisational culture of IBM would give employees so much in common that it would be easy to identify different national cultural traits, resulting in the identification of these dimensions.

2.  Cultural changes and the scores Hofstede provided were not stable since he finished his research in 1972. Hofstede (2010) rejected this criticism, emphasising that time is not an influencing factor especially that childhood acculturation is stable and difficult to change and that people adapt only superficially when faced with other cultural values.

3.  Hofstede's survey was biased since it was in English and tested in Europe and the USA, colouring the results with Western values. Hofstede admitted that this criticism is valid, especially in view of his model's omission of the people of China; however, this limitation was addressed by his work with Michael Bond in developing Confucian dynamism.

Hofstede's cultural model was initially adopted in this thesis as a representation of culture due to its stability and completeness in explaining the effect of culture; however, this decision was later reconsidered. As the focus of this study is individual-level acceptance intentions, Hofstede's National Cultural Dimensions are not applicable. As Hofstede himself posited on his official website (see www.geerthofstede.nl):



> The Hofstede dimensions of culture are group-level constructs. Dimensions of national culture are about societies; dimensions of organizational culture [are] about organizational units. Neither is about individual differences between members of society or organizations. Comparing survey responses between individuals does not yield similar patterns to the cross-population comparisons on which the Hofstede dimensions are based.

This clearly indicates that the examination of cultural impact on acceptance intention at the individual level is not possible using Hofstede's framework. Furthermore, Hofstede (2001a) emphasised that National Culture Dimensions varies and discriminates when nations are compared, but not for individuals, which makes Hofstede's model conceptually unsuitable for analysis of acceptance at the individual level. Furthermore, Hofstede's dimensions refer to work values and do not measure human values, which are related to many dimensions of life (Schwartz, 2003).

Unlike Hofstede's dimensions, Schwartz's theory of Basic Human Values explains culture at both the individual and national level and focuses on human values rather than work values. Schwartz's theory is therefore more suitable for this study's attempt to explain how culture influences individual-level decisions regarding acceptance. Schwartz's theory of Basic Human Values is discussed in the following section.

## 4.4    Theory of Basic Human Values (BHV)

Schwartz's Basic Human Values (BHV) theory was initially generated by Schwartz and Bilsky (1987, 1990) to recognise and categorise values that are guided by principles common across different cultures. Schwartz asserted that these values are universal and were developed to include all core values known in cultures around the globe (Schwartz, 1992, 2003; Schwartz & Bilsky, 1990). These identified values vary in their importance and how they motivate behaviour for different individuals and cultures. The theory also defines the structure and the relation between the comprehensive core values (Schwartz et al., 2001). Schwartz and others



continuously validated the theory until he reached a clear view of a universal typology of values (Schwartz, 1992, 1994b; Schwartz, et al., 2001). The BHV theory is deeply rooted and has been extensively used in the social psychology discipline. Schwartz's BHV is considered a comprehensive theory that is able to explain individual and national values (Alkindi, 2009). Schwartz (1994b) differentiated between Basic Personal Values (BPV) and national-level analysis, an important distinction that made BPV the most suitable theory for the goals of this thesis since cultural values can be analysed at the individual level.

To quantitatively capture these values, many instruments were developed, which are discussed later in section 4.4.1. Empirical validations of this cultural theory were conducted by collecting 210 samples from 67 countries, resulting in approximately 65,000 respondents (Schwartz, 2003).

Schwartz has adopted the following view of values: (1) they are beliefs or concepts; (2) they indicate a behaviour or relate to a desirable end state; (3) they transcend specific circumstances; (4) they direct criteria for evaluation and selection of behaviours and actions; and (5) they are hierarchically ordered, based on their importance (Schwartz, 1992, 1994b). Each value type was developed based on three universal requirements of human existence: the biological needs of individuals, the fundamentals of social interaction, and the group's welfare and survival needs. For example, the value of conformity is derived from the social interaction of the group, which requires the restrain of desires and behaviours that could harm others within the group (Schwartz, 1992). These values are listed and the motivational goals that underline each value are described in Table 4.1. Describing the central motivational goal of each basic value enables its definition and classification. Furthermore, each single value item that bears a representation of the core values is included in parentheses in the table below. These items represent actions which lead to the achievement of the goal of each basic value (Schwartz, 2003).



Table 4.1

Motivational Goals for Each Value *(Adopted from Schwartz & Boehnke, 2004)*

| Power (P) | Social status and prestige, control or dominance over people and resources *(social power, authority, wealth, preserving my public image).* |
|---|---|
| Achievement (A) | Personal success through demonstrating competence according to social standards *(successful, capable, ambitious, influential).* |
| Hedonism (H) | Pleasure and sensuous gratification for oneself *(pleasure, enjoying life).* |
| Stimulation (ST) | Excitement, novelty, and challenge in life *(daring, a varied life, an exciting life).* |
| Self-direction (SD) | Independent thought and action choosing, creating, exploring (creativity, freedom, independent, curious, choosing own goals). |
| Universalism (U) | Understanding, appreciation, tolerance, and protection for the welfare of all people and for nature (broad-minded, wisdom, social justice, equality, a world at peace, a world of beauty, unity with nature, protecting the environment). |
| Benevolence (B) | Preservation and enhancement of the welfare of people with whom one is in frequent personal contact *(helpful, honest, forgiving, loyal, responsible).* |
| Tradition (T) | Respect, commitment, and acceptance of the customs and ideas that traditional culture or religion provide the self *(humble, accepting my portion in life, devout, respect for tradition, moderate).* |
| Conformity (C) | Restraint of actions, inclinations, and impulses likely to upset or harm others and violate social expectations or norms *(politeness, obedient, self-discipline, honouring parents and elders).* |
| Security (SE) | Safety, harmony, and stability of society, of relationships, and of self (family security, national security, social order, clean, reciprocation of favours). |

*Note.* From. "Evaluating the structure of human values with confirmatory factor analysis" From S. H. Schwartz and K. Boehnke, 2004, *Journal of Research in Personality*, 38(3), p. 239. Copyright 2004 by Elsevier Inc. [adopted] with permission.

In addition to the classification of these values, Schwartz's Basic Personal Values theory claims that relationships between values can be conflicting or congruent. An individual's performance of an action motivated by a value can have social, psychological, and practical implications. These consequences can conflict or concur with the realisation of another value in that person's value system (Schwartz, 2003, 2006). For example, actions associated with the achievement values may lead to a conflict with tradition or the attainment of benevolence values. Conversely, seeking achievement values is compatible with power values. Schwartz (2003) explained this potential of conflict using an example in which the search for personal success might hinder efforts aimed at helping others, whereas seeking personal success can be enhanced by aiming for authority.



Thus, individual-level values are grouped into higher-order values as follows: self-enhancement (power, achievement, and hedonism), openness to change (hedonism, stimulation, and self-direction), self-transcendence (universalism and benevolence), and conservation (tradition, conformity, and security). Furthermore, Schwartz grouped the values of self-enhancement and openness to change as an orientation toward individualism and self-transcendence, and conservation as an orientation toward collectivism (Schwartz, 1992, 1994a, 1994b). Schwartz (1994b, 1999) analysed data at the individual case level and organised the personal values with the structure described in Table 4.2.

Table 4.2
*Structure of Schwartz's Basic Personal Values*

| Personal Values | Higher Order Values | Orientation |
|---|---|---|
| Power | | |
| Achievement | Self-Enhancement | |
| Hedonism | | *Individualism* |
| Hedonism | | |
| Stimulation | Openness to Change | |
| Self-direction | | |
| Universalism | Self-Transcendence | |
| Benevolence | | *Collectivism* |
| Tradition | | |
| Conformity | Conservation | |
| Security | | |

Analysis of the national level has empirically resulted in different national dimensions. These dimensions include seven values, as follows: conservatism, hierarchy, intellectual autonomy, affective autonomy, mastery/competency, harmony, and egalitarianism. Schwartz's national-level cultural values is not used in this study as the unit of analysis is individuals, not nations (Schwartz, 1994b).

The circular arrangement which depicts concurrency and conflict between the values of this theory can be clearly viewed in Figure 4.2 below. In the figure, the closer one value is to another, the more congruent their goals and underlying motives. Conversely, the wider the gap is between two values, the more opposed is their underlying motivation. Also, hedonism is



placed between openness to change and self-enhancement in the figure, as it has elements of

both (Schwartz, 2003).

Figure 4.2
*Values' circular structure (adapted from Schwartz, 2003).*

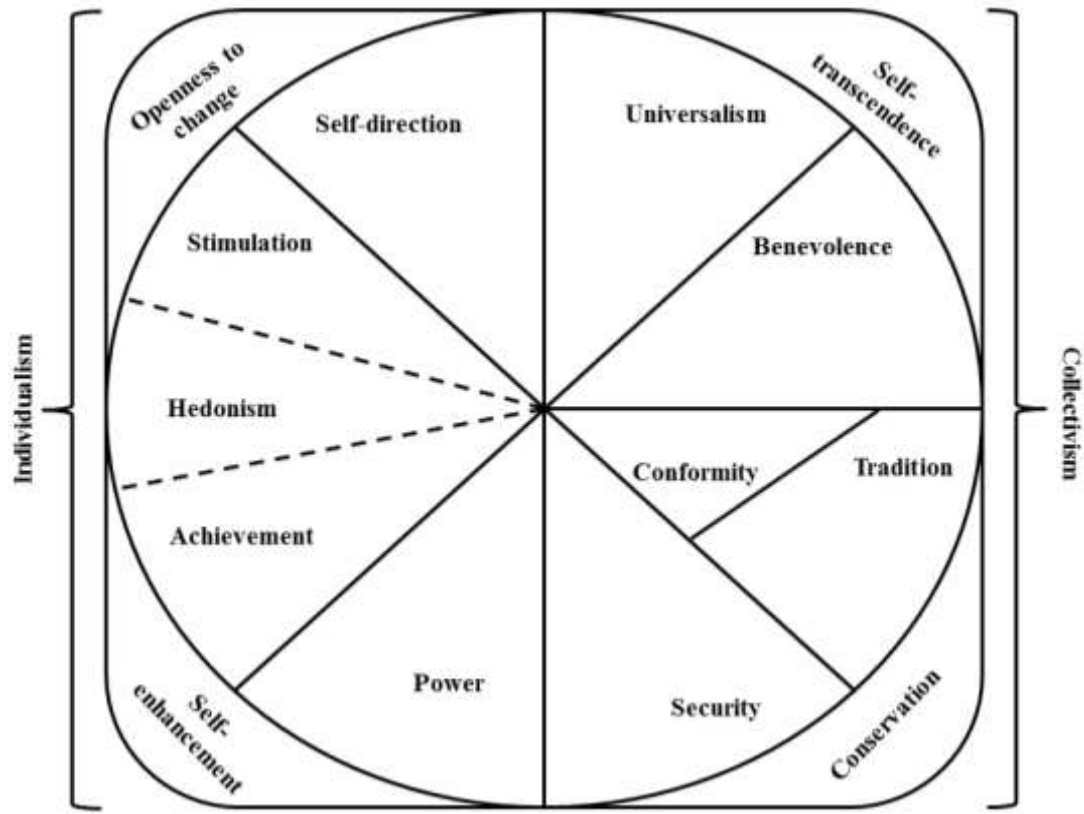

Schwartz (2003) posited that these 10 values are comprehensive and that any additional

measurement item would have a high correlation with the current structure of these values,

indicating that additions to the current value structure would be redundant.

The BHV theory has been operationalised many times since it was first introduced by

Schwartz and Bilsky (1987, 1990). Many instruments have been developed to capture the values

identified by this theory. These instruments are discussed in the following section.



### 4.4.1    *BHV Instruments*

Among the various instruments used to measure Schwartz's 10 values, the Portrait Values Questionnaire (PVQ) was the one selected for this study (Schwartz, 2003). Although a shorter version of this questionnaire is available (used by the European Social Survey), the comprehensive PVQ was adopted for this study due to the reported lack of discriminant validity of the 10 values when the shorter version is used (Knoppen & Saris, 2009). The PVQ includes 40 items, each with a 6-point Likert scale, in which the respondent identifies how closely each statement reflects to her/his values. No study was found that used PVQ as an instrument leading to an explanation of culture within Saudi society.

Many other surveys are available to assess respondents' values using the BHV theory, including but not limited to the following: the Schwartz Value Survey (SVS), the Short Schwartz's Value Survey (SSVS), which is an improvement on the SVS survey, PVQ, and the shortened version of PVQ used for the purposes of the European Social Survey (ESS). SVS is the instrument that Schwartz initially developed to capture human value theory at both the national and individual level. It included 56 items, each containing a scale followed by a statement to enable the respondents to rate a value that is important in life (Schwartz, 2003). Evaluating each statement (item) requires self-consciousness with respect to values and a high level of abstract thinking (Schwartz, et al., 2001). Due to the length of SVS and the difficulties that participants with low or no education might face in completing it, PVQ was produced as a replacement. These methods, as Schwartz argued, assess only the level of agreement without capturing real conviction. To overcome this issue, PVQ provides a statement that includes a description of an individual, and participants are required to assess their similarity to that individual on a scale of six choices: "very much like me," "like me," "somewhat like me," "a little like me," "not like me," "not like me at all." These scales are selected based on statements similar to the following: "Thinking up new ideas and being creative is important. He/She likes to do things in his/her own original way." The questionnaire starts with a declaration: "Here people are briefly described. Please read each description and think about how much each



person is or is not like you. Choose what shows how much the person in the description is like you." Appendix G contains the full questionnaire and Figure 4.3 gives an example of how items are designed in this study instrument.

Figure 4.3
*An example of the PVQ online questionnaire used in this study.*

The short version of PVQ, called the ESS scale, included 21 of the 40 original items. PVQ's 40 items were reduced to overcome space limitations in the European Social Survey. However, Davidov (2008) and Knoppen and Saris (2009) empirically concluded that the 21-item ESS is not able to capture all the distinctive elements of the BPV theory. Thus, this survey was also considered inappropriate for this research.

Although BHV is theoretically rich, critics have argued that student and teacher samples might be representative only of a given country. Furthermore, Arab samples used in Schwartz's (2006) research on Arab Israelis might not be representative of the larger Arab population. One study by Alkindi (2009) on the influence of values on management practices and styles in Oman used BHV as an explanation of culture and implemented three phases using SVS and PVQ. In the first phase, SVS and PVQ were included in one survey to collect a sample of 511 Omani participants. Phase two included the other part of the study, which analysed the dependent variables that captured managerial practices for 287 managers in Oman. The final phase concluded that values had a significant impact on the managers' actions and performance.



Alkindi's study resulted in many findings; however, the finding relevant to this thesis is identification of the cultural values for individuals in Oman using the BHV theoretical model. The results showed that Omani individuals have a stronger value priority for self-transcendence and conservation values and a weaker priority for self-enhancement and openness to change (Alkindi, 2009).

## 4.5    Summary

Cultural values and various models of culture were presented in this chapter. Through the discussion, it was found that Schwartz's Human Values theory (specifically BPV) enables the study of the influence of values at the individual level. An approach using BPV has been adopted for this study as researches have argued that technology adoption behaviour is best explained at the individual level of culture. Furthermore, Portrait Values Questionnaire (PVQ) was considered the most suitable instrument for this study because of its structure and simplicity and because it was the most accurate of the relevant examples of surveys developed to capture culture. The next chapter outlines how the theories discussed in this and previous chapter have been used to develop the research model.



# 5     RESEARCH MODEL

Chapters 2, 3, and 4 discussed and reviewed relevant literature. This chapter continues by building the research model and hypotheses. The relationship between the three main components of cultural values, e-transactions, and their acceptance by the Saudi public is empirically investigated later using the research model and research hypotheses. Perceived characteristics of e-government are selected based on their relevance in the context of the Saudi culture. The research model incorporates empirically validated and frequently cited cultural and acceptance models found in the literature according to their relevance to e-government and Saudi culture. Selected elements are chosen mainly from the Perceived Characteristics of Innovation (PCI). The cultural model is based on Schwartz's Basic Personal Values (BPV).

As this study is focused on the demand side of e-transaction acceptance, the model and hypotheses is synthesised accordingly. This chapter discusses and justifies the selection of specific constructs and their inclusion in the research model. Furthermore, the development of the research hypotheses is explicated. After developing the research model from the literature, it was established further with the assistance of other academics who have published research related to this study. The utilisation of feedback from academics enabled supporting the research model appropriateness for the goals of this thesis (see Appendix A).

The purpose of this study is to understand the influence of culture on e-government acceptance. According to Kumar et al. (2007), the average number of citizens who adopted e-government initiatives globally was only 30%. Even though Kumar's study dates back to 2007, this number is expected to be low within emerging economies generally and the KSA even in 2012 (UNDESA, 2012). This is attributable to many factors, including lack of infrastructural, educational, economic, social, and cultural readiness for electronic services. Creating a model that involves the cultural factors most relevant to these countries is important in predicting and investigating technology acceptance in voluntary situations.



Transferring technology from its nation of origin to another nation involves infusing the technology and its related cultural methods to the hosting culture. Transferred technology is usually suitable for and biased on the socio-cultural systems of its creators, which is why challenges to acceptance arise in the hosting nations (Straub, et al., 2003). Zakaria et al. (2003) argued that the cultural background of the hosting environment is usually not considered, causing delays, difficulties or failure in the process of implementing and accepting of technology. Thus, cultural factors are considered an important determinant of the intended usage of e-government services in many emerging economies, including the KSA (AlAwadhi & Morris, 2008; Baker, Al-Gahtani, & Hubona, 2010). Understanding the relevance of culture for e-transaction acceptance enabled viewing the phenomenon of acceptance from a perspective lacking in the literature, and many cultural factors are considered in detail to address low e-government acceptance and usage. This cultural perspective also enabled a better understanding of why e-government is being accepted or rejected by individuals because of their personal assumptions and perceptions on technology. A generic view of the relationship is shown below.

Figure 5.1 shows the theoretical research framework that proposes a relationship between acceptance of e-transactions and Basic Personal Values (BPV) and Perceived Characteristics of E-Transactions (PCET).



Figure 5.1
*Theoretical framework.*

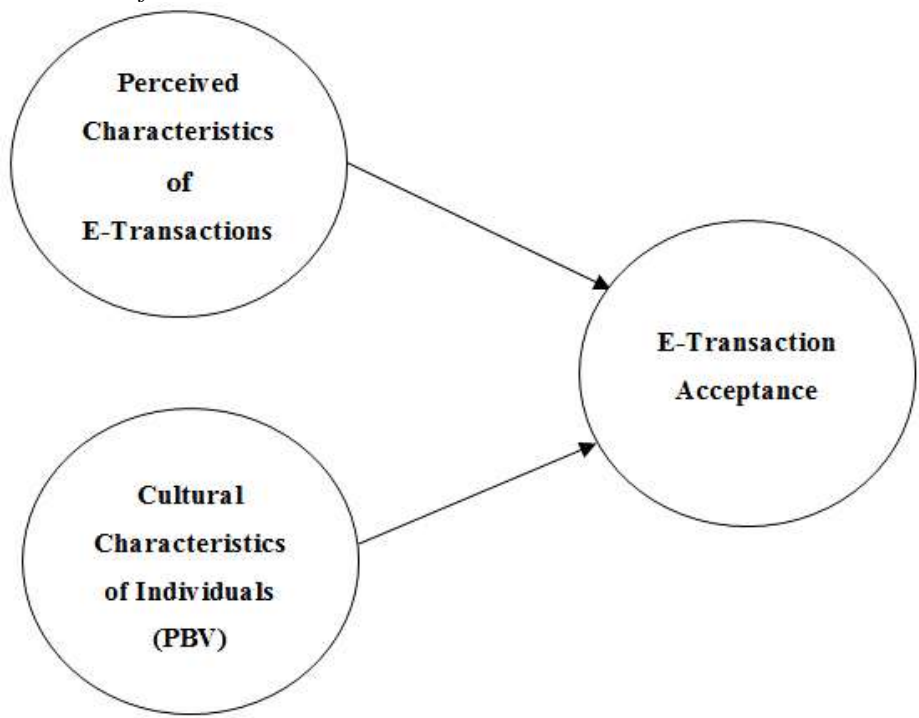

This framework was used to develop the research model. In the next section, the PCET model is described. The following section discusses the perceived characteristics of e-transactions model which includes perceptions constructs, social influence and the perspective of communication constructs.

## 5.1    Perceived Characteristics of e-Transaction (PCET)

A set of constructs has been developed based on extant literature to aid in the development of a conceptual model that enabled the understanding of e-transaction acceptance in the KSA. This set of constructs provides a preliminary framework for the development of the research instrument. This framework, PCET, is used to develop a research model that explains and predicts e-transaction acceptance. PCET is principally based on Moore and Benbasat's (1991) PCI model, which has relevance to e-government transaction acceptance in the KSA as discussed in this section.

Whetten's (1989) approach of balancing and choosing theoretical factors is used to synthesis different model into the creation of PCET. Whetten (1989) posited that researchers:



"should err in favor of including too many factors, recognizing that over time their ideas will be refined" (p. 490). Whetten (1989, 2002) emphasised that while comprehensiveness is important when selecting all factors relevant to the context, irrelevant factors need to be eliminated as well. It is therefore important to note that the research model used in this study applies only to the specific context of the KSA, as recommended by Seddon and Scheepers (2006) for information systems (IS) quantitative models.

Moore and Benbasat's (1991) PCI model was originally developed to "tap respondents' reactions in an 'initial adoption' environment where the individual acceptance decision is voluntary" (p. 194), which is also the case in this study. Moreover, PCI was chosen as this study focuses on understanding the differences between adopters and non-adopters. Further justification is that PCI was developed in a theoretically rich approach and was tested rigorously which should enrich the explanation of e-transaction acceptance (Moore & Benbasat, 1991; Plouffe, Hulland, & Vandenbosch, 2001).

The PCI model was developed to study individual-level acceptance decisions, which is also the focus of this study; nevertheless, the PCI instrument measures were applied at the organisational setting (Moore & Benbasat, 1991). Therefore, many procedures were conducted to alter the definitions and measures used in the original study (Moore & Benbasat, 1991). These procedures are discussed in the next chapter. The following table presents a list of constructs borrowed from the relevant research. The constructs were redefined so that they fit better into the present research context. Table 5.1 describes how each construct is related to the study focus, and indicates the source of the instrument items which were reworded.



Table 5.1
*Constructs and Relevance to the Research Context*

| PCI Constructs (Moore & Benbasat, 1991) | Original Definition | PCET Definition | Relevance to this context |
|---|---|---|---|
| *Relative advantage* Initially, five items were adopted from Carter and Bélanger (2005) wording altered slightly in accordance with the research context. | Perceptions that the new e-government system accomplishes a task more effectively or efficiently than the current system (Carter & Bélanger, 2005). | The degree to which using e-government transactions is perceived to be better than traditional methods. | Relative advantage construct is important for this study, as it would be able to capture whether e-transactions is perceived preferable or not to traditional methods. |
| *Compatibility* Initially, four items were adopted from Carter and Bélanger (2005) wording altered slightly in accordance with the research context. | The degree to which e-government usage is perceived seamless natural and compatible with needs (Carter & Bélanger, 2005). | The degree to which using e-transactions is consistent with the potential adopter's needs, past experiences and values. | It would be expected that compatibility would have a strong influence on acceptance. Citizens might find that e-transaction usage is also incompatible with their values (AL-Shehry, et al., 2006). |
| *Complexity* Initially, four items were adapted from Thompson, Higgins and Howell (1991), altered later in accordance with the research context. | "the degree to which an innovation is perceived as relatively difficulty to understand and use" (Rogers, 2003, p. 257). | The degree to which using e-transactions is perceived as being relatively difficulty to understand and use. | How e-transactions are perceived in terms of complexity might determine acceptance. Complexity was used as it was expected that the negative terms found in the complexity items would be more representative of Saudi users than ease of use. |
| *Result demonstrability* Initially, four items were adapted from Moore and Benbasat (1991), altered later in accordance with the research context. | "the tangibility of the results of using the innovation, including their observability and communication" (Moore & Benbasat, 1991, p. 194). | Communicability of the results of using e-transactions. | Communication between members of Saudi society on the results of using e-transactions might have an influence on transacting intentions. |
| **Acceptance of e-Government (Carter & Bélanger, 2005)** | **Original Definition** | **PCET Definition** | **Relevance to this context** |
| *Intention to use e-transactions* Initially, five items were adopted from Carter and Bélanger (2005) wording altered slightly in accordance with the research | Intention to decide to use e-government public services (Carter & Bélanger, 2005). | Intention to decide engagement with SGA using transactions available on the Internet. Citizen engagement includes usage of | These items enable capturing the essential components of e-transaction acceptance; the use of online services and inquiring, gathering and providing information (Carter & |



| context. | | service, sharing, acquisition and gathering information. | Bélanger, 2005; Pavlou, 2003). |
|---|---|---|---|
| *Trust in government agencies* Initially, four items were adopted from Carter and Bélanger (2005), later altered slightly in accordance with the research context. | "trust in the state government agency providing the service" (Lee & Turban, 2001, as cited in Carter & Bélanger, 2005, p. 10). | The perceptions of trust in the provider of e-transactions, the SGA. | The decision to engage in e-government transactions requires citizens' trust in the government agency providing the service (Lee & Turban, 2001, as cited in Carter & Bélanger, 2005). |
| *Trust in the Internet* Initially, three items were adopted from Carter and Bélanger (2005), later altered slightly in accordance with the research context. | Reliability and security of the media which e-government services are being provided, the Internet (Carter & Bélanger, 2005). | Perceptions of security and reliability of the means which e-transactions are being delivered, the Internet. | It is expected that acceptance of e-transactions would be difficult if citizens' trust in the Internet is not high. |
| **UTAUT (Venkatesh, et al., 2003)** | **Original Definition** | **PCET Definition** | **Relevance to this context** |
| *Social influence* Initially, three items were adapted from Venkatesh et al. (2003), altered later in accordance with the research context. | "the degree to which an individual perceives that important others believe he or she should use the new system" (Venkatesh, et al., 2003, p. 451). | The degree to which an individual perceives that important others believe e-transactions should be used. | Saudi society is a collective culture where individuals affect the opinions of others; citizens, therefore, might influence others when relating to acceptance intention (Al-Gahtani, et al., 2007). |
| **Accounting Information System Acceptance Model (Aoun, et al., 2010)** | **Original Definition** | **PCET Definition** | **Relevance to this context** |
| *Perspective on communication* Initially, three items were adopted from the main authors of Aoun et al. (2010). These items were altered in accordance with the research context. | The degree to which high or low context cultures prefer using Accounting Information Systems to communicate with business stakeholders (Aoun, et al., 2010). | The degree to which using e-government transactions would enable adequate communication with government agencies. | Usage of e-transactions can be a form of communication between the government and the citizens. In this context, it might be significant to understand whether e-transactions would be able to sustain an adequate level of communication that citizens might seek. |

Moore and Benbasat (1991, as cited in Plouffe et al., 2001, p. 210) defined visibility as

"the degree to which an innovation is visible during its diffusion through a user community."



Image was defined as "the degree to which use of innovation is perceived to enhance one's image or status in one's social system" (Moore & Benbasat, 1991, p. 195). Both constructs were excluded from this framework as usage of e-transactions is conducted at home, meaning that usage is not socially visible. For example, the image perceptions of e-transactions users would not be enhanced as their usage of electronic government transactions is usually unobserved by others. With regard to image, there are additional reasons for exclusion. Firstly, the social influence construct is more relevant to this study and was developed to theoretically replace image construct (Venkatesh, et al., 2003). Secondly, Carter & Bélanger (2005) have empirically found that image is not significant in determining e-government usage intention. Thus, image was excluded from the research model.

Trialability was defined as "the degree to which an innovation may be experimented with before adoption" (Moore & Benbasat, 1991, p. 195). Carter and Bélanger (2005) asserted that trialability is irrelevant to e-government adoption and would not provide enough explanatory power regarding its acceptance. Many e-government researchers (Alzahrani, 2011; Shajari & Ismail, 2010; Tung & Rieck, 2005) have therefore excluded trialability from their studies. As this argument is also applicable to the present study, the construct of trialability was excluded.

Considering the high uncertainty avoidance of the Saudi culture, it was expected that Rogers' (2003) construct of complexity would be more reflective of the sample than Moore and Benbasat's (1991) ease of use. Additionally, many studies did not find a significant relationship between ease of usage and e-government acceptance (Alomari, Woods, et al., 2009; Carter & Bélanger, 2004b; Gilbert, et al., 2004). The suitability of complexity (rather than ease of use) was also determined as a result of interviews with academics who have publications related to this study. These academics were consulted at an early stage of the present research (see Appendix A). Some of the constructs previously described in Table 5.1 were adopted from backgrounds that might not necessarily be related to the present research context. These constructs, namely result demonstrability and perspective on communication, were not found in



previous e-government adoption research models. Table 5.1 contains definitions of each construct as shown in the column titled "PCET definitions." The research model and hypothesis are discussed in the following section.

## 5.2    Research Model and Hypothesis

The research model integrates the previously discussed PCET and Schwartz's 10 personal values which enabled attending the research questions. The resulting model was integrated from various well-established models to enable deeply examining the case of Saudi culture and e-transactions technology. Thus, the model is considered as a significant contribution in the explanation of the influence of culture over technology acceptance (e-transaction) within a developing country context (Saudi Arabia). Figure 5.2 below illustrate the research model which is mainly based on Schwartz's Basic Personal Values (BPV) and Moore's and Benbasat's (1991) PCI. This model has extended PCI by incorporating cultural values and including trust, communication and social influence. The first research question 'How do perceived characteristics of e-transactions affect e-transaction acceptance?' is answered by examining the PCI elements (PCET) of the research model and their relation with the acceptance of e-government. The second question 'How does trust in the Internet and government agencies influence acceptance?' is be answered by examining the hypothesised connection between trust in the government and the Internet and e-transaction intention of usage. The third question 'How does the social influence of existing e-transaction users affect the acceptance of e-transactions?' is investigated by examining the relationship between such e-transaction acceptance and social influence. The fourth research question 'How does using e-transactions as a communication method affect acceptance of e-transactions?' is addressed by studying the association between the association between perspective on communication and acceptance of e-transactions. The fifth research question 'How do cultural values influence the acceptance of e-transactions?' is determined by examining the relation between Schwartz's BPV and transacting intention. The following figure illustrates the research model.



Figure 5.2
*Research model.*

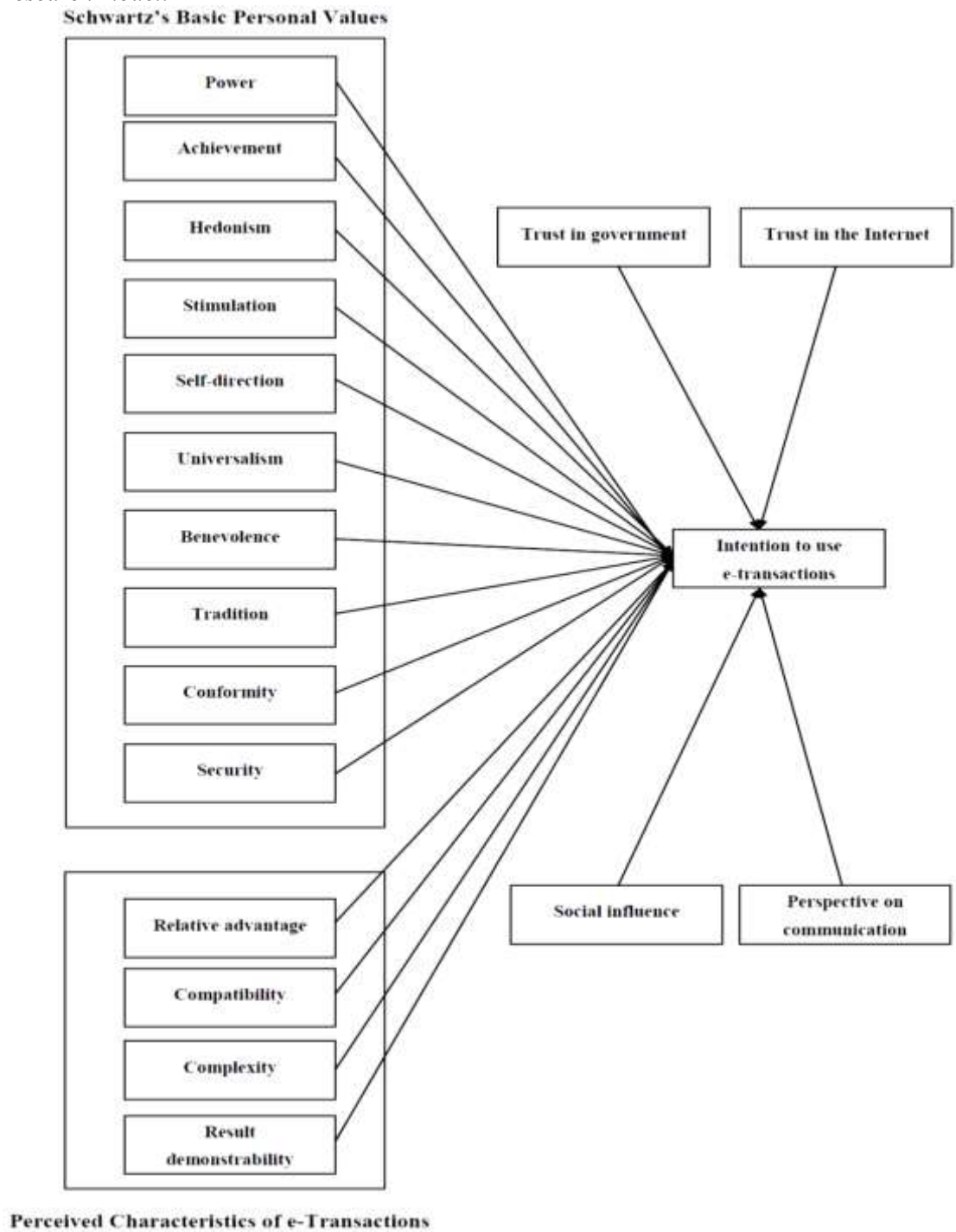

As shown in Figure 5.2 above, it was expected that factors pointing to intention to use e-transactions would be influential in the acceptance of e-transactions (usage intentions). The following sections describe how these factors play a role in the acceptance of e-transactions in the KSA. It is noted that no study used Schwartz's values to explain culture in the KSA. Hence,



research (Alshaya, 2002; Bjerke & Al-Meer, 1993) that measured the Saudi culture using Hofstede's dimensions are used a basis for explaining the research hypotheses. The following section starts with a discussion of the research hypothesis.

### 5.2.1   *Relative Advantage, Compatibility and Intention to Use e-Transactions*

It might be important to consider the perceived relative advantages for Saudi citizens when intending to use e-transactions where current social traditional methods might be preferred. Additionally, it was expected that the higher the levels of compatibility of e-transactions with cultural needs, values, and previous experiences the more higher the acceptance of e-transactions would be. The researcher took into consideration that many citizens have created many social methods in acquiring government services which was expected to hinder intention to use e-transactions (Al-Solbi & Mayhew, 2005).

Al-Gahtani et al. (2007) suggested that acceptance of technology will be hindered if a given technology clashes with the individual's affinity for certain cultural values. In conservative cultures such as Saudi Arabia, technologies developed in Western cultures are subjected to a process of selection. Technologies that best suit the adopting culture are selected based on the cultural values of that society. However, this is only one of the cultural implications of the acceptance of e-government, which requires further investigation (Baker, et al., 2010).

Collectivism would explain how the cultural norm of *wasta* (use of personal acquaintances or family members to acquire a favour or service) might affect e-government acceptance. Given that Saudi citizens are considered to represent a collective society that has developed many social norms in acquiring government services (Alshaya, 2002; Bjerke & Al-Meer, 1993), one needs to investigate whether citizens consider e-transactions as a relative advantage over the use traditional methods instead e-transactions, and whether e-transactions would be considered consistent with their needs, past experiences, and values and if these perceptions actually promote acceptance. This is examined by studying the hypothesised



relationship between relative advantage and compatibility and intention to use e-transactions, as shown below:

*H1: Relative advantage has a positive significant influence on intention to use e-transactions.*

*H2: Compatibility has a positive significant influence on intention to use e-transactions.*

Table 5.2
*Hypotheses 1 and 2*

| Code | Hypothesis | Supporting references |
|------|-----------|----------------------|
| H1 | Relative advantage has a positive significant influence on intention to use e-transactions. | (Carter & Bélanger, 2004b) (Carter & Weerakkody, 2008) (Sang, Lee, & Lee, 2009) |
| H2 | Compatibility has a positive significant influence on intention to use e-transactions. | (Carter & Bélanger, 2005) (Carter & Bélanger, 2004b) (Sang, et al., 2009) |

Previous research, as shown in Table 5.2 above, has investigated these hypotheses and assured their saliency in determining intention of usage. The next section discusses the construct complexity and its relation with intention to use e-transactions.

### 5.2.2 Complexity and Intention to Use e-Transactions

The complexity of using novel technologies is a determining factor of acceptance for many people (Al-Ghatani, 2003; Mejias, Shepherd, Vogel, & Lazaneo, 1996; Rogers, 2003). e-Transactions represent one technology which is perceived as complicated and difficult to use by many citizens around the world especially if citizens are not familiar with these services (Al-Gahtani, 2011). The novelty of e-transactions for citizens is particularly the case for Saudi citizens, taking into consideration that many SGA have recently initialised their e-transactions services (some SGA initialised online services in 2010 or 2011) (Alfarraj, Drew, & AlGhamdi, 2011). Moreover, the high uncertainty avoidance of Saudi citizens raises problems concerning the perceived complexity of technology (Al-Gahtani, 2011; Al-Gahtani, et al., 2007; Alshaya, 2002; Bjerke & Al-Meer, 1993). The complexity construct is aimed at measuring how citizens perceive the difficulty of using e-transactions. As shown in Table 5.3, e-government adoption



research negatively linked usage intention with the complexity of e-transactions websites. The

following is the research hypothesis for this relationship:

*H3: Complexity has a negative significant influence on intention to use e-transactions.*

Table 5.3
*Hypothesis 3*

| Code | Hypothesis | Supporting references |
|------|------------|-----------------------|
| H3 | Complexity has a negative significant influence on intention to use e-transactions. | (Carter & Bélanger, 2005) |

Result demonstrability is a component that is not commonly used in e-government

adoption research. However, due to the collective nature of Saudi society and the high level of

information sharing after experiencing a service, it was analysed in this study (Liu, Furrer, &

Sudharshan, 2008). The next section elaborates on this issue.

### 5.2.3    Result Demonstrability and Intention to Use e-Transactions

The extent to which the benefits of e-transactions are perceived to be sharable,

communicable, tangible, or observable might influence the actual intention to conduct a

transaction (Hussein, Mohamed, Ahlan, & Mahmud, 2011). Taking these perceptions further, it

was expected that citizens would communicate the results of their transactions to determine

whether these services are useful. This is particularly important for the nurturing, collective

cultures where the sharing of positive experiences by word of mouth is a significant factor in

determining usage (Liu, et al., 2008). Although the issue of influencing others might be more

closely related to the construct of social influence than the construct result demonstrability, how

e-transactions present themselves in terms of results might be particularly central to the Saudi

case. Thus, it is expected that perceptions of result demonstrability of e-transactions will

promote intention to use e-transactions, whether these perceptions come from the individuals

themselves or from the society. This study makes the following hypothesis, taking into

consideration the support from the literature, as shown in Table 5.4 below:



*H4: Result demonstrability has a positive significant impact on intention to use e-transactions.*

Table 5.4
*Hypothesis 4*

| Code | Hypothesis | Supporting references |
|------|-----------|----------------------|
| H4 | Result demonstrability has a positive significant impact on intention to use e-transactions. | (Baumgartner & Green, 2011) (Hussein, et al., 2011) |

Taking into consideration the high conservatism of Saudi society, the link between perceptions of trustworthiness and intention is expected to be important (Gallagher & Searle, 1985). The following section discusses this in more detail.

### 5.2.4    *Trust in the Internet and Government and Intention to Use e-Transactions*

In a conservative society such as that of Saudi Arabia, trust is an important determinant of the usage of introduced technologies. Saudi citizens' concern regarding the impact of e-government on society might affect their usage of e-transactions. Citizens might not trust the Internet as a medium for e-transactions, nor the provider of service, namely government agencies (Al-Solbi & Mayhew, 2005). Perceptions of trusting the government and the Internet might affect the number of citizens who accept e-transactions as a means to acquiring services from SGA. Even though the Internet is the means by which e-transactions are conducted, citizens might not accept the use of e-transaction unless it is trusted. Trust issues with the Internet arise especially as the private information that e-transactions require might be compromised. Citizens might doubt the relevance and the reliability of the information available on e-transaction websites, especially when this information is outdated or incorrect (Pavlou, 2003). In addition, e-transactions might require users to accept and trust the provider of the service (the SGA) to be able to adequately provide the service. However, if for example the citizen's file is lost, the SGA employees might incorrectly process the transaction, or other mistakes could occur. Hence, trusting the provider of services (the SGA) might also be important for encouraging acceptance of e-transactions, taking into consideration the distance that occurs due to the impersonal and online nature of e-transactions (Carter & Bélanger, 2005).



This relationship between the intention to use e-transactions and trust in the government and Internet has been found to be empirically significant in the positive direction by research, as shown in Table 5.5. The following hypotheses address the association between trust in the Internet and government and usage intentions:

*H5: Trust in the Internet has a positive significant influence on intention to use e-transactions.*

*H6: Trust in government agencies has a positive significant influence on intention to use e-transactions.*

Table 5.5
*Hypotheses 5 and 6*

| Code | Hypothesis | Supporting references |
|------|-----------|----------------------|
| H5 | Trust in the Internet has a positive significant influence on intention to use e-transactions. | (Carter & Bélanger, 2005) (Al-Sobhi, Weerakkody, & Kamal, 2010) (Bélanger & Carter, 2008) (Carter & Bélanger, 2004a) (Carter & Weerakkody, 2008) (Alomari, Sandhu, et al., 2009) (Alhujran, 2009) |
| H6 | Trust in government agencies has a positive significant influence on intention to use e-transactions. | (Carter & Bélanger, 2005) (Al-Sobhi, et al., 2010) (Bélanger & Carter, 2008) (Carter & Weerakkody, 2008) (Alomari, Sandhu, et al., 2009) (Alhujran, 2009) (Carter & Bélanger, 2004a) |

### 5.2.5   *Social Influence and Intention to Use e-Transactions*

Loch et al. (2003, p. 46) claimed that: "the closer the affinity of the individuals with their reference group, the more likely the individuals are to perform according to reference group expectations". It would be expected that such social influence would be especially higher in a collective society such as Saudi Arabia (Alshaya, 2002; Bjerke & Al-Meer, 1993). It was expected that when influential individuals within the society would pressure others to use e-transactions, this would increase acceptance (Loch, et al., 2003). Hence, this relationship between social influence and intention to use e-transactions is depicted in Table 5.6 below, with the references that support this hypothesis.



*H7: Social influence has a positive significant impact on intention to use e-transactions.*

Table 5.6
*Hypothesis 7*

| Code | Hypothesis | Supporting references |
|------|-----------|----------------------|
| H7 | Social influence has a positive significant impact on intention to use e-transactions. | (Al-Shafi & Weerakkody, 2009) (Gefen, Warkentin, Pavlou, & Rose, 2002) |

### 5.2.6    *Perspective on Communication and Intention to Use e-Transactions*

Hakken (1991, as cited in Straub et al. (2003), argued that technology is an establishing factor of human communications and networks. Nevertheless, the online e-government environment does not allow the natural benefits of face-to-face communications (Harfouche, 2010). Since Arabic culture is a high-context one where a significant part of meaning and information is conveyed implicitly within a conversation (Hall & Hall, 1990), it is important to study how usage of e-transactions affects their acceptance as a communication tool between the government and citizens. No actual support for this construct in relation to e-government has been found in the literature. Thus, this study has introduced the perspective on communication construct into the e-government adoption research domain. However, Aoun et al. (2010) found that this construct positively influences intention of usage. Since this study focus on Arabic culture which is considered high context, it was expected that acceptance would be favoured when citizens perceive e-transactions as a suitable means of communication with the government. Thus, the following is hypothesised:

*H8: Perspective on communication has a negative significant impact on intention to use e-transactions.*

As a major part of this research, cultural values are expected to play a key role in determining acceptance. This is explained in the following section.



### 5.2.7    *Cultural Values and Intention to Use e-Transactions*

Cultural values are associated with shaping and predicting behaviour (Schwartz, 2003). Previous researchers studying Internet adoption (Dwivedi & Weerakkody, 2007) and e-commerce (Sait, et al., 2004) have posited that cultural values strongly affect the adoption of these technologies. Therefore, it was expected that cultural values influence the acceptance of e-government in the Saudi context (AL-Shehry, et al., 2006; Webber, Leganza, & Baer, 2006).

Self-enhancement values (power and achievement values), in the sense of achievement within social expectations and authority in a collectivist and tribal culture, are gained through personal connections, power is gained from friends and family and success in utilising social relations by gaining authority or prestige. Online transactions are expected to cause a disintermediation between citizens and employees which compromises the loop of favours gained through *wasta* and personal connections within SGA (AL-Shehry, et al., 2006; Alshaya, 2002; Bjerke & Al-Meer, 1993; Mackey, 2002). Therefore, the following is hypothesised:

*H9: Power has a negative significant impact on intention to use with e-transactions.*

*H10: Achievement has negative significant impact on intention to use e-transactions.*

It was also expected that pleasure seeking, which is motivated by the hedonism value, would not be related to e-transactions in the KSA, as these transactions are used only to acquire necessary services and not for enjoyment (Abu Nadi, 2010). Based on the previous discussion it is hypothesised that:

*H11: Hedonism does not have a significant impact on intention to use e-transactions.*

Openness to change values (stimulation and self-direction values), i.e. related to novelty and independence, can be associated with e-transactions in the sense that these online transactions are rapidly changing and constantly improving in the KSA. Changing from the traditional methods of contacting the government to conducting services with SGA should be



positively related to the levels of novelty and independence of the respondents. Therefore, it is hypothesised that:

*H12: Stimulation has positive significant impact on intention to use e-government.*

*H13: Self-direction has positive significant impact on intention to use e-government.*

Self-transcendence values (universalism and benevolence values) are associated with the collectivism aspects of a society should, thus, be negatively associated with the use of e-transactions. The universalism value is concerned with sustaining the welfare of others and tolerance. Because e-transactions in the KSA cause social isolation in their in its current forms and do not provide opportunities for citizens to interact with each other and enable the social aspects of this value, it was expected that universalism plays a negative role in the acceptance of e-transactions. The benevolence value is related to assistance and loyalty to others. Within this study context, practicing *wasta* within SGA is considered by many Saudis as a way of helping others and being loyal to the family and tribe. e-Government transactions disconnect this level of social interaction between citizens (Abu Nadi, 2010; Smith, et al., 2011). This leads to the following hypothesis:

*H14: Universalism has a negative significant impact on intention to use e-transactions.*

*H15: Benevolence has a negative significant impact on intention to use e-transactions.*

Conservatism and affiliation with a tribe is a Saudi characteristic that affects Saudi society in many ways (Abu Nadi, 2010; Dwivedi & Weerakkody, 2007). Therefore, it was expected that conservation values (security, tradition and conformity values) have a strong influences on acceptance in the Saudi case. The value of security has an element of the reciprocation of favours (see Table 4.1) which may be related to *wasta,* as this practice is considered a form of favour exchange within governmental agencies (Smith, et al., 2011). *Wasta* is an Arabic word that has a very similar meaning to *nepotism* in English or *Guanxi* in Chinese. Therefore, abiding by the rules of a tribal society (conformity) and in keeping with the



tradition of *wasta* (tradition and security), it is expected that conservation values are negatively associated with e-government transaction. Thus, it is hypothesised that:

*H16: Conservation values have negative significant impact on intention to use e-transactions.*

Table 5.7
*Hypotheses Regarding Cultural Values and Intention to Use e-Transactions*

| # | Hypothesis | Supporting references |
|---|---|---|
| H9 | Power has a negative significant impact on intention to use e-transactions. | (Alhujran, 2009) (Bagchi & Kirs, 2009) (Choden, Bagchi, Udo, & Kirs, 2010) |
| H10 | Achievement has a negative significant impact on intention to use e-transactions. | |
| H11 | Hedonism does not have a significant impact on intention to use e-transactions. | |
| H12 | Stimulation has a positive significant impact on intention to use e-transactions. | |
| H13 | Self-direction has a positive significant impact on intention to use e-transactions. | |
| H14 | Universalism has a negative significant impact on intention to use e-transactions. | |
| H15 | H15: Benevolence has a negative significant impact on intention to use e-transactions. | |
| H16 | H1: Conservation values have negative impact on intention to use e-transactions. | |

BPV theoretical elements are captured using PVQ, which was added to the instrument to capture the cultural characteristics of individuals. Utilisation of PVQ and the research instrument is discussed in the subsequent chapter.

## 5.3    Summary

This chapter explained the development of the research model and hypotheses. The research model utilised elements from PCI, UTAUT and DOI and from Carter and Bélanger's (2005) e-government adoption model. Development of the model has followed Whetten's (1989, 2002) recommendation of conciseness and comprehensiveness. Thus, the constructs relevant to the research context are relative advantage, compatibility, complexity, result demonstrability, trust in the Internet, trust in government agencies, perspective on communication, and Schwartz's BPV. On the other hand, the constructs (found in PCI) not



related to the context were excluded, namely trialability, image, visibility, and ease of use (which was replaced with its opposite, complexity). Additionally, this chapter outlined the hypothesis development and the postulated significance and direction of influence for each construct on intention to use e-transactions. This study emphasis on culture and contextualisation to technology or e-transaction acceptance especially for developing countries such as the KSA was not found in the literature (Abu Nadi, 2010; Baker, et al., 2010; Udo & Bagchi, 2011). The following chapter explains the operationalisation of the research model and the methodology used to collect the data.



# 6    RESEARCH METHODOLOGY

The literature review in the previous chapters has enabled the development of the research model. The research instrument enables the empirical capturing of the latent concepts represented in the research model. The study's main goal is to explore the influences of culture on the acceptance of e-government within the KSA. To achieve this goal, methodological conventions from the field of social science and information systems are utilised to: describe this study's philosophical assumptions, its justification as valid, development of the research instrument by customising or contextualising the instrument to meet the research scope and the design of the research methods appropriate for achieving the objectives and goals of this study (Gregor, 2006). To ascertain its suitability for the goals of this thesis as discussed in Appendix A, the research methodology was further reviewed and investigated after being designed.

The following sections of this chapter present the research paradigm, research design, instrument development, sampling techniques, initial test of the developed instrument (pre-test and pilot study), design of the full-scale study, and relevant ethical considerations.

## 6.1    Research Paradigm

Describing the philosophical position underlying the research is essential as it directs and justifies the research activities (Creswell, 2009). Firstly, terms associated with this section are described to clear the way for a discussion of the philosophical stance that guides this thesis. Simply put, the research paradigm is basically the worldview adopted by the study (Creswell, 2009; Tashakkori & Teddlie, 1998). According to Neuman (2006), a research paradigm is the general philosophy for research that includes key issues, approaches to seeking answers, and basic assumptions underlying the research methods. Kuhn (1970), who is primarily associated with the term, defined a research paradigm as "a set of values and techniques which is shared by members of a scientific community, which acts as a guide or map, dictating the kinds of problems scientists should address and the types of explanations that are acceptable to them" (p.



175). Of the multiple paradigms that guide research, one is positivism or scientific research. Positivism is an epistemological position that proposes that objective reality exists and thus can be numerically measured and described independently of the researcher's and instrument's biases (Crotty, 1998; Neuman, 2006; Tashakkori & Teddlie, 1998). This study adopts the soft-positivism paradigm which is similar to what is explained by Seddon and Scheepers (2006). Seddon and Scheepers (2006) argued that this position on positivism allows capturing objective reality with caution in the context of different environments. Thus, the ontological stance of this research is that objective reality exists beyond the human mind, but that how it is perceived depends on culture and life experiences. Epistemologically, this reality can be captured empirically; however, acquired knowledge is context-bound by culture, time, and circumstances (Jupp, 2006; Seddon & Scheepers, 2006). Positivism examines causal relationships between objects within the world, which is important in answering certain research questions, in this case those involving the relationship between culture and technology acceptance (D. Byrne, 2002; Seddon & Scheepers, 2006). The 'soft' position on positivism also overcomes the rigid stance of extreme positivism by looking differently at different situations where generalisation is not always applicable (Seddon & Scheepers, 2006).

Accordingly, it was expected in this study that current research and theories on e-government acceptance suit the situation and the context they were designed to address. Hence, to enable a better understanding of Saudi e-government acceptance, this study delved deeper into the Saudi context by contextualising the research model and design. For instance, the original PCI model did not include trust, which is an important component of online transactions; consequently, this addition was essential for the context of e-transactions (Carter & Bélanger, 2005). Furthermore, the Saudi society's high-context culture considers the details which surround the communication and the communicator. For this reason, the perspective on communication construct was introduced to the final research model to take into consideration the Saudi cultural context (Hall, 2000; Hofstede & Hofstede, 2005). An example relevant to the research design is the contextualisation of the instrument's questions (which were originally



derived from other research) with the help of Saudi citizens and others who have a research background relevant to this study. Therefore, soft positivism was considered to be appropriate to guide this research. The following section discusses the research design.

## 6.2    Research Design

The research design underpinning this thesis is drawn from soft-positivism paradigm (Seddon & Scheepers, 2006) . Thus, this study has adopted a quantitative cross-sectional survey to enable capturing and studying the influence of perceived characteristics of e-transactions and cultural values on acceptance by studying a sample of a larger population. Perceptions, values (independent constructs), and acceptance (dependent construct) can be quantitatively measured using a survey design (Fink, 1995). Measured or collected data enables hypothesis testing through statistical analysis, resulting in determination of the causality between independent and dependent constructs (Neuman, 2006). Ultimately, the sample can provide a good representation of the population under consideration. Findings from the sample can be generalised to the population. The study population was Saudi citizens who have Internet access, and the sample was acquired through four email newsgroups. Although four email newsgroups were used, the survey was still considered cross-sectional and not longitudinal because each participant was approached only once for the full-scale study (Creswell, 2009). The study population and sample are described in more detail in section 7.1 and section 7.2. The sampling technique was not considered completely random. However, the usage of general topic email groups would interest a random and wide section of the society. Van Selm and Jankowski (2006) identified email newsgroups as a method for reaching samples using the Internet. General topic email newsgroups discuss and share information about any political, economic, social, and environmental news updates related to the KSA; such newsgroups are of special interest to a wide section of the society with a general interest in KSA-related news and need to use Saudi e-transactions. Other studies have found that samples collected through email newsgroups can be considered a very close representation of the studied online populations but not of offline populations (Andrews, Nonnecke, & Preece, 2003; Preece, Nonnecke, & Andrews, 2004; Teo &



Pok, 2003). Thus, an email newsgroup was considered suitable for use in this study (this point is discussed in more detail in section 6.2.5).

This study is designed as follows. Firstly, constructs and related items (questions) were determined for the instrument and then translated. Secondly, these initial items were reviewed and pre-tested by a sample of nine Saudi participants for clarity and accuracy of intended meaning. At this same stage, the participants tested the usability of Qualtrics.com as an online questionnaire tool. The third phase was to ascertain the content validity of contextualised items and constructs. This study applied Lewis et al.'s (1995) questionnaire development and content validity procedures (described later in section 6.2.3). Fourthly, the resulting instrument was pilot tested with 113 participants; feedback was collected and feasible recommendations were adopted. Finally, the full-scale questionnaire was sent to the sampling frame (100,000 online users). The following section discusses the questionnaire development and translation.

### 6.2.1    *Phase One: Questionnaire Development and Translation*

Measuring the influence of culture on e-transaction acceptance would be difficult if the sampled citizens did not understand the online environment. Those who have Internet access would be more experienced in the usage of online transactions and would be closer to representing current or potential adopters of e-transactions (Van Selm & Jankowski, 2006). To make sure that the sampled Saudi citizens have Internet access and hence the ability to use or to have used e-transactions, the questionnaire was distributed using online software. Furthermore, the usage of an online sampling method would save time and money in comparison to traditional data collection methods (such as telephone or postal mail) and enable wide geographical access to current or potential e-transaction adopters in a large country such as the KSA (Sheehan & McMillan, 1999; Van Selm & Jankowski, 2006). All of the instrument's items (questions) were originally published in English and were adopted in this thesis, with the exception of the demographic questions. However, since Arabic is the language spoken by most if not all Saudi citizens, a translation was needed (Vassiliev, 1998).



Thus, an online questionnaire was developed using Qualtrics.com software, and a modified version of back-to-back translation was conducted to enable greater clarity and accuracy in the English-to-Arabic translation (Douglas & Craig, 2007; Triandis, 1972). Items from previous research were adopted and slightly reworded in accordance with the research context. Items adopted from the literature are presented in Table 6.1 below.

Table 6.1
*Original Items from Previous Research for PCET*

| Construct | Items |
|---|---|
| ***Relative advantage*** (Carter & Bélanger, 2005) .<br><br>VA TAX (Virgina Department of Taxation). | - Using the web would enhance my efficiency in gathering information from the VA TAX.<br>- Using the web would enhance my efficiency in interacting with the VA TAX.<br>- Using the web would not make it easier to gather information from the VA TAX.<br>- Using the web would make it easier to interact with the VA TAX.<br>- Using the web would give me greater control over my interaction with the VA TAX. |
| ***Compatibility*** (Carter & Bélanger, 2005) . | - I think using the web would fit well with the way that I like to gather information from the VA TAX.<br>- I think using the web would fit well with the way that I like to interact with the VA TAX.<br>- Using the web to interact with the VA TAX would fit into my lifestyle.<br>- Using the web to interact with the VA TAX would be incompatible with how I like to do things. |
| ***Complexity*** Thompson, Higgins & Howell (1991). | - Using the system takes too much time from my normal duties.<br>- Working with the system is so complicated, it is difficult to understand what is going on.<br>- Using the system involves too much time doing mechanical operations (e.g., data input).<br>- It takes too long to learn how to use the system to make it worth the effort. |
| ***Result demonstrability*** (Moore & Benbasat, 1991).<br><br>PWS (Personal Work Station). | - I would have no difficulty in telling others about the results of using a PWS.<br>- I believe I could communicate to others the consequences of using a PWS.<br>- The results of using a PWS are apparent to me.<br>- I would have difficulty in explaining why using a PWS may or may not be beneficial. |
| ***Social influence*** (Venkatesh, et al., 2003). | - People who influence my behaviour think that I should use the system.<br>- People who are important to me think that I should use the system.<br>- The senior management of this business has been helpful in the use of the system. |
| ***Intention to use e-*** | - I would use the web for gathering information from VA TAX. |
| ***transactions*** (Carter & Bélanger, 2005). | - I would use VA TAX services provided over the web. <br> - Interacting with the VA TAX over the web is something that I would do. <br> - I would not hesitate to provide information to the VA TAX website. <br> - I would use the web to inquire about VA TAX services. |
| ***Trust in government agencies*** (Carter & Bélanger, 2005). | - I think I can trust the VA TAX. <br> - The VA TAX can be trusted to carry out online transactions faithfully. <br> - In my opinion, VA TAX is trustworthy. <br> - I trust VA TAX to keep my best interests in mind. |
| ***Trust in the Internet*** (Carter & Bélanger, 2005). | - The internet has enough safeguards to make me feel comfortable using it to interact with the VA TAX online. <br> - I feel assured that legal and technological structures adequately protect me from problems on the internet. <br> - In general, the internet is now a robust and safe environment in which to transact with the VA TAX. |
| ***Perspective on communication*** (Aoun, et al., 2010). | - My ability to communicate is enhanced when using accounting information systems. <br> - Communications through the systems enhance my ability to interpret business issues. <br> - Textual, verbal and visual information is important for business communication. |

These items were reworded according to the present research context and were used to develop the first draft of the English survey. Nevertheless, there was a need for translation into Arabic as it was expected that many participants would not fully understand the English version.

The process of translating the Perceived Characteristics of E-Transactions (PCET) and Portrait Values Questionnaire (PVQ) instruments was similar to Triandis' (1972) "back-to-back" translation methods, with additions made to increase and ascertain the translation's accuracy, as illustrated in Figure 6.1. This research included two necessary steps: assessing the accuracy of translation and amending the translation where necessary with the help of Saudi participants. These steps were conducted for both the PCET and PVQ instruments. The difference between the translations of the two instruments is that the PCET instrument was translated in full as no previous translation was found in the literature, whereas an Arabic translation of PVQ was found in Alkindi (2009). This translation was assessed and amended where necessary, translated using the back-to-back method, and then reassessed and amended



where necessary. Both PCET and PVQ translations were examined in the pre-test stage, and

minor changes to item words were made.

Figure 6.1
*PCET and PVQ instrument translation process.*

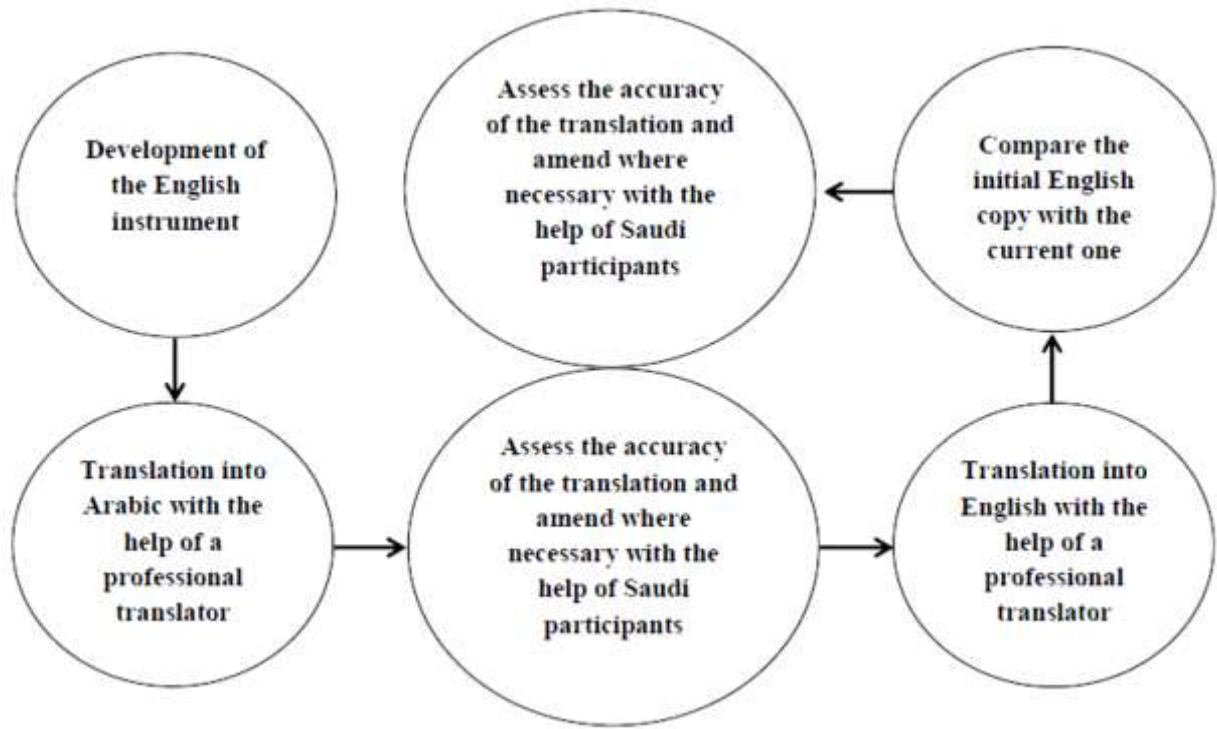

The back-to-back translation method has been widely used in research to check the

accuracy of instrument translations (Douglas & Craig, 2007). The back translation process starts

with translating to the required language, translating the words back to the original language,

and then comparing the two translations and checking for discrepancies and correcting them

(Triandis, 1972). This method helps in evaluating the accuracy of translations. However, if

participants from pre-tests were not included in this translation evaluation, the targeted

participants might not comprehend the terms being used as the pre-test participants were highly

educated and had experience with the Saudi government and e-transactions which enabled

simplifying terms used in the questionnaire (Douglas & Craig, 2007). Hence, this research has

included participants from the pre-test phase (described below) to assist in assessing the

accuracy and comprehension of the translations. As shown in Figure 6.1 above, the first

English-to-Arabic translation was assessed for accuracy and comprehension. Then, the final



draft was checked again for its clarity and accuracy. This level of participants' involvement in the development of the questionnaire is not included in the well-known back-to-back translation, which is further explained below. The following section describes the pre-test phase in which the instrument translation and usability was tested for the first time.

### 6.2.2    Phase Two: Pre-Tests for PCET and PVQ

The second phase included repeated interviews (totalling 14), conducted with only nine Saudi citizens (native Arabic speakers) who had been using e-transactions for at least two years and had at least three years of experience in acquiring services from the government by using traditional methods. All of those met were between 25 and 45 years of age and were highly educated. They were specialists in IS, IT, engineering, and business, which provided a range of perspectives and opinions. These participants were chosen because they were highly educated, native Arabic speakers, and experienced in using e-transactions, which would enable them to detect impreciseness in the meaning of the survey questions, identify technical issues with the online questionnaire, or perceive any problems with the instrument design. Therefore, the purpose of this phase was to ascertain the accuracy and understandability of the translated items. Furthermore, the usability and accessibility of online tools were noted during the meeting, and any difficulties of usage were recorded. After each meeting, suggested changes were made directly to the instrument, and any difficulty of usage related to the design of the questionnaire was dealt with. Aggregately, the following changes were applied:

- requiring a response to all questions;
- preventing participants from taking the survey more than once by changing ballot options in the Qualtrics.com software;
- placing pictures and examples illustrating e-transactions in the invitation email;
- improving the usability of the online survey;
- including 10 questions on each page, explaining the number of questions, and
- sequentially numbering the questions.

After the first English-to-Arabic translation was conducted (as illustrated in the bigger circle in Figure 6.1 above), four out of the nine participants reviewed the translation, and minor



changes to the wording of the Arabic translation item were made to clarify meaning. Then, the final copy was checked for clarity, the accuracy of translation was checked by the five pre-test participants, and minor changes were made to the Arabic instrument.

This phase was important to ascertain that the questions would be comprehensible to participants and that the Qualtrics.com online tool was easy to use. The final form of the instrument was utilised in both the English and Arabic versions to ascertain content validity and contextualisation was conducted appropriately, as described in the next stage.

### 6.2.3    *Phase Three: Content Validity of the Instrument*

The third stage was conducted to guarantee the content validity of the PCET constructs and items. Validating the content of a construct means ensuring that the instrument questions (items) represent their corresponding construct (Cronbach, 1971; Lewis, et al., 1995; Straub, Boudreau, & Gefen, 2004). Hence, content validity was not required for PVQ items which were adopted without rewording. However, to guarantee that reworded items represent the PCET constructs, content validity procedures were conducted at this stage (Straub, et al., 2004).

Content validity procedures were not conducted for PVQ items at this phase. The English and Arabic versions of the questionnaire items were adopted completely from Schwartz's (2003) and Alkindi's (2009) questionnaires, respectively. However, some minor rewordings were made in the Arabic translation, as discussed previously in the pre-test phase. The validity and reliability of PVQ was confirmed by previous research (Schwartz, 2003; Schwartz, et al., 2001; Smith, Peterson, & Schwartz, 2002). This phase continued from the previous phases on the development of PCET instrument.

After implementing the recommended changes from the previous stages, a description of the research context, construct definitions and items were emailed to 40 'highly published' academics (more than 10 publications related to the research focus) in the field of IS acceptance, This includes Saudi academics who published work in e-government acceptance (Palvia, Pinjani, & Sibley, 2007). Of these, 17 responded, including 10 Saudis. These academics were



emailed rather than met in person as they were from different countries around the world; therefore, meeting them in person was not an option due to financial and time limitations. These experts were probed, firstly, to read the research description (see Appendix C), secondly, to comment on the definitions and the instrument, and finally, to assess the content validity of the items (see Appendix D).

The method used at this stage is similar to what Straub et al. (2004, p. 387) described as "[a] good example of content validation." This was originally implemented by Lewis et al. (1995), where panellists compared the definition of each construct with the items and scored them using a scale from 1 to 3 (not relevant, important, essential). However, as an addition to Lewis et al. (1995), the panellists for this study were asked to insert any comments or suggestions that would improve the wording of the items of the constructs by making them more relevant to the context.

Data received from the three-point scale was used to compute the Content Validity Ratio (CVR) for all the items using the following formula (Lawshe, 1975, p. 567):

$$CVR = (n - \frac{N}{2})/(\frac{N}{2})$$

In the above equation, n is the number of panellists who rated the items as either "important" or "essential," while N is the total number of panellists. This equation enables measuring the percentage of panellists who indicated that the item is "important" or "essential" to the construct. Responses of "important" and "essential" were considered positive indicators of the items' relevance to the PCET constructs (Lewis, et al., 1995). Table 6.2 below includes both the average response for each item and the calculated CVR as well as the number of panellists who evaluated every item. The CVR for each item was tested for statistical significance at 0.05 (Lawshe, 1975). Significance at this level (0.05) meant that more than 50% of the panellists rated the items as "essential" or "important" (Lawshe, 1975; Lewis, et al.,



1995). Lawshe (1975) argued that when more than half the panellists indicate that an item is relevant to the construct, the item has some content validity.

Table 6.2
*Calculated Mean and CVR for Instrument Items*

| Item code | Average | CVR | Number of panellist |
|---|---|---|---|
| *Relative advantage (RA)* | | | |
| RA1 | 2.39 | 0.89 | 19 |
| RA2 | 2.16 | 0.79 | 19 |
| RA3 | 2.37 | 0.79 | 19 |
| RA4 | 2.24 | 1.00 | 17 |
| RA5 | 1.79 | 0.45 | 18 |
| *Compatibility (CT)* | | | |
| CT1 | 2.50 | 0.78 | 18 |
| CT2 | 2.35 | 0.88 | 17 |
| CT3 | 2.29 | 0.78 | 18 |
| CT4 | 2.17 | 0.79 | 18 |
| *Complexity (CMX)* | | | |
| CMX1 | 2.16 | 0.79 | 19 |
| CMX2 | 2.47 | 1.00 | 17 |
| CMX3 | 2.29 | 0.88 | 17 |
| CMX4 | 2.22 | 0.89 | 18 |
| *Result demonstrability (RED)* | | | |
| RED1 | 2.33 | 0.89 | 18 |
| RED2 | 2.12 | 0.65 | 17 |
| RED3 | 2.22 | 0.78 | 18 |
| RED4 | 2.12 | 0.65 | 17 |
| *Perspective on communication (POC)* | | | |
| POC1 | 2.22 | 0.89 | 18 |
| POC2 | 2.00 | 0.78 | 18 |
| POC3 | 2.29 | 0.78 | 18 |
| Intention to use e-transactions (USE) | | | |
| USE1 | 2.17 | 0.78 | 18 |
| USE2 | 2.00 | 0.56 | 18 |
| USE3 | 2.12 | 0.53 | 17 |
| USE4 | 1.89 | 0.56 | 18 |
| USE5 | 2.17 | 0.79 | 18 |
| *Trust in the Internet (TI)* | | | |
| TI1 | 2.61 | 0.89 | 18 |
| TI2 | 2.50 | 0.89 | 18 |
| TI3 | 2.50 | 0.89 | 18 |
| *Trust in government agencies(TG)* | | | |
| TG1 | 2.50 | 1.00 | 18 |
| TG2 | 2.59 | 0.88 | 17 |
| TG3 | 2.28 | 0.67 | 18 |
| TG4 | 2.29 | 0.76 | 17 |
| *Social influence (SI)* | | | |
| SI1 | 2.35 | 0.88 | 17 |
| SI2 | 2.24 | 0.65 | 17 |



| SI3 | 2.24 | 0.88 | 17 |

*Note.* Appendix D has the questionnaire items that were used in this phase

The average for each item (as shown in Table 6.2 above) indicates that the panellists considered all the items (except for two, RA5 and USE4) important for their corresponding constructs for the calculated mean. Furthermore, all items were significant at 0.05, and the two disputed items (RA5 and USE4) have an average mean very close to 2 (Lawshe, 1975). Therefore, no items were excluded, and the instrument was considered to have acceptable content validity.

### 6.2.4    *Phase Four: Pilot Study*

Before sending the questionnaire to a large group of participants, a pilot study was conducted on 113 participants from mid-December 2010 to mid-January 2011. This questionnaire was sent to one Saudi general topic email newsgroup. The total listed emails were 20,000; however, only 0.57% of the participants fully completed the questionnaire. The general topic email newsgroup included a wide sector of the society, which enabled a diversity of opinions and comments on the questionnaire. Conducting a pilot study at this stage provided feedback on the response rate and was an initial test for the reliability of items, test usability, and the comprehensibility on the instrument for a large group of participants.

The participants were provided with a space to comment on the questions and instrument design, usability, and understandability. Moreover, these tests enabled identifying any missing options from the demographic questions (e.g., "retiree" was added to the occupation question), and some minor changes were made to the wording of the questions (in the Arabic version). Another important addition of this phase was to include an open-ended question at the beginning of the instrument that sought to identify how citizens deal with SGA. Traditional methods (transacting with the government without using the Internet) identified by the participants were listed and included as check boxes, including an open selection for other suggestions for the next phase of the full-scale study. According to the pilot study participants



the following traditional methods are identified: face-to-face interviews with government officials; with the help of a relative or friend; mail; phone; fax; and with the services of a paid agent. In general, participants were satisfied with the clarity of the questions and the questionnaire design; however, some complained about the questionnaire length.

Cronbach's alpha reliability test was conducted as all adopted items and constructs for this research were considered 'reflective' (Cronbach, 1990). A reflective construct and items can be distinguished when a change (e.g., increase) in the construct causes or reflects a change in its items (Jarvis, Mackenzie, & Podsakoff, 2003). Therefore, reflective items are expected to be consistent to enable measurement of the construct they it represents. Cronbach's alpha provides this assessment of consistency and is therefore used here (Jarvis, et al., 2003). Table 6.3 below presents alpha reliabilities for all the constructs responded to by the 113 participants, which indicate acceptable levels for all constructs, except result demonstrability. The forth item of the construct result demonstrability (RED4) was excluded from the instrument, and the reliability of result demonstrability was improved as shown in Table 6.3 below. PCET items' wordings are shown in Appendix O.

Table 6.3
*Pilot Study Reliabilities for Reflective Items*

| Construct Scale | Items | Cronbach's Alpha | Construct Reliability Status |
|---|---|---|---|
| Relative advantage | 5 | 0.91 | Excellent |
| Compatibility | 4 | 0.91 | Excellent |
| Complexity | 4 | 0.74 | Acceptable |
| Result demonstrability | 4 | 0.48 | Unacceptable |
| Result demonstrability after RA4 is removed | 3 | 0.79 | Acceptable |
| Trust in the Internet | 3 | 0.85 | Good |
| Trust in government agencies | 4 | 0.95 | Good |
| Social influence | 3 | 0.88 | Good |
| Perspective on communication | 3 | 0.80 | Acceptable |
| Intention to use e-transactions | 5 | 0.90 | Excellent |
| Self-direction | 4 | 0.56 | Acceptable |
| Power | 3 | 0.53 | Acceptable |
| Universalism | 6 | 0.66 | Acceptable |
| Achievement | 4 | 0.66 | Acceptable |
| Security | 5 | 0.66 | Acceptable |
| Stimulation | 3 | 0.63 | Acceptable |
| Conformity | 4 | 0.55 | Acceptable |



| Tradition | 4 | 0.51 | Acceptable |
|-----------|---|------|------------|
| Hedonism | 3 | 0.71 | Acceptable |
| Benevolence | 4 | 0.67 | Acceptable |

Cronbach's alpha values above 0.9 were given the status "excellent" in the in Table 6.3 above (Creswell, 2009). Alpha reliabilities above 0.80 were considered "good," while those below this value were considered acceptable. However, alpha values less than 0.50 were considered unacceptable, which was the case for result demonstrability (Nunnally, 1967). Therefore, the only change to the instrument after the reliability test was to remove one item (RED4). After this test, minor changes were made to the Arabic wording of the instrument and other demographic options were added. The questionnaire was considered ready for the full-scale study, which is discussed in the following section. The English version of the questionnaire is shown in Appendix G and the Arabic version is shown in Appendix H.

### 6.2.5   *Phase Five: Full-Scale Study*

For the full-scale study (which took place from late January to February 2011), the improved questionnaire was sent by email using Qualtrics.com to four Saudi general topic email newsgroups. Only four large newsgroups (each containing approximately 25,000 users) were identified by the researcher as general topic email newsgroups. These emails reached approximately 100,000 participants who were considered to be the sample frame. These email newsgroups were used because they discuss topics from various areas of interest and should therefore represent a broad cross-section of Saudi society. A second email was sent as reminder after three weeks. This reminder encouraged both citizens and non-citizens to participate in the survey to permit the identification and exclusion of non-citizens from the current study. The response rate was 2.31% (2,308 participants); only 0.78% of the 100,000 contacted respondents completed the questionnaire (775 participants, including non-Saudis). The response rate was calculated using the following formula (Neuman, 2006):

$$Response\ Rate = \frac{Total\ Number\ of\ Responses}{Total\ number\ of\ participants\ in\ the\ sampling\ frame}$$



Response rate improved in comparison to the pilot study since the following procedures were conducted following Rogelberg and Luoung's (1998; 2001) recommendations:

- Sending a reminder letter.
- Encouraging potential participants by explaining that results of this questionnaire indirectly improve e-transaction services.
- Explaining that any future publications will be shared with participants.
- Activating vision- and hearing-impaired accessibility options in Qualtrics.com software.
- Enabling participants to continue the questionnaire later by saving results.
- Enabling compatibility of the questionnaire with many operating systems (e.g., Microsoft Windows and Macintosh operating systems), including mobile phones.

Nevertheless, a low response rate was expected because of the length of the questionnaire (88 items, including demographic questions). Comments on the usability of the questionnaire were positive, except for very few notifications regarding the wording and the way in which the instrument was designed. Many participants complained about its length. Furthermore, a topic related to e-government is usually of low interest to respondents (Dwivedi, Papazafeiropoulou, Gharavi, & Khoumbati, 2006). Thus for the purposes of measuring the influence of perceived characteristics of e-transactions and cultural values on acceptance, the large sample (775 participants) was considered acceptable.

### 6.2.6    *Handling Common Method Bias in the Pilot and Full-Scale Studies*

The questionnaire has implemented psychological and temporal separation between the independent and dependent constructs. This was done by locating PVQ between PCET constructs and the only dependent construct, intention to use e-transactions, in the instrument. PVQ consists of 40 item questions and is very different from PCET as it asks about values rather than e-transaction. This would create a psychological separation as many respondents in the pilot and full-scale study mentioned that PVQ questions were very different from the other questions and that these questions sometimes diverted their attention from the purpose of this



survey. Additionally, PVQ items required 40 minutes to 1 hour to be completed. This design

enabled psychological and temporal separation by keeping the participants from mentally

linking PCET and transaction intention. Therefore, this reduced the bias that might arise from

answering transaction intention items immediately after reading PCET items (Podsakoff, 2003).

The previously described questionnaire design was implemented for the pilot and the full-scale

study which were discussed previously.

## 6.3    Study Population and Sample

The study's original purpose was to measure e-transaction acceptance by Saudi citizens

who have Internet access. This is especially important since e-transactions services are available

for Saudis within or outside the KSA. In the first half of 2011 (date which the questionnaire was

distributed), about 12.5 million (or 44%) of Saudi residents (including non-Saudis had Internet

access (Communications and Information Technology Commission, 2011). An exact number of

Saudis both within and outside the KSA who use the Internet was not found. Because this study

focused only on Saudi citizens to enable measurement of the extent to which Saudi culture

influence acceptance, the inclusion of other nationalities would not lead to an accurate

measurement of Saudi cultural values. Therefore, all non-Saudi respondents were excluded from

analysis. From the sampling frame of 100,000 (expected to be mostly Saudi citizens), a total of

2,308 participants responded to the survey, including non-Saudis and those who did not fully

complete the questionnaire. Of these, 674 Saudi and 101 non-Saudi citizens completed the

survey.

This study does not claim that the study's sample represents all members of the Saudi

online population. Nonetheless, the researcher sought to gain as representative sample as

possible of Saudi Internet users using all the available ethical methods. As there is difficulty in

acquiring a representative sample from the online population, the author submitted the

questionnaire to general topic email newsgroups. Usage of multiple email newsgroups was

necessary to enable as much representation as possible. The author stopped the online survey

after five weeks, when the deadline given in the questionnaire was reached where the response



rate started to become very weak. This method might not have guaranteed a completely representative sample of the Internet population; yet it might have led to a close level of representativeness, especially after collecting a large sample of more than 600 participants and making a concerted effort to reach as many online users as possible. Furthermore, this method was considered after the researcher contacted ICT government agencies, Saudi Telecom Company, and the company responsible for the Yellow Pages guide to seek representative email lists, phone numbers, mobile phone numbers, and postal addresses. However, these lists were inaccessible. Furthermore, the researcher pursued citizens' contact databases from the White Pages, but these were also inaccessible.

The sampling method was considered optimal for this study, considering the difficulties in accessing government address databases or citizens' phone numbers. Therefore, the best possible method for this study was to directly reach Saudi citizens where they gather online to discuss or share information on non-specialised topics. Hence, it is expected that this sample is an approximate representation of the Saudi online population. Web 2.0 tools or social networking websites were not used in the sampling for the full-scale study and analysis since these methods would make sampling technique closer to snow-ball non-probability sampling rather than probably sampling (Brickman-Bhutta, 2008).

To ascertain that all participants had an equal chance of receiving the questionnaire, a participation email (see Appendix E for the English version and Appendix F for the Arabic version) and another reminder was sent after two weeks. The required sample size was calculated using the following statistical equation:

$$ n = p(1-p)(\frac{z}{E})^2 $$

In the equation, n is the sample size, p is the expected proportion sample, z is the confidence level and E is the margin of error. The calculated sample size from the equation is 384. This means that 384 participants were needed achieve 95% confidence in the results (p =



0.5 and z = 1.96), with an error margin of 0.05. The research aimed at and acquired a higher number (674), which enabled stronger validity and reliability of the results.

All the phases of the study were conducted in accordance with ethical considerations, which are described in the following section.

**6.4    Ethical Conduct**

This study was conducted in accordance with the Australian National Statement on Ethical Conduct in Human Research, which Griffith University has adopted as a set of guidelines for the ethical conduct of research. The Griffith University Human Research Ethics Committee (HREC) has approved the data collection methods of this research. The approval protocol number was ICT/04/09/HREC. The certificate of approval is in Appendix B. This approval implies the following for this study:

- Research importance: This study provides knowledge for the improvement of e-government transaction which is a method that simplifies acquiring services from the KSA government for citizens.
- Integrity in research conduct: The researcher reported truthfully to and from the research participants to maintain the originality of knowledge.
- Respect for participants: Non-disclosure of private and personal identified information by participants, not causing any kind of psychological and physical harm, ensuring that the participants are fully informed, ensuring that there is no form of coercion used, avoiding any phrases or words that are insulting in any way in the questionnaire.
- Fair treatment of participants: The questionnaire questions were the same for all participants of the study population (Saudi citizens using the Internet).
- Care for participants: There are always risks; however, they were very low in this study. Care was taken, for example, to avoid any questions which might stress participants.
- Consent from participants to for their inclusion in this study: No data was recorded without the participants' consent. Informed consent was implied when participants returned a completed survey as described in the questionnaire introduction (see Appendix G) (National Health and Medical Research Council, Australian Research Council, & Australian Vice-Chancellors' Committee, 2007).



**6.5    Summary**

        Based on the extant literature related to the acceptance of e-transactions, a set of

constructs was developed. These construct were integrated into a questionnaire design for the

development, contextualisation and validation process, which was described in detail. It was

noted that, the framework and instrument are applicable only to the KSA cultural setting and an

e-transaction context. The following chapter details the demographics and data analysis of the

full-scale study.



# 7    DATA EXAMINATION AND PREPARATION

It is important to provide an overview of the data to determine the extent to which the sampled data represents the study population, detect and manage outliers (extreme responses) by data screening, and test the data assumptions, all of which are required by parametric analysis techniques such as SEM (Hair, Black, Babin, & Anderson, 2010). Descriptive data analysis, data screening, and parametric data assumptions assessment were conducted using SPSS Version 19.0 (Pallant, 2011).

The following section provides an overview of the Saudi population to enable a comparison with the sampled data. This is followed by the preparation of the data, a process conducted by detecting and removing outliers. Finally, data assumptions are assessed to determine the suitability of the data analysis method.

## 7.1    Study Population

At this stage, it is essential to compare the demographics of the research sample with the corresponding Saudi population to determine the level of the sample's representativeness. According to the latest census which is available as an online report from the Saudi Central Department of Statistics and Information (2010), the Saudi citizens' population in the KSA is about 18.7 million, with 49.1% females. Other demographic information, such as education and age, was difficult to acquire from reliable sources. However, demographic information about the study population, Saudi citizens who have access to the Internet, is more relevant to this study. In 2011, the number of Internet users in the KSA (Saudis and non-Saudis) was approximately 44% of the population or 12.5 million (Communications and Information Technology Commission, 2011). Saudi Internet users are expected to be demographically different from the overall Saudi population. However, no official information was found online about Saudi Internet users. Application was made to the relevant government departments for current demographic information on Saudi Internet users (information was provided on 6 March 2012).



Information about the study population is presented in Table 7.1 and Table 7.2. The comparison shows that the sample's demographic characteristics are comparable to the study population. Demographic information about the sample is discussed in the following section.

## 7.2    Demographic Information on the Sample

As shown in Table 7.1 below, more males than females completed the questionnaire. The imbalance between females and males was recognised by the researcher as the pilot study also had a lower number of responses from females. Thus, the invitation letter for the full-scale study encouraged both females and males to participate in the study (see Appendix E). In comparison to the study population (Table 7.1), the sample's imbalance in the number of males and females might be related to the responses given by many females who were invited to participate in the study. They stated that they were not interested in e-transactions because their spouse, father, or brother would undertake such transactions for them. Other demographic information such as age, education, and employment showed a very close resemblance of the sample to the study population. One of the few exceptions was related to the percentage of Saudi citizens who were 17 or under in the study population (19%), which was higher than the percentage of those who participated in the questionnaire (3.71%), as it might not interest them. Furthermore, the percentage of public sector employees who participated in the questionnaire (42.43%) was higher than the percentage of such employees in the study population (9%). On the other hand, the percentage of private sector employees who participated in the questionnaire (14.54%) was lower than the percentage of such employees in the study population (42%). This may relate to the curiosity public sector employees might have about such a questionnaire, since it is about services they provide. Table 7.2 also shows that Internet and e-transaction usage in the study population is similar to the sample. However, the sample had more participants who frequently used the Internet and e-transactions.



Table 7.1

*Demographic Information about the Participants in the Full-Scale Study*

| Information | | Number of participants | Percentage in sample | Percentage in study population |
|---|---|---|---|---|
| Gender | Female | 159 | 23.59% | 42% |
| | Male | 515 | 76.41% | 58% |
| | Total number of participants | 674 | 100% | 100% |
| Age | 17 year or under | 25 | 3.71% | 19% |
| | Between 18 and 22 years | 102 | 15.13% | 16% |
| | Between 23 and 30 years | 251 | 37.24% | 32% |
| | Between 31 and 40 years | 176 | 26.11% | 19% |
| | Between 41 and 50 years | 92 | 13.65% | 8% |
| | Between 51 and 60 years | 25 | 3.71% | 5% |
| | 60 years or over | 3 | 0.45% | 1% |
| | Total number of participants | 674 | 100% | 100% |
| Education | No formal education | 1 | 0.15% | 2% |
| | Primary or secondary school education | 32 | 4.75% | 4% |
| | High school | 162 | 24.04% | 17% |
| | Technical or professional degree (No Bachelor degree) | 125 | 18.55% | 19% |
| | Bachelor degree | 239 | 35.46% | 43% |
| | Graduate certificate | 42 | 6.23% | 6% |
| | Master's degree | 57 | 8.46% | 6% |
| | Doctorate or higher | 16 | 2.37% | 3% |
| | Total number of participants | 674 | 100% | 100% |
| Employment status | Not employed and not a student | 76 | 11.28% | 12% |
| | Student | 153 | 22.70% | 28% |
| | Government sector employee | 286 | 42.43% | 9% |
| | Private sector employee | 98 | 14.54% | 42% |
| | Freelancer | 44 | 6.53% | 6% |
| | Other | 17 | 2.52% | 3% |
| | Total selections | 674 | 100% | 100% |
| Current country of residence | Australia | 29 | 4.30% | Information not available |



| | | | |
|---|---|---|---|
| | Canada | 3 | 0.45% | |
| | Egypt | 1 | 0.15% | |
| | Malaysia | 1 | 0.15% | |
| | New Zealand | 2 | 0.30% | |
| | Saudi Arabia | 628 | 93.18% | |
| | United Kingdom | 2 | 0.30% | |
| | United States | 8 | 1.19% | |
| | Total number of participants | 674 | 100% | |

As shown in Table 7.1 above, most of the participants (63.3%) were between 18 and 40 years of age, and this was expected for online users in the KSA. Nevertheless, a substantial percentage (17.81%) of the respondents was older than 41 years of age. Most of the participants completed at least high school education (78.05%), with a lower percentage (17.06%) having a postgraduate education. Government employees comprised 42.43% of the sample, and the second largest group was made up of students (22.70%). Finally, most of the participants were living in the KSA when they completed this questionnaire and had been born in the KSA (98.2%). Place of residence and birth provides an indirect indication that the participants were brought up with Saudi cultural values and that the sample is therefore an accurate representation of Saudi culture (Hofstede & Hofstede, 2005).

Table 7.2 below shows that most participants were daily users of the Internet (60.24%). Furthermore, 68.70% of the participants consider themselves either excellent or very good Internet users, which indicate high confidence in Internet usage. Most importantly, most of the participants (68.40%) had used e-transactions before completing the questionnaire. In addition, most had used e-transactions recently and frequently, implying that participants had up-to-date experience of such transactions.



Table 7.2

*Internet and e-Transactions Usage Demographic Information*

| Information | | Number of participants | Percentage in sample | Percentage in study population |
|---|---|---|---|---|
| Frequency of Internet usage | Few times a year | 1 | 0.15% | 6% |
| | Few times a month | 6 | 4.75% | 5% |
| | Few times a week | 35 | 24.04% | 6% |
| | Once a day only | 37 | 18.55% | 19% |
| | Few hours a day | 263 | 35.46% | 51% |
| | Many hours a day | 332 | 6.23% | 13% |
| | Total number of participants | 674 | 100% | 100% |
| Proficiency in the use of the Internet | Very low | 2 | 0.30% | Information not available |
| | Low | 3 | 0.45% | |
| | Satisfactory | 56 | 8.31% | |
| | Good | 150 | 22.26% | |
| | Very good | 258 | 38.28% | |
| | Excellent | 205 | 30.42% | |
| | Total number of participants | 674 | 100% | |
| Conducted e-transactions | Yes | 461 | 68.40% | 58% |
| | No | 213 | 31.60% | 42% |
| | Total number of participants | 674 | 100% | 100% |
| Last time to conduct e-transaction (percentages for 461 participants only) | Three years ago | 14 | 3.04% | Information not available |
| | Last year | 45 | 9.76% | |
| | This year (2011–2010) | 22 | 4.77% | |
| | Within the last six months | 98 | 21.26% | |
| | This month | 125 | 27.11% | |
| | This week | 121 | 26.25% | |
| | Today | 36 | 7.81% | |
| | Total number of participants | 461 | 100% | |
| Frequency of times to conduct e-transactions with the government (percentages for 461 participants only) | Once a year | 48 | 10.41% | |
| | Few times a year | 159 | 34.49% | |
| | Once a month | 49 | 10.63% | |
| | Few times a month | 120 | 26.03% | |
| | Once a week | 22 | 4.77% | |
| | Few times a week | 48 | 10.41% | |
| | Daily | 15 | 3.25% | |
| | Total number of participants | 461 | 100% | |



## 7.3     Data Screening

This section's examination of the data includes the following: identifying which cases or participants were not considered as part of the study group, describing that there was no missing data or incorrectly entered data points and managing of outliers.

Those who completed the questionnaire but did not identify themselves as Saudi nationals and those who did not complete the questionnaire were excluded from the study sample (Hair, et al., 2010). The exclusion of all other nationalities (101 participants) enabled PVQ to measure only Saudi cultural values. In addition, the data obtained from PCET focused only on the opinions and attitudes of Saudis, thereby streamlining the analysis results and conclusions. The exclusion of non-Saudi nationalities is associated with the research goal of conducting a cross-sectional study of Saudi Internet users. Moreover, respondents who did not complete all questions in the questionnaire were excluded as incomplete questionnaires would be missing some or all values from the dependent construct questions (items). The dependent construct (intention to use e-transactions) is the main focus of this study; therefore, participants were required to complete the questions associated with its items. The dependent construct items were located at the end of the questionnaire and preceded by PVQ to avoid common method biases, as discussed in section 6.2.6. Many of those who did not complete the survey did not complete the intention to use e-transactions construct items. Intention to use e-transactions is the dependent construct and the main focus of this study. Thus, their questionnaires were excluded from the sample (Hair, et al., 2010).

There were no missing data points in the data sheet because answering all questions was mandatory in the instrument. Additionally, there were no errors or mis-specified data points because the data file was downloaded directly from Qualtrics.com rather than being manually entered into the analysis software.

Outliers are defined as the cases (participants) who have specific characteristics that, for the purposes of the research, are considered to be distinctly different from other participants in



the study sample (Tabachnick & Fidell, 2007). A participant is identified as a univariate outlier if it has an extreme score on a single item. A multivariate outlier is detected when multiple (two or more) items have extreme scores (Kline, 2010). Firstly, outliers were detected using a 5% trimmed mean (univariate outlier detection) as well as the Mahalanobis distance (multivariate outlier detection). Secondly, the outliers detected were studied case by case as recommended by Pallant (2011). Only three cases were excluded (Field, 2009; Hair, et al., 2010; Pallant, 2011).

Univariate outliers were detected using a 5% trimmed mean for each item. Firstly, 5% of the extreme cases for each item were excluded, whether large or small, and then the average (trimmed mean) was calculated. Pallant (2011) argued that if the trimmed mean values for all variables (items) are very different from the mean (average for all cases), then cases within the 5% edges of these distributions should be investigated for exclusion. The difference between the trimmed mean (5%) and the mean of all of the items was at minimum 0.14 and at maximum 0.5. This very small difference between the calculated trimmed mean and the standard mean indicates that the extreme values were not very different in the remaining distribution. Hence, there were no outliers to exclude at this stage (Tabachnick & Fidell, 2007).

The Mahalanobis distance ($D^2$) is a multivariate measure that determines the distance of each case from the calculated centroid of the remaining cases. The calculated centroid is the point created by the means of all of the variables or the constructs. The higher the number, the further the case is from the other cases (Hair, et al., 2010; Tabachnick & Fidell, 2007). Hair et al. (2010) recommended a threshold $D^2$ for $D^2$/df of 3 or 4 for large samples and 2 for small samples. In this equation, df is the degrees of freedom or the number of constructs. After calculating the Mahalanobis distances for each case, all cases yielded less than 3.84, with the exception of three cases having the following $D^2$/df values: 4.40, 4.53, and 5.36. Further analyses of these cases showed that in all three instances, the respondents either selected 'strongly disagree' or 'strongly agree' consecutively for many questions, indicating that these respondents might not have completed the questionnaire properly (Hair, et al., 2010). Therefore, these three cases were removed, leaving a sample size of 671 for further analysis.



## 7.4      Parametric Data Assumptions

Parametric tests such as SEM require a set of distribution assumptions to ascertain the accuracy of the results. Parametric tests are tests of statistical significance based on certain distribution assumptions (Jupp, 2006). However, when the data assumptions are violated, parametric test results might not be fully accurate, and the use of nonparametric statistical tests (robust methods) is recommended (Tabachnick & Fidell, 2007). Since SEM was used as an analysis method, the data was tested for the extent to which it meets these assumptions.

Field (2009) suggested assessing the following parametric assumptions: normality of data, homoscedasticity, interval data, and independence. In this study the normality assumption was violated, but the homoscedasticity assumption was met. With regard to interval data, the 7- or 6-point Likert scale is considered an interval scale. Thus, this assumption was met in this study (Field, 2009; Kline, 2010). In terms of independence, each participant was expected to have completed the questionnaire individually, taking into consideration that it was taken online. Field (2009) stated that participants should not influence one another's opinions, and this is assumed to have been the case with this questionnaire.

Although the normality distribution assumption for parametric multivariate techniques was violated to an extent, this research still adopted SEM analysis technique. Statistical advice from Griffith University SEM statistical advisor suggested that the violations of the distribution assumption will not affect the results when SEM is used, especially for large samples (larger than 500). A review of the literature revealed some empirical evidence for this statement. For example, sample simulations conducted by Glass, Peckham, and Sander (1972) showed that parametric techniques are not significantly affected by violations of the distribution assumption. Furthermore, the widely cited work of Hair et al. (2010, p. 663) noted that the SEM estimation technique Maximum Likelihood (ML) "has proven fairly robust to violations of the normality assumption." ML is used in this research as an estimation technique for SEM. Prior to a more detailed discussion of this issue, the normality assumption for the sampled data is explored.



### 7.4.1    *Normality*

Normality refers to the shape of the data distribution for each variable (construct) in comparison to the benchmark, which is a bell-shaped normal distribution (Hair, et al., 2010). Hair et al. (2010) stated that an extremely large disparity between a variable's distribution and the normal distribution invalidates all statistical analyses, including F- and *t*-tests that use the normal distribution. Nevertheless, Hair et al. (2010) explained that large samples (200 participants or more) would minimise undesirable effects of skewed or non-normal distributions. Tabachnick and Fidell (2007) and Mendenhall, Beaver, and Beaver (2009) reiterated that, large samples' distribution of means have normal distributions, regardless of deviations from normality. Irrespective of the presence of a large sample, Hair et al. (2010) recommended reporting normality assumption tests.

The assessment of the normality violation can be based on the shape of the offending distribution and the sample size. The violation of the normality distribution can be measured by the kurtosis 'peakedness' or 'flatness' and the skewness of the distribution. The kurtosis indicates the height of the distribution, and the skewness refers to the balance. A skewed distribution is unbalanced and shifts to the right (positive skewness) or to the left (negative skewness), whereas a balanced distribution is centred and equally symmetrical at the edges (Hair, et al., 2010; Pallant, 2011). In a discussion of the robustness of test statistics, Curran, West, and Finch (1996) advised researchers to use values of skewness and kurtosis approaching the absolute values of 2 and 7, respectively. These values of skewness and kurtosis represent an appropriate guide for assessing acceptable non-normality to enable the robust use of parametric test statistics (Curran, et al., 1996; Fabrigar L. R., Wegener D. T., MacCallum R. C., & Strahan, 1999). As shown in Table 7.3, most of the items have skewness and kurtosis below 2 and 7, respectively. This condition was met for all of the items except four: RA3, SE14, USE3 and USE5. With the exception of USE5, all of these four items had a tolerable absolute value of kurtosis but an absolute value of skewness slightly higher than 2 (2.19, 2.3, 2.21 and 2.13, respectively). Appendix O and Appendix P shows item codes and wording. The closeness of



these values to the suggested skewness and kurtosis levels for all items suggests acceptable non-normality (Curran, et al., 1996).

Another test of normality was suggested by Hair et al. (2010), who stated that the z value of kurtosis and skewness is another method for measuring normality levels. The z-skewness and the z-kurtosis were calculated using the formulas found in Hair et al. (2010), where z-skewness is discussed on p. 72 and z-kurtosis on p. 73:

$$\text{Z kurtosis} = \frac{\text{kurtosis}}{\sqrt{\frac{24}{N}}}, \text{Z skewness} = \text{skewness}/\sqrt{(\frac{6}{N})}$$

In the formula, N is the sample size; the kurtosis and skewness are detailed in Table 7.3. The z-kurtosis for all of the items ranged from -5.97 to 39.25, and the z-skewness ranged from -24.31 to 9.59. Hair et al. (2010) noted that the critical value for a normal z-kurtosis and z-skewness is between $\pm 2.58$ for the 0.01 significance level or between $\pm 1.96$ for the 0.05 significance level. The z-kurtosis and the z-skewness values in Table 7.3 indicate violation of the normality assumption for most of the items.

Hair et al.'s (2010) assessment method proves the violation of normality. Nevertheless, taking into consideration these acceptable values of skewness and kurtosis and that the sample size was significantly larger than 200, the influence of this assumption violation on the results of the parametric analysis would be minimal (Curran, et al., 1996; Mendenhall, et al., 2009; Tabachnick & Fidell, 2007). Furthermore, Shah and Goldstein (2006) and Jöreskog and Sörbom (1989) mentioned that non-normality can be tolerable when the SEM Maximum Likelihood (ML) estimation technique is used. Therefore, violation of the normality assumption was considered to be at an acceptable level for the parametric statistical method (B. M. Byrne, 2010; Curran, et al., 1996; Mendenhall, et al., 2009). For more detailed discussion about SEM see section 8.5.1.

Table 7.3 includes a summary of the responses, the mean, the standard deviation (Std. Dev.), skewness, kurtosis and calculated z-skewness and z-kurtosis of each construct. The



descriptive statistics for the PCET items (from RA1 to USE5) are for the 7-point Likert scale.

The statistics for the PVQ items (from SD1 to B33) are for the 6-point Likert scale.

Table 7.3
*Descriptive Statistics for the Items*

| Construct | Mean | Std. Dev. | Skewness | Kurtosis | Z-skewness | Z-kurtosis |
|-----------|------|-----------|----------|----------|------------|------------|
| RA1 | 6.12 | 1.21 | -1.72 | 3.30 | -18.23 | 17.45 |
| RA2 | 5.91 | 1.27 | -1.36 | 1.83 | -14.39 | 9.66 |
| RA3 | 6.28 | 1.06 | -2.19 | 5.99 | -23.10 | 31.65 |
| RA4 | 6.09 | 1.11 | -1.64 | 3.27 | -17.32 | 17.31 |
| RA5 | 5.97 | 1.24 | -1.53 | 2.40 | -16.15 | 12.71 |
| CT1 | 6.18 | 1.15 | -1.98 | 4.46 | -20.90 | 23.58 |
| CT2 | 5.99 | 1.30 | -1.71 | 2.99 | -18.07 | 15.81 |
| CT3 | 6.10 | 1.13 | -1.74 | 3.66 | -18.44 | 19.34 |
| CT4 | 6.18 | 1.10 | -1.96 | 4.84 | -20.67 | 25.59 |
| CMX1 | 2.96 | 2.07 | 0.91 | -0.59 | 9.58 | -3.14 |
| CMX2 | 2.91 | 1.69 | 0.69 | -0.54 | 7.34 | -2.83 |
| CMX3 | 4.90 | 1.68 | -0.78 | -0.36 | -8.27 | -1.89 |
| CMX4 | 3.34 | 1.74 | 0.32 | -1.13 | 3.34 | -5.97 |
| RED1 | 5.49 | 1.31 | -1.19 | 1.27 | -12.61 | 6.69 |
| RED2 | 5.72 | 1.18 | -1.40 | 2.35 | -14.78 | 12.44 |
| RED3 | 5.71 | 1.24 | -1.27 | 1.72 | -13.47 | 9.11 |
| TI1 | 4.86 | 1.68 | -0.70 | -0.41 | -7.38 | -2.17 |
| TI2 | 4.46 | 1.73 | -0.41 | -0.82 | -4.34 | -4.33 |
| TI3 | 4.67 | 1.66 | -0.53 | -0.59 | -5.63 | -3.13 |
| TG1 | 4.77 | 1.64 | -0.69 | -0.35 | -7.31 | -1.84 |
| TG2 | 4.76 | 1.67 | -0.69 | -0.38 | -7.30 | -2.02 |
| TG3 | 4.84 | 1.64 | -0.82 | -0.17 | -8.69 | -0.88 |
| TG4 | 4.69 | 1.69 | -0.64 | -0.44 | -6.75 | -2.35 |
| SI1 | 5.14 | 1.41 | -0.90 | 0.56 | -9.51 | 2.96 |
| SI2 | 5.31 | 1.35 | -0.99 | 0.74 | -10.43 | 3.92 |
| SI3 | 5.15 | 1.40 | -0.81 | 0.28 | -8.57 | 1.46 |
| POC1 | 5.93 | 1.22 | -1.71 | 3.68 | -18.06 | 19.45 |
| POC2 | 6.04 | 1.05 | -1.62 | 3.93 | -17.09 | 20.75 |
| POC3 | 6.04 | 1.10 | -1.69 | 3.88 | -17.91 | 20.51 |
| USE1 | 6.06 | 0.98 | -1.56 | 4.10 | -16.50 | 21.67 |
| USE2 | 6.24 | 0.91 | -1.86 | 5.29 | -19.69 | 27.98 |
| USE3 | 6.18 | 1.08 | -2.21 | 6.44 | -23.38 | 34.07 |



| USE4 | 6.07 | 1.10 | -1.74 | 3.91 | -18.39 | 20.67 |
| USE5 | 6.31 | 0.87 | -2.13 | 7.42 | -22.53 | 39.25 |
| SD1 | 4.61 | 1.11 | -0.88 | 0.78 | -9.31 | 4.12 |
| SD11 | 4.91 | 1.12 | -1.27 | 1.72 | -13.38 | 9.10 |
| SD22 | 4.89 | 1.01 | -1.05 | 1.26 | -11.08 | 6.69 |
| SD24 | 4.74 | 1.20 | -0.93 | 0.47 | -9.83 | 2.48 |
| P2 | 3.31 | 1.42 | 0.13 | -0.90 | 1.40 | -4.76 |
| P17 | 3.28 | 1.49 | 0.23 | -0.95 | 2.38 | -5.03 |
| P39 | 4.11 | 1.39 | -0.40 | -0.74 | -4.24 | -3.91 |
| U3 | 5.11 | 1.18 | -1.51 | 1.83 | -15.95 | 9.68 |
| U8 | 4.92 | 1.01 | -1.09 | 1.48 | -11.56 | 7.83 |
| U23 | 4.88 | 1.19 | -1.20 | 1.02 | -12.67 | 5.41 |
| U29 | 5.49 | 0.79 | -1.99 | 5.04 | -21.08 | 26.64 |
| U19 | 5.15 | 1.05 | -1.47 | 2.31 | -15.54 | 12.20 |
| U40 | 4.90 | 1.14 | -1.13 | 1.09 | -11.95 | 5.75 |
| A4 | 3.97 | 1.50 | -0.34 | -0.95 | -3.60 | -5.05 |
| A13 | 4.66 | 1.31 | -0.89 | 0.04 | -9.46 | 0.21 |
| A24 | 4.97 | 0.98 | -1.07 | 1.28 | -11.26 | 6.75 |
| A32 | 4.99 | 1.06 | -1.11 | 1.04 | -11.72 | 5.52 |
| SE5 | 5.28 | 0.96 | -1.80 | 3.99 | -19.05 | 21.08 |
| SE14 | 5.36 | 1.01 | -2.30 | 6.25 | -24.31 | 33.07 |
| SE21 | 5.16 | 0.99 | -1.63 | 3.17 | -17.24 | 16.78 |
| SE31 | 4.97 | 1.12 | -1.17 | 0.99 | -12.33 | 5.26 |
| SE25 | 5.14 | 1.03 | -1.57 | 2.84 | -16.60 | 15.02 |
| ST6 | 4.97 | 1.10 | -1.26 | 1.55 | -13.32 | 8.21 |
| ST15 | 3.63 | 1.53 | -0.07 | -1.07 | -0.73 | -5.64 |
| ST30 | 4.22 | 1.42 | -0.50 | -0.67 | -5.26 | -3.53 |
| C7 | 4.62 | 1.38 | -1.00 | 0.22 | -10.58 | 1.17 |
| C16 | 4.90 | 1.13 | -1.26 | 1.47 | -13.31 | 7.79 |
| C28 | 5.49 | 0.80 | -1.88 | 4.18 | -19.93 | 22.08 |
| C36 | 5.35 | 0.83 | -1.74 | 4.29 | -18.35 | 22.66 |
| T9 | 4.46 | 1.39 | -0.88 | -0.03 | -9.26 | -0.16 |
| T20 | 4.92 | 1.08 | -1.23 | 1.63 | -12.97 | 8.62 |
| T25 | 3.64 | 1.43 | -0.09 | -0.89 | -0.93 | -4.71 |
| T38 | 4.87 | 1.12 | -1.02 | 0.75 | -10.76 | 3.95 |
| H10 | 4.45 | 1.35 | -0.76 | -0.25 | -8.00 | -1.31 |
| H26 | 4.15 | 1.39 | -0.43 | -0.66 | -4.52 | -3.48 |
| H37 | 5.00 | 1.04 | -1.19 | 1.40 | -12.61 | 7.40 |
| B12 | 5.12 | 0.91 | -1.17 | 2.04 | -12.35 | 10.80 |



| B18 | 5.06 | 1.03 | -1.22 | 1.60 | -12.91 | 8.46 |
| B27 | 5.03 | 0.88 | -1.24 | 2.79 | -13.14 | 14.76 |
| B33 | 4.97 | 1.11 | -1.37 | 1.93 | -14.53 | 10.19 |

*Note.* RA=relative advantage; CT=compatibility; CMX=complexity; RED=result demonstrability; TI=trust in the Internet; TG=trust in government agencies; SI=social influence; POC=perspective on communication; USE=intention to use e-transactions. SD=self-direction; P=power; U=universalism; A=achievement; SE=security; ST=stimulation; C=conformity; T=tradition; H=hedonism; B=benevolence. Appendix O and Appendix P have item wording and codes.

From the statistics (PCET variables means) in Table 7.3 above, it was apparent that, in general, the participants appeared to have positive perceptions of the use of e-transactions. The highest or lowest values of skewness and kurtosis levels for some items reflected the nature of the sampled population and also the underlying latent variable (construct) that was measured (Pallant, 2011). Additionally, the relatively small standard deviation for all of the items indicated similarity in the perceptions, values, and opinions among the participants in the sample.

### 7.4.2   *Homoscedasticity*

Homoscedasticity is the assumption that the dependent construct(s) demonstrates an equal level of variance across the set of independent construct(s). Homoscedasticity can be desirable when the variance in the dependent construct in the posited relationship disperses in a balanced way across the independent constructs. A heteroscedastic relationship occurs when fthere is unequal variance in the dependent construct across the values of the independent construct (Hair, et al., 2010). According to Field (2009), Levene's test is the most reliable method to measure homoscedasticity. This test examines the null hypothesis if the difference in the variances between the constructs is zero. A significant ($p < 0.05$) result of this test indicates violation of the homoscedasticity assumption. The results of this test were insignificant for all of the constructs, indicating that the homoscedasticity assumption was met (see Appendix I).

Therefore, from the previously discussed tests, it was revealed that the homoscedasticity assumption was met, whereas the normality assumption was violated. However, the large



sample size is expected to nullify the effect of this violation. Hair et al. (2010) and Tabachnick and Fidell (2007) noted that it is usual for large datasets of participants to tend to have strong opinions on specific issues. Therefore, normality violation is expected. Data transformation is a possible remedy to address non-normality. The data was transformed using a range of transformation techniques, and the normality assumption was tested on each transformed dataset. The results suggested that none of the possible transformations resolved the violation of the normality assumption.

## 7.5    Summary

This chapter provided demographic information on the sampled data, data screening, tests of parametric data assumptions, and the assessment of the research model. It was difficult to fully ascertain that the sampled dataset was actually representative of the Saudi population that uses the Internet. This study sampled a wide sector of the study population with the usage of on online survey (Karahanna, Evaristo, & Srite, 2002). Thus, from the comparison between the study population and the sampled data, the sampled data can be considered to be an approximate representation of the online population. Only minimal outliers were excluded from the analysis after a case-by-case investigation. Normality, which was one of the parametric data assumptions, was acceptably non-normal for use of the SEM analysis, whereas the homoscedasticity assumption was met. These tests were considered to qualify the sampled data for a parametric analysis technique (SEM). Therefore, it was decided that the parametric statistical test, i.e., SEM, would be used to analyse the data. The decision was based on statistical advice, a brief review of relevant literature, and empirical outcomes from the sampled data assumptions tests. The following chapter provides an analysis of the data revealing the direction and significance of the factors influencing the acceptance e-transactions.



# 8    DATA ANALYSIS

Straub et al. (2004) stated that quantitative research instruments, especially when positivist epistemology is applied, are used to capture and empirically measure abstract concepts (such as perceptions and culture) in the real world. These latent constructs need to be captured to posit, confirm, or reject previously proposed causality between different concepts and to draw relevant conclusions and findings. However, these findings cannot be corroborated without applying a set of heuristics to ascertain the validity of the instrument that was used to capture these concepts (Straub & Carlson, 1989).

Boudreau, Ariyachandra, Gefen, and Straub (2004) argued that the implementation of statistical validation heuristics increases the reliability, validity, and significance of research results. As noted by Straub and Carlson (1989), validation procedures for instruments centre around the concepts of validity and reliability. Validity refers to the measurement accuracy of the instrument and the extent to which the data that are obtained represent the measured constructs. Reliability refers to the evaluation of the instrument's internal consistency (Zikmund, Babin, Carr, & Griffin, 2010). Both of these validation aspects are explored further in this chapter.

Instrument validation and data analysis studies commonly cited in the field of information systems are adopted as guidelines in this chapter. These include the studies of (B. M. Byrne, 2010; Gefen, Rigdon, & Straub, 2011; Gefen, Straub, & Boudreau, 2000; Hair, et al., 2010; Kline, 2010; Straub, et al., 2004). The terms used in this chapter are mostly based on Straub et al.'s (2004) instrument validation and on Gefen et al.'s (2000) (Structural Equation Modelling) SEM data analysis guidelines.

The overall research model was considered to be large, with 19 constructs and 74 items remaining after the fourth item (RED4) of the result demonstrability construct was dropped in the pilot study. After the validation procedures, the number of constructs (16) and items (56)



was reduced but still considered large, which presented potential difficulties for the model estimation. Consequently, the assessment of the validity of the constructs was conducted at the construct level and the submodel level (instrument or scale) (Hair, et al., 2010). Confirmatory factor analysis was conducted at the construct individual level (Ahire & Devaraj, 2001) and at the scale or model level (PVQ and PCET) (Hair, et al., 2010). It is important to conduct exploratory and confirmatory factor analysis for the PCET model to establish and confirm the model's validation. This is especially important because PCET items were reworded (contextualised) to make them relevant to the research focus; therefore, both exploratory and confirmatory factor analysis were needed to ascertain the model's reliability and validity.

In contrast, Schwartz's BPV items were adopted without changes from previously validated instruments. The PVQ instrument is considered publicly available for use by researchers, so no permissions were needed for its usage (Alkindi, 2009; Schwartz, 2003; Schwartz, et al., 2001). Furthermore, Schwartz (2009) did not recommend the usage of EFA to determine the underlying item relations of the BPV model because the EFA solution cannot reveal the quasi-circumplex structure of BPV. Thus, only CFA was conducted for BPV to confirm the validity of this model for the sampled data.

The main purpose of evaluating constructs' validity is to ascertain that a group of items actually measure their underlying constructs. Thus, assessment of the submodels (PCET and BPV) would include all constructs and items in the overall research model (causal model). This scale assessment enabled a simpler evaluation and modification of each submodel separately (Ahire & Devaraj, 2001; Hair, et al., 2010). The outcomes of these assessments were combined into the final research model, which showed a good level of fit to the data obtained for the measurement model. Finally, the structural model was assessed, and the potential direction and significance between the dependent constructs and the independent construct (research hypothesis) was determined using SEM. Hair et al. (2010) suggested that the constructs' reliability should be evaluated prior to conducting any validation or analysis procedures. Therefore, the reliability assessment of all constructs is described in the following section.



## 8.1    Reliability of the Constructs

The use of reliability heuristics for reflective constructs and items include the measurement of the construct's internal consistency and further investigation of the total correlations of the items when there is a lack of homogeneity between the items. A change in reflective items is caused by the change in their underlying construct. Since all items in a group reflect one construct, the assessment of their consistency is necessary (Jarvis, et al., 2003). To assess the internal consistency of each construct, Cronbach's alpha was calculated using SPSS Version 19.0 (Pallant, 2011). The reliability of each construct was classified using a method similar to that employed in the pilot study. Cronbach's alpha values above 0.90 were considered 'excellent,' above 0.8 were considered 'good,' above 0.5 were considered 'acceptable,' and below 0.5 were considered as unacceptable (Nunnally, 1967). The overall reliability of the 74-item instrument was very high (Cronbach's alpha = 0.93). Table 7.3 below presents the Cronbach's alpha values and illustrates how these differed compared with the alpha coefficients for all of the constructs and the items' intercorrelations in the pilot study (pilot study described in section 6.2.4). The internal consistency of all of the constructs was between acceptable and excellent. The tradition construct, which had an unacceptable Cronbach's alpha value, was the exception (Cronbach, 1990). The column labelled "Differences in the alpha values compared with the pilot" in Table 8.1 highlights the slight difference between the values for the constructs in the full-scale study and those in the pilot study. In addition, the alphas of the following constructs improved in the full-scale study: relative advantage, trust in the Internet, self-direction, power, universalism, achievement, security, stimulation, conformity, hedonism, and benevolence.

Table 8.1
*Reliabilities of All of the Constructs*

| Construct | Items | Cronbach's alpha (internal consistency) | Differences in the alpha values compared with the pilot | Construct's reliability status | Item–total correlation |
|---|---|---|---|---|---|
| Relative advantage (RA) | 5 | 0.92 | +0.01 | Excellent | 0.75–0.83 |



| Compatibility (CT) | 4 | 0.90 | -0.01 | Excellent | 0.73–0.87 |
| Complexity (CMX) | 4 | 0.68 | -0.06 | Acceptable | 0.40–0.53 |
| Result demonstrability (RED) | 3 | 0.71 | -0.08 | Acceptable | 0.43–0.60 |
| Trust in the Internet (TI) | 3 | 0.89 | +0.04 | Good | 0.76–0.81 |
| Trust in the government (TG) | 4 | 0.93 | -0.02 | Excellent | 0.83–0.93 |
| Social influence (SI) | 3 | 0.86 | -0.02 | Good | 0.65–0.80 |
| Perspective on communication (POC) | 3 | 0.72 | -0.08 | Acceptable | 0.35–0.71 |
| Intention to use e-transactions (USE) | 5 | 0.88 | -0.02 | Good | 0.80–0.70 |
| Self-direction (SD) | 4 | 0.61 | +0.05 | Acceptable | 0.33–0.43 |
| Power (P) | 3 | 0.58 | +0.05 | Acceptable | 0.20–052 |
| Universalism (U) | 6 | 0.70 | +0.04 | Acceptable | 0.27–0.50 |
| Achievement (A) | 4 | 0.72 | +0.06 | Acceptable | 0.43–0.64 |
| Security (SE) | 5 | 0.69 | +0.03 | Acceptable | 0.41–0.50 |
| Stimulation (ST) | 3 | 0.63 | 0 | Acceptable | 0.34–0.59 |
| Conformity (C) | 4 | 0.64 | +0.09 | Acceptable | 0.39–0.47 |
| Tradition (T) | 4 | 0.45 | -0.06 | Unacceptable | 0.23–0.30 |
| Hedonism (H) | 3 | 0.72 | +0.01 | Acceptable | 0.54–0.57 |
| Benevolence (B) | 4 | 0.66 | +0.01 | Acceptable | 0.35–0.58 |

Pallant (2011) recommended further investigation of a construct's corrected items-total correlations if its alpha value was unacceptable. Of the four items in the tradition construct, three—T9 (0.23), T20 (0.23), and T38 (0.27)—showed a correlation lower than the acceptable value of 0.3. As shown in Table 8.2 below, the elimination of an item would not bring the tradition construct's alpha value above the acceptable level (0.3).

Table 8.2

*Inter-Item Correlations for Tradition Construct*

| Item | Corrected item-total correlation | Cronbach's alpha if item was deleted |
| --- | --- | --- |
| T9 | .230 | .408 |
| T20 | .225 | .408 |
| T25 | .301 | .329 |
| T38 | .272 | .367 |

*Note.* See Appendix P for wording of items



Schwartz and colleagues (2003; 2001) reported low alpha reliabilities for PVQ constructs. They stated that PVQ captures a wide range of content in while using few items for each construct which result in reducing Cronbach's alpha. Taking Schwartz's (2003) point into consideration, no items from the BPV model were eliminated at this stage, despite the low values for universalism item (U3; 0.27) and power item (P2; 2.0) in the total-item correlation.

Another indication of the constructs' reliability is positive values of the corrected item-total correlation, i.e., where each construct items measure the same "underlying characteristics" (Pallant, 2011, p. 100). Table 8.1 shows that all items' corrected item-total correlation values were positive, indicating good reliability for each construct. Having good reliability enables the assessment of validity. The concepts of construct's validity and related terms are firstly clarified in the following section.

## 8.2    Construct Validity

A construct's validity empirically defines the extent to which items within an instrument reflect the theoretical construct they are intended to measure (Bagozzi, Yi, & Phillips, 1991). Inferences from data would not be reliable if the validity of the construct was not confirmed. The two most important aspects of a construct's validity are discriminant and convergent validity (Campbell & Fiske, 1959; Straub, et al., 2004). These aspects complement each other in determining the validity of the construct (Bagozzi, et al., 1991). Discriminant validity is the degree to which the constructs within one instrument are actually distinct from each other. Convergent validity is the degree to which a construct's items resemble one concept (construct) (Gefen, et al., 2000; Hair, et al., 2010; Straub, et al., 2004). The validity of the constructs in both models, PCET (exploratory and confirmatory assessment) and BPV (confirmatory assessment), was assessed. The following section explores the validity of the PCET model.



### 8.3    Exploratory Factor Analysis (EFA) of the PCET Model

EFA is typically used as a theory development tool, especially when the underlying structure of the variables of the model needs to be defined. Byrne (2010) stated that EFA is frequently used when the associations between observed variables (items) and unobserved variables (constructs) are uncertain. However, when the theory is already established, EFA is usually not required, and the structure of the items and the constructs are confirmed using only CFA (Hair, et al., 2010). These associations between the constructs and the items are called loadings in both CFA and EFA. One purpose of EFA and CFA is to assess these loadings to determine whether the items measure their underlying unobservable variables (constructs) (B. M. Byrne, 2010).

Although previous research and theory (Aoun, et al., 2010; Carter & Bélanger, 2005; Moore & Benbasat, 1991; Venkatesh, et al., 2003) have already established the underlying structure of the adopted constructs, additional measures were taken in this study to confirm the validity of the contextualised PCET constructs and reworded items. EFA was not conducted for Schwartz's Basic Personal Values (BPV) model because a PVQ instrument was adopted that has been validated many times in previous research (Alkindi, 2009; Cohen, 2010; Schwartz & Bardi, 2001; Schwartz, et al., 2001). Exploratory factor analysis was used to investigate the underlying structure of the PCET model, since the adopted items and constructs were contextualised and altered to suit this research (Aoun, et al., 2010; Carter & Bélanger, 2005; Moore & Benbasat, 1991; Venkatesh, et al., 2003). Although EFA is not usually required for a well-established theory —such as PCI or UTAUT— EFA was conducted to corroborate the model's structure and to ascertain the discriminant and convergent validity of the PCET model (Hair, et al., 2010; Straub, et al., 2004). The Statistical Package for Social Science (SPSS) Version 19.0 was used to conduct EFA (Pallant, 2011).

Before describing the EFA results, the Kaiser-Meyer-Olkin (KMO) measure and Bartlett's test of sphericity is detailed. The KMO measure and Bartlett's test are used to determine the sample's appropriateness for running factor analysis. In addition, Bartlett's test of



sphericity is used to assess sampling adequacy, with a p-value below 0.05 considered significant. The KMO measure diverges between 0 and 1. Kaiser (1974, as cited in Field (2009), stated that KMO values greater than 0.9 are evidence of excellent sampling adequacy, whereas those less than 0.5 are unacceptable. The KMO for the sampled data was 0.93, indicating excellent sampling adequacy. The results of Bartlett's test of sphericity were also highly significant (p < 0.001). Therefore, the sampled data was considered to be adequate for the use of factor analysis (Field, 2009).

Straub et al. (2004) noted that items that load cleanly together on one factor and do not cross-load on other factors demonstrate convergent validity. On the other hand, items that do not cross-load on other factors provide evidence of discriminant validity. EFA is sometimes used to explore the number of factors to which items are supposed to load. The eigenvalue is, in simple terms, a condition that is used to retain the number of factors according to items' loading (Field, 2009). Kaiser (1960) defined what Jolliffe (1972, 2002) described as a strict rule in order to limit and define the number of factors that items are supposed to be loaded onto based on an eigenvalue that is equal to or greater than 1 (Field, 2009, p. 640). However, based on previous knowledge of the number of factors and of item loading, the use of this strict condition is not important. Therefore, as suggested by Field (2009) and Jolliffe (1972, 2002), a less strict condition was adopted, with factors retained that had an eigenvalue equal to or greater than 0.7.

A first run of factor analysis with principal axis factoring and Oblimin rotation resulted in clean loadings, except for three items: CMX1, RED3, and POC3. As RED3 and POC3 did not load highly on any factor, they were excluded. Table 8.3 shows the results of the factors for an eigenvalue over 0.75.

Table 8.3
*PCET Item Loadings*

| Item | Factor | | | | | | | | |
|------|------|------|------|------|------|------|------|------|------|
|      | 1    | 2    | 3    | 4    | 5    | 6    | 7    | 8    | 9    |
| RA1  | 0.69 | 0.02 | .03  | 0.11 | 0.01 | 0.00 | 0.06 | 0.03 | 0.04 |
| RA2  | 0.72 | 0.04 | .01  | 0.02 | 0.07 | 0.02 | 0.04 | 0.02 | 0.06 |
| RA3  | 0.77 | 0.05 | .00  | 0.03 | 0.05 | 0.00 | 0.01 | 0.10 | 0.03 |



| | | | | | | | | |
|---|---|---|---|---|---|---|---|---|
| RA4 | 0.76 | 0.03 | .04 | 0.05 | 0.01 | 0.08 | 0.04 | 0.05 | 0.05 |
| RA5 | 0.60 | 0.05 | .03 | 0.00 | 0.05 | 0.03 | 0.05 | 0.15 | 0.04 |
| CT1 | 0.17 | 0.03 | .02 | 0.05 | 0.03 | 0.01 | 0.00 | 0.66 | 0.03 |
| CT2 | 0.05 | 0.06 | .01 | 0.04 | 0.04 | 0.03 | 0.05 | 0.72 | 0.02 |
| CT3 | 0.06 | 0.01 | .03 | 0.01 | 0.00 | 0.02 | 0.00 | 0.85 | 0.11 |
| CT4 | 0.11 | 0.04 | .04 | 0.13 | 0.03 | 0.03 | 0.00 | 0.63 | 0.01 |
| CMX1 | 0.08 | 0.01 | .47 | 0.06 | 0.03 | 0.01 | 0.01 | 0.01 | 0.07 |
| CMX2 | 0.04 | 0.07 | .66 | 0.02 | 0.04 | 0.03 | 0.03 | 0.18 | 0.04 |
| CMX3 | 0.07 | 0.03 | .54 | 0.10 | 0.04 | 0.04 | 0.07 | 0.16 | 0.08 |
| CMX4 | 0.04 | 0.09 | .71 | 0.01 | 0.08 | 0.03 | 0.08 | 0.04 | 0.07 |
| RED1 | 0.06 | 0.01 | .03 | 0.03 | 0.02 | 0.03 | 0.75 | 0.07 | 0.01 |
| RED2 | 0.03 | 0.00 | .05 | 0.07 | 0.01 | 0.01 | 0.73 | 0.06 | 0.02 |
| TI1 | 0.00 | 0.03 | .02 | 0.04 | 0.82 | 0.01 | 0.03 | 0.04 | 0.02 |
| TI2 | 0.00 | 0.07 | .01 | 0.00 | 0.81 | 0.02 | 0.00 | 0.02 | 0.02 |
| TI3 | 0.01 | 0.05 | .00 | 0.03 | 0.86 | 0.02 | 0.02 | 0.01 | 0.00 |
| TG1 | 0.00 | 0.80 | .04 | 0.00 | 0.10 | 0.03 | 0.00 | 0.01 | 0.03 |
| TG2 | 0.03 | 0.90 | .03 | 0.02 | 0.07 | 0.03 | 0.04 | 0.03 | 0.03 |
| TG3 | 0.02 | 0.93 | .01 | 0.01 | 0.02 | 0.02 | 0.01 | 0.00 | 0.05 |
| TG4 | 0.05 | 0.81 | .02 | 0.00 | 0.04 | 0.05 | 0.04 | 0.06 | 0.03 |
| SI1 | 0.02 | 0.00 | .01 | 0.01 | 0.00 | 0.84 | 0.03 | 0.00 | 0.00 |
| SI2 | 0.04 | 0.05 | .00 | 0.00 | 0.01 | 0.97 | 0.04 | 0.02 | 0.00 |
| SI3 | 0.06 | 0.09 | .02 | 0.00 | 0.01 | 0.61 | 0.12 | 0.04 | 0.03 |
| POC1 | 0.14 | 0.07 | .05 | 0.17 | 0.05 | 0.06 | 0.04 | 0.07 | 0.55 |
| POC2 | 0.08 | 0.05 | .04 | 0.10 | 0.00 | 0.07 | 0.08 | 0.05 | 0.72 |
| USE1 | 0.02 | 0.03 | .03 | 0.64 | 0.04 | 0.05 | 0.06 | 0.02 | 0.14 |
| USE2 | 0.02 | 0.02 | .07 | 0.71 | 0.07 | 0.06 | 0.03 | 0.07 | 0.08 |
| USE3 | 0.11 | 0.01 | .02 | 0.66 | 0.01 | 0.04 | 0.02 | 0.02 | 0.02 |
| USE4 | 0.06 | 0.01 | .08 | 0.66 | 0.10 | 0.06 | 0.01 | 0.02 | 0.04 |
| USE5 | 0.05 | 0.03 | .03 | 0.83 | 0.00 | 0.01 | 0.02 | 0.01 | 0.03 |

*Note.* RA=relative advantage; CT=compatibility; CMX=complexity; RED=result demonstrability; TI=trust in the Internet; TG=trust in government agencies; SI=social influence; POC=perspective on communication; USE=intention to use e-transactions. See Appendix O for item codes and wording.

A closer look at Table 8.3 above shows that all items have factor loadings above 0.5, except CMX1, which has a value very close to the condition value (0.47). According to Hair et al. (2010), these loading values are particularly considered significant, especially when they are greater than 0.5; values above 0.3 are also considered acceptable. CMX1 item wording relevance to time explains its lower loading on the construct complexity: 'Using e-government transactions would consume too much of my time'. Straub, et al. (2003) explain that Arabs sense of time is different from westerners. Delays occurring while using a technology is acknowledged part of the process and not highly regarded as a hindrance while conducting a



technology specific task (Straub, et al., 2003). Therefore, lower loading of CMX1 item on complexity is culturally expected in comparison to other studies in western countries (e.g. USA) which reported higher loading (Moore & Benbasat, 1991; Taylor & Todd, 1995). Taking into consideration Straub, et al. (2003) argument on Arab culture and time, low cross-loading of the item on any other construct and that CMX1 loading (0.47) is higher than Hair et al. (2010) second threshold (0.3), CMX1 was decided to be retained and not deleted.

A clean structure of item loadings is shown in the table, with highly insignificant cross-loading of items on other factors, especially after the removal of the items RED3 and POC3, both of which did not load significantly on any factor. Therefore, these results confirmed the discriminant and convergent validity of the PCET model in the exploratory phase (Field, 2009; Hair, et al., 2010; Straub, et al., 2004). Common Method Bias (CMB) is assessed to ascertain that constructs really measure underlying concepts and not the setting in which these constructs were measured (Podsakoff, 2003).

**8.4     Assessment of Common Method Bias (CMB)**

CMB is considered a threat to construct validity (Boudreau, et al., 2004; Straub, et al., 2004). Harman's single-factor test was conducted to determine whether CMB was present (Podsakoff, 2003). CMB would be assumed to be present if the results of principal component analysis (part of EFA) indicated that only one factor accounted for all of the variance detected (Gefen, et al., 2011). EFA of all 74 items was conducted. The results (presented in Table 8.4 below) showed 17 factors when the eigenvalues (including BPV constructs) were greater than 1.0.

Table 8.4
*Results of Harman's CMB Assessment*

| Component | Total eigenvalue | % of variance |
|---|---|---|
| 1 | 15.38 | 20.51 |
| 2 | 6.2 | 8.28 |
| 3 | 3.54 | 4.72 |
| 4 | 3.16 | 4.21 |
| 5 | 2.47 | 3.30 |



| 6  | 2.04 | 2.72 |
|----|------|------|
| 7  | 1.79 | 2.39 |
| 8  | 1.71 | 2.29 |
| 9  | 1.66 | 2.22 |
| 10 | 1.59 | 2.12 |
| 11 | 1.37 | 1.83 |
| 12 | 1.30 | 1.74 |
| 13 | 1.18 | 1.58 |
| 14 | 1.14 | 1.52 |
| 15 | 1.04 | 1.39 |
| 16 | 1.02 | 1.36 |
| 17 | 1.00 | 1.34 |

Table 8.4 indicates that the largest factor (component) accounted for only 20.51% of the variance. As no single factor accounted for all of the variance, it was concluded that CMB was not an issue (Gefen, et al., 2011; Podsakoff, 2003). The following section is the assessment of the research model.

**8.5     Assessment of the Research Model**

The reliability and validity of each construct and each model (PCET and BPV) were confirmed individually using the procedures outlined in the following sections. Conducting these measures ascertained the validity and reliability at multiple analytical levels. PCET and BPV were carefully modified to enhance the validity and reliability of the overall research model.

For the overall model assessment, the SEM technique was adopted to assess the overall model and the research hypothesis. An overview of SEM is provided below, followed by an assessment of the measurement model. Finally, the structural model was assessed to determine the significance of the hypothesis.

*8.5.1     Overview of SEM*

SEM is a family of statistical techniques that are used to analyse and empirically explain relationships among constructs (Hair, et al., 2010; Kline, 2010). Covariance-based SEM or covariance structure analysis are different terms used in information systems and social



science research to describe SEM analysis (Diamantopoulos & Winklhofer, 2001; Gefen, et al., 2011; Gefen, et al., 2000; Kline, 2010). The focus of SEM as an analysis technique is the covariance or correlation (correlations refer to the standardisation of the covariance) parameters between the constructs (Hair, et al., 2010). This is a distinguishing characteristic of SEM analysis techniques (B. M. Byrne, 2010).

Kline (2010) explained that SEM is a statistical technique that is applied to a large sample size (larger than 200). As a general rule, SEM requires at least five participants for each item. To reach an acceptable CFA model fit, PCET and BPV CFA models were modified, resulting in a 56-item structural model. Thus, the research sample size (671 participants) exceeds the requirement of five participants per item (>280) (Hair, et al., 2010). SEM was adopted because it can measure relationships (structures) between variables (constructs or items) more accurately than other statistical techniques, such as regression analysis or factor analysis (Hair, et al., 2010).

The theory underpinning the research model, which is derived from the literature, predefines items and their underlying constructs and hypothesises relations between the constructs. SEM is used to assess the research model by identifying, estimating, and evaluating constructs-to-items and the constructs-to-constructs relations (B. M. Byrne, 2010; Hair, et al., 2010). However, when the hypothesised model does not fit the data, the model can be modified. In other words, rather than just confirming a theory, the measurement model can be modified to yield significant findings which should be theoretically meaningful correlations (Kline, 2010).

SEM as an analysis tool includes the evaluation of the measurement model and structural model (Kline, 2010). To illustrate these principles, the measurement and the structural models are depicted in



Figure 8.1 and Figure 8.2 respectively.

Figure 8.1
*Example of SEM measurement model.*

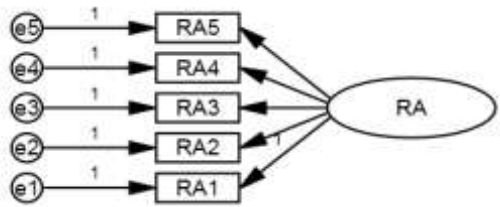

The measurement model reflects the relationship between the constructs (oval shape) and the items (square shape) as shown in

Figure 8.1 above. The measurement error (small circle) reflects the adequacy of the measuring item in measuring its underlying construct. The path (one-sided arrow) represents the path coefficient (i.e., standardised factor loading) between the item and the construct. The paths labelled "1" are fixed parameters, which are a requirement of AMOS software (B. M. Byrne, 2010). A congeneric measurement model is considered a good practice in identifying measurement models. A measurement model is considered congeneric if an item loads only on one underlying construct and there are no correlations between error terms. This practice was adopted in this study because it provides a good measure of construct validity; it was therefore applied for all measurement models (B. M. Byrne, 2010; Hair, et al., 2010).



Figure 8.2
*Example of SEM structural model.*

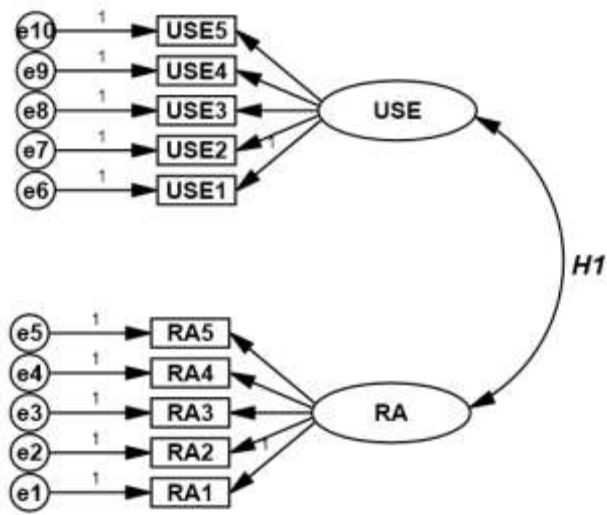

The structural model represents the hypothesised relationships between the constructs, as shown in Figure 8.2 above (B. M. Byrne, 2010). The structural model can be used to represent the interrelationships between the constructs (Hair, et al., 2010). The constructs' interrelationships (double-edged arrow) represent the covariance or the correlation between a pair of constructs (B. M. Byrne, 2010; Kline, 2010).

One of the main goals of model assessments is to determine the goodness-of-fit between the proposed overall structural model and the sampled data. Adequate goodness-of-fit increases the plausibility of the posited research model (B. M. Byrne, 2010). However, a perfect fit of the data to the structural model is very rare and almost impossible in research (Kline, 2010). Hence, the model-fitting approach in SEM includes the term 'residual,' which denotes the difference between the model's estimated parameters and the observed data (B. M. Byrne, 2010; Hair, et al., 2010).



Table 8.5
*Measurement of Fit Indices*

| Fit indices | Description | Rationale for using measure |
|---|---|---|
| CMIN (minimum discrepancy) or Chi-Square ($\chi^2$) | Signifies the difference between the covariance matrices in the posited model to the data where the difference is the ratio of $\chi^2$ to the degrees of freedom (B. M. Byrne, 2010; Hair, et al., 2010). | CMIN/df (Normed $\chi^2$) was originally introduced to reduce $\chi^2$ inflation in large sample sizes. Taking into consideration the large sample size (671), this index was included as a measure instead of $\chi^2$ (B. M. Byrne, 2010; Shah & Goldstein, 2006). |
| Goodness-of-fit (GFI) and adjusted goodness of fit (AGFI) | The GFI index is also another replacement for $\chi^2$. AGFI adjusts GFI based on the degrees of freedom. These statistical indices range between 0 and 1. However, both of these indices cannot be used alone because of their sensitivity to sample size (Hooper, Coughlan, & Mullen, 2008). | Both indices have historical importance in the literature and have been widely used in information systems and other disciplines. Therefore, GFI and PGFI were included to enable the results of this research to be compared with other studies (Gefen, et al., 2011; Hooper, et al., 2008). |
| Comparative fit index (CFI) | CFI is a popularly reported fit index in quantitative SEM research. It is considered to be a replacement of fit indices that are influenced by sample size (such as the normative fit index). This index varies between 0 and 1 (B. M. Byrne, 2010; Hooper, et al., 2008; Tabachnick & Fidell, 2007). | CFI is relatively unaffected by sample size and, therefore, is most important for the purposes of this study than the aforementioned fit indices in this table (Hooper, et al., 2008). |
| Incremental fit index (IFI) | The IFI addresses the issue of model parsimony and sample size sensitivity in other indices. Similarly to GFI, AGFI and CFI, this index ranges between 0 and 1 (B. M. Byrne, 2010). | IFI is also somewhat insensitive to sample size. Thus, it was included as a measure of the goodness of fit (B. M. Byrne, 2010). |
| Standardised root mean square residual (SRMR) | Byrne (2010) recommends the use of SRMR to measure the goodness of fit of the model. SRMR is the square root of the difference between the residuals of the sampled covariance matrix, and the posited covariance model SRMR varies between the value of 0 and 1 (Hooper, et al., 2008). | SRMR addresses the issue of having two instruments with different scales (6-point and 7-point Likert scales for PVQ and PCET, respectively) by standardising the value of RMR. As two different scales were used in this research, the inclusion of this measure was considered important for the purposes of this study (Hooper, et al., 2008). |
| Root Mean Square Error Approximation (RMSEA) | RMSEA estimates the lack of model fit per a degree of freedom (Gefen, et al., 2011). Usage of this index is preferable as it tends to be consistent when the maximum likelihood (ML) | This index favours model parsimony as it is sensitive to the number of estimated parameters in the model. For this reason, (and others), it is reported here (Gefen, |



| | is the method of estimation. Furthermore, it can detect model mis-specifications and provide an indication of the model's quality (B. M. Byrne, 2010). | et al., 2011; Hooper, et al., 2008). |
|---|---|---|

The main purpose of the assessment of the model is to confirm the constructs' validity by measuring the overall model fit, standardised factor loadings, constructs' reliability, critical ratio (CR), and correlation between the constructs (Bagozzi, et al., 1991; Hair, et al., 2010; Hooper, et al., 2008; Kline, 2010). Model fit indices, including the ratio of Chi-square ($\chi^2$ or CMIN) to the degrees of freedom (df) or $\chi^2/df$, GFI, AGFI and IFI are used in CFA to ascertain convergent and discriminant validity (Gefen, et al., 2000; Hair, et al., 2010; Straub, et al., 2004). These indices of fit jointly measure the level to which the data matches the theoretical model (Weston & Gore, 2006). To assess the model in this study, the following fit indices were used: $\chi^2/df$, GFI, AGFI, CFI, IFI, SRMR, and RMSEA. These fit indices provide a range of approaches to the assessment of the fit of the measurement model (Hair, et al., 2010). Table 8.5 provides a description and the rationale for the use of these fit indices, and Table 8.6 describes the required assessment conditions for the indices.

The model fit indices are affected by large sample sizes and the number of items in the theoretical models. Therefore, the use of fit indices thresholds is debatable, and they are used as a guideline, rather than a confirmation of the model fit (Barrett, 2007; B. M. Byrne, 2010; Hair, et al., 2010; Kline, 2010). Hair et al. (2010, p. 671) stated that "[i]t is simply not practical to apply a single set of cutoff rules that apply for all SEM models of any type." Simple models and smaller sample sizes should apply stricter model indices cut-offs than larger and more complex models. This consideration should be noted, especially with regard to sample sizes greater than 250 and models with more than 30 items higher than 30 (Hair, et al., 2010). For example, Dawes, Faulkner, and Sharp (1998) and Greenspoon and Saklofske (1998) noted that 0.8 is an acceptable GFI cut-off level for complex models with large sample sizes. Blunch (2008) indicated that only CFIs below 0.8 should be seriously considered for model modification. As a



general rule, a GFI, CFI, and IFI closer to 1 provides a better fit of the model to the data.

Conversely, the closer χ2/df, SRMR and RMSEA are to zero, the better (Hooper, et al., 2008).

Lastly, it is not advisable to eliminate more items just to increase the model fit at the expense of

the theoretical integrity (Hair, et al., 2010).

Table 8.6

*SEM Assessment Requirements and Conditions*

| Measure | Recommended criteria | Assessment | Suggested by author(s) |
|---|---|---|---|
| χ2/df (CMIN/df) | <3 is good, <5 is acceptable | Convergent, discriminant validity and model fit | (Brown, 2006; B. M. Byrne, 2010; Hair, et al., 2010) |
| GFI, AGFI, IFI and CFI | GFI, IFI and CFI >0.95 is superior, >0.90 is good, > 0.80 is tolerable. AGFI > 0.8 is good | | (Barrett, 2007; Dawes, et al., 1998; Gefen, et al., 2000; Greenspoon & Saklofske, 1998; Hair, et al., 2010) |
| SRMR | <0.05 is good <0.1 is acceptable | | (B. M. Byrne, 2010; Gefen, et al., 2011) |
| RMSEA | <0.05 superior fit <0.08 good fit <0.1 acceptable fit | | (B. M. Byrne, 2010; Hooper, et al., 2008) |
| CR | > ∓1.96, significant at the level of p <0.001 | Convergent validity | (B. M. Byrne, 2010; Hooper, et al., 2008) |
| Standardised factor loading for each item | > |0.7| is superior, > |0.50| is good | | (Brown, 2006; B. M. Byrne, 2010; Hair, et al., 2010; Kline, 2010) |
| Correlation between the constructs | <0.85 | Discriminant validity | (Kline, 2010; Weston & Gore, 2006) |

*Note.* GFI=goodness of fit index; AGFI=adjusted goodness of fit; CFI=comparative fit index; IFI=incremental fit index; SRMR=standardised root mean square residual; RMSEA=root mean square error approximation; CR=critical ratio.

Standardised factor loading (a range between -1 and 1) indicates the level to which each

item converges with the specified construct (Hair, et al., 2010). Significant loadings of items

(above |0.5|, or better, above |0.7|) on their designated constructs indicate the convergent validity

of the construct. CR is an additional indicator of convergent validity. CR represents an

estimated parameter (e.g., item loading) divided by the standardised error. The z-statistic CR



has a mean of 0 and a standard deviation of 1 (Kline, 2010). A CR of $>\pm1.96$ indicates that a parameter estimate is significant ($p < 0.05$). For the purposes of convergent validity, a significant CR provides further support that the loading of an item on its specified construct is noteworthy (B. M. Byrne, 2010).

Discriminant validity can be confirmed by ensuring that there are no correlations above 0.85 between the constructs (Kline, 2010). The presence of a pair of constructs with a high correlation means that they represent the same concept (i.e., they are redundant). Such redundancy would weaken the results of the analysis. Thus, it is advisable that these constructs are eliminated or merged (Kline, 2010; Tabachnick & Fidell, 2007). Table 8.6 summarises the required threshold values for the statistical concepts discussed above. These validation assessment criteria were applied to each construct and submodel and to the overall structural model.



Figure 8.3

*Structural equation model for the hypothesised overall research model.*

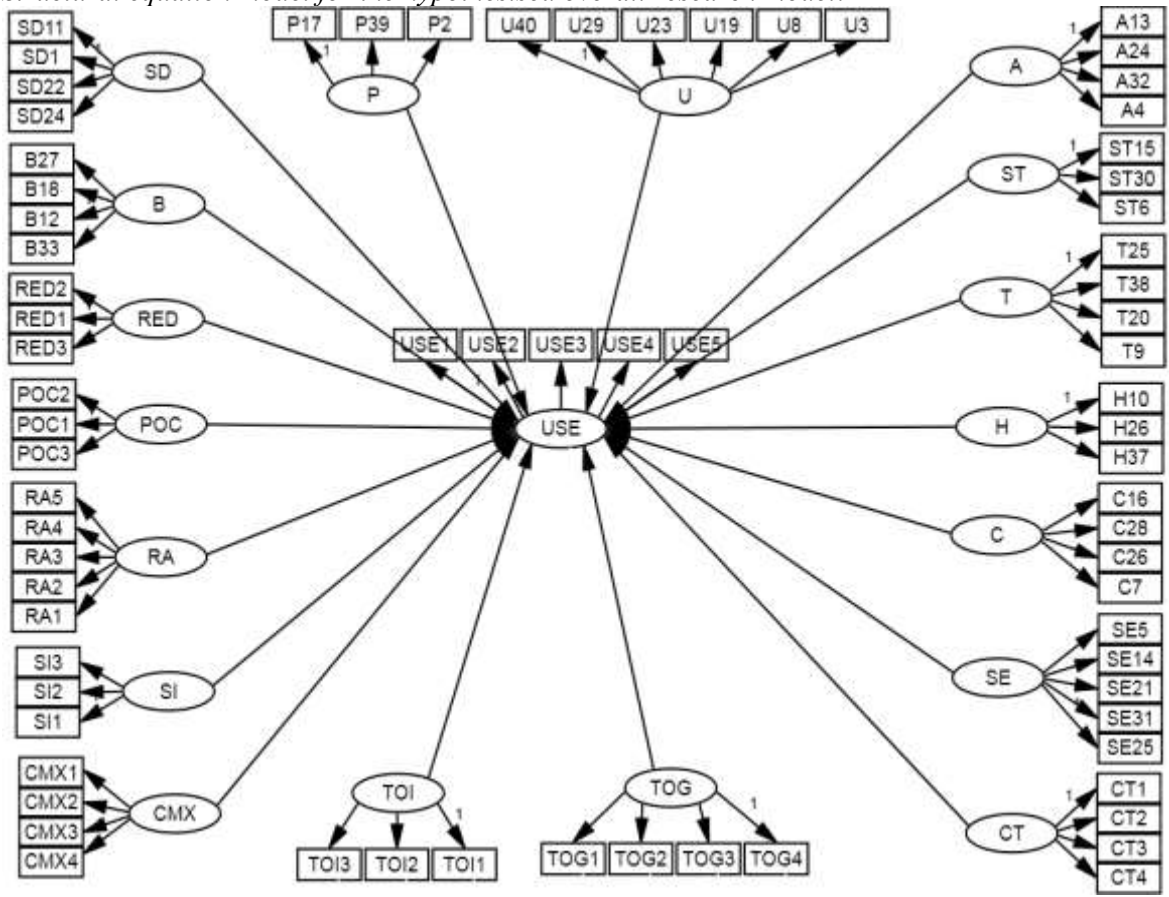

*Note:* RA=relative advantage; CT=compatibility; CMX=complexity; RED=result demonstrability; TI=trust in the Internet; TG=trust in government agencies; SI=social influence; POC=perspective on communication; USE=intention to use e-transactions. SD=self-direction; P=power; U=universalism; A=achievement; SE=security; ST=stimulation; C=conformity; T=tradition; H=hedonism; B=benevolence. Appendix O and Appendix P lists item wording and codes.

The preliminary structural and measurement models for the overall research model are shown in Figure 8.3, with the error and residual terms omitted for the sake of clarity. The following section describes the CFA assessment.

### 8.5.1.1    Confirmatory Factor Analysis (CFA)

CFA is part of the SEM statistical analysis technique (B. M. Byrne, 2010; Hair, et al., 2010). Hence, the same principles and assessment conditions of SEM apply to CFA. Therefore,



the assessment requirements outlined in Table 8.6 were used in CFA. SEM, its principles and its assessment conditions were already described in detail in section 8.5.1.

The use of CFA models enables confirmation of potential loadings between items and their corresponding constructs (Ahire & Devaraj, 2001). Many studies of unexplored domains use EFA to define the construct and its related items where a theory does not exist or has not been empirically validated. Hair et al. (2010) noted that EFA relies on statistical evidence to relate items (observable variables) and constructs (latent variables) where knowledge is lacking, whereas CFA relies on theory and then confirms or rejects it. The latter approach was adopted in this study at this stage, with empirically validated constructs and their items adopted from previous studies.

CFA models were used to confirm proposed relationships between constructs and their items (measurement theory) using the statistical Analysis of Moment Structure (AMOS) software Version 19.0 (B. M. Byrne, 2010). The CFA model, the measurement model, which evaluates the relationship between the items and the constructs, was assessed. The structural model, which assesses the relationship between the constructs, is not assessed in the CFA. SEM was applied to the structural and measurement models, with Maximum Likelihood (ML) analysis used to estimate the model's parameters. ML is the default and most widely used estimation technique for SEM in AMOS (Blunch, 2008; B. M. Byrne, 2010; Kline, 2010). ML was adopted because it is considered a reliable, stable, and robust technique for reducing the effect of the violations of normality assumptions (Hair, et al., 2010).

CFA assesses the measurement theory by utilising empirical evidence of the validity of items based on the model's overall fit and the construct's validity (Hair, et al., 2010). CFA can assess the validity of the measurement theory based on empirical evidence of the model's fit. A measurement theory is defined as the set of theoretical, logical, and systematic assumptions that suggest paths between latent constructs and items. In this study, the measurement theory is the research model, which is based on the theories discussed in Chapter 5. The paths between the



constructs and the items can be assessed using standardised loadings estimates. If a CFA model

fits the data poorly, then the proposed model can be re-specified and re-estimated. Model

modification is a common CFA practice, in which the model is modified to better represent and

fit the data (B. M. Byrne, 2010; Hair, et al., 2010; Kline, 2010). CFA model for each construct

is assessed in the following section.

### 8.5.1.1.1    *CFA at the Individual-Construct Level*

Ahire and Devaraj (2001) recommended that, when using CFA, the assessment of each

construct should be undertaken separately to enable the evaluation of each construct's GFI; if

the GFI is lower than 0.9, item(s) with the lowest loadings should be eliminated. As shown in

Table 8.7 below, all of the constructs appeared to have a good or an acceptable fit to the data

and to have convergent and discriminant validity. However, the $\chi 2/df$ for some of the constructs

(RA, CT, CMX, TG, A, and SE) was larger than the conditional level (B. M. Byrne, 2010).

Additionally, the RMSEA of the constructs RA, CT, CMX, TG, and A was above the

acceptable level. It is important to note that when the GFI = 1, AGFI cannot be calculated

(Schumacker & Lomax, 1996). Therefore, the AFGI values are not shown in the AMOS output

report for the following constructs: RED, TI, SI, POC, P, ST, and H. The fit indices for these

constructs reflected a perfect fit (Hair, et al., 2010).

Table 8.7
*Goodness of Fit Measurement for All of the Individual Constructs (N = 671)*

| Construct | $\chi2/df$ (CMIN/df) <5 | GFI <0.9 | AGFI <0.8 | CFI <0.9 | IFI <0.9 | SRMR <0.1 | RMSEA <0.1 |
|---|---|---|---|---|---|---|---|
| Relative advantage (RA) | 10.25 | 0.97 | 0.91 | 0.98 | 0.98 | 0.02 | 0.11 |
| Compatibility (CT) | 20.70 | 0.97 | 0.85 | 0.98 | 0.98 | 0.03 | 0.17 |
| Complexity (CMX) | 10.00 | 0.99 | 0.93 | 0.96 | 0.96 | 0.03 | 0.11 |
| Result demonstrability (RED) | 0.00 | 1.00 | - | 1.00 | 1.00 | 0.00 | - |
| Trust in the Internet (TI) | 0.00 | 1.00 | - | 1.00 | 1.00 | 0.00 | - |
| Trust in government agencies (TG) | 20.51 | 0.97 | 0.85 | 0.98 | 0.98 | 0.01 | 0.17 |
| Social influence (SI) | 0.00 | 1.00 | - | 1.00 | 1.00 | 0.00 | - |
| Perspective on communication (POC) | 0.00 | 1.00 | - | 1.00 | 1.00 | 0.00 | - |
| Intention to use e-transactions (USE) | 3.55 | 0.99 | 0.97 | 0.99 | 0.99 | 0.01 | 0.06 |



| Self-direction (SD) | 3.90 | 0.99 | 0.97 | 0.97 | 0.97 | 0.02 | 0.07 |
|---|---|---|---|---|---|---|---|
| Power (P) | 0.00 | 1.00 | - | 1.00 | 1.00 | 0.00 | - |
| Universalism (U) | 3.45 | 0.99 | 0.97 | 0.97 | 0.96 | 0.03 | 0.06 |
| Achievement (A) | 47.53 | 0.93 | 0.64 | 0.87 | 0.87 | 0.07 | 0.26 |
| Security (SE) | 6.91 | 0.98 | 0.94 | 0.94 | 0.94 | 0.04 | 0.09 |
| Stimulation (ST) | 0.00 | 1.00 | - | 1.00 | 1.00 | 0.00 | - |
| Conformity (C) | 2.59 | 0.99 | 0.98 | 0.98 | 0.99 | 0.01 | 0.05 |
| Tradition (T) | 2.81 | 0.99 | 0.98 | 0.98 | 0.98 | 0.02 | 0.05 |
| Hedonism (H) | 0.00 | 1.00 | - | 1.00 | 1.00 | 0.00 | - |
| Benevolence (B) | 1.72 | 0.99 | 0.99 | 0.99 | 0.99 | 0.01 | 0.03 |

*Note.* N=the number of participants; χ2=Chi-square; df=degrees of freedom; GFI=goodness of fit index; AGFI=adjusted goodness of fit; CFI=comparative fit index; IFI=incremental fit index; SRMR=standardised root mean square residual; RMSEA=root mean square error approximation

Taking into consideration how $\chi2/df$ inflates with a larger sample size and less constructs and items, all of the other indicators were checked (Hair, et al., 2010). For constructs with an RMSEA above 0.1, all of the other fit indices for these constructs indicated superior fit levels (approaching 1).Overall, the fit indices for all of the constructs appeared to exhibit an excellent level of fit. Thus, no changes were required for any construct (Gefen, et al., 2011).

Achievement was the only construct that had multiple fit indices lower than the cut-off values: CFI = 0.87, AGFI = 0.64, IFI = 0.87 and RMSEA = 0.26. Considering that the GFI value (0.93) was high and that there was only a small difference between the threshold and the construct's CFI, IFI and RMSEA, no items were removed from the achievement construct at this stage. According to Ahire and Devaraj (2001), GFI is the most important condition for construct-level CFA, and the GFI for all of the constructs was good. Therefore, all of the constructs were considered to reflect the fit indices' criteria for acceptable convergent and discriminant validity. In addition to assessments at the level of the construct, the CFA assessment was conducted at the level of the model (PCET and BPV).

### 8.5.1.1.2   *CFA for PCET model*

CFA was conducted at the model level to ascertain that the PCET model was adequately validated. The results revealed that this model showed a good fit and good validity: $\chi2/df$ = 2.18, GFI = 0.92, AGFI = 0.90, CFI = 0.96, IFI = 0.96, SRMR = 0.04 and, RMSEA = 0.04. In



addition to the fit indices, to further assess the convergent and discriminant validity, CR, standardised factor loadings and correlations between the constructs were used. As shown in Figure 8.4 and Table 8.8 (in the column labelled 'construct correlation'), all of the correlations were less than the threshold (0.85), indicating acceptable discriminant validity. Although less than the threshold, the close correlation (0.83) between RA and CT is acceptable, taking into consideration the conceptual closeness between the two constructs. Additionally, the discriminant validity is not a concern, especially given that the results of the EFA (Table 8.3) revealed strong loadings for items on their corresponding construct and insignificant cross-loading of items between the two constructs. The CMX items were negatively worded; therefore, CMX was negatively correlated with all other constructs. The negative correlations of CMX can be considered an indication that the CFA model's parameters are viable and estimated correctly as theorised (B. M. Byrne, 2010).



Figure 8.4
*CFA model of the PCET model.*

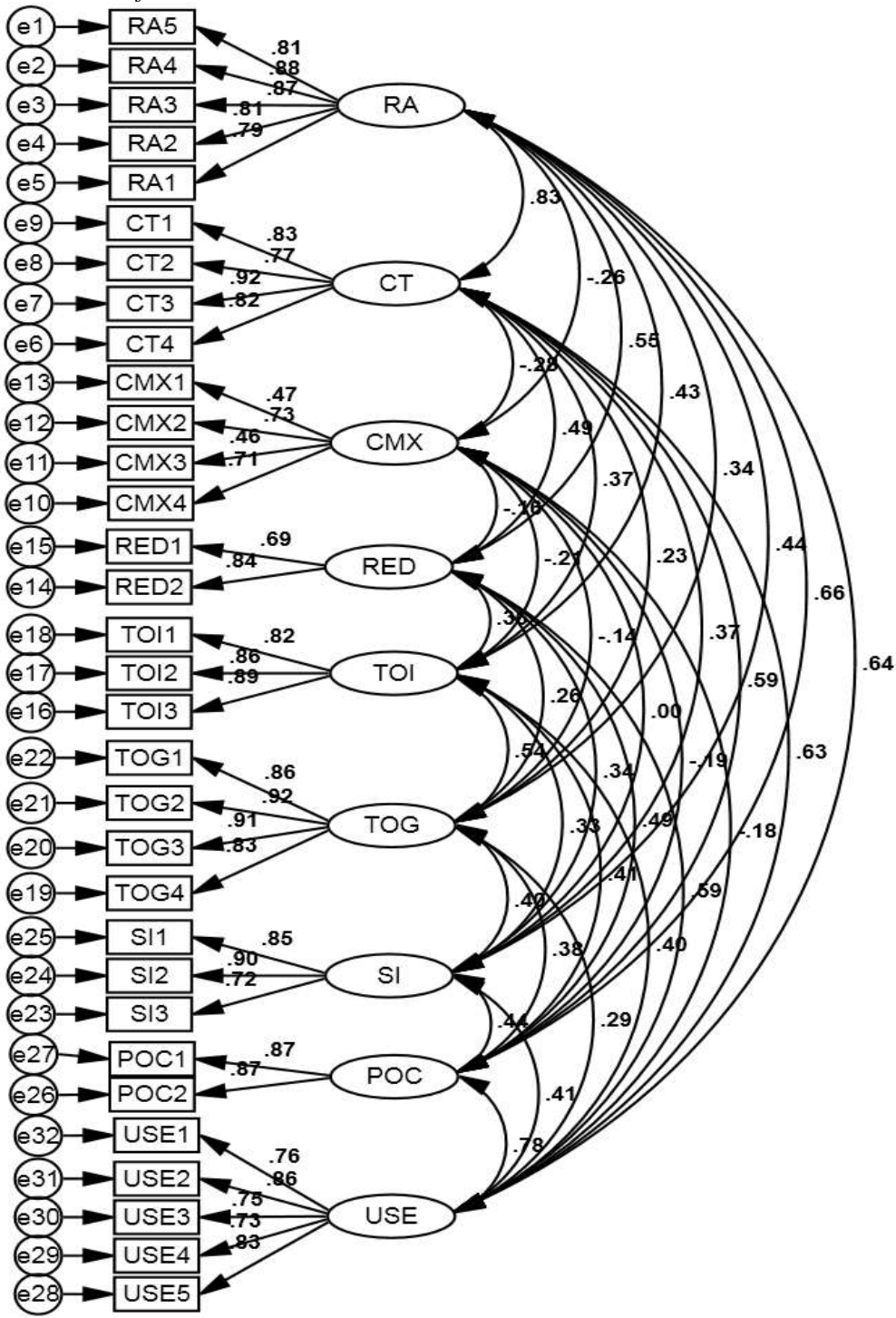

*Note:* RA=relative advantage; CT=compatibility; CMX=complexity; RED=result demonstrability; TI=trust in the Internet; TG=trust in government agencies; SI=social influence; POC=perspective on communication; USE=intention to use e-transactions. Item wording and codes are specified in Appendix O.

Table 8.8 includes the standardised factor loading for items on their corresponding construct, in addition to CR, CR significance, construct correlations, and the model fit indices.



Table 8.8
*Results of the CFA of PCET Model*

| Item | Loading | CR | P | Constructs' correlations | | | |
|------|---------|-----|-----|------|-----|------|------|
| **Relative advantage** | | | | RA | ←→ | CT | 0.83 |
| RA5 | 0.81 | 25.04 | *** | RA | ←→ | CMX | -0.26 |
| RA4 | 0.88 | 27.25 | *** | RA | ←→ | RED | 0.55 |
| RA3 | 0.87 | 26.80 | *** | RA | ←→ | TI | 0.43 |
| RA2 | 0.81 | 24.26 | *** | RA | ←→ | TG | 0.34 |
| RA1 | 0.79 | 23.31 | *** | RA | ←→ | SI | 0.44 |
| | | | | RA | ←→ | POC | 0.66 |
| | | | | RA | ←→ | USE | 0.64 |
| **Compatibility** | | | | CT | ←→ | CMX | -0.28 |
| CT4 | 0.83 | 25.64 | *** | CT | ←→ | RED | 0.49 |
| CT3 | 0.92 | 29.83 | *** | CT | ←→ | TI | 0.37 |
| CT2 | 0.77 | 22.98 | *** | CT | ←→ | TG | 0.23 |
| CT1 | 0.83 | 25.48 | *** | CT | ←→ | SI | 0.37 |
| | | | | CT | ←→ | POC | 0.59 |
| | | | | CT | ←→ | USE | 0.63 |
| **Complexity** | | | | CMX | ←→ | RED | -0.16 |
| CMX4 | 0.71 | 16.69 | *** | CMX | ←→ | TI | -0.21 |
| CMX3 | 0.46 | 9.47 | *** | CMX | ←→ | TG | -0.14 |
| CMX2 | 0.73 | 12.03 | *** | CMX | ←→ | SI | 0.00 |
| CMX1 | 0.47 | 9.72 | *** | CMX | ←→ | POC | -0.19 |
| | | | | CMX | ←→ | USE | -0.18 |
| **Result demonstrability** | | | | RED | ←→ | TI | 0.36 |
| RED2 | 0.84 | 20.31 | *** | RED | ←→ | TG | 0.26 |
| RED1 | 0.70 | 13.19 | *** | RED | ←→ | SI | 0.34 |
| | | | | RED | ←→ | POC | 0.49 |
| | | | | RED | ←→ | USE | 0.59 |
| **Trust in the Internet** | | | | TI | ←→ | TG | 0.54 |
| TI3 | 0.89 | 28.20 | *** | TI | ←→ | SI | 0.33 |
| TI2 | 0.86 | 28.31 | *** | TI | ←→ | POC | 0.41 |
| TI1 | 0.82 | 26.27 | *** | TI | ←→ | USE | 0.41 |
| **Trust in government agencies** | | | | TG | ←→ | SI | 0.40 |
| TG4 | 0.83 | 25.85 | *** | TG | ←→ | POC | 0.38 |
| TG3 | 0.91 | 30.27 | *** | TG | ←→ | USE | 0.29 |
| TG2 | 0.92 | 30.75 | *** | | | | |
| TG1 | 0.86 | 27.22 | *** | | | | |
| **Social influence** | | | | SI | ←→ | POC | 0.44 |
| SI3 | 0.72 | 20.60 | *** | SI | ←→ | USE | 0.41 |
| SI2 | 0.90 | 20.95 | *** | | | | |
| SI1 | 0.85 | 20.53 | *** | | | | |
| **Perspective on communication** | | | | | | | |
| POC2 | 0.87 | 26.77 | *** | | | | |
| POC1 | 0.87 | 25.56 | *** | Shown in other arrangements in the | | | |
| **Usage intention** | | | | | | | |



| USE5 | 0.83 | 25.73 | *** | column: 'Constructs' correlations' |
| USE4 | 0.73 | 21.04 | *** | |
| USE3 | 0.75 | 21.84 | *** | |
| USE2 | 0.86 | 26.67 | *** | |
| USE1 | 0.76 | 22.36 | *** | |
| Model Fit Indices: $\chi2/df$ = 2.18; GFI = 0.92; AGFI=0.90; CFI = 0.96; IFI = 0.96; SRMR = 0.04; RMSEA = 0.04, *** significant at the level p <0.001 | | | | |

*Note.* RA=relative advantage; CT=compatibility; CMX=complexity; RED=result demonstrability; TI=trust in the Internet; TG=trust in government agencies; SI=social influence; POC=perspective on communication; USE=intention to use e-transactions; GFI=goodness of fit index; AGFI=adjusted goodness of fit; CFI=comparative fit index; IFI=incremental fit index; SRMR=standardised root mean square residual; RMSEA=root mean square error approximation. Items codes and wording are shown in Appendix O.

All construct-to-items loadings were above the previously identified required threshold (0.5) except for CMX1 and CMX3 which had loadings very close to 0.5. The CR of each item was more than ±1.96 with high significance (p < 0.001). Both of these indicators suggest convergent validity of the model. As all of the indicators discussed were acceptable, the PCET model was considered to establish discriminant and convergent validity. The following section discusses the empirical requirements for discriminant and convergent validity of the BPV model.

### 8.5.1.1.3   CFA for Schwartz's BPV model

The proposed CFA model was developed according to the BPV model measurement theory. Determining whether the measurement model fit the Saudi sample data was a concern because Schwartz's BPV model has not been validated using a sample of Saudi citizens (Bardi & Guerra, 2010; Piurko, Schwartz, & Davidov, 2011; Schwartz, 2003). The CFA model of the BPV shown in Figure 8.5 indicated an unacceptable fit: $\chi2/df$ = 3.5, GFI = 0.82, AGFI = 0.79, CFI = 0.79, IFI = 0.79, SRMR = 0.06, and RMSEA = 0.06. Additionally, the model exhibited a lack of validity due to the high (> 0.85) correlations between several constructs and a lower-than-acceptable standardised loading for the other items (< 0.5). A cut-off point of 0.5 was considered acceptable for item loading, considering that each value's "items sought



coverage of the conceptual breadth of each value rather than homogeneity of the items that operationalized each value" (Beierlein et al., 2012, p. 34).

The CFA model was then re-specified to fit the data, in accordance with BPV theory (B. M. Byrne, 2010; Hair, et al., 2010). Table 8.9 points out (in italics) high correlations and low item loadings. The CFA model for BPV is shown in Figure 8.5.

Table 8.9
*Results of the CFA of BPV Model*

| Item | Loading | CR | P | Constructs' correlations | | | |
|------|---------|-----|-----|-----|-----|-----|-----|
| **Achievement** | | | | A | ←→ | SD | 0.77 |
| *A4* | *0.46* | *11.39* | *** | A | ←→ | ST | 0.50 |
| A13 | 0.68 | 10.40 | *** | A | ←→ | SE | 0.65 |
| A24 | 0.73 | 10.66 | *** | A | ←→ | H | 0.55 |
| A32 | 0.68 | 10.40 | *** | A | ←→ | B | 0.65 |
| | | | | A | ←→ | T | 0.34 |
| | | | | A | ←→ | C | 0.54 |
| **Benevolence** | | | | B | ←→ | T | 0.70 |
| B12 | 0.64 | 16.76 | *** | B | ←→ | C | 0.80 |
| B18 | 0.57 | 12.14 | *** | | | | |
| B27 | 0.61 | 12.81 | *** | | | | |
| B33 | 0.52 | 11.35 | *** | | | | |
| **Conformity** | | | | | | | |
| C7 | 0.50 | 13.05 | *** | Shown in other arrangements in the column 'Constructs' correlations' | | | |
| C16 | 0.55 | 10.58 | *** | | | | |
| C36 | 0.70 | 12.10 | *** | | | | |
| C28 | 0.57 | 10.86 | *** | | | | |
| **Hedonism** | | | | H | ←→ | B | 0.42 |
| H10 | 0.66 | 16.76 | *** | H | ←→ | T | 0.24 |
| H26 | 0.65 | 12.86 | *** | H | ←→ | C | 0.33 |
| H37 | 0.76 | 13.78 | *** | | | | |
| **Power** | | | | P | ←→ | A | 0.61 |
| *P2* | *0.24* | *5.56* | *** | P | ←→ | SD | 0.57 |
| P17 | 0.72 | 5.44 | *** | P | ←→ | ST | 0.51 |
| P39 | 0.83 | 5.44 | *** | P | ←→ | SE | 0.29 |
| | | | | P | ←→ | H | 0.32 |
| | | | | P | ←→ | B | 0.25 |
| | | | | P | ←→ | T | 0.25 |
| | | | | P | ←→ | C | 0.24 |
| **Self-direction** | | | | SD | ←→ | ST | 0.61 |
| SD1 | 0.52 | 13.00 | *** | SD | ←→ | SE | 0.73 |
| SD11 | 0.51 | 9.80 | *** | SD | ←→ | H | 0.53 |
| SD22 | 0.61 | 10.97 | *** | SD | ↔ | B | 0.72 |
| SD24 | 0.47 | 9.33 | *** | SD | ↔ | T | 0.59 |
| | | | | SD | ↔ | C | 0.60 |
| **Security** | | | | SE | ↔ | H | 0.38 |
| SE5 | 0.52 | 13.76 | *** | SE | ↔ | B | 0.71 |
| SE14 | 0.55 | 10.88 | *** | SE | ↔ | T | 0.80 |
| SE21 | 0.57 | 11.15 | *** | *SE* | ↔ | *C* | *0.95* |
| SE25 | 0.62 | 11.76 | *** | | | | |
| SE31 | 0.52 | 10.47 | *** | | | | |
| **Stimulation** | | | | ST | ↔ | SE | 0.19 |
| ST6 | 0.46 | 10.95 | *** | ST | ↔ | H | 0.61 |
| ST15 | 0.70 | 9.83 | *** | ST | ↔ | B | 0.47 |
| ST30 | 0.67 | 9.71 | *** | ST | ↔ | T | 0.20 |
| | | | | ST | ↔ | C | 0.07 |
| **Tradition** | | | | *T* | ↔ | *C* | *0.89* |
| *T9* | *0.35* | *7.77* | *** | | | | |
| *T20* | *0.42* | *6.36* | *** | | | | |
| *T25* | *0.31* | *5.43* | *** | | | | |
| T38 | 0.53 | 6.94 | *** | | | | |
| **Universalism** | | | | U | ↔ | P | 0.25 |
| *U3* | *0.31* | *7.77* | *** | U | ↔ | A | 0.56 |
| U8 | 0.54 | 7.08 | *** | U | ↔ | SD | 0.78 |
| U19 | 0.58 | 7.23 | *** | U | ↔ | ST | 0.29 |
| U23 | 0.56 | 7.17 | *** | *U* | ↔ | *SE* | *0.97* |
| U29 | 0.64 | 7.42 | *** | U | ↔ | H | 0.34 |
| U40 | 0.60 | 7.29 | *** | *U* | ↔ | *B* | *0.85* |
| | | | | U | ↔ | T | 0.80 |
| | | | | *U* | ↔ | *C* | *0.96* |

Model Fit Indices: $\chi^2/df$ = 3.50; GFI = 0.82; AGFI = 0.79; CFI = 0.79; IFI = 0.79; SRMR = 0.06; RMSEA = 0.06, *** $p < 0.001$

*Note.* SD=self-direction; P=power; U=universalism; A=achievement; SE=security; ST=stimulation; C=conformity; T=tradition; H=hedonism; B=benevolence; CR=Critical Ratio; $\chi^2$=Chi-square; df=degrees of freedom; GFI=goodness of fit index; AGFI=adjusted goodness of fit; CFI=comparative fit index; IFI=incremental fit index; SRMR=standardised root mean square residual; RMSEA=root mean square error approximation. Items codes and wordings are shown in Appendix P.

Considering the problematic values in Table 8.9 above, the Generalised Least Squares (GLS) estimation technique was tested in addition to ML using AMOS. It was noticed that the GLS solutions improved $\chi^2/df$, GFI, and AGFI, while CFI, IFI, and SRMR were poorer than the previous run using the ML estimation technique. Running the model using GLS provided the following fit indices: $\chi^2/df$ = 2.3; GFI = 0.88; AGFI = 0.88; CFI = 0.33; IFI = 0.47; SRMR =



0.07; RMSEA = 0.04. Therefore, GLS was not an optimal solution and using ML as an estimation technique was sustained for further modifications of the model and for its previously described advantages (section 8.5.1.1). Other estimation techniques such as Unweight Least Square or Asymptotically distribution-free techniques were not considered as a they require extremely large sample sizes (in thousands) (Forero, Maydeu-Olivares, & Gallardo-Pujol, 2009; Gefen, et al., 2011).



Figure 8.5
*Preliminary CFA model of BPV.*

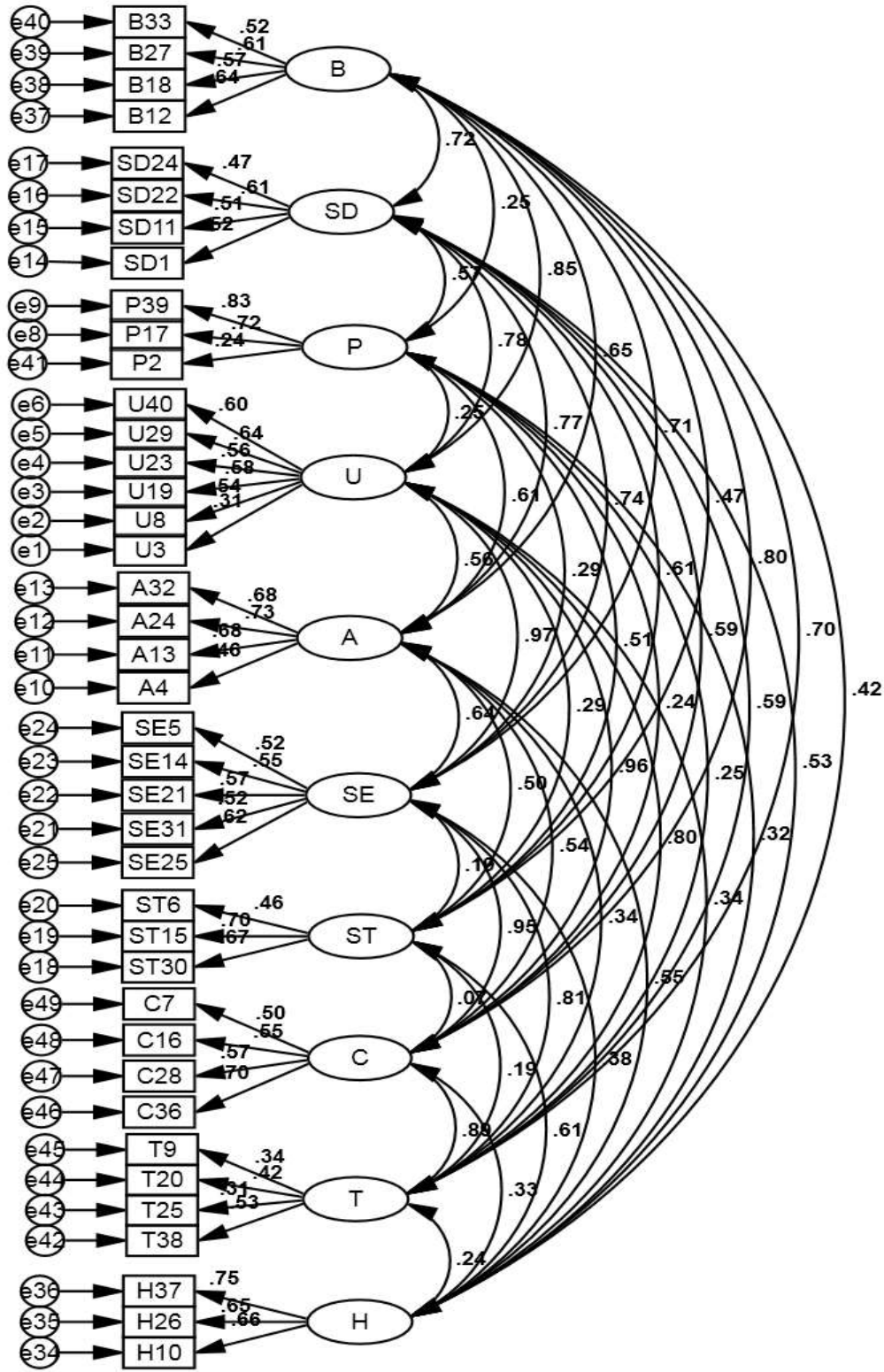

*Note:* SD=self-direction; P=power; U=universalism; A=achievement; SE=security;
ST=stimulation; C=conformity; T=tradition; H=hedonism; B=benevolence. Appendix P shows
item wording and codes.



A postulated (Confirmatory Factor Analysis) CFA model should be modified only when there is a lack of fit and multiple problematic parameter estimates (Blunch, 2008; B. M. Byrne, 2010). Based on the aforementioned empirical evidence (i.e., fit indices, standardised estimates and the construct's correlations), it was decided to modify and re-estimate the model. To overcome the lack of model validity, the approach adopted was similar to model re-specification procedures suggested by Hair et al. (2010), Byrne (2010), and Kline (2010). The re-specification of the CFA model was undertaken in accordance with accepted theory and based on empirical evidence (B. M. Byrne, 2010; Hair, et al., 2010; Kline, 2010). Twenty-five iterations of the model re-estimation were conducted for each modification to the CFA, and the standardised residuals, correlations and loadings were noted where they improved at each re-estimation. The resulting modifications to and adjustments of the CFA model are highlighted below.

One approach that can be employed to improve CFA model fit indices is to use modification indices (MIs) to correlate measurement errors, such as, (e4) and (e5) in Figure 8.5. MIs are possible relationships that are not estimated in the model. Additionally, the value of the MIs represents the expected drop in the chi-square value, which indicates a better fit if the parameter is identified by the MIs (B. M. Byrne, 2010). However, this approach is not recommended for modifying models (Gefen, et al., 2011; Hair, et al., 2010; Kline, 2010). Adding these correlations between measurement errors would only mask the lack of validity in the model (Hair, et al., 2010; Kline, 2010). Therefore, no measurement errors were correlated for any of the models.

A high intercorrelation between independent constructs is referred to as multicollinearity. Addressing multicollinearity is the first priority in model modification because it causes redundancy between highly correlated constructs and produces spurious relationships between dependent and independent constructs (Blunch, 2008; Grewal, Cote, & Baumgartner, 2004; Hair, et al., 2010). Constructs are considered highly correlated if correlations are greater than 0.85 (Kline, 2010). According to Tabachnick and Fidell (2007),



multicollinearity causes highly correlated constructs to be redundant and, as a result, weakens the analysis. Multicollinearity suggests that at least one of the redundant constructs is dispensable. Table 8.9 displays high correlations between the following constructs: universalism, benevolence, security, conformity, and tradition. The correlation between benevolence and universalism was expected, given the conceptual similarity between these two constructs and because they both belong to the higher-order self-transcendence value. However, the high correlation between universalism and the other constructs (conservation values) was unexpected, taking the theoretical difference between universalism and conservation values (security, tradition, and conformity) into consideration. These very high correlations between the benevolence, universalism and conservation values are not desirable because they cause redundancy. There were also high correlations between the conservation values.

There are three possible approaches to reduce the likelihood of multicollinearity: (1) remove the universalism construct from the CFA model only, (2) combine the three values (security, tradition, and conformity) into the conservation higher-order value without removing universalism, or (3) do both.

Option (1), the removal of universalism, slightly improved some of the fit indices (Table 8.11). However, as shown in Table 8.10 below, high correlations remained between the conservation values.

Table 8.10
*Correlations between Conservation Values when Universalism is removed*

| Security | ←→ | Conformity | 0.951 |
| Tradition | ←→ | Conformity | 0.885 |
| Security | ←→ | Tradition | 0.811 |

Option (2), combining the three values into the conservation higher order value without removing universalism, also did not solve the problem; a high correlation (0.96) remained between the higher order value of conservation (CON) and universalism. The fit indices



improved only slightly, in contrast to the original CFA model. Table 8.11 below lists the fit

indices of the preliminary CFA models based on options (1), (2), and (3) for comparison.

Table 8.11
*Fit Indices for Each Proposed Option to Reduce Multicollinearity*

| CFA model | Fit Indices | Notes |
|---|---|---|
| Preliminary CFA model | χ2/df = 3.50; GFI = 0.82; AGFI = 0.79; CFI = 0.79; IFI =0.79; SRMR =0.06; RMSEA = 0.06 | Initial full BPV model |
| Option (1) solution | χ2/df = 3.79; GFI = 0.84; AGFI = 0.80; CFI =0.79; IFI = 0.79; SRMR =0.06; RMSEA = 0.07 | Deleting universalism |
| Option (2) solution | χ2/df = 3.81; GFI = 0.81; AGFI = 0.79; CFI =0.75; IFI = 0.76; SRMR =0.08; RMSEA = 0.07 | Combination of conformity, tradition and security |
| Option (3) solution | χ2/df = 4.18; GFI = 0.83; AGFI = 0.80; CFI =0.75; IFI = 0.76; SRMR =0.08; RMSEA = 0.07 | Combination of conformity, tradition and security and deleting universalism. |

*Note.* χ2=Chi-square; df=degrees of freedom; GFI=goodness of fit index; AGFI=adjusted
goodness of fit; CFI=comparative fit index; IFI=incremental fit index; SRMR=standardised
root mean square residual; RMSEA=root mean square error approximation

The third solution (3) was the optimal in terms of eliminating multicollinearity. High

intercorrelations between the constructs were absent; all the correlations were less than 0.77.

Although the fit indices were slightly different and there was no clear improvement using any of

the solutions, the elimination of multicollinearity was the main concern (B. M. Byrne, 2010;

Hair, et al., 2010). Thus, the three values (conformity, tradition, and security) were combined

(as shown in Figure 8.6) into what was defined by Schwartz and colleagues (1992; 2001) as a

higher value: conservation. In addition, the universalism construct was removed from this CFA

model. These two adjustments to the model took into account both the existing theory and the

empirical evidence (B. M. Byrne, 2010). The elimination of the value universalism can be

compensated by the inclusion of the value benevolence, both of which have underlying concept

of self-transcendence and caring for the welfare of others. The value benevolence is preferable

to universalism in this instance because it can provide better parameter estimates. This was

empirically verified firstly by comparing the standardised loadings for benevolence and

universalism constructs (as shown in Table 8.9 above) and secondly by comparing the fit



indices from the results of the CFA model for each construct (as shown in Table 8.7, section

8.5.1.1.1). For the purposes of comparison, these results are compiled in Table 8.12.

Table 8.12

*Comparison between Universalism's and Benevolence's Standardised Loadings and Fit Indices*

| Universalism (U) | | Benevolence (B) | |
|---|---|---|---|
| Standardised item loadings: | | Standardised item loadings: | |
| U3 | 0.31 | B12 | 0.64 |
| U8 | 0.54 | B18 | 0.57 |
| U19 | 0.58 | B27 | 0.61 |
| U23 | 0.56 | B33 | 0.52 |
| U29 | 0.64 | | |
| U40 | 0.60 | | |
| Fit indices: <br> $\chi 2/df$ = 3.45, GFI = 0.99, <br> AGFI = 0.97, CFI = 0.97, <br> IFI = 0.96; SRMR = 0.03; <br> RMSEA = 0.06 | | Fit Indices: <br> $\chi 2/df$ = 1.72 , GFI = 0.99, <br> AGFI = 0.99, CFI = 0.99, <br> IFI = 0.99; SRMR= 0.01; <br> RMSEA = 0.03 | |

Table 8.12 shows that the benevolence value provides better fit indices and standardised

loadings, with no item being lower than the threshold (0.5). In contrast, the universalism value

has one item below the acceptable the threshold (0.5). The combination of the values (tradition,

conformity, and security) into the higher-order value of conservation is also theoretically

justified as Schwartz likewise grouped these values to form the conservation value (Schwartz &

Boehnke, 2004; Schwartz, et al., 2001). Although multicollinearity was eliminated, additional

re-specification of the CFA model was necessary.

Items with standardised loadings lower than the threshold (0.5) were eliminated,

including A4, P2, T9, and T25. However, the item T20 (loading 0.42) was retained to capture

the conceptual essence of the tradition value. Even after these modifications, the fit indices were

still subpar ($\chi 2/df$ = 3.99, GFI = 0.85, AGFI = 0.81, CFI = 0.80, IFI = 0.80, SRMR = 0.06, and

RMSEA = 0.07). Therefore, additional modifications were considered to improve the fit indices

to a tolerable level while adhering to the BPV theory.



An exploratory diagnostic measure that involves the use of standardised residuals offers another method of improving the model (Hair, et al., 2010). These residuals refer to the difference between the estimated covariance terms and the observed covariance terms. The standardised residuals are obtained by dividing the residuals by the standard error of the residuals. These standardised residuals can be negative or positive, depending on whether the estimated covariance is larger or smaller than the observed covariance. Items with standardised residuals lower than |2.5| are acceptable, whereas items with standardised residuals more than |4| indicate higher error levels. Therefore, as shown in the covariance matrix in Appendix J, the items (in italics in the table) with the highest residual values were eliminated. The eliminated items included SE5, ST6, SD22, and SD24 (Hair, et al., 2010).

After eliminating the problematic items, at least two items remained for each construct. This approach can again be theoretically justified, as Schwartz (2003) recommended maintaining at least two items for each BPV constructs to ascertain an optimal conceptual coverage of the values. Therefore, there was no possibility that the values with two items (self-direction, power, and stimulation, as shown in Figure 8.6) would not fully capture the broader sense of each value. The value conservation is hereafter abbreviated as CON, as shown in Figure 8.6.



Figure 8.6
*Modified CFA model for BPV.*

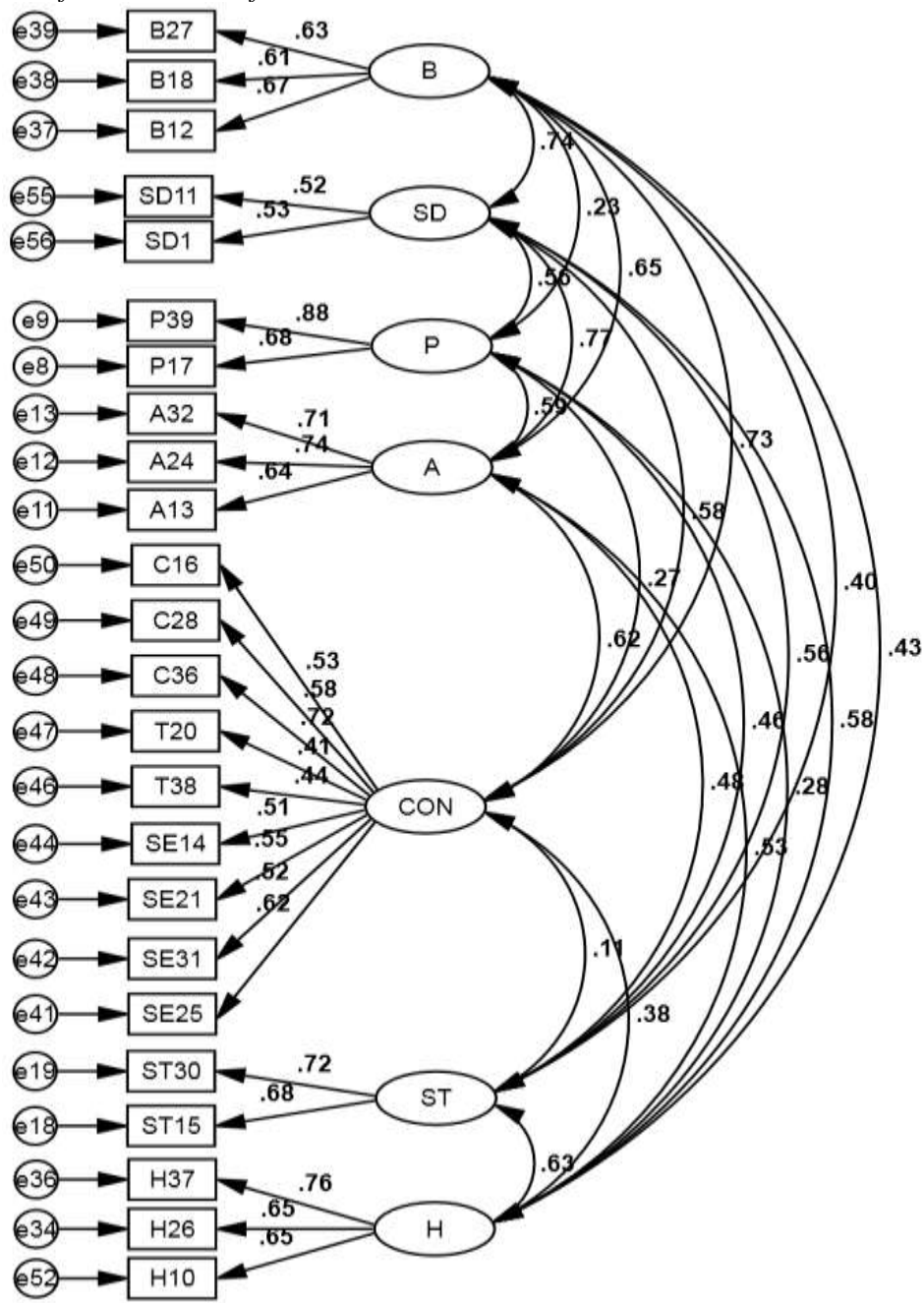

*Note.* P=power; A=achievement; SE=security; ST=stimulation; C=conformity; T=tradition; H=hedonism; B=benevolence CON=conservation values. Appendix P shows item wording and codes

After making the aforementioned changes, the modified model was re-estimated. The

loading (0.48) of item B33 was below the threshold (0.5); consequently, it was eliminated

considering that there are other items to capture this concept (B12, B18 and B28). Although the

loadings of the items T20 ( 0.41) and T38 (0.44) were lower than the threshold, they were



retained since it is the loading values are higher than Hair et al. (2010) second suggested threshold (0.3). Additionally, it was important to retain at least these two items to capture the tradition concept, which is part of the conservation value. The final modified model presented acceptable fit indices: $\chi 2/df$ = 3.5, GFI = 0.90, AGFI = 0.87, CFI = 0.8, IFI = 0.88, SRMR = 0.05, and RMSEA = 0.06. The model fit indices were acceptable, considering that GFI and AGFI surpassed the threshold value (CFI = 0.9 and AGFI > 0.8). The values of CFI and IFI were also acceptable, and those of SRMR and RMSEA were better than the acceptable threshold (Gefen, et al., 2011). In addition, based on the satisfactory items' loadings, CRs, p-values and interconstruct correlations (shown in Table 8.13), this model was believed to offer the best possible discriminant and convergent validity for the Saudi sample. The model fit indices can be improved by removing more items or eliminating outliers. However, it was decided to retain both and to accept the validation level of this model. The elimination of additional items might compromise the theory, and the deletion of outliers might affect the generalisability of the findings to the population. Therefore, no further changes were made to the CFA model (B. M. Byrne, 2010; Hair, et al., 2010; Kline, 2010).

Table 8.13
*Outcome of the Modified CFA Model for BPV*

| Item | Loading | CR | P | Constructs' correlations | | | |
|------|---------|------|-----|------|------|------|------|
| **Achievement** | | | | A | ←→ | SD | 0.77 |
| A13 | 0.64 | 16.68 | *** | A | ←→ | ST | 0.48 |
| A24 | 0.74 | 14.60 | *** | A | ←→ | H | 0.53 |
| A32 | 0.71 | 14.23 | *** | A | ←→ | B | 0.65 |
| | | | | A | ←→ | CON | 0.62 |
| **Benevolence** | | | | B | ←→ | SD | 0.74 |
| B12 | 0.67 | 16.86 | *** | B | ←→ | CON | 0.73 |
| B18 | 0.61 | 12.34 | *** | | | | |
| B27 | 0.63 | 12.76 | *** | | | | |
| **Conservation** | | | | | | | |
| C16 | 0.53 | 13.66 | *** | | | | |
| C36 | 0.72 | 12.38 | *** | | | | |
| C28 | 0.58 | 10.92 | *** | | | | |
| SE14 | 0.51 | 10.10 | *** | Shown in other arrangements in the column 'Constructs' correlations' | | | |
| SE21 | 0.55 | 10.58 | *** | | | | |
| SE25 | 0.62 | 11.48 | *** | | | | |
| SE31 | 0.52 | 10.20 | *** | | | | |



| | | | | | | | | |
|---|---|---|---|---|---|---|---|---|
| T20 | 0.41 | 8.60 | *** | | | | | |
| T38 | 0.44 | 9.12 | *** | | | | | |
| **Hedonism** | | | | H | ←→ | SD | 0.58 | |
| H10 | 0.66 | 16.59 | *** | H | ←→ | B | 0.43 | |
| H26 | 0.65 | 12.83 | *** | H | ←→ | CON | 0.38 | |
| H37 | 0.76 | 13.80 | *** | | | | | |
| **Power** | | | | P | ←→ | A | 0.59 | |
| P17 | 0.68 | 16.66 | *** | P | ←→ | SD | 0.56 | |
| P39 | 0.87 | 12.65 | *** | P | ←→ | ST | 0.46 | |
| | | | | P | ←→ | H | 0.28 | |
| | | | | P | ←→ | B | 0.23 | |
| | | | | P | ←→ | CON | 0.27 | |
| **Self-direction** | | | | SD | ←→ | CON | 0.58 | |
| SD1 | 0.68 | 11.19 | *** | | | | | |
| SD11 | 0.88 | 9.28 | *** | | | | | |
| **Stimulation** | | | | ST | ←→ | SD | 0.56 | |
| ST15 | 0.68 | 16.35 | *** | ST | ←→ | H | 0.63 | |
| ST30 | 0.72 | 12.30 | *** | ST | ←→ | B | 0.40 | |
| | | | | ST | ←→ | CON | 0.11 | |
| Model fit indices: χ2/df = 3.50; GFI = 0.90, AGFI = 0.87; CFI = 0.88; IFI = 0.88; SRMR= 0.05; RMSEA = 0.06, *** p <0.001 | | | | | | | | |

*Note.* P=power; A=achievement; SE=security; ST=stimulation; C=conformity; T=tradition; H=hedonism; B=benevolence χ2=Chi-square; df=degrees of freedom; GFI=goodness of fit index; AGFI=adjusted goodness of fit; CFI=comparative fit index; IFI=incremental fit index; SRMR=standardised root mean square residual; RMSEA=root mean square error approximation. Appendix O and Appendix P show item codes and wording.

Previous researches using PVQ have reported that a translated version of the instrument yielded lower-than-acceptable fit indices and problematic discriminant and convergent validity (e.g., Davidov, 2008). For example, Steinmetz, Baeuerle, and Isidor (in press) reported a CFI of lower than a good or acceptable level for five multinational samples (CFI = 0.47, 0.81, 0.80, 0.82, and 0.84). Another study reported a GFI of 0.89 for a Slovakian sample (Oreg et al., 2008). BPV was considered adequately validated, taking into consideration that other studies have reported lower-than-acceptable fit indices and a lack of discriminant and convergent validity.

The reliability of the BPV constructs was reassessed since there were many modifications to the CFA model. The overall reliability of the complete 56-item instrument



(PCET and the modified BPV) was very high (Cronbach's alpha = 0.92). As shown in Table

8.14, the reliabilities of the modified constructs improved in terms of the Cronbach's alpha

values with the exception of self-direction.

Table 8.14
*Reliability of the Modified CFA Model for the BPV Constructs*

| Construct | Items | Cronbach's alpha (internal consistency) | Difference in the alpha values between the initial and the modified model | Construct's reliability status | Item-total correlation |
|---|---|---|---|---|---|
| Self-direction (SD) | 2 | 0.45 | -0.16 | unacceptable | 0.28-0.28 |
| Power (P) | 2 | 0.75 | +0.17 | Acceptable | 0.60-0.60 |
| Achievement (A) | 3 | 0.72 | 0 | Acceptable | 0.56-0.59 |
| Stimulation (ST) | 2 | 0.66 | +0.03 | Acceptable | 0.50-.50 |
| Conservation (CON) | 9 | 0.78 | - | Acceptable | 0.37-0.61 |
| Hedonism (H) | 3 | 0.72 | 0 | Acceptable | 0.54-0.57 |
| Benevolence (B) | 3 | 0.67 | +0.01 | Acceptable | 0.50-0.51 |

Cronbach's alpha values for the P, ST and B constructs were improved, whereas the

values for H and A did not change. Conservation's reliability (Cronbach's alpha = 78) improved

in comparison with the values in the initial model (conformity: Cronbach's alpha = 0.64;

tradition: Cronbach's alpha = 0.45; and security: Cronbach's alpha = 0.69). In contrast, the

reliability of the value SD did not improve but actually decreased by 0.16. This is tolerable

considering the relatively small difference (0.05) with the acceptable level (0.50). Regarding the

reliability of the 40-item PVQ, Schwartz et al. (2010, p. 433) stated that "[i]nternal reliabilities

of the basic personal values are necessarily low because the few items that measure each one are

intended to cover the conceptual breadth of the value rather than a core idea." Therefore, the

modified CFA model for BPV was considered to reflect acceptable levels of reliability

(Schwartz, et al., 2001).

Based on previously discussed CFA and Cronbach's alpha assessments, the modified

CFA model was considered to offer acceptable levels of validity and reliability for inclusion in



the overall structural (causal) model. Therefore, the modified CFA model shown in Figure 8.6 provides a better fit to the sampled data.

The theoretical structure of the Basic Personal Value (BPV) model could support the use of second-order CFA models because all of the values can be represented as higher order values (shown in Figure 4.2 p.67) (Krystallis, Vassallo, Chryssohoidis, & Perrea, 2008; Schwartz, 1994b, 1999). Basically, second-order models contain multiple layers (mostly two but sometimes more) of constructs. This type of representation of BPV was not considered for the following reasons. Firstly, in discussing conditions on the use of second-order CFA models, Hair et al. (2010, p. 757) stated that "a minimum of three first-order (first-level) constructs is required to assess a single second-order construct." Fewer than three first-order constructs is not enough to represent a second-order (higher order) construct. For instance, the higher order value, self-transcendence (B and U), is represented by two values only. Secondly, the use of a second-order CFA model for BPV would complicate the abstraction of the model. Thirdly, it would make it unfeasible to create direct paths between each individual value when testing the previously proposed research hypotheses (Hair, et al., 2010; Kline, 2010).

Modification of the CFA models (PCET and BPV) separately helped in simplifying this process which could have been difficult if conducted together for the overall model. Furthermore, it provided an enhanced view of the assessment of discriminant and convergent validity for both models. Combining both CFA models provided a measurement model for the overall research model. The restructured PCET and BPV models replaced the model postulated originally (see Appendix L). The following section evaluates the measurement model of the overall research model.

### 8.5.2    *Evaluation of Measurement Theory for the Overall Research Model*

The measurement model for the overall research model (see Appendix L) served as the postulated model to test the fit of the data. The model exhibited good levels of fit to the data: $\chi2/df = 1.93$, GFI = 0.88, AGFI = 0.85, CFI =0.93, IFI = 0.94, SRMR =0.04, and RMSEA =



0.04. The standardised loadings for all of the items were acceptable (see Appendix L). They had

a significant p-value (p < 0.05), and the CR was above 1.96. Furthermore, all of the constructs'

correlations were lower than the threshold (0.85), pointing to the discriminant validity of the

model. Therefore, the overall measurement model was considered suitable for path analysis, and

assessment of the structural model was performed. Details of the assessment of the overall

structural model are provided in the following section.

### 8.5.3    *Structural Evaluation of the Overall Research Model*

The evaluation of the structural model can be conducted by the assessment of (1) the

structural model fit indices and (2) the structural parameter estimates for the hypothesised

relationships (Hair, et al., 2010). Structural assessment is conducted for the postulated relations

in Table 8.15.

Table 8.15
*Assessed Hypotheses*

| Code | Hypothesis |
|------|------------|
| USE←RA | H1: Relative advantage has a positive significant influence on intention to use e-transactions. |
| USE←CT | H2: Compatibility has a positive significant influence on intention to use e-transactions. |
| USE←CMX | H3: Complexity has a negative significant influence on intention to use e-transactions. |
| USE←RED | H4: Result demonstrability has a positive significant impact on intention to use e-transactions. |
| USE←TI | H5: Trust in the Internet has a positive significant influence on intention to use e-transactions. |
| USE←TG | H6: Trust in government agencies has a positive significant influence on intention to use e-transactions. |
| USE←SI | H7: Social influence has a positive significant impact on intention to use e-transactions. |
| USE←POC | H8: Perspective on communication has a positive significant impact on intention to use e-transactions. |
| USE←P | H9: Power has a negative significant impact on intention to use e-transactions. |
| USE←A | H10: Achievement has a negative significant impact on intention to use e-transactions. |
| USE←H | H11: Hedonism does not have a significant impact on intention to use e-transactions. |
| USE←ST | H12: Stimulation has a positive significant impact on intention to use e-transactions. |
| USE←SD | H13: Self-direction has a positive significant impact on intention to use e-transactions. |
| USE←U | H14: Universalism has a negative significant impact on intention to use e-transactions. |



| USE←B | H15: Benevolence value has a negative significant impact on intention to use e-transactions. |
| USE←CON | H16: Conservation values have a negative significant impact on intention to use e-transactions. |

The hypothesised structural model is shown in Figure 8.7. The structural model was developed by replacing the correlations (double-headed arrows) between the independent constructs and the dependent construct (intention to use e-transactions) with causal paths (one-headed arrows). These arrows graphically represent the research hypotheses. Again, for the sake of clarity, the correlations between the independent constructs, the measurement errors and the residual terms are not included in the figure.

Hair et al. (2010) suggested that theoretically similar constructs should be correlated in a structural model. Thus, the constructs from the PCET model were correlated together, and the BPV model constructs were correlated together. The fit indices were changed only after the hypothesised relationships were freed for estimation, and all of the other relations were set to zero: $\chi 2/\text{df} = 2.0$, GFI = 0.86, AGFI = 085, CFI = 0.92, IFI = 0.9, RMSEA = 0.04, and SRMR = 0.09. The structural model fit indices were similar to the measurement model fit indices, indicating that the overall model has a good fit (B. M. Byrne, 2010; Hair, et al., 2010).



Figure 8.7
*Revised structural model for the overall research model.*

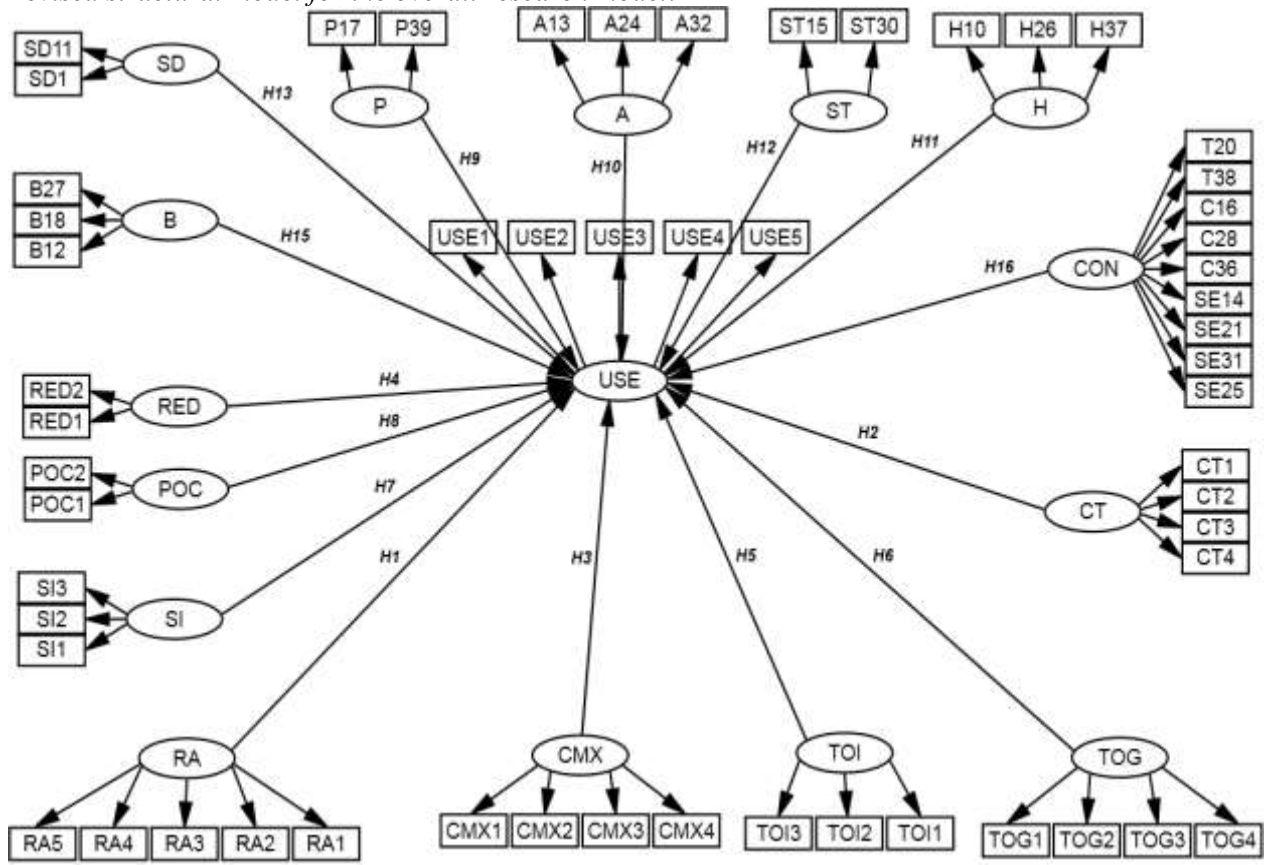

*Note.* RA=relative advantage; CT=compatibility; CMX=complexity; RED=result
demonstrability; TI=trust in the Internet; TG=trust in government agencies; SI=social influence;
POC=perspective on communication; USE=intention to use e-transactions. SD=self-direction;
P=power; A=achievement; ST=stimulation; CON=conservation values; H=hedonism;
B=benevolence. Appendix O and Appendix P shows item wording and codes.

        As discussed in section 8.5.1.1.3, the construct universalism was removed because of

multicollinearity. Therefore, the hypothesis H14 (USE<-U) was not assessed. The second set of

evaluation criteria for the structural model comprised the structural parameter estimates, which

included the standardised regression weights (SRW), p-values, and squared multiple

correlations (SMC) or $R^2$ for the dependent construct (Gefen, et al., 2000; Hair, et al., 2010).

The direction (positive or negative) of the structural path is shown by the positive or negative

sign of the standardised regression weight. A p-value less than 0.05 signified that the relation is

significant. Kline (2010) noted that SRW (path coefficients) with an absolute value less than



0.10 signify a small effect on the dependent construct. He further noted that a medium effect is

indicated by standardised path coefficients around 0.30 and those coefficients larger than or

equal to 0.50 indicate a large effect or significance. Finally, SMC determines whether the

overall model is able to predict acceptance (USE) by calculating the percentage of variance that

the independent constructs explain in the dependent construct (B. M. Byrne, 2010; Hair, et al.,

2010; Kline, 2010).

Table 8.16 illustrates the outcome of the structural model estimation using AMOS. As

shown in the table below, 7 of the 15 hypotheses were significant (p < 0.05) and 6 of out of 15

were in the hypothesised direction and significance; hypothesis 14 was not assessed because of

multicollinearity.

Table 8.16
*Structural Parameter Estimates for the Structural Model*

| Hypothesis | SRW | CR | P-value | Significant? | In the proposed direction? | Supported? | Conclusion |
|---|---|---|---|---|---|---|---|
| H1: USE←RA | -0.02 | -0.29 | 0.77 | No | No | No | RA has a non-significant effect on USE. |
| H2: USE←CT | 0.18 | 2.87 | *** | Yes | Yes | Yes | CT is positively related to USE at the level of p < 0.001. |
| H3: USE←CMX | -0.01 | -0.22 | 0.82 | No | Yes | No | CMX has a non-significant effect on USE. |
| H4: USE←RED | 0.19 | 4.39 | *** | Yes | Yes | Yes | RED is positively related to USE at the level of p < 0.001. |
| H5: USE←TI | 0.09 | 2.2 | 0.03 | Yes | Yes | Yes | TI is positively related to USE at the level of p < 0.05. |
| H6: USE←TG | -0.09 | -2.23 | 0.03 | Yes | No | No | TG is negatively related to USE at the level of significance of p < 0.05. |
| H7: USE←SI | 0.03 | 0.74 | 0.46 | No | Yes | No | SI has a non-significant effect on USE. |



| H8:<br>USE←POC | 0.54 | 10.31 | *** | Yes | Yes | Yes | POC is positively related to USE at the level of p < 0.001. |
|---|---|---|---|---|---|---|---|
| H9:<br>USE←P | -0.20 | -2.16 | 0.03 | Yes | Yes | Yes | P has a negative effect on USE at the level of p < 0.05. |
| H10:<br>USE←A | 0.10 | 1.02 | 0.31 | No | No | No | A has a non-significant effect. |
| H11:<br>USE←H | -0.11 | -1.31 | 0.19 | No | - | Yes | H has a non-significant effect. |
| H12:<br>USE←ST | 0.09 | 0.95 | 0.34 | No | Yes | No | ST has a non-significant effect |
| H13:<br>USE←SD | 0.28 | 1.37 | 0.17 | No | Yes | No | SD has a non-significant effect on USE. |
| H14:<br>USE←U | U influence on the intention to use e-transactions was not directly measured due to multicollinearity. However, it is expected that the influence of this value would be similar to benevolence. | | | | | | |
| H15:<br>USE←B | -0.29 | -1.78 | 0.08 | No | Yes | No | B has a non-significant effect on USE. |
| H16:<br>USE←CON | 0.27 | 2.69 | 0.01 | Yes | No | No | CON is positively related to USE with significance of p < 0.05. |

*Note.* *** Significance at the level of p < 0.001

As show in the Table 8.16 above, the SRW indicates that the only significant (p < 0.05) negative relationships were between USE and TG (-0.04) and USE and P (-0.14). The most significant construct was POC, which had a standardised regression weight (0.54) higher than 0.50, indicating a large effect on acceptance. On the other hand, the other constructs with significant influence (CT, RED, P, and CON) were considered to have a medium effect (0.18, 0.19, -0.20, and 0.27). TI and TG were significant (p < 0.05) contributors to the intention to use e-transactions (USE) and had a small influence (< 0.1) on the dependent construct USE. The overall structural model contributed 70% of the variance (SMC = 0.70) in the dependent construct (USE), indicating that the overall structural model can strongly predict acceptance.



Intercorrelations between the independent variables for each submodel (PCET and BPV) were

all less than 0.85, suggesting discriminant validity (see Appendix M).

Figure 8.8
*Hypothesised structural model with standardised regression weights for supported hypotheses.*

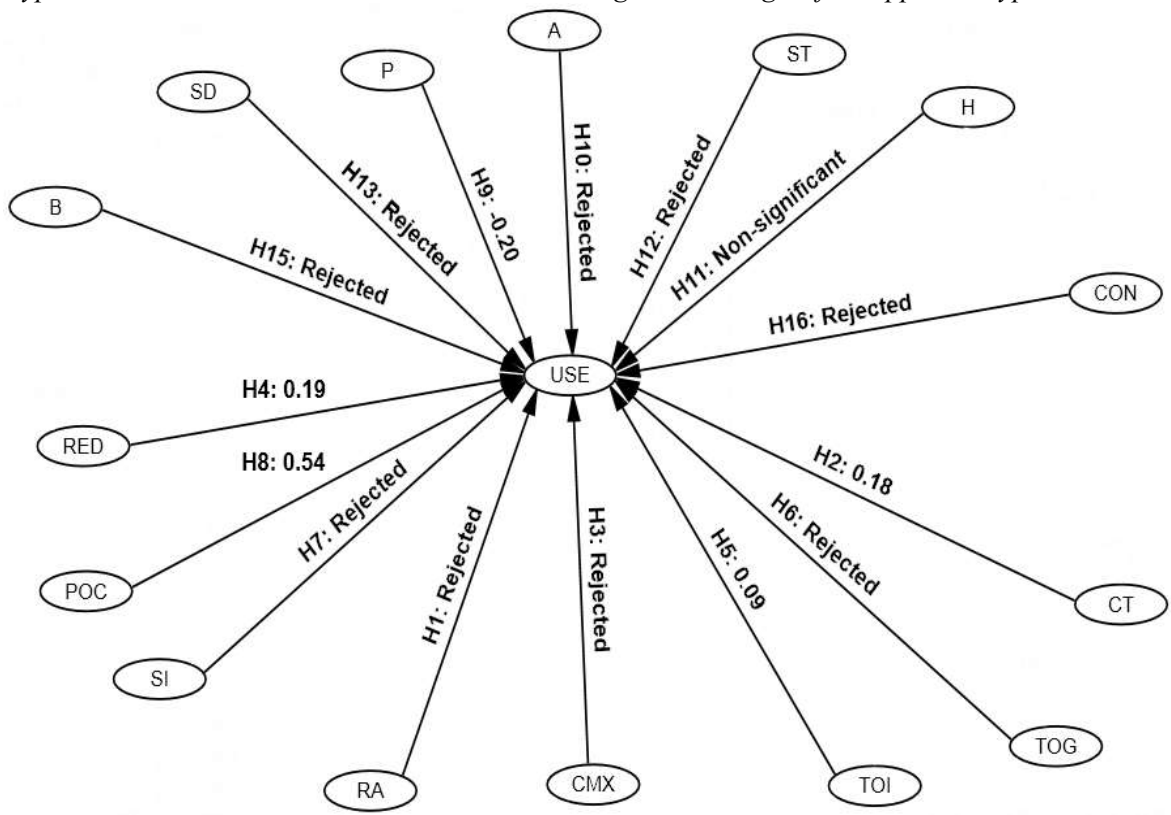

*Note.* RA=relative advantage; CT=compatibility; CMX=complexity; RED=result
demonstrability; TI=trust in the Internet; TG=trust in government agencies; SI=social influence;
POC=perspective on communication; USE=intention to use e-transactions. SD=self-direction;
P=power A=achievement; ST=stimulation; CON=conservation values; H=hedonism;
B=benevolence.

        Figure 8.8 above depicts the hypothesised relations in the structural model; the SRW are

illustrated for the supported hypothesis and the unsupported hypotheses are denoted as rejected.

The data from the structural model shows that significant relationships exit in the following

hypotheses: H2 (USE←CT), H4 (USE←RED), H5 (USE←TI), H6 (USE←TG), H8

(USE←POC), H9 (USE←P) and H16 (USE←CON). The following hypothesises were found

non-significant: H1 (USE←RA), H3 (USE←CMX), H7 (USE←SI), H10 (USE←A), H11



(USE←H), H12 (USE←ST), H13 (USE←SD), H12 (USE←ST) and H15 (USE←B). The postulated direction and significance were supported by the data for the following hypothesises: H2 (USE←CT), H4 (USE←RED), H5 (USE←TI), H8 (USE←POC), H9 (USE←P) and H11 (USE←H). Therefore, the data supported 6 of the 15 hypotheses, and H14 (USE←U) was not tested. The structural model can be further explored to identify relationships that were not theoretically hypothesised yet are empirically suggested by the data. Further exploratory analysis is described in Appendix N.

## 8.6    Summary

The assessment of the research model included Confirmatory Factor Analysis (CFA) assessment for Perceived Characterises of e-Transactions (PCET), Basic Personal Values (BPV), and all constructs. The PCET model met all of the requirements of the assessment of the CFA, whereas the BPV model was modified and reassessed for fit, validity, and reliability (conducted validation procedures are summarised in Appendix K). For the overall research model, the measurement and the structural model were found to be of an adequate fit and to be valid for causality or path analysis. Less than half of the postulated hypothesises were found to be in the expected direction and significance. Finally, the data was explored further for the purpose of identifying potential new relationships.



# 9    DISCUSSION

This study contributes to the literature in that it elucidates the influence of culture on the acceptance of e-transaction practices in the KSA. No other studies were found that addressed the association between culture and intention to use e-transactions, particularly in developing countries such as the KSA with a contextualised focus.

The main research question for this thesis is 'How does culture influence the acceptance of e-transactions?' This question is further divided into five questions designed to describe the cultural focus and context of the research, as well as add depth to the understanding of e-transaction acceptance. The relevant questions are as follows: (1) How do perceived characteristics of e-transactions affect e-transaction acceptance? (2) How does trust in the Internet and government agencies influence acceptance? (3) How does the social influence of existing e-transaction users affect the acceptance of e-transactions? (4) How does using e-transactions as a communication method affect acceptance of e-transactions? and (5) How do cultural values influence the acceptance of e-transactions? On the basis of these questions and associated studies, 16 hypotheses were developed, which were empirically tested to determine the significance and direction of the relationships that exist between intention to use e-transactions and each of the following: perceptions of e-transactions, trust, social influence, perspective on communication, and cultural values.

In this chapter, significant findings are discussed and non-significant results are explained. The discussion of the findings is organised with the list of the research questions. Significant results related to perceptions of e-transactions and intention to use e-transactions are discussed followed by significant results from the assessment of the posited hypothesis between intention to use e-transactions and trust, social influence, perspective on communication, and cultural values. Finally, non-significant findings are discussed. The discussion of these findings



addresses the main research question explicating the influence of culture on e-transaction acceptance.

**9.1    Influence of Perceptions on the Acceptance of e-Transactions**

This section discusses the significant findings related to the following research question and hypotheses:

**Research Question 1:** How do perceived characteristics of e-transactions affect e-transaction acceptance?

*H2: Compatibility has a positive significant influence on intention to use e-transactions.*

*H4: Result demonstrability has a positive significant impact on intention to use e-transactions.*

Perceptions of a technology and, more particularly, e-government transactions and services are identified in the literature as determinants of acceptance (Carter & Bélanger, 2005; Sang, et al., 2009). The perceptions related to the Saudi culture and e-transactions were included in the research model for hypothesis testing. The SEM structural assessment shows that compatibility with cultural needs, values, and previous experiences and result demonstrability of e-transactions are salient perceptions that affect the intention to use e-transactions, whereas relative advantage and complexity are not. The constructs compatibility and result demonstrability have a positive significant effect on the intention to use e-transactions. These findings are congruent with those of Carter and Bélanger (2005), AlAwadhi (2009), and Sang et al. (2009) for compatibility, as well as with the results derived by Baumgartner and Green (2011) and Hussein et al. (2011) for result demonstrability. The significant relationships between intention to use e-transactions and both compatibility and result demonstrability enhance the understanding of the role of culture in the acceptance of e-transactions.

Perceptions of the compatibility of e-transactions with users' cultural needs, values, and previous experiences significantly determine acceptance. Compatibility is an important part of acceptance in the KSA given that Saudi society is conservative, making cultural needs and



values essential influencing factors for accepting introduced technologies. Therefore, the significant positive relationship between intention and the compatibility construct indicates that Saudis who have Internet access find e-transactions with the government highly compatible with their cultural needs, values, and previous experiences.

Perceptions of the result demonstrability of e-transactions and their significant positive effect on acceptance indicate that the outcomes of using e-transactions are communicable to others. The relationship between result demonstrability and intention to use e-transactions can be understood by considering the collective orientation of Saudi culture, which is characterised by the tendency to share experiences with others (Liu, et al., 2008).

## 9.2    Influence of Trust in Internet and Government on the Acceptance of e-Transaction

The response to the following research question and hypotheses are discussed in this section:

**Research Question 2:** How does trust in the Internet and government agencies influence acceptance?

*H5: Trust in the Internet has a positive significant influence on intention to use e-transactions.*

*H6: Trust in government agencies has a positive significant influence on intention to use e-transactions.*

Trust in both the Internet and government agencies are significant contributors to the acceptance of e-transactions, but in opposite directions. As hypothesised, Trust in the Internet exhibits a significant positive influence, whereas trust in the government has a significant negative effect in the opposite direction of the hypothesis. The positive correlation between trust in the Internet and e-transaction acceptance shows that Saudis who have Internet access trust the Internet as a medium for conducting transactions with government agencies. The sampled citizens prefer e-transactions because this technology reduces the frustration of having to experience the negative consequences of visiting a government agency. Examples include long



queues, unequal treatment, and vague requirements for completing a transaction, lengthy

transaction processes and the uncooperativeness of government employees. Therefore, those

who do not trust government agencies believe that technology facilitates the provision of better

and more reliable services (Abu Nadi, 2010; AlAwadhi, 2009). Conversely, trust in government

agencies diminishes the acceptance of e-transactions because most citizens do not have faith in

the government agencies as a provider of service. This finding indicates that citizens view some

government agencies as lacking in integrity, effectiveness, efficiency, and trustworthiness in

terms of providing and completing e-transaction services. Thus, it was understood that, for most

citizens, trust in the Internet and e-transactions as a medium and technology to conduct

government transactions was accompanied by a lack of trust in government agencies to facilitate

the usage of these technologies.

### 9.3     Influence of Perspective on Communication on the Acceptance of e-Transactions

Findings related to the following research question and hypotheses are addressed in this

section:

**Research Question 4:** How does using e-transactions as a communication method affect

acceptance of e-transactions?

*H8: Perspective on communication has a positive significant impact on intention to use e-transactions.*

Perspective on communication significantly advances acceptance of e-transactions. The

significant positive effect of communication perceptions on e-transaction acceptance reflects the

inclination of citizens towards the use of e-transaction technology as a means of contacting the

government. Furthermore, the use of e-transactions to communicate with the government

improves the understanding of transaction requirements, the structure of transaction processes,

and the simplification of delivery or receipt of necessary documents for the completion of

transactions (Alhomod & Shafi, 2012). Therefore, the high level of preference for e-transactions

as a communication method is associated with increased acceptance.



## 9.4    Influence of Cultural Values on the Acceptance of e-Transactions

In this section, the following research questions and hypotheses are discussed:

**Research Question 5:** How do cultural values influence the acceptance of e-transactions?

*H9: Power has a negative significant impact on intention to use e-transactions.*

*H16: Conservation values have a negative significant impact on intention to use e-transactions.*

Previous research on cultural values has indicated their importance in determining behaviour or intentions to engage in a specific act (Hofstede, 2001a; Schwartz, 2003). Schwartz (2003) and Hofstede (2001a) support the findings on the influence of values on behavioural intentions, but determining which values are most relevant to e-transaction acceptance is difficult because no published study is directly relevant to the KSA and BPV. Empirical evidence supports the significance of conservation and power values. An unexpected result is the positive relationship between conservation values and e-transaction acceptance, which is attributed to the respect that Saudi citizens have for their leaders. Such respect extends to their willingness to comply with the government's choice to introduce e-transactions. This situation is especially true for developing nations such as Saudi Arabia, as confirmed by previous research, which has demonstrated that citizens of such nations exhibit high deference to leadership authority or score high in the Hofstede power distance index (Hofstede, et al., 2010). The tendency to grant political leaders the utmost respect is the essence of conservation values as political and economic stability is sought (Hofstede, 2001a; Schwartz, 2003). The value power which is motivated by seeking prestige and authority is a significant negative contributors to acceptance; this result was expected, due to the disintermediation that e-transactions cause for those who seek prestige and social status by using their personal relations or *wasta* within government agencies in the KSA (AL-Shehry, et al., 2006).



**9.5     Discussion of Non-Significant Results**

        Non-significant results are discussed because they highlight interesting perspectives. For perceptions of e-transactions, relative advantage and complexity are non-significant contributors to acceptance, which is similar to the findings reported by Carter and Bélanger (2005). Additionally, the non-significance of complexity in e-transaction acceptance is consistent with the results derived by Carter and Bélanger (2003). Therefore, the following hypothesis is discussed:

*H1: Relative advantage has a positive significant influence on intention to use e-transactions.*

        Relative advantage was expected to be an important and positive significant contributor to e-transaction acceptance, but the empirical assessment of this hypothesis reflects the opposite. The non-significant result of relative advantage is similar to Carter and Bélanger's (2005) study. Those who have Internet access do not consider the use of e-transactions as a relative advantage under circumstances where a more novel or more innovative method (e.g., government services provided over mobile phones) exists (Carter & Bélanger, 2005). Although e-transactions may be preferred, this preference does not translate to perceptions of superior advantage in all e-transaction features, such as better control over transactions or quality of results of conducted transactions. The following postulated hypothesis is discussed:

*H3: Complexity has a negative significant influence on intention to use e-transactions.*

        The hypothesised relationship between complexity and acceptance of e-transactions is in the proposed direction but with non-significant influence. As discussed in section 5.2.2, many Saudis were expected to consider e-transactions a novel method, consequently triggering the high uncertainty avoidance tendency, from which perceptions of complexity arise. However, high uncertainty avoidance was not found to be a related factor in the perceived complexity of e-transactions. As described in section 7.2, most of the respondents frequently use e-transactions and have used them recently (within one month or less). Compared with traditional methods, e-transactions are not perceived as complex. Thus, citizens find e-transactions considerably easier



to accomplish as a result of their availability, simplicity, and efficiency in comparison to visiting government agencies (Abu Nadi, 2010; AlAwadhi, 2009). Therefore, the relation between complexity perceptions of using e-transactions and intention to use e-transactions was found non-significant.

In the following paragraph the hypothesis which addresses the third research question: 'How does the social influence of existing e-transaction users affect the acceptance of e-transactions?' is discussed:

*H7: Social influence has a positive significant impact on intention to use e-transactions.*

Social influence is also a non-significant contributor to e-transaction acceptance—a result that agrees with those of AlAwadhi (2009) and Hung, Chang and Yu (2006). Society or peer influence is a particularly important factor when there are few adopters of a given technology because non-adopters in collective cultures tend to seek advice from others (Al-Gahtani, et al., 2007; Hung, et al., 2006). This situation, however, does not hold true for this study, in which the citizens sampled are mostly (68.4%) previous users of e-transactions. Social influence can be a particularly essential factor for non-adopters or when e-transactions have only been recently introduced (AlAwadhi, 2009; Gefen, et al., 2002; Hung, et al., 2006). Al-shafi et al. (2009) found that social influence significantly determines e-government acceptance in Qatar. The advanced e-government in Qatar provides an opportunity for citizens to discuss and share their experiences with online services (Al-Shafi, et al., 2009; UNDESA, 2012). In the KSA and Kuwait (AlAwadhi, 2009), the adoption of e-government is motivated by need, and individuals form their perceptions mostly independently of others given that no formal channel for sharing experiences is in place. The following hypotheses are discussed:

*H10: Achievement has a negative significant impact on intention to use e-transactions.*

*H11: Hedonism does not have a significant impact on intention to use e-transactions.*

*H12: Stimulation has a positive significant impact on intention to use e-transactions.*



*H13: Self-direction has a positive significant impact on intention to use e-transactions.*

*H14: Universalism has a negative significant impact on intention to use e-transactions.*

*H15: Benevolence value has a negative significant impact on intention to use e-transactions.*

The hypothesised associations between e-transaction acceptance and achievement, self-direction, stimulation, and benevolence, are also non-significant. Although these results were not expected, it was hypothesised that hedonism value would cause a non-significant influence on e-transaction acceptance. The influence of universalism was not assessed given its multicollinearity with other constructs (very high correlation with benevolence and conservation values).

This finding is supported by the results of Choden et al. (2010), who found that the aforementioned values are irrelevant in European developing countries. Specifically, achievement (in Poland and Hungary), self-direction (in Hungary), hedonism (in the Czech Republic, Poland, Hungary, and Estonia), and stimulation (in Estonia, Poland, and Hungary) were irrelevant to Internet acceptance. In Choden et al.'s (2010) study, benevolence and universalism were non-significant contributors to Internet usage for all sampled European developed (Denmark, Netherlands, Sweden, and Finland) and developing nations (Czech Republic, Poland, Hungary, and Estonia).

Except for benevolence, most collectivist values (security, conformity, and tradition) are significant contributors to the acceptance of e-transactions. Most individualist values (achievement, self-direction, hedonism, and stimulation) are non-significant, except for power value. These findings indicate the strong predisposition of Saudi society towards collectivism and the influence of this predisposition on the acceptance of technology, especially e-transactions (Al-Gahtani, et al., 2007).



## 9.6    Summary

The significant and non-significant findings were explored in this chapter. These findings addressed the sub-research questions, which consequently answered the main research question. Results indicate that the constructs perspective on communication, conservation values, result demonstrability, compatibility, and trust in the Internet have positive significant influence while power and trust in government agencies have negative significant influence on acceptance of e-transactions. These finding demonstrated that the technology of e-transactions is accepted by the culture of Saudi citizens. Relative advantage, complexity, social influence, and the following cultural values: achievement, hedonism, stimulation, self-direction, and benevolence did not have significant influence on acceptance of e-transactions.



## 10   CONCLUSIONS

This study examined the influence of culture on e-transaction acceptance for Saudis who have Internet access. Understanding the influence of culture is important theoretically and practically. Theoretically, this study provided a link between acceptance and culture with a contextualised method that has been insufficiently explored in research especially for developing countries such as the KSA. Practically, strategies were formulated on the basis of an empirical study in the KSA. These e-government design and implementation strategies would provide insight for the government of the KSA, which if implemented might lead to higher levels of acceptance of e-transactions.

The influence of culture on e-transaction acceptance was explored through understanding the impact caused by the perceived characteristics of e-transactions, trust in the Internet and government, social influence, usage of e-transactions as a communication method, and cultural values. In order to address the research questions at an initial stage, literature on culture, technology acceptance, and e-government acceptance was integrated with a focus on context-related issues to develop a model for determining the effect of perceptions, trust, and cultural values on the acceptance of e-transactions. Using questionnaires adopted from previous studies as bases, an online questionnaire was developed, which was then translated into Arabic. The developed questionnaire was contextualised by ensuring that the questionnaire's items corresponded with the focus of the research. The online questionnaire was pre-tested and pilot-tested for usability and clarity. The final modified and improved questionnaire was sent to Saudi respondents with Internet access. Data was screened for outliers and demographic information was provided. Structural Equation Modelling (SEM) was rigorously applied to determine significant relationships. As indicated by the SEM structural assessment, about half (7) of the tested hypotheses are significant or important predictors of the intention to use e-transactions. The outcomes of the SEM analysis are discussed in relation to the research questions.



Perceptions associated with e-transactions, trust in the Internet and government agencies, and cultural values explain the acceptance of e-government transactions in the KSA. These findings present important implications for the government of the KSA and the entities associated with it (such as embassies that provide online services or professional organisations that are developing e-government services in the country), as well as for other countries with cultural characteristics similar to those of Saudi Arabia. International research and consulting organisations as well as researchers interested in the link between culture and acceptance of technology (particularly e-government) will find the study relevant.

The significant factors that influence the acceptance of e-transactions are perspective on communication, conservation values, power value, result demonstrability, compatibility, trust in the Internet, and trust in the government. The non-significant factors are relative advantage, complexity, social influence, achievement, self-direction, hedonism, stimulation, and benevolence. The most interesting and unexpected finding is the positive influence of conservation values. Saudi society is religious and conservative, but these attributes have not impeded technology acceptance; rather, they enhance acceptance—a finding that is congruent with the TNS digital life report, which suggested that Saudis are the most engaged people in the world when it comes to Internet usage. In the report, engagement included attitudes towards technology and length of time spent on the Internet. TNS is a market researcher and global consulting organisation that focuses on worldwide growth indicators (TNS Digital Life, 2010).

The constructs that significantly influence e-transaction acceptance will also be relevant to citizens who did not have Internet access at the time of data collection because mobile Internet and broadband access have become more accessible in the near future in KSA (Ministry of Communications and Information Technology, 2007b). A contact database of Saudi citizens was not accessible; thus, the sample is considered as representative as possible, especially under current circumstances. The theoretical and practical contributions of this study are outlined. Based on the results, design and implementation strategies were crafted for the government of the KSA to enhance the successful implementation of current e-transaction



programmes. A presentation of limitations and future directions follow. This chapter ends with a summary.

## 10.1   Theoretical Contributions

The findings of this study expand the technology and e-government acceptance literature, add to the theoretical and cultural understanding of e-government acceptance, and highlight the importance of culture and context in the development and assessment of conceptual models.

The research model was adapted and contextualised from previously tested and approved valid and reliable models. The assessment of the research model confirms that it is reliable and valid for replication in other contexts. Whether the hypothesised relationships are significant and in the proposed direction, were determined via the SEM analysis method. In descending order of influence, the most significant factors that influence the acceptance of e-transactions are perspective on communication, conservation values, power value, result demonstrability, compatibility, trust in the Internet, and trust in government agencies. Perspective on communication is a factor that describes the level to which e-transactions would enable adequate communication with the government and it has a positive significant effect on intention to use e-transactions. Conservation values relate to the conformity of citizens to social expectations and leadership, seeking stability and security of society and nation, and humbleness and commitment to traditions and religion have positive significant influence on intention to use e-transactions. Intention to use e-transactions was negatively affected by power value which supports seeking social status and authority. Communicability of the outcomes of using e-transactions is a positive significant determinant of intention to use e-transactions. Consistency of using e-transactions with the users' needs, values and past experiences has a positive significant influence on intention to use e-transaction. Trust in the Internet as a medium for conducting transactions with the government has a positive significant influence while trust in government agencies as a provider of service has a negative effect on intention to use e-transactions.



These findings are also especially relevant to developing countries that partially share analogous cultural or contextual characteristics with the KSA. For instance, Thailand, China, and Japan are high-context cultures, in which perspective on communication can be a significant determinant (whether it has a positive or negative influence) of acceptance. The conservation values of countries such as other Arab nations are expected to influence technology acceptance. The results of this study therefore provide a theoretical explanation for the acceptance of e-transactions in developing countries, and more specifically, Arab countries where language, religion, culture, traditions, and economic interests are shared to a large extent (Aoun, et al., 2010; Barakat, 1993; Nydell, 2006).

A relatively atypical approach to contextualising the research model, questionnaire items, and instrument was implemented. Although all constructs were adopted from previous studies, methods typically used to develop constructs in research were applied in multiple phases to ascertain that the phrasing of the questionnaire items was suitable for the research context. Many participants that specialise in the field of information systems and other related fields were consulted at an early stage to discuss which factors were most relevant to the context (see Appendix A). Lewis' content validity ratio, typically used to assess the relevance of developed items, was employed to assess the level to which the wording of each questionnaire item was aligned with the research context. Within this same phase, the participants were provided opportunities to provide feedback on item wording; the most relevant and useful recommendations were adopted. These phases were intended to address the need for contextualisation and localisation when discussing the adoption of technology in the information systems field, and more specifically, in e-government research (Bolívar, Muñoz, & Hernánde, 2010; Heeks & Bailur, 2007; Orlikowski & Iacono, 2001).

The 'perspective on communication' construct was adapted from the field of accounting information systems. This construct is the most important factor (it has the largest standardised regression weight) for determining the acceptance of e-government in the KSA. The perspective on communication construct was originally developed by Aoun (2010) to determine whether



such perspectives influence the adoption of accounting information systems. The main author of this study (Aoun) was contacted to acquire the items considered in measuring this construct. After redefining the scope of the construct as 'the degree to which using e-government transactions enables adequate communication with the government,' the acquired items were also contextualised and tested for relevance to the construct definition (see Appendix A for contextualisation process and section 6.2.3 for content validity assessment). Exploratory factor analysis revealed that the third item, 'Textual, verbal and visual information is important for carrying out government transactions' is not related to the construct, and was therefore excluded. The remaining items were 'My ability to communicate with the government would be enhanced when using e-government transactions' and 'Communication through e-government enhances my ability to understand government transactions'. The perspective on communication construct facilitated the exploration of a dimension that was not previously examined in e-government research, thereby aiding the understanding of the importance of e-transactions as a communication tool between citizens and the government. This finding indicates that online transactions represent an essential communication channel between stakeholders in the KSA.

Schwartz's Basic Personal Values model and Portrait Values Questionnaire have not been previously used to explain the acceptance of e-transactions. To the best of the researcher's knowledge, this study is the first to use the aforementioned model and instrument to explain KSA culture, and the second after Alkindi, (2009) to adopt the model and instrument in elucidating the culture of Arab countries. Using PVQ, the presence of collective cultural values—especially the conservation values of security, tradition, and conformity—was found to exhibit the strongest influence over e-transaction acceptance in the country. Conservation values are the second most salient factors in determining the acceptance of e-transactions. The positive influence of conservation values (which are considered a major component of a society's culture) on the acceptance of e-transactions was unexpected, and contradicts the findings reported in other studies (C. Hill, et al., 1994; Loch, et al., 2003; Straub, et al., 2003). Power



value is the third most significant factor in e-transaction acceptance. In contrast to conservation values, however, power is a cultural inhibitor of e-transaction acceptance.

The strong correlation of universalism with benevolence and conservation values (multicollinearity) indicates a cultural finding about the Saudi society. In a conservative and collective society such as the KSA, concern for social justice and equity is more strongly related to the family, tribe, inner circles, and society than it is to other countries. This moral concern is also associated with the overall level of development in a country, which advocates consideration for assisting others 'close by' over concern for external affairs. As a developing country, therefore, the Saudis attach more importance to benevolence, social justice, and equity within the borders of the country (Schwartz, 2007). However, this situation is expected to change in the coming years, given the country's increasing participation in globalisation trends (Ramady, 2010).

The importance of cultural values in technology acceptance has not been sufficiently emphasised in the literature. A theoretical implication for research on information systems is that Schwartz's Basic Personal Values model requires further examination in studies that focus on culture would strengthen the understanding of the association between culture and technology acceptance. Only two relevant studies in the realm of information systems exist, and both investigated the influence of cultural values on Internet use (Bagchi & Kirs, 2009; Choden, et al., 2010). In these studies, the shorter-version (European Social Survey) instrument was used to determine relevant cultural values, instead of the more comprehensive Portrait Values Questionnaire. Studies that address the influence of cultural values on the acceptance of complicated systems, such as e-transactions, are needed, especially those that concentrate on developing nations and Arab countries.

## 10.2   Practical Contributions

The research methodology, questionnaire, and findings can be used as reference by researchers and decision makers in public and private organisations.



Using an online survey enabled reaching citizens located in the wide geographical scope of the KSA. As described in section 7.2, the participants in this survey were Saudi citizens living in the KSA and other countries. Although a pencil-and- paper survey is easier to develop, the costs of establishing an online survey are lower than those presented by printing paper-based surveys. The response rate in this study was low (2.31%), but a considerable number of citizens completed the questionnaire (674). The online survey facilitated easier and faster inputting of responses into the analysis software (SPSS and AMOS) and enabled the acquisition of a large sample size comprising participants who reside in many countries. Such a sample would be difficult for a PhD candidate to acquire without the use of an online survey. Furthermore, the validity and reliability of the online survey data was confirmed. Based on the large acquired sample, lower costs of questionnaire development and distribution and validity of transferring and analysing electronically collected data, the use of online surveys is recommended to researchers, governments, and organisations that intend to study a portion of a population with Internet access.

A rigorous translation method was used to translate the survey questions from English into Arabic. The translation is considered accurate on the basis of the pre-tests and pilot study. The resultant Arabic questionnaire can be used by other researchers, governments, or organisations interested in a similar research focus.

Decision makers and researchers can benefit by recognising the relevance of of cultural values, as well as their influence on the acceptance of technology. The congruence of technologies with the cultural values of citizens or employees should exhibit increased acceptance rates. The research model and developed questionnaire can be used in the KSA or other countries with similar cultural characteristics for the prediction, assessment, and determination of e-transaction acceptance. A comprehensively contextualised and customised model and questionnaire can be used in different settings (such as organisations) and in studying the acceptance of varied technologies (including e-commerce and mobile technologies).



Research and consulting organisations (such as Garter or United Nations) that do not have access to local KSA perspectives on e-government programmes can use this study as reference material. The e-government global survey report of the United Nations Department of Economic and Social Affairs (2012) and similar reports rely on domestic research in providing information and findings related to e-government acceptance. These reports are intended to assist many countries in their development initiatives. Business organisations that frequently transact with the KSA government will also find the results of this research relevant, especially those responsible for assisting the government in implementing e-government programmes. KSA government officials and decision makers can use this study's findings to understand how citizens accept and perceive e-transaction programmes and how culture and cultural values influence such acceptance.

## 10.3   Design and Implementation Strategies for Government

Here, strategies and practical solutions for developing a successful e-government programme in the KSA are provided. This study indicates that adoption of some specific strategies will facilitate the country's transition into an information society and knowledge-based economy. These implications are also relevant to countries similar to the KSA. These countries can take into consideration the strategies specific to their circumstances (e.g., strategies addressing the predominance of collective culture in a country). Suggestions include the following: (1) enhancing e-transactions as a tool of communication between the government and citizens; (2) understanding and taking advantage of conservation and power values; (3) enabling communication channels for information sharing among citizens; (4) considering the cultural needs of citizens; (5) improving the public image and perception of government agencies; and (6) developing Internet infrastructure in the KSA.

e-Transactions can be considered a tool for enhancing communication between citizens and the government. Although Saudi Arabia is a high-context culture, in which oral and face-to-face communication are favoured, most citizens prefer using e-transactions, indicating that citizens value technology as a communication platform. This finding also implies that citizens



desire improvements in the current traditional methods of communication with Saudi government agencies. To improve communication within the online environment, government agencies can expand their scope to other communication channels. The instructions provided by e-transaction systems are insufficient in terms of clarity and usability. To adequately support users, the government can add features such as online text chat, audio messaging, video conferencing, webinars, and online collaboration tools. Forums and social networking sites (e.g., Twitter and Facebook) can also improve communication. Government agencies can integrate the use of social networking sites into e-transaction processes and practices to increase their usage by government employees. However, effective usage of these innovations necessitates high levels of transparency, tolerance, patience, and understanding on the part of government employees. It is suggested that the Saudi government increase its online presence through social networks, and provide text (e.g., chat or email), voice, and video support to its citizens. The use of such tools also necessitates support from top management in government agencies, as well as training and awareness programmes for government employees who frequently interact with the public. Improving services by forming feedback-active teams (task forces) composed of programmers and information systems specialists also hastens responses to feedback. Such teams are in a unique position to adopt feasible recommendations from the public. Rapid response to issues materialises into tangible changes and developments in e-government websites, thereby encouraging trust in these communication channels and enhancing perceptions of government integrity (Baumgarten, 2009). In 2012, the Ministry of Communication and Information Technology implemented an initiative that involves citizens in the development of policies, strategies, and regulations for the next five-year communication and information technology plan (ideas.mcit.gov.sa). This initiative signals a move towards the espousal of better interaction between the government and citizens. Nevertheless, this level of interaction, in which citizens put forward suggestions for future plans, remains one sided. The establishment of two-way interactions, in which government officials actively discuss such plans with citizens, is suggested based on the findings.



The strong positive influence of conservation values shows that a large element of collective culture (within Schwartz BPV model) supports rather than impedes acceptance of e-transactions. Moreover, the negative influence of power value can be managed. Website interfaces as well as website functionality and design can be conceived of in such a way that embraces these values. Examples include the inclusion of family pictures, which reinforce the image of the government as an institution that cares about stability and family security. Leadership is well revered in Saudi Arabia; including the name or a picture of the King (for example) can encourage trust in a website. Adding the names and profiles of developers and managers as well as the story behind website development can establish a personal connection with users. The establishment of community networks within or connected to e-government websites also intensifies the social experience aspect of these websites. An example would be tribe-specific blogs or forums, in which technically capable members provide guidance or assistance on the use of e-government websites. Another promising feature is online distance assistance, in which a government employee can remotely assist a citizen in completing online transactions. The last two features are related to benevolence and power value because government officials can assist others easily and also acquire as result of extending positive legal and technical advice extended to citizens (Warkentin, et al., 2002; Zakaria, et al., 2003). Power value imposes a negative influence on acceptance. The negatively perceived prestige and authority gained through the use of connections (*wasta*) can be eliminated by constantly monitoring the behaviour of government employees (Smith, et al., 2011). Incentives such as online tributes and praise for the assistance provided to others can be provided to encourage positive behaviour. These incentives can be extended to both government employees and citizens who assist new adopters or those encountering difficulties in completing their e-transactions. Such initiatives will not only increase the level of acceptance, but also establish stronger connections with the government and build integrity via socialisation and two-way communication. Awareness programmes and advertisements with conservative themes can be launched to motivate more citizens to use e-transactions. Other awareness programmes, as well as newspaper and television advertisements, can highlight the negative aspects of obtaining



power and authority when conducting government transactions as a method to discourage such behaviour.

Social networking tools and services can be used not only between citizens and the government, but also among citizens who wish to share their experiences with e-transactions. As a collective society that places a premium on the result demonstrability of e-transactions, the KSA can benefit from the use of social networks and Web 2.0 technologies (e.g., blogs, YouTube, Facebook, and Wikis) by increasing opportunities for information sharing. These social networking tools can also serve as feedback channels between citizens and e-government developers; such feedback can drive government employees to improve e-transactions as required by citizens. Another interesting idea to explore is establishing real-time communication channels, in which government employees interact with citizens as they complete online transactions. A chat tool effectively serves this purpose. This approach will enhance trust in technologies, especially in recently introduced online services and transactions where finances are involved. The Saudi government can also consider transitioning into eGovernment 2.0 or government 2.0. This shift does not pertain merely to adopting a new technology; government 2.0 implementation would be an appropriate strategy in promoting acceptance because the Saudi collective culture is characterized by a tendency to share and discuss experiences with others. Government 2.0 is defined as the use of technology that enables the commoditisation and socialisation of the internal and external data, services, and processes of governments. Its adoption, therefore, is not restricted to using Web 2.0 and social networking tools. Government 2.0 involves transparency, inclusion of citizens in decision making, sharing processes, service conduct, and methods with citizens, and citizen and employee empowerment (Di Maio, 2009; Henman, 2010).

The shift from mere publishing and transacting with citizens to a sharing and collaborative mentality via government 2.0 elevates the compatibility of services with cultural traditions within government agencies. Compatibility increases because government 2.0 enables a high level of social interaction among citizens, and between government officials and citizens,



thereby stimulating numerous social cues and trends that are usually induced via traditional methods. Establishing e-government or government 2.0 features that promote social seamlessness and cultural compatibility is therefore a crucial objective (Carter & Bélanger, 2005).

Citizens trust the Internet as an intermediary between them and the government, but continued faith necessitates constant support. Government agencies can also exert efforts to encourage trust in those who doubt the effectiveness of the Internet. Internet security can be enhanced by regulating use; that is, policies that penalise hackers can be formulated. Authentication methods can be employed, either by using national identification card biometrics, encryption methods, or Internet-specific policies, such as public key infrastructures or complex password authentication techniques. Internet research and innovations have engendered numerous methods and techniques for authentication. These features will reinforce trust in the Internet, especially when awareness campaigns devoted to Internet safety and security methods are launched; citizens will realise that the Internet can guarantee privacy and confidentiality (Al-Gahtani, 2011).

The research findings indicate that Saudi citizens trust and accept the technology itself, but not the government agencies that oversee operations. Government agencies can enhance their public image by improving organisational culture, practices, and processes. These improvements can be effective because the transition into e-government will not entail a replacement of all traditionally provided services, even if all citizens have Internet access. An organisational culture of transparency is suggested. To increase adoption of such a culture, the members of top management can serve as role models for other employees. Government officials will also be consistent in their dealings with citizens, thereby reducing perceived unequal treatments and the *wasta* mentality. In addition, top management can provide government employees with courses and lectures that emphasise increasing transparency to enhance public trust. Another means of increasing transparency is the online posting of e-



transaction requirements, so that citizens are aware of them and realise that the prerequisites imposed are the same for all citizens.

**10.4   Limitations and Future Research**

        The limitations and shortcomings of this study can serve as opportunities for future research. Only citizens with Internet access were sampled in this study; a sample that also covers individuals who do not have Internet access will generate more comprehensive and representative results. Researchers can perform random probability sampling in the KSA to ascertain representativeness. To facilitate this process, scholars can acquire or establish contact databases. A qualitative method can also be triangulated with the research findings to deepen the understanding of the influence of culture on the acceptance of e-transactions. Instead of conducting a cross-sectional study, a longitudinal study can be carried out to accurately determine the influence of acceptance across different periods of time. This research focused on the culture of Saudi nationals. The preliminary efforts initiated in this research can be extended to individuals of different nationalities who are residing in the KSA. Such an extension would be favourable, especially when more e-transaction services become available and the Internet becomes more accessible (e.g., via mobile phones). Demographic characteristics can be incorporated as moderators of currently considered factors or can be assessed as independent determinants of acceptance. The acceptance of e-transactions in general was considered, but other researchers can enhance the focus of future efforts by narrowing this emphasis and contextualising the research based on a specific online service. It is suggested that researchers focus on the influence of culture on the acceptance of website interfaces, and elucidate online technology features and aspects in terms of different cultures. Scholars can explore the Perceived Characteristics of E-Transactions (PCET) model and Schwartz's theory of Basic Human Values (BHV) in relation to technology acceptance. The Portrait Values Questionnaire (PVQ) can be applied to the circumstances of other Arab countries given that the current work on BHV lacks an extensive encapsulation of cultural values and comparisons with other



findings. Researchers can test the validity of the PCET model within different settings and varied cultures, which will enable an understanding of acceptance in different contexts.

## 10.5 Summary

The theoretical and practical contributions of this study were presented, as well as implications for successful e-government development of in the KSA. The chapter concluded with a discussion of limitations and future research directions. As a theoretical contribution, this research focused on the influence of culture on the acceptance of e-transactions. Usage of e-transactions as a communication channel with the government, conservation and power values, result demonstrability and compatibility of e-transactions and trust in the Internet and government are influential factors in the acceptance of e-transactions. A major finding of this research is the positive impact of conservation values and preference of using e-transactions as a communication tool which contradicts with other studies within this stream of research particularly focusing on culture (Pons, 2004). Based on the theoretical findings, a set of design and implementation strategies were formulated for the government of KSA. Improvements includes enhancing the communication methods between government and citizens, implementation of government 2.0 and taking into consideration the influence of cultural values, focusing on improving the public perceptions on government agencies, and provision of enhanced Internet infrastructure within the KSA. While the focus of this research was on quantitative method, a qualitative study would provide a different perspective into the explanation of how culture could affect acceptance of e-transactions.



**APPENDIX A: EXAMINATION OF THE RESEARCH MODEL AND DESIGN**

The main purpose of this phase was to ascertain that the factors developed from the literature were relevant and important to this thesis. Another purpose was to ascertain that the research design was suitable to the research goals. After reviewing the related literature, open-ended interviews were conducted with 12 highly published academics (including Saudi citizens) in the fields of IS acceptance, cultural studies, social psychology, and e-government. Highly published academics (called 'experts' in this section) were identified as such if they had published at least 10 papers related to this study (Palvia, et al., 2007). The interviewed academics were selected based on their background and specialisation, as shown in the table below. Most of these interviews were conducted during information systems conferences, and the researcher scheduled appointments lasting for one to two hours per interview during the days following the conference to discuss the research model which was developed based on the literature. Some of the academics who were located in Australia were interviewed more than once during 2010, as shown in the table below.

Before the interviews were conducted, the research context was described to each expert, and discussion recordings were reviewed (Miles & Huberman, 1994). The table below presents information on the experts who participated in these interviews, the date of each interview, and the number of times each expert was interviewed.

*Expert Demographics*

| # | Academic position | Specialisation area | Publications | Date and number of times interviewed |
|---|---|---|---|---|
| 1 | Assistant Professor | Innovation management | Mainly on technology adoption. Has done some work on culture and e-government adoption. | Once in September 2009 |
| 2 | Professor | Information systems, technology adoption, and culture | Many publications on technology adoption and some studies on developing countries. Few publications on e- | Five times in January, February, and March 2010 |



| | | | government adoption in developing countries. | |
|---|---|---|---|---|
| 3 | Professor | Information systems | Technology adoption, information systems research, and the Arab culture. | Once in June 2010 |
| 4 | Professor | Culture, organisational culture, and social psychology | Focus on culture, organisational culture, and social psychology. | Twice in February, and March 2010 |
| 5 | Associate Professor | Information systems, and technology adoption | Many publications on technology adoption in developing countries. Also some publications on e-government adoption. | Twice in June 2010 |
| 6 | Professor | Information systems, and technology adoption | Technology adoption, e-government adoption, and information systems research. | Three times in June 2010 |
| 7 | Associate Professor | Management information systems | National culture and technology adoption. | Twice in July 2010 |
| 8 | Senior Lecturer | Culture and information systems | Many papers on technology adoption and culture in Arab countries, including Saudi Arabia. | Once in January 2010 |
| 9 | Associate Professor | Information systems and computer science | Papers on information systems and research methods. | Once in December 2010 |
| 10 | Senior Lecturer | Information systems | Technology adoption, research methods, and culture. | Twice in March and April 2010 |
| 11 | Assistant Professor | Information systems | Many papers on e-government adoption in Saudi Arabia. | Six times between January and August 2010 |
| 12 | Assistant Professor | Information systems | Many papers on the Arab culture and technology adoption. | Once in June 2010 |

Most of these experts agreed that the research model should be customised to suit the research context. Eight experts supported using the complete PCI model as the basic framework for explaining the acceptance of the participants in the study, owing to the relevance of the context described. The author discussed the usage of the social influence construct from Venkatesh et al.'s (2003) UTAUT and Aoun et al.'s (2010) perspective on communication, and both were advocated. To enable the study of culture at the individual level, the usage of



Schwartz's PVQ was also advocated. The Saudi experts mentioned that due to the recent

introduction of e-government transactions in the KSA, the complexity construct would be more

reflective of the society than ease of use. Also, the visibility, trialability and image constructs

were excluded, because they are not related to e-government acceptance in Saudi Arabia, as

discussed in the interviews. Trustworthiness perceptions were suggested for inclusion due to

their importance in determining acceptance of online transactions. The table below summarises

suggestions from experts and to whom these are attributed. This table uses the participants'

numbers from the table above to summarise the expert opinions.

*Suggestions from Experts*

| Suggestions | Experts who suggested or concurred with the suggestion |
|---|---|
| Consideration of research context. | 1, 2, 3, 5, 8, 9, 10, 11, 12 |
| Using Schwartz Basic Personal Values theory. | 4, 2 |
| Inclusion of trustworthiness perceptions. | 1, 2, 11 |
| Exclusion of visibility and image constructs as citizens would have difficulties in seeing others use e-transactions. | 3, 6, 11 |
| Inclusion of trialability is irrelevant as users of e-transactions would not be interested in trying the service. | 6, 11 |
| Using complexity instead of ease of use. The complexity construct and TAM's ease of use are very similar and would lead to the same conclusions. | 2, 11 |
| Voluntariness is not important for e-transactions, because it will not provide variance as Saudi citizens have the choice to use electronic transactions or go to SGA. | 3, 6, 11 |
| Result demonstrability construct might be important when citizens contact each other to describe the outcome of using e-transactions. | 3, 6, 11 |
| Include a construct that captures perceptions on communication. | 2, 10, 12 |
| Base the research model on PCI. | 1, 2, 3, 6, 7, 8, 10, 11 |
| Replace image construct with social influence construct as a measure of society's impact. | 6, 11 |
| Avoid common method bias. | 2, 4 |



**APPENDIX B: ETHICAL CLEARANCE CERTIFICATION**

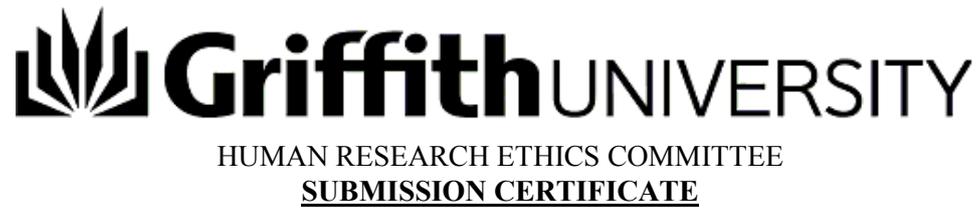

HUMAN RESEARCH ETHICS COMMITTEE
**SUBMISSION CERTIFICATE**

This certificate generated on 16-04-2012.

This certificate confirms that an application for 'Micro-Cultural Influence Modeling for E-Government Adoption' (GU Protocol Number ICT/04/09/HREC) . This application will shortly be considered by the Griffith University Human Research Ethics Committee (HREC).

The applicant will be advised of the outcome of this consideration in due course.

This correspondence will list the standard conditions of ethical clearance that apply to Griffith University protocols.

The HREC is established in accordance with the *National Statement on Ethical Conduct in Research Involving Humans*. The operation of this Committee is outlined in the HREC standard Operating Procedure, which is available from www.gu.edu.au/or/ethics.

Please do not hesitate to contact me if you have any further queries about this matter.

Dr Gary Allen
Manager, Research Ethics
Office for Research
G39 room 3.55 Gold Coast Campus
Griffith University
Phone: 3735 5585
Facsimile: 07 5552 9058
Email: g.allen@griffith.edu.au



**APPENDIX C: INVITATION LETTER FOR CONTENT VALIDITY PHASE**

Dear *****

I hope this message finds you well,

I am validating the content of my instrument as I will need your help in this phase. This instrument is part of a study titled: Micro-cultural Influence Modelling for E-Government Adoption in Saudi Arabia.

The results of this study is a set of recommendations that will enable creating electronic government transactions (government services on the Internet) which are compatible with human needs and takes into account different personal requirements and perceptions. Thus, an instrument was created from PCI (Perceived Characteristics of Innovation), DOI (Diffusion of Innovation), UTAUT (Unified theory of Acceptance and Use of Technology) and Trustworthiness models to study the citizens' perceptions on e-government transactions. The result of synthesising measures from these models is a questionnaire which includes 40 questions.

What is required is your opinion on each construct and its related items in the instrument as I am trying to get representative measures of the given constructs. Please click on this link which contains more explanation about the research.

This will only take about 5 to maximum 15 minutes

I appreciate your help

Ibrahim Abu Nadi



**APPENDIX D: CONTENT VALIDITY QUESTIONNAIRE**

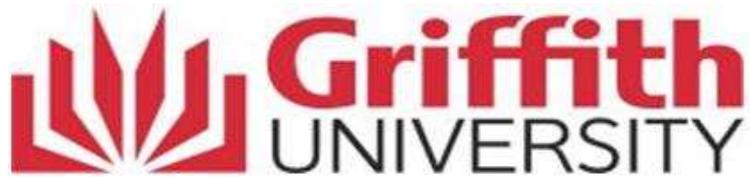

<u>**Title of this Project**</u>
Micro-cultural Influence Modelling for E-Government Adoption in Saudi Arabia

<u>**Why is the research being conducted?**</u>
This study would result in a set of recommendations that will enable creating electronic government transactions (government services on the Internet) which are compatible with human needs and takes into account different personal requirements and perceptions. Thus, the research question is: what influences do cultural values have on the perceptions of e-government diffusion different characteristics? (Nevertheless, there is another questionnaire to measure culture which is not included here) Additionally, feedback on this questionnaire provides content validity for this instrument.

<u>**What is an E-Government Transaction?**</u>
On-line e-government transactions include paying speeding infringements using the internet, applying for a government job on-line, applying or reviewing information about scholarships using the Ministry of Higher Education websites, paying bills, making a complaint.

<u>**What you will be asked to do?**</u>
Kindly please give your feedback on this instrument, which will take about 5-10 minutes.
There are 40 questions distributed in 2 pages.
Constructs of this research will be defined, each following its items.

<u>**The expected benefits of the research**</u>
By modelling cultural factors which might impede or enhance e-government transaction acceptance, a deeper understanding will be gained of the following factors: e-government cultural motivational factors, required cultural aspects in implementation and development of e-government.

<u>**Your confidentiality and risks involved**</u>
The questionnaire is completely anonymous with no other private or corporate identifying data being recorded. The research focus will be on categories drawn from the aggregated material rather than from any individual; any report or publication from this study will conceal or remove any identifying features, which might tend to connect you with any of the reported responses.

<u>**Your participation is voluntary**</u>
Your participation is voluntary and you may discontinue your participation at any time if your decision was not to participate in this study.

<u>**The ethical conduct of this research**</u>
Griffith University conducts research in accordance with the National Statement on Ethical Conduct in Human Research.  If potential participants have any concerns or complaints about the ethical conduct of the research project they should contact the Manager, Research Ethics on 3735 5585 or research-ethics@griffith.edu.au.



**Questions / further information**
For any inquiries about this research you can contact;
The researcher: Ibrahim Abu Nadi, School of ICT, Griffith University,
i.abunadi@griffith.edu.au, Mobile +61(0)413649905,
The supervisor: Dr Steve Drew, School of ICT, Griffith University, S.Drew@griffith.edu.au.

**Feedback to you**
The result of this study will be available on-line once published

Completing and returning feedback on this instrument means that you have read and understood the previously mentioned information and agreed that you would allow usage of the data in the manner described above.

Please note that your answers will be saved (to be able to continue later) if you close or leave the browser, but it will not be saved if you change the computer you are completing the questionnaire from.

Info. Please click on 'Next' to proceed.

*. Please evaluate each item based on its relevance and consistency with each give definition. Please note that version of the survey is only used for evaluating the instrument.

Def. E-Government Relative Advantage:

The degree to which usage of e-government transactions are seen as being superior to traditional method transactions with government officials.

- Traditional methods include face-to-face interaction with government officials or using the help of friend or relative.

Relative Advantage (Rogers, 2003):
The degree to which an innovation is seen as being superior to its predecessor.

*. Any comments about this definition?

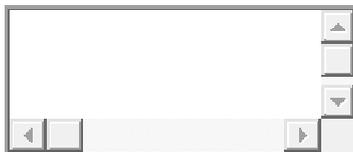

RA1. Using e-government would enable me to carry out my transactions more quickly.

  ○  Not relevant

  ○  Important

  ○  Essential

*. Any comments about RA1 (this question above) ?



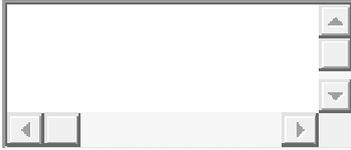

RA2. Using e-government would improve the quality of the way my transactions are conducted.

○  Not relevant

○  Important

○  Essential

*. Any comments about RA2?

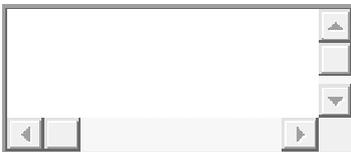

RA3. Using e-government would make it easier to carry out transactions with the government.

○  Not relevant

○  Important

○  Essential

*. Any comments about RA3?

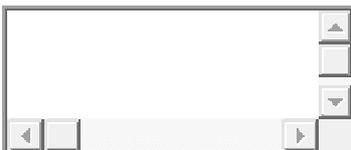

RA4. Using e-government services helps carry out transactions more effectively

○  Not relevant

○  Important

○  Essential

*. Any comments about RA4?

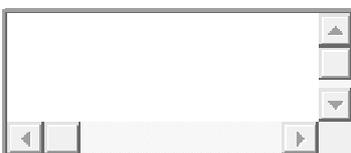



RA5. Using e-government would give me greater control over conducting transactions with the government.

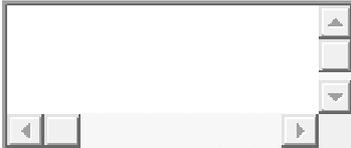 Not relevant

Important

Essential

*. Any comments about RA5?


Def. <u>E-Government Compatibility:</u>
Consistency of the usage of e-government transactions with a potential adopter's needs, past experiences with government agencies and values.

The consistency of e-government services to a potential adopter's needs, etc

<u>Compatibility (Rogers, 2003):</u>

The consistency of an innovation to a potential adopter's needs, past experiences and values.

*. Any comments about this definition?


CT1. Using e-government is compatible with how I like to conduct transactions with the government.

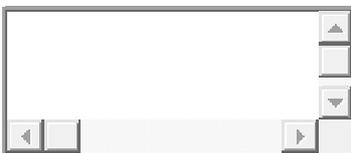 Not relevant

Important

Essential

*. Any comments about CT1?


CT2. Using e-government transactions is completely compatible with my current needs.

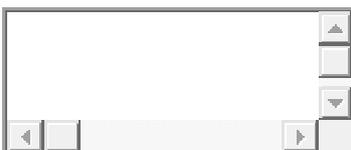



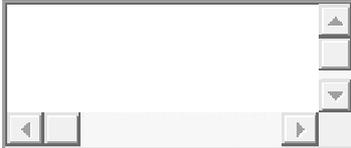

Not relevant

Important

Essential

*. Any comments about CT2?

CT3. I think that using e-government would fit well with the way that I prefer to conduct transactions with the government.

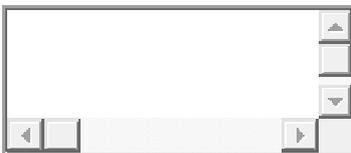

Not relevant

Important

Essential

*. Any comments about CT3?

CT4. Using e-government transactions would fit well into my lifestyle.

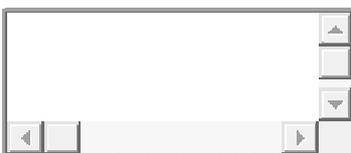

Not relevant

Important

Essential

*. Any comments about CT4?

Def. <u>E-Government Complexity</u>

The level to which using e-government transaction is perceived acceptable and effortless in usage.



Complexity (Rogers, 2003):
Level to which an innovation is perceived acceptable and effortless in terms of usage.

*. Any comments about this definition?

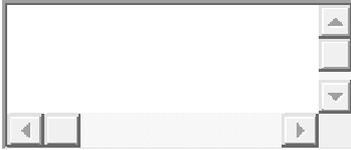

CMX1. Using e-government transactions would consume too much of my time.

   ◯  Not relevant

   ◯  Important

   ◯  Essential

      *. Any comments about CMX1?

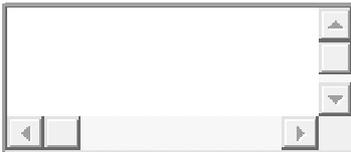

CMX2. Conducting e-government transactions would be so complicated, it would be difficult to understand what is going on.

   ◯  Not relevant

   ◯  Important

   ◯  Essential

*. Any comments about CMX2?

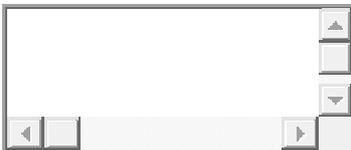

CMX3. Using e-government transactions would involve too much time doing technical operations (e.g. data entry)

   ◯  Not relevant

   ◯  Important

   ◯  Essential



*. Any comments about CMX3?

CMX4. It would take too long to learn how to use e-government transactions to make it worth the effort.

  ○ Not relevant

  ○ Important

  ○ Essential

*. Any comments about CMX4?

*. Original Perceived innovation Characteristics Model by Benbasat and Moore (1991) contained "Perceived Ease of Use" construct which was avoided in this instrument due to weak loadings and results found in previous e-government research on Saudi Arabia (Hisham et al, 2010) and Arab world (Alomari et al, 2009) and additionally due to wide criticism of TAM Perceived Ease of Use and Usefulness (Benbasat and Barki, 2007).

Any suggestions or comments?

Def. E-Government Result Demonstrability
The tangibility of the results of using e-government transactions, inducing observability and communicability of its results.

Result Demonstrability (Benbasat and Moore, 1991):

The tangibility of the results of using the innovation, including the observability and communicability.

*. Any comments about this definition?



RED1. I would have no difficulty telling others about the results of using e-government transactions.

   ○  Not relevant

   ○  Important

   ○  Essential

*. Any comments about RED1?

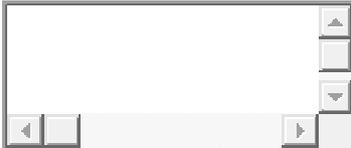

RED2. I believe I could communicate to others the consequences of using e-government transactions.

   ○  Not relevant

   ○  Important

   ○  Essential

*. Any comments about RD2?

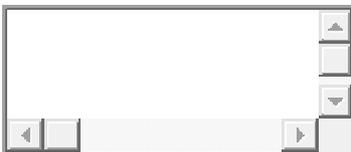

RED3. The results of using e-government transactions are apparent to me.

   ○  Not relevant

   ○  Important

   ○  Essential

*. Any comments about RD3?

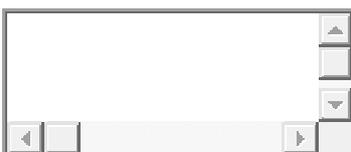

RED4. I would have difficulty explaining why using e-government transactions may or may not be beneficial.



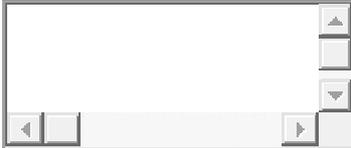 Not relevant

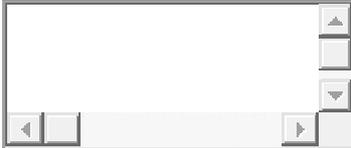 Important

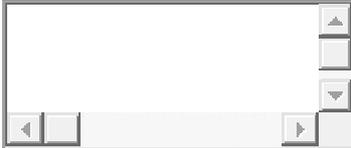 Essential

*. Any comments about RED4?

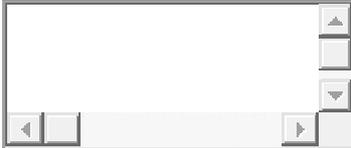

Def. Communication using E-Government Transactions
The degree to which using e-government transactions would enable adequate communication with the government.

Perspective on Communication (Chadi et al, 2010):

Preference of high or low context cultures in the communication with business stakeholders using Accounting Information Systems.

*. Any comments about this definition?

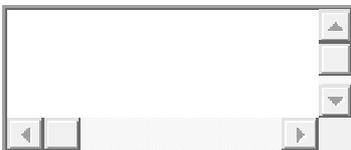

POC1. My ability to communicate with the government would be enhanced when using e-government transactions.

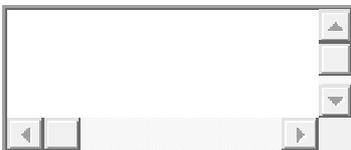 Not relevant

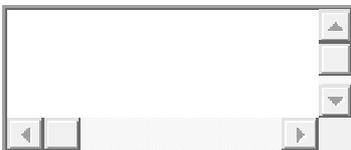 Important

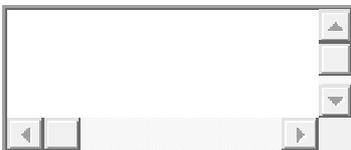 Essential

*. Any comments about POC1?

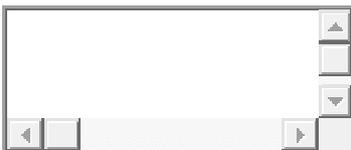

POC2. Communications through e-government transactions enhance my ability to interpret government services.

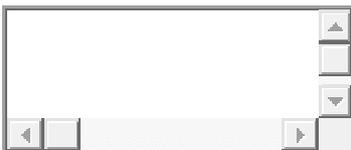 Not relevant



○ Important

○ Essential

*. Any comments about POC2?

POC3. Textual, verbal and visual information is important for carrying out e-government transactions.

○ Not relevant

○ Important

○ Essential

*. Any comments about POC3?

**Def. <u>Intention to use E-Government Transactions</u>**
Citizens' behavioural intention towards the usage of e-government transactions.

**<u>Intention to use E-Government (Carter and Bélanger, 2005):</u>**

Citizens' behavioural intention towards the usage of e-government Tax services.

*. Any comments about this definition?

USE1. I would use e-government to gather information about my required transactions in the future.

○ Not relevant

○ Important

○ Essential



*. Any comments about USE1?

USE2. I would use e-government transactions provided over the Internet.

○  Not relevant

○  Important

○  Essential

*. Any comments about USE2?

USE3. Using e-government transactions is something that I would do.

○  Not relevant

○  Important

○  Essential

*. Any comments about USE3?

USE4. I would not hesitate to provide information to e-government websites to conduct my transactions.

○  Not relevant

○  Important

○  Essential

*. Any comments about USE4?



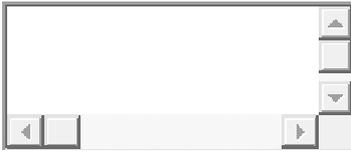

USE5. I would use e-government to inquire about my government transactions.

○ Not relevant

○ Important

○ Essential

*. Any comments about USE5?

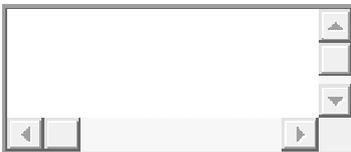

**Def. <u>Trustworthiness</u>**
Citizen trust in the state government agency providing the service and citizen trust in the technology through which electronic transactions are executed, the internet (Carter and Bélanger, 2005, P.9-10)

*. Any comments about this definition?

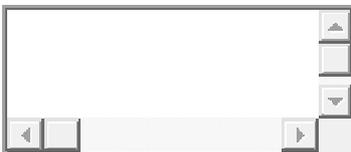

TI1. The Internet has enough safeguards to make me feel comfortable in conducting transactions using e-government.

○ Not relevant

○ Important

○ Essential

*. Any comments about TI1?

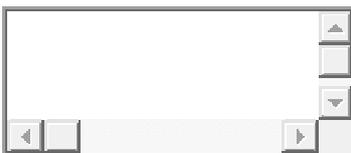

TI2. I feel assured that legal and technological structures adequately protect me from problems on the
Internet while using e-government transactions.



○ Not relevant

○ Important

○ Essential

*. Any comments about TI2?

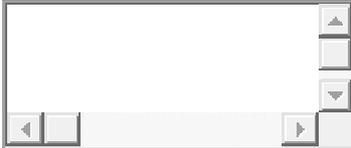

TI3. Generally, I feel that the Internet is now a robust and safe environment in which to conduct on-line transactions with the government.

○ Not relevant

○ Important

○ Essential

*. Any comments about TI3?

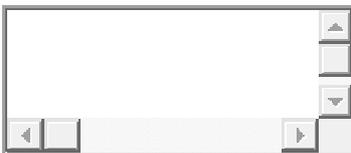

TG1. I think I can trust government agencies in delivering my transactions using e-government.

○ Not relevant

○ Important

○ Essential

*. Any comments about TG1?

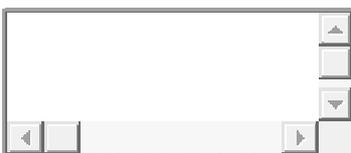

TG2. Government agencies can be trusted to carry out on-line transactions faithfully.

○ Not relevant

○ Important



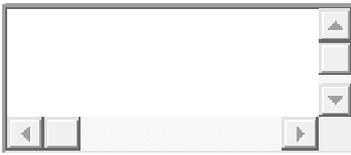 Essential

\*. Any comments about TG2?

TG3. In my opinion, government agencies are trustworthy in their ability to deliver services using e-government transactions.

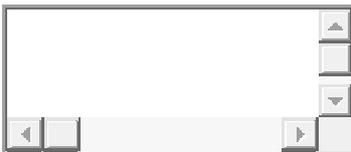 Not relevant

Important

Essential

\*. Any comments about TG3?

TG4. I trust government agencies to keep my best interest in mind while delivering on-line services using e-government transactions.

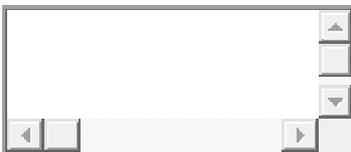 Not relevant

Important

Essential

\*. Any comments about TG4?

**Def. <u>Social Influence (Venkatash et al, 2003)</u>**
The degree to which an individual perceives that important others believe he or she should use the new system (P.451).

**Def. <u>E-Government transaction Social Influence</u>**
The degree to which a citizen perceives that important others believe he or she should use e-government transactions.

\*. Any comments about this definition?



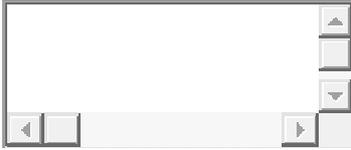

SI1. People who influence my behaviour would think that I should use e-government to conduct transactions.

  ○  Not relevant

  ○  Important

  ○  Essential

*. Any comments about SI1?

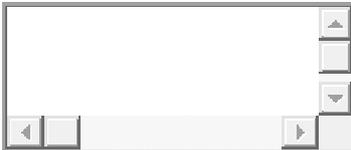

SI2. People who are important to me would think that I should use the e-government transactions.

  ○  Not relevant

  ○  Important

  ○  Essential

*. Any comments about SI2?

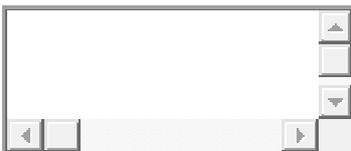

SI3. People who are in my social circle would think that I should use the e-government to conduct transactions.

  ○  Not relevant

  ○  Important

  ○  Essential

*. Any comments about SI3?



*. If you have any comments or suggestions about the complete instrument please add it below.

Info. Thank you for participating,



**APPENDIX E: INVITATION LETTER FOR THE FULL-SCALE PHASE (ENGLISH)**

Dear Sir/Madam

We would like to call upon you to participate in a study titled: "Micro-Cultural Influence Modeling for E-Government Transaction Adoption in Saudi Arabia" by completing this on-line questionnaire.

As an Internet user you are a nominee to answer this survey where only your general experience with the Internet is required.

The questionnaire might take about 15-30 minutes to complete.

The research team was not provided any personal information about you and the questionnaire has been designed to keep your privacy intact.

Ibrahim Abu Nadi

PhD Candidate

School of Information Communication Technology, Griffith University

Note: The results of this study will be available on an external link after completing the questionnaire (The results will be provided on the same link once published).

To participate in this questionnaire please click on the following link

or please copy this link to your browser

http://qualtrasia.qualtrics.com/SE/?SID=SV_agdcHOMQ8mhaMNS

Examples of e-government transactions include the following:-- Transactions and services available for citizens and non-citizens through Saudi E-Government National Portal:



www.saudi.gov.sa

- Transactions available for Saudi students via Ministry of Higher Education Portal: student.mohe.gov.sa

- Inquiry transaction of "Saher" speeding infringement:

http://www.rt.gov.sa/saher-.php



- Electronic Visa services for residences and visitors: visa.mofa.gov.sa/eDefault.asp

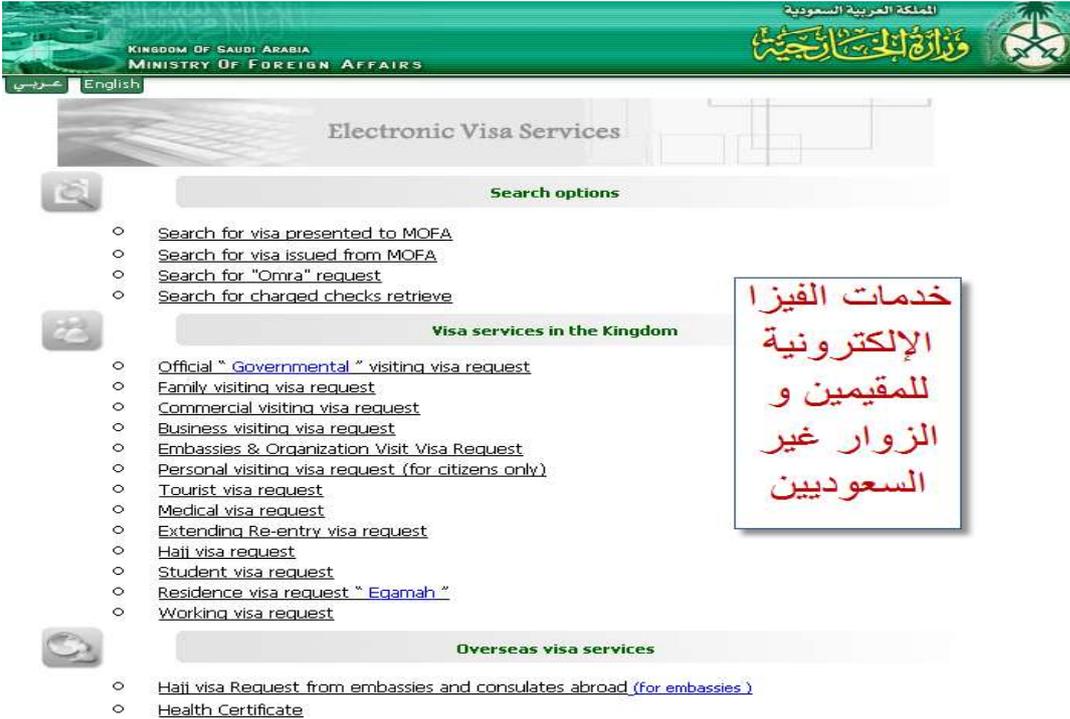

Please note that these above transaction are just examples which are used to clarify the term "E-Government Transactions" as there are many more transactions available in both languages (Arabic and English).

To participate in this questionnaire please click on the following link

or please copy this link to your browser

http://qualtrasia.qualtrics.com/SE/?SID=SV_agdcHOMQ8mhaMNS



**APPENDIX F: INVITATION LETTER FOR THE FULL-SCALE PHASE (ARABIC)**

السلام عليكم ورحمة الله وبركاته

أخي/أختي الكريمـة
ندعوكم للمشاركة في دراسة بعنوان : "تأثير الثقافة على تبني تعاملات الحكومة الإلكترونية في المملكة العربية السعودية" من خلال استكمال هذا الإستبيان.

كمستخدم للإنترنت ، فأنت مرشح للإجابة على هذا الاستبيان حيث أن المطلوب هو خبرتك العامة في استخدام الإنترنت، ذلك يشمل السعوديين والمقيمين، الرجال و النساء، من استخدم الحكومة الإلكترونية ومن لم يستخدمها
الإنتهاء من الاستبيان قد يستغرق مابين 15-30 دقيقة.
تم تصميم الإستبيان بطريقة تحافظ على معلوماتك الشخصية لذا ففريق البحث لن يحصل على أي معلومات خاصة عنك.
للمشاركة في الإستبيان الرجاء الدخول على الرابط
أو نسخ هذا الرابط على المتصفح
http://qualtrasia.qualtrics.com/SE/?SID=SV_agdcHOMQ8mhaMNS
إبراهيم أبونادي
طالب الدكتوراة في كلية الإتصالات و تقنية المعلومات في جامعة جريفيث
بعض الأمثلة على تعاملات الحكومة الإلكترونية السعودية تشمل الآتي-:
الخدمات و التعاملات الإلكترونية المتوفرة للمواطنين و غير المواطنين من خلال البوابة الوطنية للتعاملات الإلكترونية
http://www.saudi.gov.sa

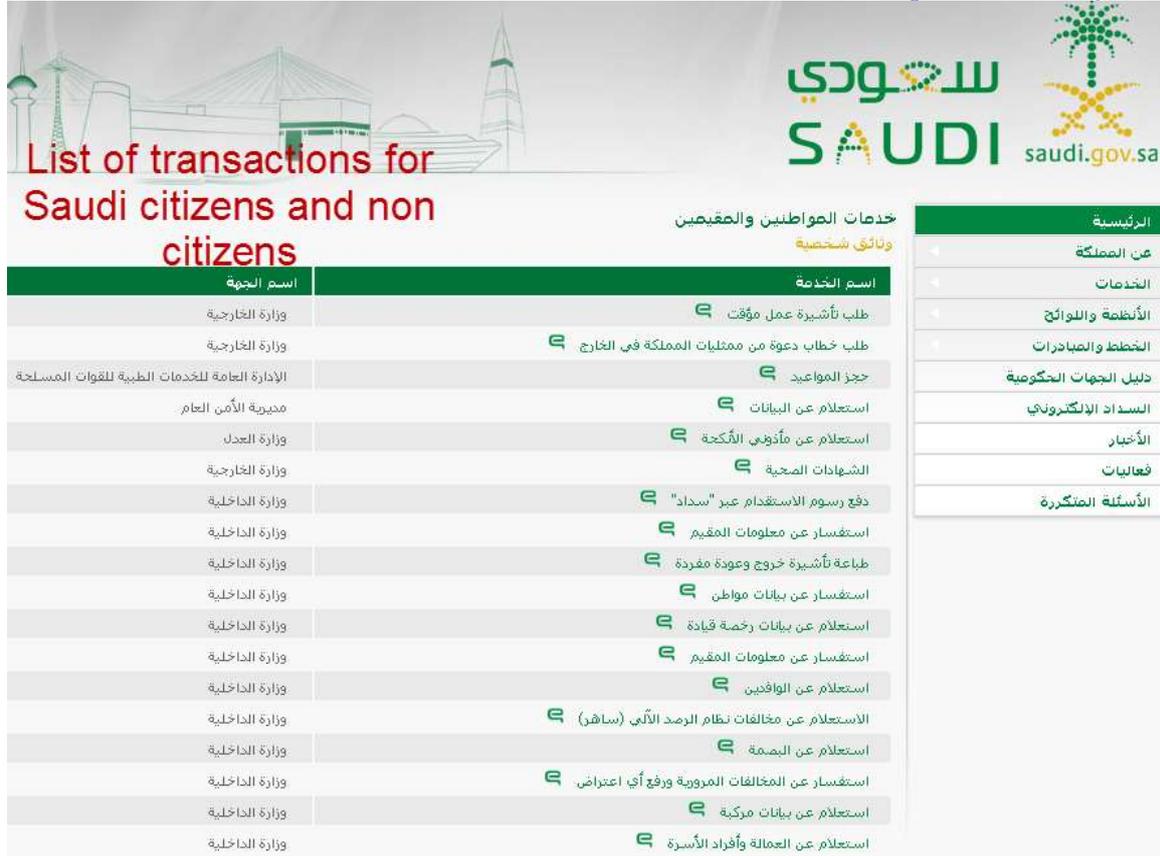

لتكبير الصورة الرجاء الضغط عليها
المعاملات الإلكترونية للمبتعثين المتوفرة من خلال بوابة وزارة التعليم العالي
student.mohe.gov.sa



بوابة
الطلبة المبتعثين

وزارة التعليم العالي
Ministry of Higher Education

تسجيل الخروج | تغيير كلمة المرور | التخصيص | دعم فني | اتصل بنا

أدوائي | الخدمات الدراسية | الخدمات المالية | المعاملات الشخصية | بياناتي | الرئيسية

متابعة الطلبات
طلب تذكرة سفر
طلب استفسار
تنبيهات
طلب مكافأة تميز
طلب إلحاق مرافق بعضوية البعثة
طلب اجازة
طلب تعريف

الخدمات الشخصية > تنبيهات

example of
e-government
transactions for
students

أمثلة على
معاملات
إلكترونية

الخدمات الشخصية

تنبيهات
لا يوجد تنبيهات في الوقت الحالي

متابعة الطلبات

طلب تذكرة سفر

لتكبير الصورة الرجاء الضغط عليها

خدمات الفيزا الإلكترونية للمقيمين و الزائرين
https://visa.mofa.gov.sa/eDefault.asp

المملكة العربية السعودية
وزارة الخارجية
Kingdom Of Saudi Arabia
Ministry Of Foreign Affairs

عربي | English

Electronic Visa Services

**Search options**

○ Search for visa presented to MOFA
○ Search for visa issued from MOFA
○ Search for "Omra" request
○ Search for charged checks retrieve

**Visa services in the Kingdom**

○ Official " Governmental " visiting visa request
○ Family visiting visa request
○ Commercial visiting visa request
○ Business visiting visa request
○ Embassies & Organization Visit Visa Request
○ Personal visiting visa request (for citizens only)
○ Tourist visa request
○ Medical visa request
○ Extending Re-entry visa request
○ Hajj visa request
○ Student visa request
○ Residence visa request " Eqamah "
○ Working visa request

**Overseas visa services**

○ Hajj visa Request from embassies and consulates abroad (for embassies )
○ Health Certificate

خدمات الفيزا
الإلكترونية
للمقيمين و
الزوار غير
السعوديين

لتكبير الصورة الرجاء الضغط عليها
الرجاء ملاحظة أن المعاملات السابقة هي عبارة عن أمثلة لتوضيح مصطلح "التعاملات الإلكترونية" حيث أن هناك العديد من
المعاملات الإلكترونية الأخرى.
بالإضافة إلى ذلك فإن أغلب التعاملات الإلكترونية السعودية متاحة باللغة الإنجليزية و العربية



## APPENDIX G: QUESTIONNAIRE (ENGLISH)

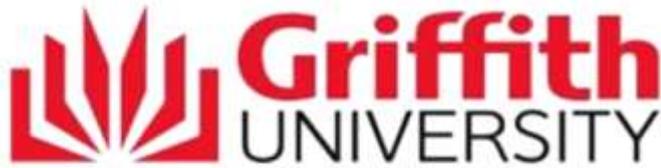

العربية | English

**Title of this Project**
Micro-Cultural Influence Modeling for E-Government Adoption

### Why is the research being conducted?
This study would result in a set of recommendations that will enable creating electronic government transactions (government services on the Internet) which are compatible with human needs and takes into account different personal requirements and perceptions. Additionally, this questionnaire forms part of my PhD study in Griffith University ICT (Information Communication Technology) school.

### What is an E-Government Transaction?
On-line e-government transactions include paying speeding infringements using the internet, applying for a government job on-line, applying or reviewing information about scholarships using the Ministry of Higher Education websites, paying bills, making a complaint.

Therefore, e-government transactions are any kind of contact with the Saudi government which involves filling out forms or contacting the government using the internet about a specific transaction.

### What you will be asked to do?
Please complete this on-line questionnaire, which will take about 15-30 minutes.
There are 88 questions distributed in 12 pages. Please note that this questionnaire will expire on 15/2/2011.

### Please give your opinion based on your general experience with the Internet OR recently conducted transactions with the Saudi e-government, which include:
**1- Using government websites to gather information about your transactions with governmental agencies.**

**2- Conducting transactions with the government via the Internet.**

### The expected benefits of the research
By modelling cultural factors which might impede or enhance e-government transaction acceptance, a deeper understanding will be gained of the following factors: e-government cultural motivational factors, required cultural aspects in implementation and development of e-government.

### Your confidentiality and risks involved
The questionnaire is completely anonymous with no other private or corporate identifying data being recorded. The research focus will be on categories drawn from the aggregated material rather than from any individual; any report or publication from this study will conceal or remove any identifying features, which might tend to connect you with any of the reported responses. The information you provided in this questionnaire will be kept secured and confidential in the office of the School of Information and Communication Technology for a period of 5 years, and will then be destroyed.

### Your participation is voluntary
Your participation is voluntary and you may discontinue your participation at any time without explanation or fear of reprisal. If your decision was not to participate in this study.

### Questions / further information
For any inquiries about this research you can contact;
The researcher: Ibrahim Abu Nadi, School of ICT, Griffith University, i.abunadi@griffith.edu.au,



**Questions / further information**
For any inquiries about this research you can contact;
The researcher: Ibrahim Abu Nadi, School of ICT, Griffith University, i.abunadi@griffith.edu.au,
Mobile +61(0)413649905,
The supervisor: Dr Steve Drew, School of ICT, Griffith University, s.drew@griffith.edu.au.

**Feedback to you**
The result of this study will be available on-line on a provided link after conducting the questionnaire.

Completing and returning the questionnaire means that you have read and understood the previously
mentioned information and agreed that you would allow usage of the data in the manner described
above.

Please note that your answers will be saved (to be able to continue later) if you close or leave the
browser, but it will not be saved if you change the computer you are completing the questionnaire from.

Q1. You are:
   Male     Female
   ○        ○

Q2. Age:
| 17 or under | 18-22 | 23-30 | 31-40 | 41-50 | 51-60 | 60 or over |
|-------------|-------|-------|-------|-------|-------|------------|
| ○ | ○ | ○ | ○ | ○ | ○ | ○ |

Q3. I was born in:

Q4. I am living in:

Q5. Your nationality:

Q6. Level of education:
○ Was not educated in schools
○ Primary or secondary school education
○ High School
○ Technical or professional degree (No Bachelor degree)
○ Bachelor degree
○ Graduate certificate
○ Master's degree
○ Doctorate or higher

Q7. Proficiency of Internet usage:
| Very low | Low | Satisfactory | Good | Very Good | Excellent |
|----------|-----|--------------|------|-----------|-----------|
| ○ | ○ | ○ | ○ | ○ | ○ |

Q8. Frequency of Internet usage:
○ Few times a year
○ Few times a month



○ Few times a week

○ Once a day only

○ Few hours a day

○ Many hours a day

Q9. Have you ever conducted an e-government transaction with the government of Saudi Arabia? (i.e. using the Internet to acquire services or information from the government).

Yes          No (goes to question 12)
○            ○

Q10. When was the last time you conducted an e-government transaction?

| Three years ago | Last year | This year | Within the last six months | This month | This week | Today |
|---|---|---|---|---|---|---|
| ○ | ○ | ○ | ○ | ○ | ○ | ○ |

Q11. How often do you conduct transactions with the Saudi government using the Internet?

| Once a year | Few times a year | Once a month | Few times a month | Once a week | Few times a week | Daily |
|---|---|---|---|---|---|---|
| ○ | ○ | ○ | ○ | ○ | ○ | ○ |

Q12. Which method do you usually use to conduct transactions with the Saudi government?

☐ Mail

☐ Phone

☐ Fax

☐ With the help of an agent

☐ Face to face meeting with government officials

☐ With the help of a friend or relative

☐ Using e-government

☐ I have never conducted any transactions with the Saudi government

☐ Other

Q13. You are:

○ Not employed and not a student

○ Student

○ Government sector employee

○ Private sector employee

○ Freelancer

○ Other

You have completed 15/88 questions, 17% of this survey.
If you have any suggestions about this page and the previous one please add it here



Please answer from your general experience with Internet OR Saudi e-government websites while comparing e-government with traditional methods of transacting with government agencies (e.g. using phone, physically visiting government agencies, using the help of a friend, etc).

Q14. Using e-government would enable me to carry out my transactions more quickly.

| Strongly Disagree | Disagree | Somewhat Disagree | Neither Agree nor Disagree | Somewhat Agree | Agree | Strongly Agree |
|---|---|---|---|---|---|---|
| ○ | ○ | ○ | ○ | ○ | ○ | ○ |

Q15. Using e-government would improve the quality of my transactions.

| Strongly Disagree | Disagree | Somewhat Disagree | Neither Agree nor Disagree | Somewhat Agree | Agree | Strongly Agree |
|---|---|---|---|---|---|---|
| ○ | ○ | ○ | ○ | ○ | ○ | ○ |

Q16. Using e-government would make it easier to carry out transactions with the government.

| Strongly Disagree | Disagree | Somewhat Disagree | Neither Agree nor Disagree | Somewhat Agree | Agree | Strongly Agree |
|---|---|---|---|---|---|---|
| ○ | ○ | ○ | ○ | ○ | ○ | ○ |

Q17. Using e-government would enhance my effectiveness in carrying out transactions with the government.

| Strongly Disagree | Disagree | Somewhat Disagree | Neither Agree nor Disagree | Somewhat Agree | Agree | Strongly Agree |
|---|---|---|---|---|---|---|
| ○ | ○ | ○ | ○ | ○ | ○ | ○ |

Q18. Using the e-government would give me greater control over conducting transactions with the government.

| Strongly Disagree | Disagree | Somewhat Disagree | Neither Agree nor Disagree | Somewhat Agree | Agree | Strongly Agree |
|---|---|---|---|---|---|---|
| ○ | ○ | ○ | ○ | ○ | ○ | ○ |

Q19. Using e-government is compatible with how I like to conduct transactions with the government.

| Strongly Disagree | Disagree | Somewhat Disagree | Neither Agree nor Disagree | Somewhat Agree | Agree | Strongly Agree |
|---|---|---|---|---|---|---|
| ○ | ○ | ○ | ○ | ○ | ○ | ○ |

Q20. Using e-government transactions is completely compatible with my current needs.

| Strongly Disagree | Disagree | Somewhat Disagree | Neither Agree nor Disagree | Somewhat Agree | Agree | Strongly Agree |
|---|---|---|---|---|---|---|
| ○ | ○ | ○ | ○ | ○ | ○ | ○ |

Q21. I think that using e-government would fit well with the way that I prefer to conduct transactions with the government.

| Strongly Disagree | Disagree | Somewhat Disagree | Neither Agree nor Disagree | Somewhat Agree | Agree | Strongly Agree |
|---|---|---|---|---|---|---|
| ○ | ○ | ○ | ○ | ○ | ○ | ○ |



Q22. Using e-government transactions would fit well into my lifestyle.

| Strongly Disagree | Disagree | Somewhat Disagree | Neither Agree nor Disagree | Somewhat Agree | Agree | Strongly Agree |
|---|---|---|---|---|---|---|
| ○ | ○ | ○ | ○ | ○ | ○ | ○ |

Q23. Using e-government transactions would consume too much of my time.

| Strongly Disagree | Disagree | Somewhat Disagree | Neither Agree nor Disagree | Somewhat Agree | Agree | Strongly Agree |
|---|---|---|---|---|---|---|
| ○ | ○ | ○ | ○ | ○ | ○ | ○ |

Q24. Conducting e-government transactions would be so complicated; it would be difficult to understand what is going on.

| Strongly Disagree | Disagree | Somewhat Disagree | Neither Agree nor Disagree | Somewhat Agree | Agree | Strongly Agree |
|---|---|---|---|---|---|---|
| ○ | ○ | ○ | ○ | ○ | ○ | ○ |

You have completed 24/88 questions, 27% of this survey.
Please add any suggestions about this page.

Please answer from your general experience with Internet OR Saudi e-government websites while comparing e-government with traditional methods of transacting with government agencies (e.g. using phone, physically visiting government agencies, using the help of a friend, etc).

Q25. Using e-government transactions would involve too much time doing technical operations (e.g. data entry)

| Strongly Disagree | Disagree | Somewhat Disagree | Neither Agree nor Disagree | Somewhat Agree | Agree | Strongly Agree |
|---|---|---|---|---|---|---|
| ○ | ○ | ○ | ○ | ○ | ○ | ○ |

Q26. It would take too long to learn how to use e-government transactions to make it worth the effort.

| Strongly Disagree | Disagree | Somewhat Disagree | Neither Agree nor Disagree | Somewhat Agree | Agree | Strongly Agree |
|---|---|---|---|---|---|---|
| ○ | ○ | ○ | ○ | ○ | ○ | ○ |

Q27. I would have no difficulty telling others about the results of using e-government transactions.

| Strongly Disagree | Disagree | Somewhat Disagree | Neither Agree nor Disagree | Somewhat Agree | Agree | Strongly Agree |
|---|---|---|---|---|---|---|
| ○ | ○ | ○ | ○ | ○ | ○ | ○ |

Q28. I believe I could communicate to others the consequences of using e-government transactions.

| Strongly Disagree | Disagree | Somewhat Disagree | Neither Agree nor Disagree | Somewhat Agree | Agree | Strongly Agree |
|---|---|---|---|---|---|---|
| ○ | ○ | ○ | ○ | ○ | ○ | ○ |

Q29. The results of using e-government transactions are apparent to me.

| Strongly Disagree | Disagree | Somewhat Disagree | Neither Agree nor Disagree | Somewhat Agree | Agree | Strongly Agree |
|---|---|---|---|---|---|---|
| ○ | ○ | ○ | ○ | ○ | ○ | ○ |



Q30. I would have difficulty explaining why using e-government transactions may or may not be beneficial.

| Strongly Disagree | Disagree | Somewhat Disagree | Neither Agree nor Disagree | Somewhat Agree | Agree | Strongly Agree |
|---|---|---|---|---|---|---|
| ○ | ○ | ○ | ○ | ○ | ○ | ○ |

You have completed 30/88 questions, 40% of this survey.
If you have any suggestions about this page please add it here

Please answer depending on your experience with the Internet.

Q31. The Internet has enough safeguards to make me feel comfortable in conducting transactions using e-government.

| Strongly Disagree | Disagree | Somewhat Disagree | Neither Agree nor Disagree | Somewhat Agree | Agree | Strongly Agree |
|---|---|---|---|---|---|---|
| ○ | ○ | ○ | ○ | ○ | ○ | ○ |

Q32. I feel assured that legal and technological structures adequately protect me from problems on the Internet while using e-government transactions.

| Strongly Disagree | Disagree | Somewhat Disagree | Neither Agree nor Disagree | Somewhat Agree | Agree | Strongly Agree |
|---|---|---|---|---|---|---|
| ○ | ○ | ○ | ○ | ○ | ○ | ○ |

Q33. Generally, I feel that the Internet is now a robust and safe environment in which to conduct on-line transactions with the government.

| Strongly Disagree | Disagree | Somewhat Disagree | Neither Agree nor Disagree | Somewhat Agree | Agree | Strongly Agree |
|---|---|---|---|---|---|---|
| ○ | ○ | ○ | ○ | ○ | ○ | ○ |

Please answer based on your experiences or expectations of the Saudi government agencies OR Saudi e-government.

Q34. I think I can trust government agencies in delivering my transactions using e-government.

| Strongly Disagree | Disagree | Somewhat Disagree | Neither Agree nor Disagree | Somewhat Agree | Agree | Strongly Agree |
|---|---|---|---|---|---|---|
| ○ | ○ | ○ | ○ | ○ | ○ | ○ |

Q35. Government agencies can be trusted to carry out on-line transactions faithfully.

| Strongly Disagree | Disagree | Somewhat Disagree | Neither Agree nor Disagree | Somewhat Agree | Agree | Strongly Agree |
|---|---|---|---|---|---|---|
| ○ | ○ | ○ | ○ | ○ | ○ | ○ |

Q36. In my opinion, government agencies are trustworthy in their ability to deliver services using e-government transactions.

| Strongly Disagree | Disagree | Somewhat Disagree | Neither Agree nor Disagree | Somewhat Agree | Agree | Strongly Agree |
|---|---|---|---|---|---|---|
| ○ | ○ | ○ | ○ | ○ | ○ | ○ |



Q37. I trust government agencies to keep my best interest in mind while delivering on-line services using e-government transactions.

| Strongly Disagree | Disagree | Somewhat Disagree | Neither Agree nor Disagree | Somewhat Agree | Agree | Strongly Agree |
|---|---|---|---|---|---|---|
| ○ | ○ | ○ | ○ | ○ | ○ | ○ |

Please answer based on the opinions you received about e-government transactions.

Q38. People who influence my behaviour would think that I should use e-government to conduct transactions.

| Strongly Disagree | Disagree | Somewhat Disagree | Neither Agree nor Disagree | Somewhat Agree | Agree | Strongly Agree |
|---|---|---|---|---|---|---|
| ○ | ○ | ○ | ○ | ○ | ○ | ○ |

Q39. People who are important to me would think that I should use the e-government transactions.

| Strongly Disagree | Disagree | Somewhat Disagree | Neither Agree nor Disagree | Somewhat Agree | Agree | Strongly Agree |
|---|---|---|---|---|---|---|
| ○ | ○ | ○ | ○ | ○ | ○ | ○ |

Q40. People who are in my social circle would think that I should use the e-government to conduct transactions.

| Strongly Disagree | Disagree | Somewhat Disagree | Neither Agree nor Disagree | Somewhat Agree | Agree | Strongly Agree |
|---|---|---|---|---|---|---|
| ○ | ○ | ○ | ○ | ○ | ○ | ○ |

You have completed 40/88 questions, 46% of this survey.
If you have any suggestions about this page please add it here

Here people are briefly described. Please read each description and think about how much each person is or is not like you. Choose what shows how much the person in the description is like you.

Q41. Thinking up new ideas and being creative is important. He/She likes to do things in his/her own original way.

| Not like me at all | Not like me | A little like me | Somewhat like me | Like me | Very much like me |
|---|---|---|---|---|---|
| ○ | ○ | ○ | ○ | ○ | ○ |

Q42. It is important to be rich. He/She wants to have a lot of money and expensive things.

| Not like me at all | Not like me | A little like me | Somewhat like me | Like me | Very much like me |
|---|---|---|---|---|---|
| ○ | ○ | ○ | ○ | ○ | ○ |

Q43. He/She thinks it is important that every person in the world be treated equally. He/She believes everyone should have equal opportunities in life.

| Not like me at all | Not like me | A little like me | Somewhat like me | Like me | Very much like me |
|---|---|---|---|---|---|
| ○ | ○ | ○ | ○ | ○ | ○ |



Q44. It is very important to show his/her abilities. He/She wants people to admire what they do.

| Not like me at all | Not like me | A little like me | Somewhat like me | Like me | Very much like me |
|---|---|---|---|---|---|
| ○ | ○ | ○ | ○ | ○ | ○ |

Q45. It is important to live in secure surroundings. He/She avoids anything that might endanger their safety.

| Not like me at all | Not like me | A little like me | Somewhat like me | Like me | Very much like me |
|---|---|---|---|---|---|
| ○ | ○ | ○ | ○ | ○ | ○ |

Q46. It is important to do many different things in life. He/She always looks for new things to try.

| Not like me at all | Not like me | A little like me | Somewhat like me | Like me | Very much like me |
|---|---|---|---|---|---|
| ○ | ○ | ○ | ○ | ○ | ○ |

Q47. He/She believes that people should do what they're told. He/She thinks people should follow rules at all times, even when no-one is watching.

| Not like me at all | Not like me | A little like me | Somewhat like me | Like me | Very much like me |
|---|---|---|---|---|---|
| ○ | ○ | ○ | ○ | ○ | ○ |

Q48. It is important to listen to people who are different from him/her. Even when disagrees with them, he/she still wants to understand them.

| Not like me at all | Not like me | A little like me | Somewhat like me | Like me | Very much like me |
|---|---|---|---|---|---|
| ○ | ○ | ○ | ○ | ○ | ○ |

Q49. He/She thinks it is important not to ask for more than what you have. He/She believes that people should be satisfied with what they have.

| Not like me at all | Not like me | A little like me | Somewhat like me | Like me | Very much like me |
|---|---|---|---|---|---|
| ○ | ○ | ○ | ○ | ○ | ○ |

Q50. He/She seeks every chance to have fun. It is important to do things that give him/her pleasure.

| Not like me at all | Not like me | A little like me | Somewhat like me | Like me | Very much like me |
|---|---|---|---|---|---|
| ○ | ○ | ○ | ○ | ○ | ○ |

Q51. It is important to make his/her own decisions about what he/she does. He/She likes to be free to plan and to choose activities.

| Not like me at all | Not like me | A little like me | Somewhat like me | Like me | Very much like me |
|---|---|---|---|---|---|
| ○ | ○ | ○ | ○ | ○ | ○ |

Q52. It is very important to help the people around him/her. He/She wants to care for their well-being.

| Not like me at all | Not like me | A little like me | Somewhat like me | Like me | Very much like me |
|---|---|---|---|---|---|
| ○ | ○ | ○ | ○ | ○ | ○ |



Q53. Being very successful is important. He/She likes to impress.

| Not like me at all | Not like me | A little like me | Somewhat like me | Like me | Very much like me |
|---|---|---|---|---|---|
| ○ | ○ | ○ | ○ | ○ | ○ |

Q54. It is very important to him/her that the country be safe. He/She thinks the state must be on watch against threats from within and without.

| Not like me at all | Not like me | A little like me | Somewhat like me | Like me | Very much like me |
|---|---|---|---|---|---|
| ○ | ○ | ○ | ○ | ○ | ○ |

Q55. He/She likes to take risks. He/She is always looking for adventures.

| Not like me at all | Not like me | A little like me | Somewhat like me | Like me | Very much like me |
|---|---|---|---|---|---|
| ○ | ○ | ○ | ○ | ○ | ○ |

Q56. It is important to him/her always to behave properly. He/She wants to avoid doing anything people would say is wrong.

| Not like me at all | Not like me | A little like me | Somewhat like me | Like me | Very much like me |
|---|---|---|---|---|---|
| ○ | ○ | ○ | ○ | ○ | ○ |

Q57. It is important to be in charge and tell others what to do. He/She wants people to do what he/she says.

| Not like me at all | Not like me | A little like me | Somewhat like me | Like me | Very much like me |
|---|---|---|---|---|---|
| ○ | ○ | ○ | ○ | ○ | ○ |

Q58. It is important to be loyal to friends. He/She wants to devote him/herself to people close to him/her.

| Not like me at all | Not like me | A little like me | Somewhat like me | Like me | Very much like me |
|---|---|---|---|---|---|
| ○ | ○ | ○ | ○ | ○ | ○ |

Q59. He /She strongly believe that people should care for nature. Looking after the environment is important to him/her.

| Not like me at all | Not like me | A little like me | Somewhat like me | Like me | Very much like me |
|---|---|---|---|---|---|
| ○ | ○ | ○ | ○ | ○ | ○ |

Q60. Being religious is important. He/She tries hard to follow religious beliefs.

| Not like me at all | Not like me | A little like me | Somewhat like me | Like me | Very much like me |
|---|---|---|---|---|---|
| ○ | ○ | ○ | ○ | ○ | ○ |

You have completed 60/88 questions, 68% of this survey.
If you have any suggestions about this page please add it here

Here people are briefly described. Please read each description and think about how much each person is or is not like you. Choose what shows how much the person in the description is like you.



Q61. It is important that things be organised and clean. He/She really does not like things to be a mess.

| Not like me at all | Not like me | A little like me | Somewhat like me | Like me | Very much like me |
|---|---|---|---|---|---|
| ○ | ○ | ○ | ○ | ○ | ○ |

Q62. He/She thinks it is important to be interested in things. He/She likes to be curious and to try to understand all sorts of things.

| Not like me at all | Not like me | A little like me | Somewhat like me | Like me | Very much like me |
|---|---|---|---|---|---|
| ○ | ○ | ○ | ○ | ○ | ○ |

Q63. He/She believes all the worlds' people should live in harmony. Promoting peace among all groups in the world is important.

| Not like me at all | Not like me | A little like me | Somewhat like me | Like me | Very much like me |
|---|---|---|---|---|---|
| ○ | ○ | ○ | ○ | ○ | ○ |

Q64. He/She thinks it is important to be ambitious. He/She wants to show how capable he/she is.

| Not like me at all | Not like me | A little like me | Somewhat like me | Like me | Very much like me |
|---|---|---|---|---|---|
| ○ | ○ | ○ | ○ | ○ | ○ |

Q65. It is best to do things in traditional ways. It is important to keep up the customs he/she has learned.

| Not like me at all | Not like me | A little like me | Somewhat like me | Like me | Very much like me |
|---|---|---|---|---|---|
| ○ | ○ | ○ | ○ | ○ | ○ |

Q66. Enjoying life's pleasures is important. He/She likes to "spoil" him/herself.

| Not like me at all | Not like me | A little like me | Somewhat like me | Like me | Very much like me |
|---|---|---|---|---|---|
| ○ | ○ | ○ | ○ | ○ | ○ |

Q67. It is important to respond to the needs of others. He/She tries to support those he/she knows.

| Not like me at all | Not like me | A little like me | Somewhat like me | Like me | Very much like me |
|---|---|---|---|---|---|
| ○ | ○ | ○ | ○ | ○ | ○ |

Q68. It is important always to show respect to parents and to older people. It is important to be obedient.

| Not like me at all | Not like me | A little like me | Somewhat like me | Like me | Very much like me |
|---|---|---|---|---|---|
| ○ | ○ | ○ | ○ | ○ | ○ |

Q69. He/She wants everyone to be treated justly, even people do not know. It is important to protect the weak in society.

| Not like me at all | Not like me | A little like me | Somewhat like me | Like me | Very much like me |
|---|---|---|---|---|---|
| ○ | ○ | ○ | ○ | ○ | ○ |



Q70. He/She likes surprises. It is important to have an exciting life.

| Not like me at all | Not like me | A little like me | Somewhat like me | Like me | Very much like me |
|---|---|---|---|---|---|
| ○ | ○ | ○ | ○ | ○ | ○ |

Q71. He/She tries hard to avoid getting sick. Staying healthy is very important.

| Not like me at all | Not like me | A little like me | Somewhat like me | Like me | Very much like me |
|---|---|---|---|---|---|
| ○ | ○ | ○ | ○ | ○ | ○ |

Q72. Getting ahead in life is important. He/She strives to do better than others do.

| Not like me at all | Not like me | A little like me | Somewhat like me | Like me | Very much like me |
|---|---|---|---|---|---|
| ○ | ○ | ○ | ○ | ○ | ○ |

Q73. Forgiving people who have hurt him/her is important. He/She tries to see what is good in them and not to hold a grudge.

| Not like me at all | Not like me | A little like me | Somewhat like me | Like me | Very much like me |
|---|---|---|---|---|---|
| ○ | ○ | ○ | ○ | ○ | ○ |

Q74. It is important to be independent. He/She likes to rely on him/herself.

| Not like me at all | Not like me | A little like me | Somewhat like me | Like me | Very much like me |
|---|---|---|---|---|---|
| ○ | ○ | ○ | ○ | ○ | ○ |

Q75. Having a stable government is important. He/She is concerned that the social order be protected.

| Not like me at all | Not like me | A little like me | Somewhat like me | Like me | Very much like me |
|---|---|---|---|---|---|
| ○ | ○ | ○ | ○ | ○ | ○ |

Q76. It is important to be polite to other people all the time. He/She tries never to disturb or irritate others.

| Not like me at all | Not like me | A little like me | Somewhat like me | Like me | Very much like me |
|---|---|---|---|---|---|
| ○ | ○ | ○ | ○ | ○ | ○ |

Q77. He/She really wants to enjoy life. Having a good time is very important.

| Not like me at all | Not like me | A little like me | Somewhat like me | Like me | Very much like me |
|---|---|---|---|---|---|
| ○ | ○ | ○ | ○ | ○ | ○ |

Q78. It is important to be humble and modest. He/She tries not to draw attention to him/her.

| Not like me at all | Not like me | A little like me | Somewhat like me | Like me | Very much like me |
|---|---|---|---|---|---|
| ○ | ○ | ○ | ○ | ○ | ○ |



Q79. Always wants to be the one who makes the decisions. He/She likes to be the leader.

| Not like me at all | Not like me | A little like me | Somewhat like me | Like me | Very much like me |
|---|---|---|---|---|---|
| ○ | ○ | ○ | ○ | ○ | ○ |

Q80. It is important to adapt to nature and to fit into it. He/She believes that people should not change nature.

| Not like me at all | Not like me | A little like me | Somewhat like me | Like me | Very much like me |
|---|---|---|---|---|---|
| ○ | ○ | ○ | ○ | ○ | ○ |

You have completed 80/88 questions, 91% of this survey.
If you have any suggestions about this page please add it here

**This is the last page of the survey.**
**Please don't forget to click on next to submit the survey after you complete these questions.**

Q81. My ability to communicate with the government would be enhanced when using e-government transactions.

| Strongly Disagree | Disagree | Somewhat Disagree | Neither Agree nor Disagree | Somewhat Agree | Agree | Strongly Agree |
|---|---|---|---|---|---|---|
| ○ | ○ | ○ | ○ | ○ | ○ | ○ |

Q82. Communications through e-government enhance my ability to understand government transactions.

| Strongly Disagree | Disagree | Somewhat Disagree | Neither Agree nor Disagree | Agree | Agree | Strongly Agree |
|---|---|---|---|---|---|---|
| ○ | ○ | ○ | ○ | ○ | ○ | ○ |

Q83. Textual, verbal and visual information is important for carrying out government transactions.

| Strongly Disagree | Disagree | Somewhat Disagree | Neither Agree nor Disagree | Somewhat Agree | Agree | Strongly Agree |
|---|---|---|---|---|---|---|
| ○ | ○ | ○ | ○ | ○ | ○ | ○ |

Q84. I would use e-government to gather information about my required transactions.

| Strongly Disagree | Disagree | Somewhat Disagree | Neither Agree nor Disagree | Agree | Agree | Strongly Agree |
|---|---|---|---|---|---|---|
| ○ | ○ | ○ | ○ | ○ | ○ | ○ |

Q85. I would use e-government transactions provided over the Internet.

| Strongly Disagree | Disagree | Somewhat Disagree | Neither Agree nor Disagree | Agree | Agree | Strongly Agree |
|---|---|---|---|---|---|---|
| ○ | ○ | ○ | ○ | ○ | ○ | ○ |



Q86. Using e-government transactions is something that I would do.

| Strongly Disagree | Disagree | Somewhat Disagree | Neither Agree nor Disagree | Somewhat Agree | Agree | Strongly Agree |
|---|---|---|---|---|---|---|
| ○ | ○ | ○ | ○ | ○ | ○ | ○ |

Q87. I would not hesitate to provide information to e-government websites to conduct my transactions.

| Strongly Disagree | Disagree | Somewhat Disagree | Neither Agree nor Disagree | Somewhat Agree | Agree | Strongly Agree |
|---|---|---|---|---|---|---|
| ○ | ○ | ○ | ○ | ○ | ○ | ○ |

Q88. I would use e-government to inquire about my government transactions.

| Strongly Disagree | Disagree | Somewhat Disagree | Neither Agree nor Disagree | Somewhat Agree | Agree | Strongly Agree |
|---|---|---|---|---|---|---|
| ○ | ○ | ○ | ○ | ○ | ○ | ○ |

If you have any suggestions about this page please add it here

*If you are interested in participating further in this study please add your* email *or leave it empty if not*

*Please add you're* mobile *if you are interested in participating further in the study or leave it empty if not.*

Please add any suggestions about the complete survey.

Thank you. You have completed the survey. To submit please click on the Next button.



**APPENDIX H: QUESTIONNAIRE (ARABIC)**

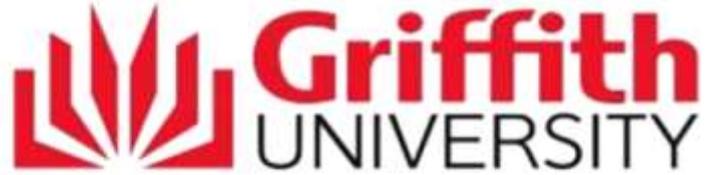

**عنوان المشروع**
تأثير الثقافة على تبني الحكومة الإلكترونية في المملكة العربية السعودية

**ماهو الهدف من هذا البحث؟**
سوف ينتج عن هذه الدراسة مجموعة من التوصيات التي من شأنها تمكين إنشاء معاملات حكومية إلكترونية (الخدمات الحكومية على شبكة الإنترنت) تتوافق مع الاحتياجات البشرية المختلفة. بالإضافة إلى ذلك فإن هذه الدراسة تشكل جزءاً من متطلبات مرحلة الدكتوراه في جامعة جريفيث كلية الإتصالات و تقنية المعلومات .

**ماهي معاملات الحكومة الإلكترونية**
تتضمن معاملات الحكومة الإلكترونية الآتي:
دفع مخالفات "ساهر" عبر الإنترنت، التقديم و مراجعة المعاملات الخاصة بالمبتعث من خلال موقع وزارة التعليم العالي، دفع الفواتير، تقديم شكاوى، إلخ.
لذلك، فإن معاملات الحكومة الألكترونية هي أي نوع من انواع التعامل مع الحكومة السعودية والتي تشمل تعبئة الاستمارات، أو التعاملات مع الحكومة باستخدام شبكة الانترنت.

**ما هو المطلوب منك؟**
الرجاء تعبئة هذا الإستبيان و الذي قد يستغرق مابين 15-30 دقيقة. هناك 88 سؤال موزعة في 12 صفحات.
يرجى ملاحظة أن هذا الإستبيان سوف يتم إيقافه في 2011/2/15 ، لذا لن يتم قبول المشاركين بعد هذا التاريخ.

الرجاء اعطاء رأيك الشخصي بكل صراحة و ذلك بناء على خبرتك العامة في استخدام الانترنت، أو بناءاً على تجربتك في آخر استخدام لك في المعاملات التي تقدمها الحكومة الإلكترونية السعودية و ذلك يشمل:
-استخدام مواقع الحكومة الإلكترونية بهدف جمع معلومات عن معاملاتك مع الدوائر الحكومية.
-المعاملات الحكومية التي تم اجراؤها مع الدوائر الحكومية باستخدام الإنترنت.

**الفوائد المتوقعة من البحث**
من خلال هذه الدراسة، سيتم اكتساب فهم أعمق للعوامل التالية :
العوامل التحفيزية لاستخدام الحكومة الإلكترونية و متطلبات إنشاء حكومة إلكترونية متوافقة مع حاجات المجتمع.

**خصوصية المشارك**
هذا الإستبيان تم تصميمه بطريقة تحافظ على خصوصية بياناتك من الباحثين أنفسهم أو من أي جهة أخرى. وسيكون التركيز في هذا البحث على البيانات بشكل عام و جماعي وليس على أي فرد، لذلك فإن أي تقرير أو بحث منشور من هذه الدراسة سوف يتم فيه إخفاء أو إزالة أي معلومات يمكن من خلالها تحديد هوية المشارك. سيتم الاحتفاظ بالمعلومات التي قدمتها في هذا الإستبيان بشكل يحافظ على خصوصيتك في مكتب كلية الإتصالات و تقنية المعلومات في جامعة جريفيث و لمدة خمسة سنوات فقط، وسوف يتم التخلص منها بعد ذلك.

**مشاركتكم طوعية**
المشاركة هي هذا الإستبيان تطوعية تماماً و بالإمكان التوقف في أي وقت دون إبداء سبب لذلك.

**الإجراءات الأخلاقية لهذا البحث**
يتم إجراء هذه الدراسة وفقاً لشروط جامعة جريفيث و التي تقدم بحوثها من خلال المعايير الأسترالية في البحوث.
إذا كان للمشاركين أية مخاوف حول السلوك الأخلاقي للمشروع البحثي ينبغي الاتصال بمدير البحوث على 3735 5585 أو research-ethics@griffith.edu.au.أو الإتصال ب د.عبدالعزيز السحيباني على هذا الإيميل:



alsehibani@cis.psu.edu.sa

أسئلة / مزيد من المعلومات
لأية استفسارات حول هذا البحث ، يمكنك الاتصال بـ :
الباحث : إبراهيم أبو نادي ،طالب الدكتوراة في كلية الإتصالات و تقنية المعلومات في جامعة جريفيث
i.abunadi@griffith.edu.au
جوال: 0061413649905

المشرف : الدكتور ستيف درو ، كلية الإتصالات و تقنية المعلومات ، جامعة جريفيث ، s.drew@griffith.edu.au

نتائج البحث
نتائج البحث ستكون متوفرة على رابط بعد الإنتهاء من الإستبيان. حيث سيتم الإعلان عن نتائج البحث على الرابط بعد نشرها.

استكمال وإرسال الإستبيان يعني أنك قرأت وفهمت المعلومات المذكورة سابقا ، و أنك موافق على أن يتم استخدام البيانات على الطريقة الموضحة أعلاه.

الرجاء ملاحظة أن إجاباتك سوف يتم حفظها في حال أغلقت المتصفح حيث أنه بالإمكان إكمال الإستبيان لاحقا على نفس الجهاز فقط.

أرجوا اختيار "التالي" للمتابعة.

Q1. الجنس:
ذكر          أنثى
○            ⊙

Q2. العمر:
17 سنة أو أقل    22-18    30-23    40-31    50-41    60-51    أكبر من 60
○              ○        ○        ○        ○        ○        ○

Q3. مكان الميلاد:

Q4. الدولة التي أعيش فيها حاليا:

Q5. الجنسية:

Q6. أعلى مؤهل تعليمي:
○ لم يتم التعلم في المدارس
○ الإبتدائي أو المتوسط
○ الثانوية
○ شهادة فنية أو مهنية (لم يتم الحصول على درجة البكالويوس)
○ درجة البكالويوس
○ شهادة تعليم عالي (لم يتم الحصول على شهادة الماجستير )
○ درجة الماجستير
○ الدكتوراة



Q7. مستوى احتراف استخدام الإنترنت:

| ممتاز | جيد جدا | جيد | لا بأس به | ضعيف | ضعيف جدا |
|--------|----------|------|-----------|-------|-----------|
| ○ | ○ | ○ | ○ | ○ | ○ |

Q8. عدد مرات استخدام الإنترنت:

○ عدة مرات في السنة

○ عدة مرات في الشهر

○ عدة مرات في الأسبوع

○ مرة في اليوم فقط

○ ساعات قليلة في اليوم

○ ساعات طويلة في اليوم

Q9. هل سبق و أن قمت بإجراء معاملات حكومية إلكترونية سعودية (مثال ، دفع المخالفات أو معاملات الجوازات أو معاملات وزارة التعليم العالي والأحوال المدنية إلخ) ؟

نعم        لا (سيتم الإنتقال إلى مجموعة الأسئلة التالية)

○          ○

Q10. متى كانت آخر مرة أجريت فيها معاملة حكومية إلكترونية سعودية؟

| اليوم | هذا الأسبوع | هذا الشهر | خلال الستة شهور الماضية | هذه السنة | قبل ثلاث سنوات السنة الماضية |
|--------|-------------|------------|--------------------------|-----------|------------------------------|
| ○ | ○ | ○ | ○ | ○ | ○          ○ |

Q11. ماهو معدل قيامك بإجراء معاملات حكومية إلكترونية سعودية؟

| يوميا | مرات قليلة في الأسبوع | مرة في الأسبوع | مرات قليلة في الشهر | مرة في الشهر | مرات قليلة في السنة | مرة في السنة |
|--------|----------------------|-----------------|---------------------|---------------|---------------------|---------------|
| ○ | ○ | ○ | ○ | ○ | ○ | ○ |

Q12. ماهي الوسيلة التي تتعامل بها عادة مع الدوائر الحكومية السعودية؟

☐ البريد

☐ الهاتف

☐ الفاكس

☐ بمساعدة المعقب

☐ باللقاء مع موظفي الدوائر الحكومية

☐ بمساعدة صديق أو قريب

☐ باستخدام الحكومة الإلكترونية

☐ لم يسبق أن تعاملت مع الحكومة السعودية

☐ طرق أخرى

Q13. يمكن وصف حالتك المهنية:

○ لست موظفا أو طالبا



<div dir="rtl">

○ طالب
○ موظف في القطاع الحكومي
○ موظف في القطاع الخاص
○ اعمال حرة
○ اخرى  [______________]

**أكملت 13 من 88 سؤال، أي 16٪ من الإستبيان.**
**اذا كان لديك أي ملاحظات على الأسئلة في هذه الصفحة الرجاء إضافتها هنا:**

أرجوا الإجابة من خلال خبرتك و تجربتك العامة مع الإنترنت أو من خلال تعاملك مع مواقع الحكومة الإلكترونية السعودية.

في إجاباتك أرجو مقارنة التعاملات الإلكترونية على الإنترنت مع الطرق التقليدية مثل (الإتصال بالهاتف ، أو الزيارة الفعلية للدوائر والهيئات الحكومية أو استخدام مساعدة صديق ... إلخ)

Q14. استخدام الحكومة الإلكترونية يمكنني من إجراء معاملاتي بشكل أسرع.

| غير موافق بشدة | غير موافق | نوعا ما غير موافق | محايد | موافق نوعا ما | موافق | موافق بشدة |
|---|---|---|---|---|---|---|
| ○ | ○ | ○ | ○ | ○ | ○ | ○ |

Q15. استخدام الحكومة الإلكترونية سوف يرفع من جودة معاملاتي.

| غير موافق بشدة | غير موافق | نوعا ما غير موافق | محايد | موافق نوعا ما | موافق | موافق بشدة |
|---|---|---|---|---|---|---|
| ○ | ○ | ○ | ○ | ○ | ○ | ○ |

Q16. استخدام الحكومة الإلكترونية سوف يجعل من الأسهل إجراء معاملات مع الدوائر الحكومية.

| غير موافق بشدة | غير موافق | نوعا ما غير موافق | محايد | موافق نوعا ما | موافق | موافق بشدة |
|---|---|---|---|---|---|---|
| ○ | ○ | ○ | ○ | ○ | ○ | ○ |

Q17. استخدام الحكومة الإلكترونية من الممكن أن يعمل على تحسين فعاليتي في إجراء معاملاتي مع الدوائر الحكومية.

| غير موافق بشدة | غير موافق | نوعا ما غير موافق | محايد | موافق نوعا ما | موافق | موافق بشدة |
|---|---|---|---|---|---|---|
| ○ | ○ | ○ | ○ | ○ | ○ | ○ |

Q18. استخدام الحكومة الإلكترونية من الممكن أن يعطيني تحكم أكبر على إجراء المعاملات مع الدوائر الحكومية.

| غير موافق بشدة | غير موافق | نوعا ما غير موافق | محايد | موافق نوعا ما | موافق | موافق بشدة |
|---|---|---|---|---|---|---|
| ○ | ○ | ○ | ○ | ○ | ○ | ○ |

Q19. استخدام الحكومة الإلكترونية هو أمر متوافق مع الطريقة التي أفضل لإجراء معاملاتي.

| غير موافق بشدة | غير موافق | نوعا ما غير موافق | محايد | موافق نوعا ما | موافق | موافق بشدة |
|---|---|---|---|---|---|---|
| ○ | ○ | ○ | ○ | ○ | ○ | ○ |

</div>



Q20. استخدام معاملات الحكومة الإلكترونية هو أمر متوافق تماما مع احتياجاتي الحالية.

| موافق بشدة | موافق | موافق نوعا ما | محايد | نوعا ما غير موافق | غير موافق بشدة غير موافق |
|---|---|---|---|---|---|
| ○ | ○ | ○ | ○ | ○ | ○ |

Q21. أعتقد أن استخدام الحكومة الإلكترونية من الممكن أن يتناسب مع الطريقة التي أفضل في إجراء معاملاتي الحكومية.

| موافق بشدة | موافق | موافق نوعا ما | محايد | نوعا ما غير موافق | غير موافق بشدة غير موافق |
|---|---|---|---|---|---|
| ○ | ○ | ○ | ○ | ○ | ○ |

Q22. استخدام معاملات الحكومة الإلكترونية قد يناسب أسلوب حياتي.

| موافق بشدة | موافق | موافق نوعا ما | محايد | نوعا ما غير موافق | غير موافق بشدة غير موافق |
|---|---|---|---|---|---|
| ○ | ○ | ○ | ○ | ○ | ○ |

Q23. استخدام معاملات الحكومة الإلكترونية قد يستهلك الكثير من وقتي.

| موافق بشدة | موافق | موافق نوعا ما | محايد | نوعا ما غير موافق | غير موافق بشدة غير موافق |
|---|---|---|---|---|---|
| ○ | ○ | ○ | ○ | ○ | ○ |

Q24. استخدام معاملات الحكومة الإلكترونية قد يكون معقدا جدا لدرجة أنه من الممكن أن يكون من الصعب فهم كيف يتم.

| موافق بشدة | موافق | موافق نوعا ما | محايد | نوعا ما غير موافق | غير موافق بشدة غير موافق |
|---|---|---|---|---|---|
| ○ | ○ | ○ | ○ | ○ | ○ |

أكملت 24 من 88 سؤال، أي ٢٧ ٪ من الاستبيان.
اذا كان لديك أي ملاحظات على الأسئلة في هذه الصفحة الرجاء إضافتها هنا:

أرجوا الإجابة من خلال خبرتك و تجربتك العامة مع الإنترنت أو من خلال تعاملك مع مواقع الحكومة الإلكترونية السعودية.

في إجاباتك أرجو مقارنة التعاملات الإلكترونية على الإنترنت مع الطرق التقليدية مثل (الاتصال بالهاتف ، أو الزيارة الفعلية للدوائر والهيئات الحكومية أو استخدام مساعدة صديق ... الخ)

Q25. استخدام معاملات الحكومة الإلكترونية قد يتضمن إجراء الكثير من العمليات التقنية (مثل إدخال البيانات من خلال لوحة المفاتيح).

| موافق بشدة | موافق | موافق نوعا ما | محايد | نوعا ما غير موافق | غير موافق بشدة غير موافق |
|---|---|---|---|---|---|
| ○ | ○ | ○ | ○ | ○ | ○ |

Q26. قد يستغرق مني وقتا طويلا لتعلم كيف استخدم معاملات الحكومة الإلكترونية و لذلك من أجل الإستفادة الفعلية منها.

| موافق بشدة | موافق | موافق نوعا ما | محايد | نوعا ما غير موافق | غير موافق بشدة غير موافق |
|---|---|---|---|---|---|
| ○ | ○ | ○ | ○ | ○ | ○ |



Q27. قد لا أواجه أي صعوبات في إخبار الأخرين عن النتائج التي حصلت عليها من خلال إجراء معاملات الحكومة الإلكترونية.

| موافق بشدة | موافق | موافق نوعا ما | محايد | نوعا ما غير موافق | غير موافق بشدة غير موافق |
|---|---|---|---|---|---|
| ○ | ○ | ○ | ○ | ○ | ○ ○ |

Q28. أعتقد أنه يمكن التواصل مع الأخرين وإخبار هم عن نتائج اجراء المعاملات الحكومية الإلكترونية.

| موافق بشدة | موافق | موافق نوعا ما | محايد | نوعا ما غير موافق | غير موافق بشدة غير موافق |
|---|---|---|---|---|---|
| ○ | ○ | ○ | ○ | ○ | ○ ○ |

Q29. إن نتائج استخدام معاملات الحكومة الإلكترونية هو أمر واضح بالنسبة لي.

| موافق بشدة | موافق | موافق نوعا ما | محايد | نوعا ما غير موافق | غير موافق بشدة غير موافق |
|---|---|---|---|---|---|
| ○ | ○ | ○ | ○ | ○ | ○ ○ |

Q30. من الممكن أن أعاني الكثير من الصعوبات في محاولة شرح فائدة أو عدم فائدة استخدام تعاملات الحكومة الإلكترونية.

| موافق بشدة | موافق | موافق نوعا ما | محايد | نوعا ما غير موافق | غير موافق بشدة غير موافق |
|---|---|---|---|---|---|
| ○ | ○ | ○ | ○ | ○ | ○ ○ |

أكملت 30 من 88 سؤال، أي 40٪ من الإستبيان.
اذا كان لديك أي ملاحظات على الأسئلة في هذه الصفحة الرجاء إضافتها هنا:

**أرجو الإجابة اعتمادا على خبرتك و تعاملك مع الانترنت**

Q31. الإنترنت يحتوي على ضمانات و حماية كافية تجعلني أثق في القيام بتعاملات الحكومة الإلكترونية.

| موافق بشدة | موافق | موافق نوعا ما | محايد | نوعا ما غير موافق | غير موافق بشدة غير موافق |
|---|---|---|---|---|---|
| ○ | ○ | ○ | ○ | ○ | ○ ○ |

Q32. أشعر بأن الضمانات القانونية و التكنولوجية الحالية كافية لحمايتي من المشاكل الموجودة على الإنترنت عند استخدام التعاملات الحكومية الإلكترونية السعودية.

| موافق بشدة | موافق | موافق نوعا ما | محايد | نوعا ما غير موافق | غير موافق بشدة غير موافق |
|---|---|---|---|---|---|
| ○ | ○ | ○ | ○ | ○ | ○ ○ |

Q33. بشكل عام أشعر بأن الإنترنت الآن أصبح بيئة آمنة و نظام متين للقيام بالمعاملات الحكومية الإلكترونية السعودية.

| موافق بشدة | موافق | موافق نوعا ما | محايد | نوعا ما غير موافق | غير موافق بشدة غير موافق |
|---|---|---|---|---|---|
| ○ | ○ | ○ | ○ | ○ | ○ ○ |



أرجوا الإجابة اعتمادا على خبرتك أو توقعاتك في التعامل مع الدوائر الحكومية في السعودية أو من خلال تعاملاتك مع خدمات الحكومة الإلكترونية السعودية.

Q34. أعتقد أنه يمكنني الوثوق بقدرة الدوائر الحكومية السعودية عند تقديم خدمات الحكومة الإلكترونية.

| موافق بشدة | موافق | موافق نوعا ما | محايد | نوعا ما غير موافق | غير موافق بشدة غير موافق |
|---|---|---|---|---|---|
| ○ | ○ | ○ | ○ | ○ | ○ ○ |

Q35. يمكن الوثوق بقدرة و أمانة الدوائر الحكومية السعودية لتقديم تعاملات حكومية إلكترونية.

| موافق بشدة | موافق | موافق نوعا ما | محايد | نوعا ما غير موافق | غير موافق بشدة غير موافق |
|---|---|---|---|---|---|
| ○ | ○ | ○ | ○ | ○ | ○ ○ |

Q36. في رأيي فإن الدوائر الحكومية السعودية محل ثقة لتقديم خدمات إلكترونية.

| موافق بشدة | موافق | موافق نوعا ما | محايد | نوعا ما غير موافق | غير موافق بشدة غير موافق |
|---|---|---|---|---|---|
| ○ | ○ | ○ | ○ | ○ | ○ ○ |

Q37. إنني أثق في أن الدوائر الحكومية سوف تعمل على الحفاظ على منفعتي و تضعها في أولوياتها عند تقديم خدمات إلكترونية.

| موافق بشدة | موافق | موافق نوعا ما | محايد | نوعا ما غير موافق | غير موافق بشدة غير موافق |
|---|---|---|---|---|---|
| ○ | ○ | ○ | ○ | ○ | ○ ○ |

أرجو الإجابة بناء على الأراء التي سمعتها أو قد تسمعها عن الحكومة الإلكترونية السعودية.

Q38. الأشخاص الذين لهم أثر على سلوكي قد يشجعوني على استخدام معاملات الحكومة الإلكترونية.

| موافق بشدة | موافق | موافق نوعا ما | محايد | نوعا ما غير موافق | غير موافق بشدة غير موافق |
|---|---|---|---|---|---|
| ○ | ○ | ○ | ○ | ○ | ○ ○ |

Q39. الأشخاص المهمين بالنسبة لي قد يشجعوني على استخدام معاملات الحكومة الإلكترونية

| موافق بشدة | موافق | موافق نوعا ما | محايد | نوعا ما غير موافق | غير موافق بشدة غير موافق |
|---|---|---|---|---|---|
| ○ | ○ | ○ | ○ | ○ | ○ ○ |

Q40. الأشخاص الذين في دائرتي الإجتماعية يظنوا أنه من الملائم بالنسبة لي أن استخدم معاملات الحكومة الإلكترونية.

| موافق بشدة | موافق | موافق نوعا ما | محايد | نوعا ما غير موافق | غير موافق بشدة غير موافق |
|---|---|---|---|---|---|
| ○ | ○ | ○ | ○ | ○ | ○ ○ |

أكملت 40 من 88 سؤال، أي 46٪ من الإستبيان.
اذا كان لديك أي ملاحظات على الأسئلة في هذه الصفحة الرجاء إضافتها هنا:



الشخصية و السلوك

أدناه نورد أوصاف و سلوكيات شخص ما، يرجى مطالعة كل وصف و سلوك لتحديد مدى مطابقته او عدم مطابقته لشخصيتك و سلوكك.

يرجى مراعاة الصراحة في إختيار المستوى الأقرب لك.

Q41. التفكير المتجدد و الإبداع أمر مهم. هو/ هي يحب القيام بالأشياء بطريقته المبتكرة.

| يشبهني كثيرا | يشبهني | يوجد تشابه بسيط | يوجد تشابه بسيط جدا | لا يشبهني | لا يوجد أي شبه |
|---|---|---|---|---|---|
| ○ | ○ | ○ | ○ | ○ | ○ |

Q42. من المهم أن تكون غنيا فهو / هي يريد أن يحصل على مبالغ طائلة ومقتنيات ثمينة.

| يشبهني كثيرا | يشبهني | يوجد تشابه بسيط | يوجد تشابه بسيط جدا | لا يشبهني | لا يوجد أي شبه |
|---|---|---|---|---|---|
| ○ | ○ | ○ | ○ | ○ | ○ |

Q43. هو / هي يعتقد أنه من المهم أن تكون هناك مساواة في التعامل مع كل البشر. فهو يؤمن أن كل شخص في هذه الحياة يجب أن يحصل على فرص متساوية.

| يشبهني كثيرا | يشبهني | يوجد تشابه بسيط | يوجد تشابه بسيط جدا | لا يشبهني | لا يوجد أي شبه |
|---|---|---|---|---|---|
| ○ | ○ | ○ | ○ | ○ | ○ |

Q44. من المهم أن يُظهر الإنسان مهاراته. ويرغب من الناس أن يمدحوا ما يقوم به.

| يشبهني كثيرا | يشبهني | يوجد تشابه بسيط | يوجد تشابه بسيط جدا | لا يشبهني | لا يوجد أي شبه |
|---|---|---|---|---|---|
| ○ | ○ | ○ | ○ | ○ | ○ |

Q45. من المهم أن يعيش الإنسان في بيئة آمنة ويتجنب أي شيء قد يعرض سلامته للخطر.

| يشبهني كثيرا | يشبهني | يوجد تشابه بسيط | يوجد تشابه بسيط جدا | لا يشبهني | لا يوجد أي شبه |
|---|---|---|---|---|---|
| ○ | ○ | ○ | ○ | ○ | ○ |

Q46. من المهم القيام بأعمال مختلفة في الحياة، وهو / هي إنسان يتطلع دائما لتجربة الأشياء أو الخبرات الجديدة.

| يشبهني كثيرا | يشبهني | يوجد تشابه بسيط | يوجد تشابه بسيط جدا | لا يشبهني | لا يوجد أي شبه |
|---|---|---|---|---|---|
| ○ | ○ | ○ | ○ | ○ | ○ |

Q47. هو / هي يعتقد أنه على الجميع الإنصياع إلى الأوامر، ويعتقد أنه على الناس أن تحترم القوانين طوال الوقت، حتى في حالة عدم وجود رقابة.

| يشبهني كثيرا | يشبهني | يوجد تشابه بسيط | يوجد تشابه بسيط جدا | لا يشبهني | لا يوجد أي شبه |
|---|---|---|---|---|---|
| ○ | ○ | ○ | ○ | ○ | ○ |



Q48. يرى أنه من المهم الاستماع إلى الأشخاص المختلفين عنه. من الضروري أن يتفهمهم حتى ولو كان يختلف معهم في وجهات النظر.

| لا يوجد أي شبه | لا يشبهني | يوجد تشابه بسيط جدا | يوجد تشابه بسيط | يشبهني | يشبهني كثيرا |
|---|---|---|---|---|---|
| ○ | ○ | ○ | ○ | ○ | ○ |

Q49. يرى أنه من الضروري أن لا يطلب أكثر مما يملك. ويعتقد أن على الناس الإقتناع بما لديهم.

| لا يوجد أي شبه | لا يشبهني | يوجد تشابه بسيط جدا | يوجد تشابه بسيط | يشبهني | يشبهني كثيرا |
|---|---|---|---|---|---|
| ○ | ○ | ○ | ○ | ○ | ○ |

Q50. يسعى خلف كل فرصة تتيح الإستمتاع بالحياة، ومن المهم أن يقوم بالأشياء التي تجلب المتعة.

| لا يوجد أي شبه | لا يشبهني | يوجد تشابه بسيط جدا | يوجد تشابه بسيط | يشبهني | يشبهني كثيرا |
|---|---|---|---|---|---|
| ○ | ○ | ○ | ○ | ○ | ○ |

Q51. من المهم أن يضع المرء قراراته بنفسه حول الأمور التي يريد القيام بها. ويحب الحرية في التخطيط و اختيار ما يريد.

| لا يوجد أي شبه | لا يشبهني | يوجد تشابه بسيط جدا | يوجد تشابه بسيط | يشبهني | يشبهني كثيرا |
|---|---|---|---|---|---|
| ○ | ○ | ○ | ○ | ○ | ○ |

Q52. يرى أنه من المهم جدا مساعدة الأشخاص حوله. ويرغب بر عاية مصالحهم.

| لا يوجد أي شبه | لا يشبهني | يوجد تشابه بسيط جدا | يوجد تشابه بسيط | يشبهني | يشبهني كثيرا |
|---|---|---|---|---|---|
| ○ | ○ | ○ | ○ | ○ | ○ |

Q53. يرى أنه من المهم جدا أن يكون ناجحا. ويحب أن يُثير إعجاب الآخرين بإنجازاته.

| لا يوجد أي شبه | لا يشبهني | يوجد تشابه بسيط جدا | يوجد تشابه بسيط | يشبهني | يشبهني كثيرا |
|---|---|---|---|---|---|
| ○ | ○ | ○ | ○ | ○ | ○ |

Q54. سلامة الوطن مهمة بالنسبة له. ويعتقد أن على الدولة مراقبة التهديدات الداخلية و الخارجية.

| لا يوجد أي شبه | لا يشبهني | يوجد تشابه بسيط جدا | يوجد تشابه بسيط | يشبهني | يشبهني كثيرا |
|---|---|---|---|---|---|
| ○ | ○ | ○ | ○ | ○ | ○ |

Q55. يحب / تحب المخاطرة ودائمُ البحث عن المغامرات.

| لا يوجد أي شبه | لا يشبهني | يوجد تشابه بسيط جدا | يوجد تشابه بسيط | يشبهني | يشبهني كثيرا |
|---|---|---|---|---|---|
| ○ | ○ | ○ | ○ | ○ | ○ |

Q56. يرى أنه من المهم جدا أن يتصرف دائما التصرف اللائق. وأن يتجنب ما يعتبره الآخرون خطأ.

| لا يوجد أي شبه | لا يشبهني | يوجد تشابه بسيط جدا | يوجد تشابه بسيط | يشبهني | يشبهني كثيرا |
|---|---|---|---|---|---|
| ○ | ○ | ○ | ○ | ○ | ○ |



Q57. يرى أنه من الضروري أن يكون في مركز القيادة و أن يأمر الآخرين. ويرغب في أن يمتثل الآخرين لما يقوله.

| يشبهني كثيرا | يشبهني | يوجد تشابه بسيط | يوجد تشابه بسيط جدا | لا يشبهني | لا يوجد أي شبه |
|---|---|---|---|---|---|
| ○ | ○ | ○ | ○ | ○ | ○ |

Q58. يرى أنه من المهم أن يكون وفيا لأصدقاءه. ويرغب أن يكرس نفسه للمقربين منه.

| يشبهني كثيرا | يشبهني | يوجد تشابه بسيط | يوجد تشابه بسيط جدا | لا يشبهني | لا يوجد أي شبه |
|---|---|---|---|---|---|
| ○ | ○ | ○ | ○ | ○ | ○ |

Q59. يؤمن بشدة أن على الناس المحافظة على البيئة. الإهتمام والحفاظ على البيئة أمر مهم بالنسبة له.

| يشبهني كثيرا | يشبهني | يوجد تشابه بسيط | يوجد تشابه بسيط جدا | لا يشبهني | لا يوجد أي شبه |
|---|---|---|---|---|---|
| ○ | ○ | ○ | ○ | ○ | ○ |

Q60. التدين امر مهم بالنسبة له. ويحاول جاهدا اتباع المعتقدات الدينية.

| يشبهني كثيرا | يشبهني | يوجد تشابه بسيط | يوجد تشابه بسيط جدا | لا يشبهني | لا يوجد أي شبه |
|---|---|---|---|---|---|
| ○ | ○ | ○ | ○ | ○ | ○ |

أكملت 60 من 88 سؤال، أي 68٪ من الإستبيان.
اذا كان لديك أي ملاحظات على الأسئلة في هذه الصفحة الرجاء اضافتها هنا:

**الشخصية و السلوك**

**أدناه نورد أوصاف و سلوكيات شخص ما، يرجى مطالعة كل وصف لتحديد مدى مشابهة او عدم مشابهته لشخصيتك و سلوكك.**

**يرجى مراعاة الصراحة في إختيار المستوى الأقرب لك.**

Q61. من المهم جدا أن تكون الأمور منظمة و نظيفة. ولايحب أن تكون الأشياء فوضوية.

| يشبهني كثيرا | يشبهني | يوجد تشابه بسيط | يوجد تشابه بسيط جدا | لا يشبهني | لا يوجد أي شبه |
|---|---|---|---|---|---|
| ○ | ○ | ○ | ○ | ○ | ○ |

Q62. شخص يعتقد أنه من الضروري أن يكون لديه اهتمام بأمور معينة. ويحب أن يكون مُتساءلا و أن يفهم ماحوله.

| يشبهني كثيرا | يشبهني | يوجد تشابه بسيط | يوجد تشابه بسيط جدا | لا يشبهني | لا يوجد أي شبه |
|---|---|---|---|---|---|
| ○ | ○ | ○ | ○ | ○ | ○ |



Q63. شخص يعتقد أن على كل سكان العالم أن يتعايشوا بإنسجام. من المهم الترويج للسلام بين شعوب الارض.

| لا يوجد أي شبه | لا يشبهني | يوجد تشابه بسيط جدا | يوجد تشابه بسيط | يشبهني | يشبهني كثيرا |
|---|---|---|---|---|---|
| ○ | ○ | ○ | ○ | ○ | ○ |

Q64. شخص يرى أن الطموح مهم و يحب اظهار مدى قدراته.

| لا يوجد أي شبه | لا يشبهني | يوجد تشابه بسيط جدا | يوجد تشابه بسيط | يشبهني | يشبهني كثيرا |
|---|---|---|---|---|---|
| ○ | ○ | ○ | ○ | ○ | ○ |

Q65. شخص يرى أن من الأفضل القيام بالأمور بالطريقة التقليدية. من المهم المحافظة على التقاليد المتوارثة.

| لا يوجد أي شبه | لا يشبهني | يوجد تشابه بسيط جدا | يوجد تشابه بسيط | يشبهني | يشبهني كثيرا |
|---|---|---|---|---|---|
| ○ | ○ | ○ | ○ | ○ | ○ |

Q66. من المهم الاستمتاع بملذات الحياة. ويحب أن "يدلل" نفسه.

| لا يوجد أي شبه | لا يشبهني | يوجد تشابه بسيط جدا | يوجد تشابه بسيط | يشبهني | يشبهني كثيرا |
|---|---|---|---|---|---|
| ○ | ○ | ○ | ○ | ○ | ○ |

Q67. من المهم التجاوب مع احتياجات الآخرين. يحاول أن يساند من يعرف.

| لا يوجد أي شبه | لا يشبهني | يوجد تشابه بسيط جدا | يوجد تشابه بسيط | يشبهني | يشبهني كثيرا |
|---|---|---|---|---|---|
| ○ | ○ | ○ | ○ | ○ | ○ |

Q68. من الضروري وعلى الدوام ، احترام الوالدين ومن هم أكبر سنا. ومن المهم أن تكون مطيعا.

| لا يوجد أي شبه | لا يشبهني | يوجد تشابه بسيط جدا | يوجد تشابه بسيط | يشبهني | يشبهني كثيرا |
|---|---|---|---|---|---|
| ○ | ○ | ○ | ○ | ○ | ○ |

Q69. يرغب في أن يتم التعامل مع كل الناس بطريقة عادلة حتى مع من لايعرف. من المهم حماية الضعيف في المجتمع.

| لا يوجد أي شبه | لا يشبهني | يوجد تشابه بسيط جدا | يوجد تشابه بسيط | يشبهني | يشبهني كثيرا |
|---|---|---|---|---|---|
| ○ | ○ | ○ | ○ | ○ | ○ |

Q70. يحب المفاجآت، من المهم أن يعيش حياة مثيرة.

| لا يوجد أي شبه | لا يشبهني | يوجد تشابه بسيط جدا | يوجد تشابه بسيط | يشبهني | يشبهني كثيرا |
|---|---|---|---|---|---|
| ○ | ○ | ○ | ○ | ○ | ○ |

Q71. يحاول أن يتجنب الأمراض. البقاء بصحة وعافية مهمة جدا بالنسبة له.

| لا يوجد أي شبه | لا يشبهني | يوجد تشابه بسيط جدا | يوجد تشابه بسيط | يشبهني | يشبهني كثيرا |
|---|---|---|---|---|---|
| ○ | ○ | ○ | ○ | ○ | ○ |



Q72. البلوغ للقمة أمر مهم في الحياة. هو / هي شخص يحاول جاهدا أن يكون أداءه افضل من الأخرين.

| لا يوجد أي شبه | لا يشبهني | يوجد تشابه بسيط | يوجد تشابه بسيط جدا | يشبهني | يشبهني كثيرا |
|---|---|---|---|---|---|
| ○ | ○ | ○ | ○ | ○ | ○ |

Q73. يرى أهمية العفو و الصفح عن من يتعرض له بالأذى. يحاول رؤية النواحي الإيجابية في الناس ويتجنب أن يحقد على أي أحد.

| لا يوجد أي شبه | لا يشبهني | يوجد تشابه بسيط جدا | يوجد تشابه بسيط | يشبهني | يشبهني كثيرا |
|---|---|---|---|---|---|
| ○ | ○ | ○ | ○ | ○ | ○ |

Q74. الإستقلالية أمر مهم جدا.يحاول الإعتماد على نفسه فقط.

| لا يوجد أي شبه | لا يشبهني | يوجد تشابه بسيط جدا | يوجد تشابه بسيط | يشبهني | يشبهني كثيرا |
|---|---|---|---|---|---|
| ○ | ○ | ○ | ○ | ○ | ○ |

Q75. من المهم وجود حكومة مستقرة، هو شخص يهتم بحماية الأنظمة الإجتماعية.

| لا يوجد أي شبه | لا يشبهني | يوجد تشابه بسيط جدا | يوجد تشابه بسيط | يشبهني | يشبهني كثيرا |
|---|---|---|---|---|---|
| ○ | ○ | ○ | ○ | ○ | ○ |

Q76. من المهم أن نكون مهذبين مع الأخرين دائما. و يحاول ألا يز عج أو يغضب الأخرين.

| لا يوجد أي شبه | لا يشبهني | يوجد تشابه بسيط جدا | يوجد تشابه بسيط | يشبهني | يشبهني كثيرا |
|---|---|---|---|---|---|
| ○ | ○ | ○ | ○ | ○ | ○ |

Q77. هو شخص يريد فعلا أن يستمتع بالحياة. قضاء أوقات ممتعة أمر مهم جدا.

| لا يوجد أي شبه | لا يشبهني | يوجد تشابه بسيط جدا | يوجد تشابه بسيط | يشبهني | يشبهني كثيرا |
|---|---|---|---|---|---|
| ○ | ○ | ○ | ○ | ○ | ○ |

Q78. التواضع أمر مهم جدا. هو شخص يحاول عدم لفت انتباه الأخرين إليه.

| لا يوجد أي شبه | لا يشبهني | يوجد تشابه بسيط جدا | يوجد تشابه بسيط | يشبهني | يشبهني كثيرا |
|---|---|---|---|---|---|
| ○ | ○ | ○ | ○ | ○ | ○ |

Q79. هو شخص ير غب دائما باتخاذ القرارات. و يحب أن يكون قياديا.

| لا يوجد أي شبه | لا يشبهني | يوجد تشابه بسيط جدا | يوجد تشابه بسيط | يشبهني | يشبهني كثيرا |
|---|---|---|---|---|---|
| ○ | ○ | ○ | ○ | ○ | ○ |

Q80. يرى أنه من المهم التكيف و التأقلم مع الطبيعة. و يرى أن على الناس عدم العبث بالطبيعة أو محاولة تغيير ها.

| لا يوجد أي شبه | لا يشبهني | يوجد تشابه بسيط جدا | يوجد تشابه بسيط | يشبهني | يشبهني كثيرا |
|---|---|---|---|---|---|
| ○ | ○ | ○ | ○ | ○ | ○ |



أكملت 80 من 88 سؤال، أي 91٪ من الاستبيان.
اذا كان لديك أي ملاحظات على الأسئلة في هذه الصفحة الرجاء إضافتها هنا:

هذه الصفحة الأخيرة من الإستبيان
الرجاء التنبه باختيار التالي في نهاية هذه الصفحة لإرسال الإستبيان بعد الإنتهاء من الأسئلة.

Q81. إن قدرتي على التواصل مع الحكومة ستتحسن من خلال استخدام معاملات الحكومة الإلكترونية.

| موافق بشدة | موافق | نوعا ما موافق | محايد | نوعا ما غير موافق | غير موافق بشدة غير موافق | موافق بشدة |
|---|---|---|---|---|---|---|
| ○ | ○ | ○ | ○ | ○ | ○ | ○ |

Q82. التواصل و الإتصال من خلال الحكومة الإلكترونية قد يحسن من قدرتي على فهم التعاملات الحكومية.

| موافق بشدة | موافق | نوعا ما موافق | محايد | نوعا ما غير موافق | غير موافق بشدة غير موافق | موافق بشدة |
|---|---|---|---|---|---|---|
| ○ | ○ | ○ | ○ | ○ | ○ | ○ |

Q83. المعلومات النصية و اللفظية و البصرية مهمة جدا لإجراء المعاملات الحكومية.

| موافق بشدة | موافق | نوعا ما موافق | محايد | نوعا ما غير موافق | غير موافق بشدة غير موافق | موافق بشدة |
|---|---|---|---|---|---|---|
| ○ | ○ | ○ | ○ | ○ | ○ | ○ |

Q84. قد استخدم معاملات الحكومة الإلكترونية لجمع معلومات عن معاملاتي الحكومية.

| موافق بشدة | موافق | نوعا ما موافق | محايد | نوعا ما غير موافق | غير موافق بشدة غير موافق | موافق بشدة |
|---|---|---|---|---|---|---|
| ○ | ○ | ○ | ○ | ○ | ○ | ○ |

Q85. قد اقوم بإجراء معاملات الحكومة الإلكترونية المتوفرة من خلال الإنترنت.

| موافق بشدة | موافق | نوعا ما موافق | محايد | نوعا ما غير موافق | غير موافق بشدة غير موافق | موافق بشدة |
|---|---|---|---|---|---|---|
| ○ | ○ | ○ | ○ | ○ | ○ | ○ |

Q86. استخدام معاملات الحكومة الإلكترونية هو أمر قد أقوم به.

| موافق بشدة | موافق | نوعا ما موافق | محايد | نوعا ما غير موافق | غير موافق بشدة غير موافق | موافق بشدة |
|---|---|---|---|---|---|---|
| ○ | ○ | ○ | ○ | ○ | ○ | ○ |

Q87. قد لا أتردد في تزويد معلوماتي إلى مواقع الحكومة الإلكترونية و ذلك لإجراء معاملاتي الحكومية.

| موافق بشدة | موافق | نوعا ما موافق | محايد | نوعا ما غير موافق | غير موافق بشدة غير موافق | موافق بشدة |
|---|---|---|---|---|---|---|
| ○ | ○ | ○ | ○ | ○ | ○ | ○ |



Q88. قد استخدم الحكومة الإلكترونية للإستفسار عن تعاملاتي الحكومية.

| موافق بشدة | موافق | موافق نوعا ما | محايد | نوعا ما غير موافق | غير موافق بشدة | غير موافق |
|---|---|---|---|---|---|---|
| ○ | ○ | ○ | ○ | ○ | ○ | ○ |

اذا كان لديك أي ملاحظات على الأسئلة في هذه الصفحة الرجاء إضافتها هنا:

اذا كنت مهتما بالمشاركة مستقبلا في هذا البحث الرجاء إضافة ايميلك أو تركه فارغا إن لم ترغب بذلك.

الرجاء إضافة جوالك اذا كنت مهتما بالمشاركة مستقبلا في هذا البحث أو تركه فارغا إن لم ترغب بذلك.

**اذا كان لديك أي ملاحظات على هذا الاستبيان الرجاء إضافتها هنا:**

شكرا لك، الرجاء إختيار "التالي" لتسليم الاستبيان



**APPENDIX I: TEST OF HOMOSCEDASTICITY**

| Information | | Levene Statistic | df1 | df2 | Sig. |
|---|---|---|---|---|---|
| USE | Based on Mean | 0.04 | 1.00 | 669.00 | 0.84 |
| | Based on Median | 0.00 | 1.00 | 669.00 | 0.97 |
| | Based on Median and with adjusted df | 0.00 | 1.00 | 665.12 | 0.97 |
| | Based on trimmed mean | 0.02 | 1.00 | 669.00 | 0.89 |
| RA | Based on Mean | 0.03 | 1.00 | 669.00 | 0.86 |
| | Based on Median | 0.04 | 1.00 | 669.00 | 0.84 |
| | Based on Median and with adjusted df | 0.04 | 1.00 | 656.70 | 0.84 |
| | Based on trimmed mean | 0.07 | 1.00 | 669.00 | 0.79 |
| CT | Based on Mean | 2.56 | 1.00 | 669.00 | 0.11 |
| | Based on Median | 1.66 | 1.00 | 669.00 | 0.20 |
| | Based on Median and with adjusted df | 1.66 | 1.00 | 642.19 | 0.20 |
| | Based on trimmed mean | 1.79 | 1.00 | 669.00 | 0.18 |
| CMX | Based on Mean | 0.47 | 1.00 | 669.00 | 0.49 |
| | Based on Median | 0.45 | 1.00 | 669.00 | 0.50 |
| | Based on Median and with adjusted df | 0.45 | 1.00 | 666.03 | 0.50 |
| | Based on trimmed mean | 0.49 | 1.00 | 669.00 | 0.49 |
| RED | Based on Mean | 1.53 | 1.00 | 669.00 | 0.22 |
| | Based on Median | 1.57 | 1.00 | 669.00 | 0.21 |
| | Based on Median and with adjusted df | 1.57 | 1.00 | 666.42 | 0.21 |
| | Based on trimmed mean | 1.86 | 1.00 | 669.00 | 0.17 |
| TI | Based on Mean | 0.50 | 1.00 | 669.00 | 0.48 |
| | Based on Median | 0.25 | 1.00 | 669.00 | 0.62 |
| | Based on Median and with adjusted df | 0.25 | 1.00 | 663.80 | 0.62 |
| | Based on trimmed mean | 0.48 | 1.00 | 669.00 | 0.49 |
| TG | Based on Mean | 2.40 | 1.00 | 669.00 | 0.12 |
| | Based on Median | 1.43 | 1.00 | 669.00 | 0.23 |
| | Based on Median and with adjusted df | 1.43 | 1.00 | 666.96 | 0.23 |
| | Based on trimmed mean | 2.22 | 1.00 | 669.00 | 0.14 |
| SI | Based on Mean | 3.77 | 1.00 | 669.00 | 0.05 |
| | Based on Median | 3.18 | 1.00 | 669.00 | 0.08 |
| | Based on Median and with adjusted df | 3.18 | 1.00 | 660.49 | 0.08 |
| | Based on trimmed mean | 3.77 | 1.00 | 669.00 | 0.05 |
| SD | Based on Mean | 0.18 | 1.00 | 669.00 | 0.67 |
| | Based on Median | 0.43 | 1.00 | 669.00 | 0.52 |
| | Based on Median and with adjusted df | 0.43 | 1.00 | 667.75 | 0.52 |
| | Based on trimmed mean | 0.31 | 1.00 | 669.00 | 0.58 |
| P | Based on Mean | 0.44 | 1.00 | 669.00 | 0.51 |
| | Based on Median | 0.38 | 1.00 | 669.00 | 0.54 |
| | Based on Median and with adjusted df | 0.38 | 1.00 | 668.54 | 0.54 |
| | Based on trimmed mean | 0.40 | 1.00 | 669.00 | 0.53 |
| U | Based on Mean | 0.01 | 1.00 | 669.00 | 0.91 |
| | Based on Median | 0.00 | 1.00 | 669.00 | 0.99 |
| | Based on Median and with adjusted df | 0.00 | 1.00 | 666.84 | 0.99 |
| | Based on trimmed mean | 0.01 | 1.00 | 669.00 | 0.93 |
| A | Based on Mean | 0.79 | 1.00 | 669.00 | 0.37 |
| | Based on Median | 0.93 | 1.00 | 669.00 | 0.34 |
| | Based on Median and with adjusted df | 0.93 | 1.00 | 668.90 | 0.34 |
| | Based on trimmed mean | 0.90 | 1.00 | 669.00 | 0.34 |
| SE | Based on Mean | 0.28 | 1.00 | 669.00 | 0.60 |
| | Based on Median | 0.37 | 1.00 | 669.00 | 0.54 |
| | Based on Median and with adjusted df | 0.37 | 1.00 | 655.23 | 0.54 |
| | Based on trimmed mean | 0.35 | 1.00 | 669.00 | 0.56 |
| ST | Based on Mean | 1.73 | 1.00 | 669.00 | 0.19 |
| | Based on Median | 1.82 | 1.00 | 669.00 | 0.18 |
| | Based on Median and with adjusted df | 1.82 | 1.00 | 662.57 | 0.18 |
| | Based on trimmed mean | 1.82 | 1.00 | 669.00 | 0.18 |
| C | Based on Mean | 0.19 | 1.00 | 669.00 | 0.66 |
| | Based on Median | 0.42 | 1.00 | 669.00 | 0.52 |
| | Based on Median and with adjusted df | 0.42 | 1.00 | 657.09 | 0.52 |
| | Based on trimmed mean | 0.33 | 1.00 | 669.00 | 0.57 |
| T | Based on Mean | 0.25 | 1.00 | 669.00 | 0.62 |
| | Based on Median | 0.13 | 1.00 | 669.00 | 0.72 |
| | Based on Median and with adjusted df | 0.13 | 1.00 | 668.99 | 0.72 |
| | Based on trimmed mean | 0.25 | 1.00 | 669.00 | 0.62 |
| H | Based on Mean | 0.00 | 1.00 | 669.00 | 0.98 |
| | Based on Median | 0.08 | 1.00 | 669.00 | 0.78 |
| | Based on Median and with adjusted df | 0.08 | 1.00 | 667.35 | 0.78 |
| | Based on trimmed mean | 0.03 | 1.00 | 669.00 | 0.87 |
| B | Based on Mean | 1.44 | 1.00 | 669.00 | 0.23 |
| | Based on Median | 1.16 | 1.00 | 669.00 | 0.28 |
| | Based on Median and with adjusted df | 1.16 | 1.00 | 668.18 | 0.28 |
| | Based on trimmed mean | 1.34 | 1.00 | 669.00 | 0.25 |
| POC | Based on Mean | 0.89 | 1.00 | 669.00 | 0.35 |
| | Based on Median | 0.29 | 1.00 | 669.00 | 0.59 |
| | Based on Median and with adjusted df | 0.29 | 1.00 | 666.84 | 0.59 |
| | Based on trimmed mean | 0.50 | 1.00 | 669.00 | 0.48 |

*Note.* RA=relative advantage; CT=compatibility; CMX=complexity; RED=result demonstrability; TI=trust in the Internet; TG=trust in government agencies; SI=social influence; POC=perspective on communication; USE=intention to use e-transactions. SD=self-direction; P=power A=achievement; ST=stimulation; C=conformity; T=tradition; SE=security; H=hedonism; B=benevolence; df=degrees of freedom; sig=significance.



# APPENDIX J: STANDARDISED RESIDUAL COVARIANCES MATRIX- CFA INCONCLUSIVE MODEL FOR BPV

| | C7 | C16 | C28 | C36 | T20 | T38 | SE5 | SE14 | SE21 | SE31 | SE25 | B33 | B27 | B18 | B12 | H37 | H26 | H10 | ST6 | ST15 | ST30 | SD24 | SD22 | SD11 | SD1 | A32 | A24 | A13 | A4 | P39 | P17 |
|---|---|---|---|---|---|---|---|---|---|---|---|---|---|---|---|---|---|---|---|---|---|---|---|---|---|---|---|---|---|---|---|
| C16 | 1.8 | 0.0 | | | | | | | | | | | | | | | | | | | | | | | | | | | | | |
| C28 | 0.1 | 0.6 | 0.0 | | | | | | | | | | | | | | | | | | | | | | | | | | | | |
| C36 | -0.9 | 0.0 | 0.4 | 0.0 | | | | | | | | | | | | | | | | | | | | | | | | | | | |
| T20 | -1.2 | -0.9 | 1.3 | -1.6 | 0.0 | | | | | | | | | | | | | | | | | | | | | | | | | | |
| T38 | -0.7 | 0.2 | -0.8 | 3.3 | 0.7 | 0.0 | | | | | | | | | | | | | | | | | | | | | | | | | |
| SE5 | 2.7 | 2.0 | -0.6 | -1.8 | -0.1 | -1.3 | 0.0 | | | | | | | | | | | | | | | | | | | | | | | | |
| SE14 | 1.1 | 1.7 | -1.0 | -1.6 | 0.2 | -2.2 | 3.0 | 0.0 | | | | | | | | | | | | | | | | | | | | | | | |
| SE21 | 0.5 | -1.2 | 1.4 | -0.4 | 0.0 | -0.2 | 0.2 | -1.0 | 0.0 | | | | | | | | | | | | | | | | | | | | | | |
| SE31 | -0.4 | -1.3 | 0.1 | 0.0 | 1.3 | 0.7 | 0.7 | -1.7 | 1.3 | 0.0 | | | | | | | | | | | | | | | | | | | | | |
| SE25 | 0.3 | -1.6 | -1.5 | 0.8 | 0.1 | 0.0 | -0.8 | 1.5 | -0.4 | 0.0 | 0.0 | | | | | | | | | | | | | | | | | | | | |
| B33 | 0.0 | 0.1 | 3.6 | 3.4 | 0.6 | 4.5 | -1.7 | -0.9 | 0.8 | 2.5 | 0.0 | 0.0 | | | | | | | | | | | | | | | | | | | |
| B27 | -2.3 | -0.6 | 1.0 | 2.1 | 0.0 | 0.1 | -2.4 | -1.0 | -0.6 | 0.4 | 1.2 | -0.4 | 0.0 | | | | | | | | | | | | | | | | | | |
| B18 | -0.7 | 0.2 | 0.5 | -1.9 | 2.1 | -1.2 | -0.1 | 0.0 | -1.5 | -0.9 | -1.4 | -0.9 | 0.9 | 0.0 | | | | | | | | | | | | | | | | | |
| B12 | 0.3 | 1.1 | -0.5 | -0.9 | 0.7 | -1.5 | 0.0 | 3.1 | -1.8 | -1.7 | -0.8 | -0.8 | -0.7 | 1.3 | 0.0 | | | | | | | | | | | | | | | | |
| H37 | -2.9 | 0.4 | 0.8 | 3.7 | 0.1 | 2.2 | 0.4 | -0.3 | 1.1 | 3.6 | 1.8 | 1.0 | 1.7 | 0.3 | -0.1 | 0.0 | | | | | | | | | | | | | | | |
| H26 | -3.8 | -2.8 | -0.5 | -1.6 | -2.0 | -1.8 | -0.6 | -3.1 | -0.6 | 2.3 | -1.1 | -3.4 | 2.7 | -1.0 | -2.8 | 0.1 | 0.0 | | | | | | | | | | | | | | |
| H10 | -0.5 | -0.3 | -1.0 | -1.3 | -1.2 | -2.3 | 1.5 | 0.9 | -0.6 | 0.3 | -2.0 | -1.6 | 0.4 | -0.5 | 0.6 | -0.5 | 0.7 | 0.0 | | | | | | | | | | | | | |
| ST6 | 1.7 | 3.3 | 1.9 | 2.8 | 2.8 | 2.2 | 6.0 | 3.6 | 2.4 | 3.4 | 3.3 | 2.7 | 0.4 | 4.1 | 4.9 | -1.1 | -2.7 | 1.4 | 0.0 | | | | | | | | | | | | |
| ST15 | -2.2 | -1.8 | -2.4 | -3.5 | 0.5 | -1.7 | -2.8 | -0.6 | -2.3 | -0.4 | -1.5 | -2.7 | -1.0 | -0.7 | 0.1 | -2.5 | -1.1 | 0.0 | 0.9 | 0.0 | | | | | | | | | | | |
| ST30 | -2.0 | -1.0 | 1.5 | 0.6 | 0.9 | 0.8 | 0.5 | -1.8 | 1.5 | 5.2 | 1.2 | 0.3 | 0.6 | -0.4 | -1.8 | 2.3 | 3.0 | 0.6 | -2.1 | 0.4 | 0.0 | | | | | | | | | | |
| SD24 | -1.3 | -2.0 | -1.7 | 0.9 | 0.0 | 3.1 | -1.0 | -2.1 | 0.4 | 0.8 | 4.1 | 1.1 | 1.4 | -2.0 | -2.4 | 1.0 | -0.8 | -2.3 | 1.5 | -1.9 | -1.1 | 0.0 | | | | | | | | | |
| SD22 | -0.6 | -0.5 | -0.2 | 1.2 | 1.5 | -0.5 | -0.3 | 0.0 | 4.2 | 0.7 | 1.0 | 1.7 | 0.6 | -0.9 | -0.5 | 1.0 | -2.0 | -1.4 | 2.2 | -2.7 | 0.3 | 0.6 | 0.0 | | | | | | | | |
| SD11 | -2.5 | -1.6 | -2.5 | -1.9 | 0.0 | -1.8 | 0.1 | -0.2 | -1.5 | -1.2 | -0.5 | -2.5 | -0.6 | -1.0 | 0.7 | 1.8 | 0.4 | 4.5 | 3.8 | 0.5 | -0.4 | 2.6 | -0.5 | 0.0 | | | | | | | |
| SD1 | 0.1 | -0.1 | 0.3 | -1.0 | 0.2 | -1.1 | 0.9 | 0.2 | -0.4 | 0.8 | 1.6 | 0.9 | -0.1 | 1.9 | 1.5 | -1.6 | -1.5 | 0.0 | 7.4 | -0.3 | -0.4 | -1.4 | -0.9 | 0.2 | 0.0 | | | | | | |
| A32 | -0.1 | -0.2 | -0.5 | 0.4 | 0.9 | -0.8 | -1.2 | 0.0 | 2.5 | 3.6 | 1.1 | 0.3 | 0.4 | -0.5 | -0.2 | 0.3 | -2.4 | -3.1 | 0.7 | -1.6 | 0.1 | 1.0 | 1.1 | -0.2 | 0.6 | 0.0 | | | | | |
| A24 | -1.4 | -0.5 | 0.8 | 1.0 | 0.2 | -2.1 | -0.3 | 0.7 | 1.2 | 1.6 | 1.5 | 0.8 | 1.1 | -0.4 | 0.3 | -0.2 | -1.0 | 0.2 | 1.7 | -1.1 | 1.1 | -1.0 | 2.0 | -0.9 | 0.9 | 0.8 | 0.0 | | | | |
| A13 | -2.2 | -0.3 | -1.7 | -1.4 | 1.3 | -4.7 | 0.4 | 0.9 | -0.3 | 0.0 | -0.8 | -2.3 | -1.0 | 0.7 | 1.6 | 1.4 | 0.5 | 1.8 | 0.8 | -0.7 | 1.4 | -3.1 | -0.9 | 1.3 | -1.4 | -0.9 | -0.6 | 0.0 | | | |
| A4 | -0.6 | -0.3 | -1.2 | -1.8 | -0.6 | -5.9 | 2.5 | 0.1 | 0.1 | -0.1 | -1.1 | -2.7 | -1.3 | -1.0 | 0.9 | 1.3 | 2.4 | 0.3 | -1.2 | 0.5 | -1.5 | -0.4 | -1.3 | -0.9 | -2.4 | -1.2 | 5.6 | 0.0 | | | |
| P39 | -0.3 | 0.2 | -0.5 | -0.7 | 1.4 | -0.3 | -1.6 | -2.5 | 1.1 | 3.0 | 1.8 | 1.0 | -0.1 | 0.5 | -0.8 | -0.3 | 0.7 | -0.6 | 0.2 | -0.7 | 0.0 | 2.1 | -0.8 | 0.4 | 0.0 | 1.8 | -0.4 | -1.0 | -1.0 | 0.0 | |
| P17 | 1.5 | 1.7 | 0.0 | -2.5 | 2.5 | -3.7 | -1.2 | -0.9 | -0.7 | 1.3 | 0.4 | -1.5 | 0.3 | 1.0 | -0.5 | -0.3 | 0.5 | 1.3 | 0.3 | 1.7 | -0.1 | -1.3 | -1.5 | 0.6 | -0.4 | -0.4 | -1.3 | 1.7 | 1.1 | 0.0 | 0.0 |



**APPENDIX K: SUMMARY OF INSTRUMENT VALIDATION PROCEDURES**

| Validity measure | Definition | Conducted Test or Method |
|---|---|---|
| Content Validity | The degree to which instrument questions represent the conceptual domain being measured (Straub, et al., 2004). | • Literature review of the related area of study (Ahire & Devaraj, 2001).<br>• Expert Opinion.<br>• Content Validity Ratio (Lawshe, 1975; Straub, et al., 2004). |
| Construct Reliability | Level of consistency between different items of a construct (Creswell, 2009; Cronbach, 1951). | • Internal consistency assessment using Cronbach's alpha (Cronbach, 1990; Nunnally, 1967; Straub & Carlson, 1989). |
| Construct Validity | Group of heuristics measuring the level to which an instrument is able to capture underlying construct(s) (Straub, et al., 2004; Straub & Carlson, 1989). | • Discriminant and convergent validity heuristics (Straub, et al., 2004; Straub & Carlson, 1989). |
| Discriminant Validity | The degree to which a construct and its items are distinguished from other constructs and their items in an instrument (Campbell & Fiske, 1959; Straub, et al., 2004). | • Exploratory factor analysis.<br>• Fit indices assessment of models at different levels of analysis (Submodels, construct level and overall research model).<br>• Avoiding multicollinearity by assuring there is no high correlations between constructs (Kline, 2010; Straub, et al., 2004). |
| Convergent Validity | Assessment of the degree to which items of the same construct are related to the construct being measured (Hair, et al., 2010; Zikmund, et al., 2010). | • Exploratory factor analysis.<br>• Fit indices assessment for all models and at construct level.<br>• Assuring high loadings of items to corresponding constructs in the CFA and EFA methods (Straub, et al., 2004; Straub & Carlson, 1989). |



**APPENDIX L: MEASUREMENT MODEL AND CFA OUTCOME OF THE OVERALL RESEARCH MODEL**

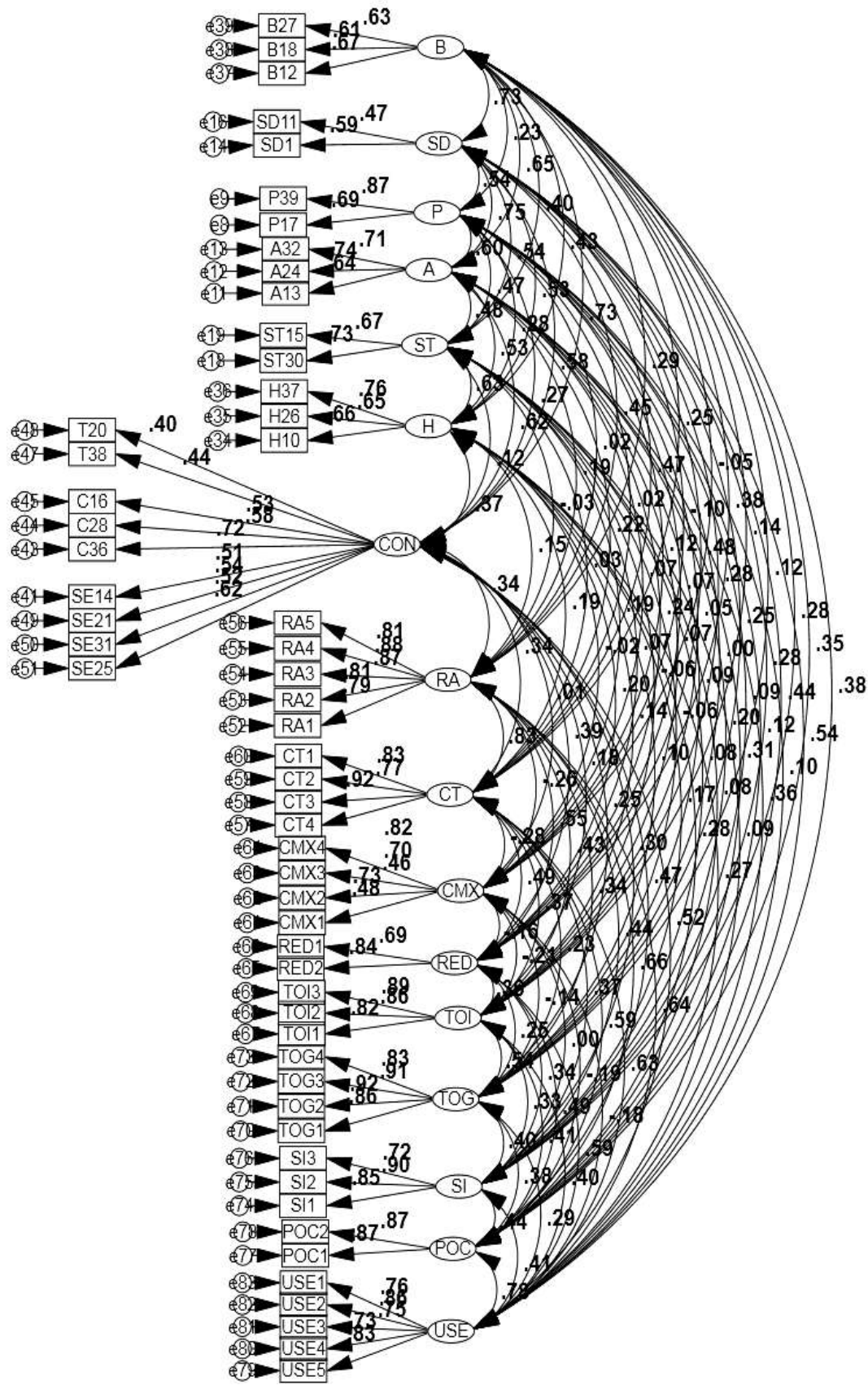



| Item | Loading | CR | P | Construct Correlation | | | |
|------|---------|-----|---|------|---|------|------|
| *Relative Advantage* | | | | RA | ←→ | CT | 0.83 |
| RA5 | 0.81 | 25.04 | *** | RA | ←→ | CMX | -0.26 |
| RA4 | 0.88 | 27.25 | *** | RA | ←→ | RED | 0.55 |
| RA3 | 0.87 | 26.80 | *** | RA | ←→ | TI | 0.43 |
| RA2 | 0.81 | 24.26 | *** | RA | ←→ | TG | 0.34 |
| RA1 | 0.79 | 23.31 | *** | RA | ←→ | SI | 0.44 |
| | | | | RA | ←→ | POC | 0.66 |
| | | | | RA | ←→ | USE | 0.64 |
| *Compatibility* | | | | CT | ←→ | CMX | -0.28 |
| CT4 | 0.83 | 25.64 | *** | CT | ←→ | RED | 0.49 |
| CT3 | 0.92 | 29.83 | *** | CT | ←→ | TI | 0.37 |
| CT2 | 0.77 | 22.98 | *** | CT | ←→ | TG | 0.23 |
| CT1 | 0.83 | 25.48 | *** | CT | ←→ | SI | 0.37 |
| | | | | CT | ←→ | POC | 0.59 |
| | | | | CT | ←→ | USE | 0.63 |
| *Complexity* | | | | CMX | ←→ | RED | -0.16 |
| CMX4 | 0.71 | 16.69 | *** | CMX | ←→ | TI | -0.21 |
| CMX3 | 0.46 | 9.47 | *** | CMX | ←→ | TG | -0.14 |
| CMX2 | 0.73 | 12.03 | *** | CMX | ←→ | SI | 0.00 |
| CMX1 | 0.47 | 9.72 | *** | CMX | ←→ | POC | -0.19 |
| | | | | CMX | ←→ | USE | -0.18 |
| *Result demonstrability* | | | | RED | ←→ | TI | 0.36 |
| RED2 | 0.84 | 20.31 | *** | RED | ←→ | TG | 0.26 |
| RED1 | 0.70 | 13.19 | *** | RED | ←→ | SI | 0.34 |
| | | | | RED | ←→ | POC | 0.49 |
| | | | | RED | ←→ | USE | 0.59 |
| *Trust in the Internet* | | | | TI | ←→ | TG | 0.54 |
| TI3 | 0.89 | 28.20 | *** | TI | ←→ | SI | 0.33 |
| TI2 | 0.86 | 28.31 | *** | TI | ←→ | POC | 0.41 |
| TI1 | 0.82 | 26.27 | *** | TI | ←→ | USE | 0.40 |
| *Trust in government agencies* | | | | TG | ←→ | SI | 0.40 |
| TG4 | 0.83 | 25.85 | *** | TG | ←→ | POC | 0.38 |
| TG3 | 0.91 | 30.27 | *** | TG | ←→ | USE | 0.29 |
| TG2 | 0.92 | 30.75 | *** | | | | |
| TG1 | 0.86 | 27.22 | *** | | | | |
| *Social influence* | | | | SI | ←→ | POC | 0.44 |
| SI3 | 0.72 | 20.60 | *** | SI | ←→ | USE | 0.41 |
| SI2 | 0.90 | 20.95 | *** | | | | |
| SI1 | 0.85 | 20.53 | *** | | | | |
| *Perspective on communication* | | | | POC | ←→ | USE | 0.78 |
| POC2 | 0.87 | 26.77 | *** | | | | |
| POC1 | 0.87 | 25.56 | *** | | | | |
| *Usage intention* | | | | | | | |
| USE5 | 0.83 | 25.73 | *** | | | | |
| USE4 | 0.73 | 21.04 | *** | Shown in other arrangements in the column 'Constructs' correlations' | | | |
| USE3 | 0.75 | 21.84 | *** | | | | |
| USE2 | 0.86 | 26.67 | *** | | | | |
| USE1 | 0.76 | 22.36 | *** | | | | |
| *Achievement* | | | | A | ←→ | SD | 0.77 |
| A13 | 0.64 | 16.68 | *** | A | ←→ | ST | 0.48 |
| A24 | 0.74 | 14.60 | *** | A | ←→ | H | 0.53 |
| A32 | 0.71 | 14.23 | *** | A | ←→ | B | 0.65 |
| | | | | A | ←→ | CON | 0.62 |
| | | | | A | ←→ | RA | 0.19 |
| | | | | A | ←→ | CT | 0.22 |
| | | | | A | ←→ | CMX | 0.07 |
| | | | | A | ←→ | RED | 0.24 |
| | | | | A | ←→ | TI | 0.07 |
| | | | | A | ←→ | TG | 0.10 |
| | | | | A | ←→ | SI | 0.20 |
| | | | | A | ←→ | POC | 0.31 |
| | | | | A | ←→ | USE | 0.36 |
| | | | | A | ←→ | RA | 0.19 |
| | | | | A | ←→ | CT | 0.22 |
| | | | | A | ←→ | CMX | 0.07 |
| | | | | A | ←→ | RED | 0.24 |
| | | | | A | ←→ | TI | 0.07 |
| | | | | A | ←→ | TG | 0.10 |
| | | | | A | ←→ | SI | 0.20 |
| | | | | A | ←→ | POC | 0.31 |
| | | | | A | ←→ | USE | 0.36 |
| *Benevolence* | | | | B | ←→ | CON | 0.73 |
| B12 | 0.67 | 16.86 | *** | B | ←→ | RA | 0.29 |
| B18 | 0.61 | 12.34 | *** | B | ←→ | CT | 0.25 |
| B27 | 0.63 | 12.76 | *** | B | ←→ | CMX | -0.05 |
| | | | | B | ←→ | RED | 0.38 |
| | | | | B | ←→ | TI | 0.14 |
| | | | | B | ←→ | TG | 0.12 |
| | | | | B | ←→ | SI | 0.28 |
| | | | | B | ←→ | POC | 0.35 |
| | | | | B | ←→ | USE | 0.38 |
| *Conservation* | | | | CON | ←→ | RA | 0.34 |
| C16 | 0.53 | 13.66 | *** | CON | ←→ | CT | 0.34 |
| C36 | 0.72 | 12.38 | *** | CON | ←→ | CMX | 0.01 |
| C28 | 0.58 | 10.92 | *** | CON | ←→ | RED | 0.39 |
| SE14 | 0.51 | 10.10 | *** | CON | ←→ | TI | 0.18 |
| SE21 | 0.55 | 10.58 | *** | CON | ←→ | TG | 0.25 |
| SE25 | 0.62 | 11.48 | *** | CON | ←→ | SI | 0.30 |
| SE31 | 0.52 | 10.20 | *** | CON | ←→ | POC | 0.47 |



| | | | | | | | | |
|---|---|---|---|---|---|---|---|---|
| T20 | 0.41 | 8.60 | *** | CON | ←→ | USE | 0.52 |
| T38 | 0.44 | 9.12 | *** | | | | |
| *Hedonism* | | | | H | ←→ | B | 0.43 |
| H10 | 0.66 | 16.59 | *** | H | ←→ | CON | 0.38 |
| H26 | 0.65 | 12.83 | *** | H | ←→ | RA | 0.15 |
| H37 | 0.76 | 13.80 | *** | H | ←→ | CT | 0.19 |
| | | | | H | ←→ | CMX | -0.02 |
| | | | | H | ←→ | RED | 0.20 |
| | | | | H | ←→ | TI | 0.14 |
| | | | | H | ←→ | TG | 0.10 |
| | | | | H | ←→ | SI | 0.17 |
| | | | | H | ←→ | POC | 0.28 |
| | | | | H | ←→ | USE | 0.27 |
| *Power* | | | | P | ←→ | A | 0.60 |
| P17 | 0.68 | 16.66 | *** | P | ←→ | SD | 0.54 |
| P39 | 0.87 | 12.65 | *** | P | ←→ | ST | 0.47 |
| | | | | P | ←→ | H | 0.28 |
| | | | | P | ←→ | B | 0.23 |
| | | | | P | ←→ | CON | 0.27 |
| | | | | P | ←→ | RA | 0.02 |
| | | | | P | ←→ | CT | 0.02 |
| | | | | P | ←→ | CMX | 0.12 |
| | | | | P | ←→ | RED | 0.07 |
| | | | | P | ←→ | TI | 0.05 |
| | | | | P | ←→ | TG | 0.00 |
| | | | | P | ←→ | SI | 0.09 |
| | | | | P | ←→ | POC | 0.12 |
| | | | | P | ←→ | USE | 0.10 |
| *Self-direction* | | | | SD | ←→ | ST | 0.54 |
| SD1 | 0.68 | 11.19 | *** | SD | ←→ | H | 0.53 |
| SD11 | 0.88 | 9.28 | *** | SD | ←→ | B | 0.73 |
| | | | | SD | ←→ | CON | 0.58 |
| | | | | SD | ←→ | RA | 0.45 |
| | | | | SD | ←→ | CT | 0.47 |
| | | | | SD | ←→ | CMX | -0.10 |
| | | | | SD | ←→ | RED | 0.48 |
| | | | | SD | ←→ | TI | 0.28 |
| | | | | SD | ←→ | TG | 0.25 |
| | | | | SD | ←→ | SI | 0.28 |
| | | | | SD | ←→ | POC | 0.44 |
| | | | | SD | ←→ | USE | 0.54 |
| *Stimulation* | | | | ST | ←→ | H | 0.63 |
| ST15 | 0.68 | 16.35 | *** | ST | ←→ | B | 0.40 |
| ST30 | 0.72 | 12.30 | *** | ST | ←→ | CON | 0.12 |
| | | | | ST | ←→ | RA | -0.03 |
| | | | | ST | ←→ | CT | 0.03 |



| | | | | ST | ←→ | CMX | 0.19 |
|---|---|---|---|---|---|---|---|
| | | | | ST | ←→ | RED | 0.07 |
| | | | | ST | ←→ | TI | -0.06 |
| | | | | ST | ←→ | TG | -0.06 |
| | | | | ST | ←→ | SI | 0.08 |
| | | | | ST | ←→ | POC | 0.08 |
| | | | | ST | ←→ | USE | 0.09 |
| Model Fit Indices: χ2/df = 3.50; GFI = 0.90; AGFI = 0.87; CFI = 0.88; IFI = 0.94; SRMR = 0.04; RMSEA = 0.04, *** p<0.001 | | | | | | | |

*Note.* RA=relative advantage; CT=compatibility; CMX=complexity; RED=result demonstrability; TI=trust in the Internet; TG=trust in government agencies; SI=social influence; POC=perspective on communication; USE=intention to use e-transactions; SD=self-direction; P=power; U=universalism; A=achievement; SE=security; ST=stimulation; C=conformity; T=tradition; H=hedonism; B=benevolence; CR=Critical Ratio; χ2=Chi-square; df=degrees of freedom; GFI=goodness of fit index; AGFI=adjusted goodness of fit; CFI=comparative fit index; IFI=incremental fit index; SRMR=standardised root mean square residual; RMSEA=root mean square error approximation. Appendix O and P show item codes and wording.



**APPENDIX M: CORRELATIONS FOR THE STRUCTURAL MODEL (INCLUDING INTERCORRELATIONS ONLY BETWEEN BPV CONSTRUCTS AND BETWEEN PCET CONSTRUCTS)**

| Correlations | | | Estimate |
|---|---|---|---|
| *Correlations between Basic Personal Values (BPV) constructs* | | | |
| Hedonism | ←→ | Conservation values | .375 |
| Stimulation | ←→ | Conservation values | .111 |
| Achievement | ←→ | Conservation values | .616 |
| Power | ←→ | Conservation values | .267 |
| Self-direction | ←→ | Conservation values | .583 |
| Benevolence | ←→ | Conservation values | .729 |
| Stimulation | ←→ | Hedonism | .630 |
| Achievement | ←→ | Hedonism | .527 |
| Power | ←→ | Hedonism | .282 |
| Self-direction | ←→ | Hedonism | .572 |
| Benevolence | ←→ | Hedonism | .428 |
| Achievement | ←→ | Stimulation | .480 |
| Power | ←→ | Stimulation | .465 |
| Self-direction | ←→ | Stimulation | .563 |
| Benevolence | ←→ | Stimulation | .398 |
| Power | ←→ | Achievement | .594 |
| Self-direction | ←→ | Achievement | .766 |
| Benevolence | ←→ | Achievement | .653 |
| Self-direction | ←→ | Power | .556 |
| Benevolence | ←→ | Power | .233 |
| Benevolence | ←→ | Self-direction | .740 |
| *Correlations between Perceived Characteristics of e-Transactions (PCET) constructs* | | | |
| Trust in government agencies | ←→ | Social influence | .402 |
| Trust in the Internet | ←→ | Social influence | .329 |
| Result demonstrability | ←→ | Social influence | .345 |
| Complexity | ←→ | Social influence | -.002 |
| CT | ←→ | Social influence | .368 |
| RA | ←→ | Social influence | .436 |
| Social influence | ←→ | Perspective on communication | .440 |
| Trust in the Internet | ←→ | Trust in government agencies | .538 |
| Result demonstrability | ←→ | Trust in government agencies | .256 |
| Complexity | ←→ | Trust in government agencies | -.142 |
| Compatibility | ←→ | Trust in government agencies | .232 |
| Relative advantage | ←→ | Trust in government agencies | .343 |
| Trust in government agencies | ←→ | Perspective on communication | .380 |
| Result demonstrability | ←→ | Trust in the Internet | .361 |
| Complexity | ←→ | Trust in the Internet | -.212 |
| Compatibility | ←→ | Trust in the Internet | .371 |
| RA | ←→ | Trust in the Internet | .428 |
| Trust in the Internet | ←→ | Perspective on communication | .406 |
| Complexity | ←→ | Result demonstrability | -.163 |
| Compatibility | ←→ | Result demonstrability | .495 |
| Relative advantage | ←→ | Result demonstrability | .548 |
| Result demonstrability | ←→ | Perspective on communication | .494 |



| Compatibility | ←→ | Complexity | -.277 |
|---|---|---|---|
| Relative advantage | ←→ | Complexity | -.257 |
| Complexity | ←→ | Perspective on communication | -.194 |
| Relative advantage | ←→ | Compatibility | .826 |
| Compatibility | ←→ | Perspective on communication | .593 |
| Relative advantage | ←→ | Perspective on communication | .658 |



**APPENDIX N: EXPLORATORY ANALYSIS OF THE STRUCTURAL MODEL**

In this section, the data is used to identify relationships that were not hypothesised in this study. Knowledge of these relationships might be of use for future studies to confirm or reject. There are two approaches to structural model re-specification: model building and model trimming. Model building starts with a "bare-boned" model where no relationships are identified in the structural model and relations are defined based on empirical evidence (Kline, 2010). Model trimming is the opposite: the initial structural model is fully saturated (all possible relations are freed for estimation) and insignificant relations are eliminated. In this study, model building was used to improve the model fit.

The goal of both model trimming and model building is to find a model that is parsimonious and fits the data relatively well (Kline, 2010). Kline (2010) recommended that insignificant paths should not be removed from the model; instead, they should be retained until the research is replicated and the insignificance is assured. Therefore, to identify relationships, modification indices were used to guide the process, and the modification indices were checked for each freed estimate or relationship contained in the model. However, to increase the fit of the model to the data, the model building commenced with the proposed research hypothesis, rather than with a bare-boned model. Thirteen re-estimations were conducted until there were no modification indices. The model was re-specified according to the largest modification index and re-estimated (B. M. Byrne, 2010). Each proposed alteration to the model, the modification index, fit indices and the parameter significance are noted in the table below. In the table, the Standardised Root Mean Root (SRMR) is not reported for some of the altered structural models (e.g., solution 1) because the AMOS program did not calculate this parameter due to the lack of fit of these particular models. However, the SRMR was calculated for solution 7. In addition, the final solution (number 13) yielded a good fit.



*Model Alterations Suggested by the Modification Indices*

| # | Proposed relation | Largest modification index | Estimated parameter change | Fit indices after proposed change was specified | Actual parameter after re-estimation |
|---|---|---|---|---|---|
| 1 | Relative advantage → Compatibility | 385.119 | 0.73 | χ2/df = 3.60; GFI = 0.73; AGFI = 0.70; CFI = 0.81; IFI = 0.81; RMSEA = 0.06 | 0.82 |
| 2 | Relative advantage → Perspective on communication | 235.35 | 0.65 | χ2/df = 3.21; GFI = 0.76; AGFI = 0.74; CFI = 0.83; IFI = 0.83; RMSEA = 0.06 | 0.76 |
| 3 | Benevolence → Conservation values | 179.95 | 0.18 | χ2/df = 3.60; GFI = 0.73; AGFI = 0.70; CFI = 0.81; IFI = 0.81; RMSEA = 0.06 | 0.55 |
| 4 | Benevolence → Achievement | 180.98 | 0.33 | χ2/df = 3.06; GFI = 0.78; AGFI = 0.76; CFI = 0.84; IFI = 0.84; RMSEA = 0.06 | 0.99 |
| 5 | Trust in the Internet → Trust in government agencies | 162.85 | 1.00 | χ2/df = 2.94; GFI = 0.79; AGFI = 0.77; CFI = 0.85; IFI = 0.85; RMSEA = 0.05 | 0.55 |
| 6 | Relative advantage → Result demonstrability | 149.05 | 0.45 | χ2/df = 2.81; GFI = 0.80; AGFI = 0.78; CFI = 0.86; IFI = 0.86; RMSEA = 0.05 | 0.54 |
| 7 | Stimulation → Hedonism | 117.59 | 0.19 | χ2/df = 2.71; GFI = 0.81; AGFI = 0.79; CFI = 0.87; IFI = 0.87; SRMR = 0.15; RMSEA = 0.05 | 0.61 |
| 8 | Relative advantage → Trust in the Internet | 114.68 | 0.59 | χ2/df = 2.63; GFI = 0.82; AGFI = 0.80; CFI = 0.88; IFI = 0.88; SRMR = 0.13; RMSEA = 0.05 | 0.65 |
| 9 | Relative advantage → Social influence | 112.06 | 0.50 | χ2/df = 2.54; GFI = 0.83; AGFI = 0.81; CFI = 0.88; IFI = 0.88; SRMR = 0.12; RMSEA = 0.05 | 0.57 |
| 10 | Benevolence → Self-direction | 105.46 | 0.29 | χ2/df = 2.43; GFI = 0.84; AGFI = 0.82; CFI = 0.89; IFI = 0.89; SRMR = 0.11; RMSEA = 0.05 | 0.91 |
| 11 | Benevolence → Stimulation | 79.34 | 0.22 | χ2/df = 2.36; GFI = 0.84; AGFI = 0.83; CFI = 0.90; IFI = 0.90; SRMR = 0.11; RMSEA = 0.05 | 0.81 |
| 12 | Benevolence→ Power | 78.62 | 0.31 | χ2/df = 2.29; GFI = 0.84; AGFI = 0.83; CFI = 0.90; IFI = 0.90; SRMR= 0.10; RMSEA = 0.04 | 0.92 |



| 13 | Benevolence→ Relative advantage | 62.448 | 0.18 | χ2/df = 2.25; GFI = 0.85; AGFI = 0.83; CFI = 0.91; IFI = 0.91; SRMR= 0.07; RMSEA = 0.04 | 0.62 |

*Note.* χ2=Chi-square; df=degrees of freedom; GFI=goodness of fit index; AGFI=adjusted goodness of fit; CFI=comparative fit index; IFI=incremental fit index; SRMR=standardised root mean square residual; RMSEA=root mean square error approximation.

The final solution (i.e., 13) provided a good fit to the data: χ2/df = 2.25, GFI = 0.85, AGFI = 0.83, CFI = 0.91, IFI = 0.91, SRMR= 0.07, and RMSEA = 0.04. Furthermore, none of the modification indices pointed to relationships between the constructs. Therefore, all of the suggested relationships based on the modification and fit indices were correlated (B. M. Byrne, 2010). The figure below represents the proposed structural model by the data with standardised regression weights for each relationship (the items, error and residual means are not shown for the sake of clarity).

*Re-specified structural model with standardised regression weights.*

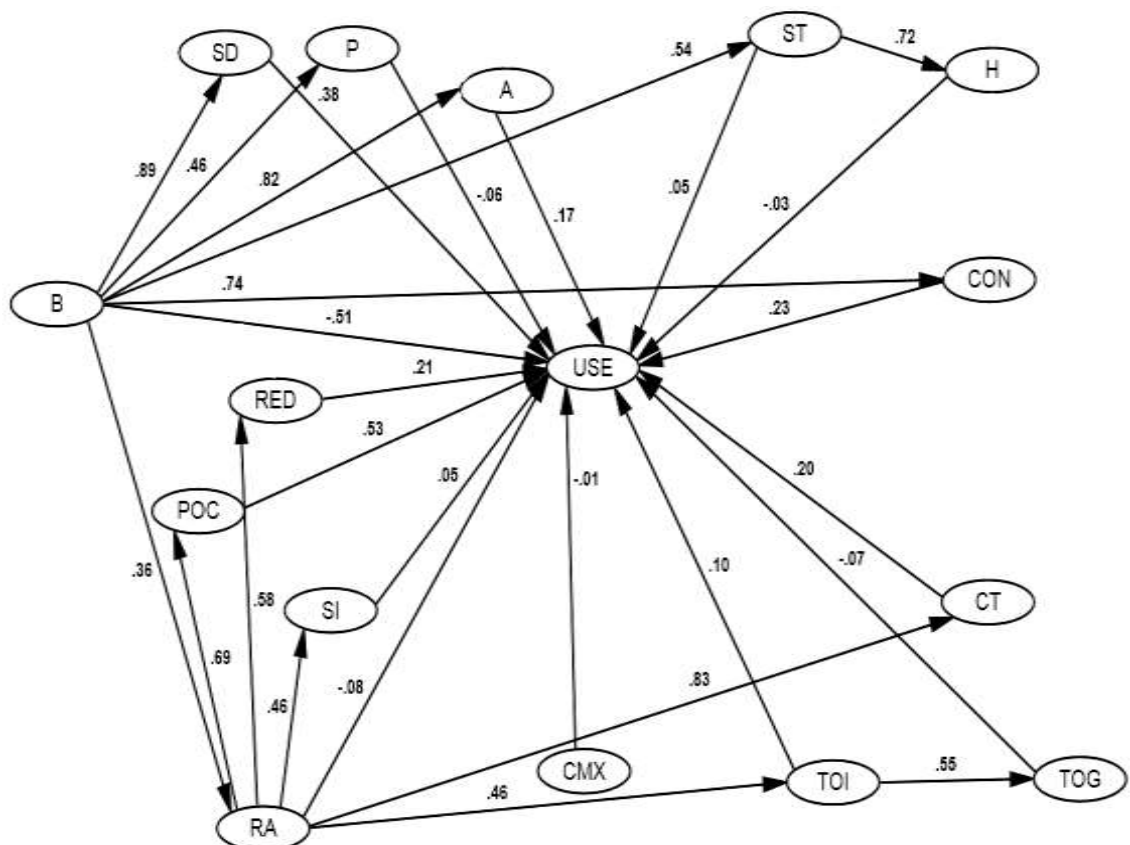

*Note.* RA=relative advantage; CT=compatibility; CMX=complexity; RED=result demonstrability; TI=trust in the Internet; TG=trust in government agencies; SI=social influence;



POC=perspective on communication; USE=intention to use e-transactions. SD=self-direction; P=power A=achievement; ST=stimulation; CON=conservation values; H=hedonism; B=benevolence.

       Only relationships related to the construct (intention to use e-transactions) USE are of particular interest for this study. Additionally, these proposed relationships are not of substantive interest for the current study because they have not been corroborated for the literature and are only based on the data. However, the usefulness of identify relationships based on the data is that these relationships might be helpful for future studies (Hair, et al., 2010).



**APPENDIX O: ITEM CODING FOR PCET MODEL**

| Code | Item |
|------|------|
| *Relative advantage (RA)* | |
| RA1 | Using e-government would enable me to carry out my transactions more quickly. |
| RA2 | Using e-government would improve the quality of my transactions. |
| RA3 | Using e-government would make it easier to carry out transactions with the government. |
| RA4 | Using e-government would enhance my effectiveness in carrying out transactions with the government. |
| RA5 | Using the e-government would give me greater control over conducting transactions with the government. |
| *Compatibility (CT)* | |
| CT1 | Using e-government is compatible with how I like to conduct transactions with the government. |
| CT2 | Using e-government transactions is completely compatible with my current needs. |
| CT3 | I think that using e-government would fit well with the way that I prefer to conduct transactions with the government. |
| CT4 | Using e-government transactions would fit well into my lifestyle. |
| *Complexity (CMX)* | |
| CMX1 | Using e-government transactions would consume too much of my time. |
| CMX2 | Conducting e-government transactions would be so complicated; it would be difficult to understand what is going on. |
| CMX3 | Using e-government transactions would involve too much time doing technical operations (e.g. data entry). |
| CMX4 | It would take too long to learn how to use e-government transactions to make it worth the effort. |
| *Result demonstrability (RED)* | |
| RED1 | I would have no difficulty telling others about the results of using e-government transactions. |
| RED2 | I believe I could communicate to others the consequences of using e-government transactions. |
| RED3 | The results of using e-government transactions are apparent to me. |
| RED4 | I would have difficulty explaining why using e-government transactions may or may not be beneficial. |
| *Trust in the Internet (TI)* | |
| TI1 | The Internet has enough safeguards to make me feel comfortable in conducting transactions using e-government. |
| TI2 | I feel assured that legal and technological structures adequately protect me from problems on the Internet while using e-government transactions. |
| TI3 | Generally, I feel that the Internet is now a robust and safe environment in which to conduct on-line transactions with the government. |
| TI4 | I think I can trust government agencies in delivering my transactions using e-government. |
| *Trust in government agencies (TG)* | |
| TG1 | Government agencies can be trusted to carry out on-line transactions faithfully. |
| TG2 | In my opinion, government agencies are trustworthy in their ability to deliver services using e-government transactions. |
| TG3 | I trust government agencies to keep my best interest in mind while delivering on-line services using e-government transactions. |



| Social influence (SI) | |
|---|---|
| SI1 | People who influence my behaviour would think that I should use e-government to conduct transactions. |
| SI2 | People who influence my behaviour would think that I should use e-government to conduct transactions. |
| SI3 | People who are in my social circle would think that I should use the e-government to conduct transactions. |
| Perspective on communication (POC) | |
| POC1 | My ability to communicate with the government would be enhanced when using e-government transactions. |
| POC2 | Communications through e-government enhance my ability to understand government transactions. |
| POC3 | Textual, verbal and visual information is important for carrying out government transactions. |
| Intention to use e-transactions (USE) | |
| USE1 | I would use e-government to gather information about my required transactions. |
| USE2 | I would use e-government transactions provided over the Internet. |
| USE3 | Using e-government transactions is something that I would do. |
| USE4 | I would not hesitate to provide information to e-government websites to conduct my transactions. |
| USE5 | I would use e-government to inquire about my government transactions. |



**APPENDIX P: ITEM CODING FOR BPV MODEL**

| Code | Item |
|------|------|
| *Self-direction (SD)* | |
| SD1 | Thinking up new ideas and being creative is important. He/She likes to do things in his/her own original way. |
| SD11 | It is important to make his/her own decisions about what he/she does. He/She likes to be free to plan and to choose activities. |
| SD22 | He/She thinks it is important to be interested in things. He/She likes to be curious and to try to understand all sorts of things. |
| SD24 | It is important to be independent. He/She likes to rely on him/herself. |
| *Power (P)* | |
| P2 | It is important to be rich. He/She wants to have a lot of money and expensive things. |
| P17 | It is important to be in charge and tell others what to do. He/She wants people to do what he/she says. |
| P39 | Always wants to be the one who makes the decisions. He/She likes to be the leader. |
| *Universalism (U)* | |
| U3 | He/She thinks it is important that every person in the world be treated equally. He/She beliefs everyone should have equal opportunities in life |
| U8 | It is important to listen to people who are different from him/her. Even when disagrees with them, he/she still wants to understand them. |
| U23 | He/She believes all the worlds' people should live in harmony. Promoting peace among all groups in the world is important. |
| U29 | He/She wants everyone to be treated justly, even people do not know. It is important to protect the weak in society. |
| U19 | He /She strongly believe that people should care for nature. Looking after the environment is important to him/her. |
| U40 | It is important to adapt to nature and to fit into it. He/She believes that people should not change nature. |
| *Achievement (A)* | |
| A4 | It is very important to show his/her abilities. He/She wants people to admire what they do. |
| A13 | Being very successful is important. He/She likes to impress. |
| A24 | He/She thinks it is important to be ambitious. He/She wants to show how capable he/she is. |
| A32 | Getting ahead in life is important. He/She strives to do better than others do. |
| *Security (SE)* | |
| SE5 | It is important to live in secure surroundings. He/She avoids anything that might endanger their safety. |
| SE14 | It is very important to him/her that the country be safe. He/She thinks the state must be on watch against threats from within and without. |
| SE21 | It is important that things be organised and clean. He/She really does not like things to be a mess. |
| SE31 | He/She tries hard to avoid getting sick. Staying healthy is very important. |
| SE25 | Having a stable government is important. He/She is concerned that the social order be protected. |
| *Stimulation (ST)* | |
| ST6 | It is important to do many different things in life. He/She always looks for new things to try. |
| ST15 | He/She likes to take risks. He/She is always looking for adventures. |
| ST30 | He/She likes surprises. It is important to have an exciting life. |



| Conformity (C) | |
|---|---|
| C7 | He/She believes that people should do what they're told. He/She thinks people should follow rules at all times, even when no-one is watching |
| C16 | It is important to him/her always to behave properly. He/She wants to avoid doing anything people would say is wrong. |
| C28 | It is important always to show respect to parents and to older people. It is important to be obedient. |
| C36 | It is important to be polite to other people all the time. He/She tries never to disturb or irritate others. |
| Tradition (T) | |
| T9 | He/She thinks it is important not to ask for more than what you have. He/She believes that people should not change nature. |
| T20 | Being religious is important. He/She tries hard to follow religious beliefs. |
| T25 | It is best to do things in traditional ways. It is important to keep up the customs he/she has learned. |
| T38 | It is important to be humble and modest. He/She tries not to draw attention to him/her. |
| Hedonism (H) | |
| H10 | He/She seeks every chance to have fun. It is important to do things that give him/her pleasure. |
| H26 | Enjoying life's pleasures is important. He/She likes to "spoil" him/herself. |
| H37 | He/She really wants to enjoy life. Having a good time is very important. |
| Benevolence (B) | |
| B12 | It is very important to help the people around him/her. He/She wants to care for their well-being. |
| B18 | It is important to be loyal to friends. He/She wants to devote him/herself to people close to him/her. |
| B27 | It is important to respond to the needs of others. He/She tries to support those he/she knows. |
| B33 | Forgiving people who have hurt him/her is important. He/She tries to see what is good in them and not to hold a grudge |